\documentclass[11pt, oneside]{article}   	
\usepackage{geometry}                		
\pdfoutput=1						

\usepackage[parfill]{parskip}    			
\usepackage{graphicx}				
\usepackage{amssymb}
\usepackage{amsmath}

\usepackage{bm} 					
\usepackage{natbib}					
\usepackage[ngerman, english]{babel}
\usepackage{booktabs} 				
\usepackage[table]{xcolor}			
\usepackage{rotating}	
\usepackage{float}

\usepackage{url} 
\usepackage{multirow} 
\usepackage[bookmarks]{hyperref}
\hypersetup{colorlinks=true, citecolor=black, linkcolor=black, urlcolor=black} 

\usepackage{sectsty} 
\allsectionsfont{\sffamily}

\title{\bfseries\sffamily{A Comparison of Approaches for Imbalanced Classification Problems in the Context of Retrieving Relevant Documents for an Analysis}}
\author{\textsf{Sandra Wankm\"uller} \\
\textsf{\emph{Ludwig-Maximilians-University Munich}} \\
\textsf{sandra.wankmueller@gsi.lmu.de}}
\date{}							

\begin{document}
\maketitle

{\bfseries\sffamily{{Abstract.}}} One of the first steps in many text-based social science studies is to retrieve documents that are relevant for the analysis from large corpora of otherwise irrelevant documents. The conventional approach in social science to address this retrieval task is to apply a set of keywords and to consider those documents to be relevant that contain at least one of the keywords. But the application of incomplete keyword lists risks drawing biased inferences. More complex and costly methods such as query expansion techniques, topic model-based classification rules, and active as well as passive supervised learning could have the potential to more accurately separate relevant from irrelevant documents and thereby reduce the potential size of bias. Yet, whether applying these more expensive approaches increases retrieval performance compared to keyword lists at all, and if so, by how much, is unclear as a comparison of these approaches is lacking. This study closes this gap by comparing these methods across three retrieval tasks associated with a data set of German tweets \citep{Linder2017}, the Social Bias Inference Corpus (SBIC) \citep{Sap2020}, and the Reuters-21578 corpus \citep{Reuters1997}. Results show that query expansion techniques and topic model-based classification rules in most studied settings tend to decrease rather than increase retrieval performance. Active supervised learning, however, if applied on a not too small set of labeled training instances (e.g.~1,000 documents), reaches a substantially higher retrieval performance than keyword lists.


\section{Introduction} \label{sec:intro} 


When conducting a study on the basis of textual data, at the very start of an analysis researchers are often confronted with a difficulty: Online platforms and other sources from which textual data are generated usually cover multiple topics and hence tend to contain textual references toward a huge number of various entities. Social scientists, however, are typically interested in text elements referring to a single entity
, e.g.~a specific person, organization, object, event, or issue. 

Imagine, for example, that a study seeks to examine how rape incidents are framed in newspaper articles \citep{Baum2018}, or that a study seeks to detect electoral violence based on social media data \citep{Muchlinski2021}, or that a study seeks to measure attitudes expressed towards further European integration in speeches of political elites \citep{Rauh2019}. In all these studies, one of the first steps is to extract documents that refer to the entity of interest from a large, multi-thematic corpus of documents.\footnote{A corpus is a set of documents. A document is the unit of observation. A document can be a very short to a very long text (e.g.~a sentence, a speech, a newspaper article). Here, the term corpus refers to the set of documents a researcher has collected and from which he or she then seeks to retrieve the relevant documents.} This is, researchers have to separate the relevant documents that refer to the entity of interest from the documents that focus on entities irrelevant for the analysis at hand. Newspaper articles reporting about rape incidents have to be parted from those articles that do not. Tweets relating to electoral violence have to be extracted from the stream of all other tweets. And speech elements about the European integration have to be separated from elements in which the speaker talks about other entities.

Given a corpus comprising many diverse topics, it is likely that only a small proportion of documents relate to the entity of interest. Hence, the proportion of relevant documents is substantively smaller than the proportion of irrelevant documents in the data and the task of separating relevant from irrelevant documents turns into an imbalanced classification problem (\citeauthor{Manning:2008vf}, \citeyear{Manning:2008vf}, p.~155). How researchers address this imbalanced classification problem is highly important as the selection of documents affects the inferences drawn. More precisely: If there is a systematic bias in the selection of documents such that the value on a variable of interest is related to the question of whether a document is selected for analysis or not, the inferences that are made on the basis of the documents that have been selected for analysis are likely to be biased. Selection biases can be induced when the corpus 
is collected in the first place.
And selection biases can be induced when from the already collected corpus documents that refer to relevant entities are selected for analysis. This work focuses on the second step. The more accurately a method can separate relevant from irrelevant documents, the less the potential size of the bias resulting from this second selection step.

Despite the relevance of this problem, the question of how best to retrieve documents from large, heterogenous corpora so far has received little attention in social science research. In many applications, researchers have relied on applying human-created sets of keywords and regard those documents as relevant that comprise at least one of the keywords \citep[see e.g.][]{Burnap:2016kv, Jungherr2016, Beauchamp:2017cp, Baum2018, Stier2018, FogelDror2019, Rauh2019, Watanabe2020, Muchlinski2021}. Yet, research indicates that humans are not good at generating comprehensive keyword lists and are highly unreliable at the task \citep[p.~973-975]{King:2017io}. This is, the keyword list generated by one human is likely to contain only a small amount of the universe of terms one could use to refer to a given entity of interest \citep[p.~973-975]{King:2017io}. Moreover, the list of keywords that one human comes up with is likely to show little overlap with the keyword list generated by another human \citep[p.~973-975]{King:2017io}. Joining forces by combining keyword lists that researchers have created independently may alleviate the problem somewhat. But still, the conventional approach of using keywords to identify relevant documents is likely to be unreliable and thus is likely to lead to very different (and potentially biased) conclusions depending on which set of keywords the researchers have used 
\citep[p.~974-976]{King:2017io}.

Other approaches for identifying relevant documents---such as passive and active supervised learning, query expansion techniques, or the construction of topic model-based classification rules---are less frequently employed in social science applications. These approaches also require human input but they detect patterns or keywords the researchers do not have to know beforehand. Except for query expansion, these methods require the researchers to \emph{recognize} documents or terms related to the entity of interest rather than requiring the researchers to \emph{recall} such information a priori \citep[p.~972]{King:2017io}. This does not preclude these techniques from generating selection biases. A supervised learning algorithm, for example, may systematically misclassify some documents as not being relevant based on word usage that could be correlated with a main variable of the analysis. Yet, as these approaches have the potential to extract patterns beyond what a team of researchers may come up with, these methods have the potential to more precisely separate relevant from non-relevant documents. And the higher the retrieval performance of a method, the smaller the potential for strongly biasing effects due to selection biases. 

These other techniques, however, also have a disadvantage: they are much more resource intensive to implement. Supervised learning algorithms require labeled training documents, query expansion techniques depend on a data source to operate on, and topic model-based classification rules hinge on estimating a topic model. As the identification of relevant documents from a large heterogeneous corpus is likely to only constitute an early small step in an elaborate text analysis, considerations regarding the costs and benefits of a retrieval method have to be taken into account.

Hence, an ideal procedure reliably achieves a high retrieval performance such that it reduces the risk of incurring large selection biases and simultaneously is cost-effective enough to be conducted as a single step of an extensive study. In practice, the performance and the cost-effectiveness of a procedure is likely to depend on the characteristics of an application (such as the length and textual style of documents, the type of the entity of interest, or the heterogeneity vs.~homogeneity of the documents in the corpus). If the entity of interest is a person or organization and there is only a small set of expressions that is usually used to refer to this entity, then a list of keywords may lead to a similar performance than the resource intensive application of a supervised learning algorithm. If on the other hand the entity of interest is not easily denominated (e.g.~a policy issue such as the set of restrictions implemented to address the COVID-19 pandemic), then an acceptable retrieval performance may only be achieved by training a supervised learning algorithm.

So far, however, a systematic comparison of the performances of these different retrieval methods across social science applications is lacking. Thus, it is unclear what, if anything, could be gained in terms of retrieval performance by applying a more elaborate procedure. 
This study seeks to answer this question by comparing the retrieval performance of a small set of predictive keywords to (1) query expansion techniques extending this initial set, (2) topic model-based classification rules as well as (3) passive and active supervised learning. The procedures are compared on the basis of three retrieval tasks: (1) the identification of tweets referring to refugees, refugee policies, and the refugee crisis from a dataset of 24,420 German tweets \citep{Linder2017}, (2) the retrieval of posts that are offensive toward mentally or physically disabled people from the Social Bias Inference Corpus (SBIC) \citep{Sap2020} that covers 44,671 potentially toxic and offensive posts from various social media platforms, and (3) the extraction of newspaper articles referring to crude oil from the Reuters-21578 corpus \citep{Reuters1997} that comprises economically focused newspaper articles of which 10,377 are assigned to a topic.

The results show that with the model settings studied here query expansion techniques as well as topic-model-based classification rules tend to decrease rather than increase retrieval performance compared to sets of predictive keywords. They only yield minimal improvements or acceptable results in specific settings. By contrast, active supervised learning---if implemented with a not too small number of labeled training documents---achieves relatively high retrieval performances across contexts. Moreover, in each application active learning substantively improves upon the mediocre to fair results reached by the best performing lists of predictive keywords. The observed differences of the mean $F_1$-Scores achieved by active learning with 1,000 labeled training documents to the mean $F_1$-Scores of keyword lists range between 0.194 and 0.476. Although active learning is designed to reduce the number of training documents that have to be annotated by human coders, it is nevertheless particularly resource intensive. Yet, the achieved performance enhancements are so considerable (and the consequences of selection biases potentially so severe) that researchers should consider spending more of their available resources on the step of separating relevant from irrelevant documents.

In the following Section \ref{sec:imblearn} basic concepts relevant for discussing imbalanced classification problems in retrieval contexts are introduced. Afterward the benefits and disadvantages of the usage of keyword lists, query expansion techniques, topic model-based classification rules, and passive as well as active supervised learning techniques in the context of identifying documents relevant for further analyses are discussed (\ref{sec:approaches}). Then the procedures are applied on the datasets and their retrieval performances are inspected (\ref{sec:comparison}). The final discussion in Section \ref{sec:conclusio} summarizes what has been learned and points toward aspects that merit further study.

Before continuing, note that the vocabulary used in this study often makes use of the term \emph{retrieval}. As this study focuses on contexts in which the task is to retrieve relevant documents from corpora of otherwise irrelevant documents, the usage of the term \emph{retrieval} seems adequate. Yet, the task examined in this study is different from the task that is typically examined in \emph{document retrieval}. Document retrieval is a subfield of information retrieval in which the task usually is to rank documents according to their relevance for an explicitly stated user query \citep[p.~14, 16]{Manning:2008vf}. In this study, in contrast, the aim is to classify---rather than rank---documents as being relevant vs.~not relevant. 
Moreover, not all of the approaches evaluated here require the query, that states the information need, to be expressed explicitly in the form of keywords or phrases.

\section{Imbalanced Classification, Precision, Recall} \label{sec:imblearn} 

Imbalanced classification problems are common in information retrieval tasks \citep[p.~155]{Manning:2008vf}. They are characterized by an imbalance in the proportions made up by one vs.~the other category. When retrieving relevant documents from large corpora typically only a small fraction of documents falls into the positive relevant category whereas an overwhelming majority of documents is part of the negative irrelevant category \citep[p.~155]{Manning:2008vf}.

When evaluating the performance of a method in a situation of imbalance, the accuracy measure that gives the share of correctly classified documents is not adequate \citep[p.~155]{Manning:2008vf}. The reason is that a method that would assign all documents to the negative irrelevant category would get a very high accuracy value \citep[p.~155]{Manning:2008vf} Thus, evaluation metrics that allow for a refined view, such as precision and recall, should be employed \citep[p.~155]{Manning:2008vf}. Precision and recall are defined as: 

\begin{table}
\begin{center}
{\renewcommand{\arraystretch}{1.5}
\renewcommand{\tabcolsep}{0.2cm}
\begin{tabular}{c|c|c|c}
 & \textbf{truly positive} & \textbf{truly negative} & \\
 \hline
 \textbf{predicted positive} & True Positives ($TP$) & False Positives ($FP$) & $TP + FP$\\
  \hline
 \textbf{predicted negative} & False Negatives ($FN$) & True Negatives ($TN$) & $FN + TN$\\
   \hline
    & $TP + FN$ & $FP + TN$ & $N$ \\
\end{tabular}}
\footnotesize
\caption[Confusion Matrix]{\textbf{Confusion Matrix.}}
\label{tab:precrecall}
\end{center}
\end{table}

\begin{equation}
Precision = \frac{TP}{TP + FP}
\end{equation}
\begin{equation}
Recall = \frac{TP}{TP + FN}
\end{equation}

whereby $TP$, $FP$ and $FN$ are defined in Table 1. Precision and recall are in the range $[0, 1]$. However, if none of the documents is predicted to be positive, then $TP + FP = 0$ and precision is undefined. If there are no truly positive documents in the corpus, then $TP + FN = 0$ and recall is undefined. The higher precision and recall, the better.

Precision exclusively takes into account all documents that have been assigned to the positive relevant category by the classification method and informs about the share of truly positive documents among all documents that are predicted to fall into the positive category. Recall, on the other hand, exclusively focuses on the truly relevant documents and informs about the share of documents that has been identified as relevant among all truly relevant documents.

There is a trade-off between precision and recall \citep[p.~156]{Manning:2008vf}. A keyword list comprising many terms or a classification algorithm that is lenient in considering documents to be relevant will likely identify many of the truly relevant documents (high recall). Yet, as the hurdle for being considered relevant is low, they will also classify many truly irrelevant documents into the relevant category (low precision). A keyword list consisting of few specific terms or a classification algorithm with a high threshold for assigning documents to the relevant class will likely miss out many relevant instances (low recall), but among those considered relevant many are likely to indeed be relevant (high precision).

In this study's context of identifying relevant documents to be used for further analyses from a set of otherwise irrelevant documents, recall as well as precision should be as high as possible; but recall is the slightly more important metric: Recall operates on the set of all truly relevant documents and focuses on the inclusion vs.~exclusion of relevant documents into the analysis---the analytic step at which selection biases may arise. If there is a correlation between the documents identified as relevant vs.~not relevant and the value of the variable of interest, a selection bias is generated. This is, if truly relevant documents are systematically misclassified in the sense that the higher (or lower) the value on the variable of interest, the higher (or lower) the probability of being assigned to the negative irrelevant category, inferences made on the basis of the set of instances classified into the positive category are biased. High recall values do not guarantee that there are no systematic misclassifications. But the higher recall, the smaller the maximum size of the bias that arises from systematic misclassifications of truly relevant documents can become.

 
Because of its exclusive focus on true and false positives, precision provides no information on the potential of selection bias due the missing out of truly relevant documents. Nevertheless, precision should also be high. The lower precision, the less documents among those considered to be relevant by the classification method are indeed relevant. A considerable share of false positives among the set of documents classified to be relevant also has the potential to severely bias the inferences drawn or can impede the researcher from conducting any analysis at all because the retrieved documents are not those documents he or she seeks to analyze. Yet, whereas low precision can be handled by a researcher in subsequent steps, low recall implies that a substantial proportion of truly relevant documents are never to be considered for analysis. Hence, falsely classifying a truly relevant document as irrelevant can be considered to be more severe than falsely predicting an irrelevant document to be relevant.

The trade-off between precision and recall is incorporated in the $F_{\omega}$-measure, which is the weighted harmonic mean of precision and recall \citep[p.~156]{Manning:2008vf}:

\begin{equation}
F_{\omega} = \frac{(\omega^2 + 1) \cdot Precision \cdot Recall}{\omega^{2} \cdot Precision + Recall}
\end{equation}

The $F_{\omega}$-measure also is in the range $[0, 1]$. $\omega$ is a real-valued factor balancing the importance of precision vs.~recall \citep[p.~156]{Manning:2008vf}. For $\omega > 1$ recall is considered more important than precision and for $\omega < 1$ precision is weighted more than recall \citep[p.~156]{Manning:2008vf}. A very common choice for $\omega$ is 1 \citep[p.~156]{Manning:2008vf}. In this case, the $F_1$-measure (or synonymously: $F_1$-Score) is the harmonic mean between precision and recall \citep[p.~156]{Manning:2008vf}. 

\begin{equation}
F_1 = \frac{2 \cdot Precision \cdot Recall}{Precision + Recall}
\end{equation}

The $F_1$-Score is a widely used measure to evaluate the performance of classification tasks. Although recall here is considered the slightly more important measure, the $F_1$-Score---because it is the measure nearly always reported---will be employed to assess the performances of the retrieval approaches evaluated in the following.

\section{Retrieval Approaches} \label{sec:approaches} 

\subsection{Keyword Lists}

In social science, a very commonly used approach to identify documents on relevant entities is to set up a set of keywords and to consider those documents as relevant that contain at least one of the keywords (see for example the studies listed in Table \ref{tab:literatureoverview}). 
This procedure in fact is a keyword-based boolean query in which the keywords are connected with the OR operator \citep[p.~4]{Manning:2008vf}. Slightly more advanced are boolean queries in which in addition to the OR operator also the AND operator is used. Using the AND operator is important in situations in which expressions denoting the entity of interest are composed of more than a single term (e.g.~\emph{`United States'}).

\begin{table}
\small
\begin{tabular}{llll}
\textbf{Study} &  \multirow{2}{2.5cm}{\textbf{Number of keywords}} & \multirow{2}{4cm}{\textbf{How are the keywords selected?}} &  \multirow{2}{2.75cm}{\textbf{Operators in boolean query}}   \\
 &&  &   \\ 
 \hline
\citet{Puglisi2011} & 11+ & likely by the authors & OR, AND \\
\citet{King2013} & unspecified & likely by the authors & unclear \\
\citet{Burnap:2016kv} & 33 & likely by the authors & OR  \\
\citet{Jungherr2016} & 86 & by the authors & OR  \\
\citet{Beauchamp:2017cp} & 36 & likely by the authors & OR \\
\citet{vanAtteveldt2017} & 1 & by the authors &  - \\
\citet{Baum2018} & 2 & likely by the authors & OR   \\
\citet{Stier2018} & 218 & by the authors & OR   \\
\citet{FogelDror2019} & 27-170 & by the authors & OR \\
\citet{Katagiri2019} & unspecified & from COPDAB data bank & OR, AND \\
\citet{Zhang2019} & 50 & empirically; frequency-based & OR  \\
\citet{Rauh2019} & 14 & likely by the authors & OR  \\
\citet{Uyheng2020} & 1 & likely by the authors & - \\
\citet{AbdulReda2021} & 57 & by the authors & OR, AND  \\
\multirow{2}{2.5cm}{\citet{Gessler2021}} & \multirow{2}{2.5cm}{94} & \multirow{3}{4cm}{by the authors; re-usage of lists created by other authors} & \multirow{2}{2.75cm}{OR}  \\
&  & &  \\
&  & &  \\
\citet{Muchlinski2021} & 30-38 & by the authors & OR  \\
\citet{Watanabe2020} & 2-4 & by the authors & OR  \\
\end{tabular}
\footnotesize
\caption[Social Science Studies Applying Keyword Lists]{\textbf{Social Science Studies Applying Keyword Lists.} \footnotesize{This table exemplary lists social science studies that employ lists of keywords to retrieve documents or text elements that are relevant for (a part of) their analysis. A similar but older list of studies can be found in \citet[p.~5]{Linder2017}. Note that the column \emph{`Number of keywords'} gives the number of keywords the authors in the listed studies use to extract documents relating to one entity of interest. If the authors are interested in several entities, then typically several keyword lists are applied which is why here for some articles a range rather than a single number is given. Note also that \citet[p.~161]{Katagiri2019} state that the keywords they use come from the Conflict and Peace Data Bank (COPDAB) \citep{Azar2009}. They do not specify how they extract keywords from this data bank.}}
\label{tab:literatureoverview}
\end{table}

The ways in which the authors come up with a set of keywords range from simply using the most obvious terms \citep[e.g.][]{Baum2018}, to collecting a set of typical denominations for the entity of interest \citep[e.g.][]{Burnap:2016kv, Jungherr2016, Beauchamp:2017cp}, to carefully thinking about, testing, and revising sets of keywords \citep[e.g.][]{Stier2018, AbdulReda2021, Gessler2021}, to collecting keywords empirically based on word-usage in texts known to be about the entity \citep[e.g.][]{Zhang2019}. Though these approaches vary in their complexity and costs, they are all still very cheap and relatively fast procedures. Another advantage of the usage of keyword lists for the extraction of relevant documents is that a researcher has full control over the terms that are included---and not included---as keywords. 

Yet, research suggests that the human construction of keyword lists is not reliable \citep[p.~973-975]{King:2017io}. If a 
researcher generates a keyword list, then another researcher or the same researcher at another point in time is likely to construct a very different set of keywords. This is problematic: Depending on which human-generated set of search terms is used to identify relevant documents, inferences drawn may vary greatly \citep[p.~974-976]{King:2017io}.

Moreover, this conventional procedure of human keyword list generation might lead to biased inferences if the terms that are used to denote an entity correlate with the values of the variable of interest. To illustrate: Imagine that a researcher is interested in attitudes toward Joe Biden as expressed in comments on an online platform during a given time period. The researcher analyzes the sentiments of all comments that contain the search term \emph{`Biden'}. The obtained results can be biased if the attitudes expressed in comments that refer to Joe Biden as `Biden' or \emph{`Joe Biden'} differ from the attitudes in comments that refer to him as \emph{`Sleepy Joe'}. For keyword-based approaches to avoid such types of selection bias, a researcher thus has to set up a set of keywords that fully captures the universe of terms and expressions that is used to refer to the entity of interest in the given corpus.\footnote{Such a comprehensive list of keywords implies low precision and thus would come with another problem: a large share of false positives. Nevertheless, a comprehensive list would imply perfect recall and thus would preclude selection bias due to false negatives.} But humans tend to perform very poorly when it comes to constructing an extensive set of search terms \citep[p.~973-975]{King:2017io}.

There are several likely reasons for the problems human researchers encounter when trying to set up an extensive list of keywords. First, language is highly varied \citep{Durrell2008}. There are numerous ways to refer to the same entity---and entities also can be referred to indirectly without the usage of proper names or well-defined denominations \citep[p.~167]{Baden2020}. Especially if the entity of interest is abstract and/or not easily denominated, the universe of terms and expressions referring to the entity is likely to be large and not easily to be captured \citep[p.~167]{Baden2020}. Such entities are abundant in social science. Typical entities of interest, for example, are policies (e.g.~the policies implemented to address the COVID-19 pandemic), concepts (e.g.~European integration or homophobia), and occurrences (e.g.~the 2015 European refugee crisis or the 2021 United States Capitol riot).

A second likely reason for the human inability to come up with a comprehensive keyword list are inhibitory processes (\citealp{Baeuml2007}; \citealp[p.~974]{King:2017io}). After a set of concepts has been retrieved from memory, inhibitory processes suppress the representation of related, non-retrieved concepts in memory and thereby reduce the probability of those concepts to be recovered \citep{Baeuml2007}. One method that has the potential to alleviate this second aspect are query expansion methods, which are discussed next. 

\subsection{Query Expansion}
By being able to move beyond keywords that researchers are able to recall a priori, query expansion methods can be employed to create a more comprehensive set of search terms. 
Query expansion techniques expand the original query (i.e.~the original set of keywords) by appending related terms \citep[p.~1699-1700]{Azad2019}. Here, the focus is on similarity-based automatic query expansion methods, that add new terms automatically---i.e.~without interactive relevance feedback from the user---and make use of the similarity between the set of query terms to potential expansion terms \citep[p.~1700, 1706]{Azad2019}. The underlying hypothesis used here is the association hypothesis formulated by van Rijsbergen stating that ``[i]f one index term is good at discriminating relevant from non-relevant documents, then any closely associated index term is also likely to be good at this'' \citep[p.~11]{vanRijsbergen2000}. The specific methods differ regarding
\begin{itemize}
\item the data source to extract candidate terms for the expansion,
\item how candidate terms from this data source are ranked (such that the ranks reflect the relatedness to the original query), and
\item how (many) additional terms are selected and integrated into the original query
\end{itemize}
\citep[p.~1701]{Azad2019}. Data sources from which expansion terms are identified can be the corpus from which relevant documents are to be retrieved, the documents retrieved by the initial query, human-created thesauri such as WordNet, knowledge bases as Wikipedia, external corpora such as a collection of web texts, or a combination of these \citep[p.~1701-1704]{Azad2019}. If thesauri such as WordNet are employed as a data source, terms the thesaurus encodes to be related to the query terms can be considered candidate terms for expansion \citep[p.~1702]{Azad2019}. Path lengths between the synsets (word senses) in a thesaurus can be used to compute a similarity score between a query term and potential expansion terms \citep[p.~1705]{Azad2019}. In Wikipedia, the network of hyperlinks between articles can be used to extract articles about concepts related to the query terms \citep{ALMasri2013}. A similarity score, for example, can be computed based on shared ingoing and outgoing hyperlinks between articles \citep[][p.~6]{ALMasri2013}. If the data source for query expansion is the local corpus from which documents are to be retrieved or if the data source is an external global corpus, then the similarity between terms can be assessed via similarity measures that are computed based on the terms' vector representations \citep[p.~1706]{Azad2019}. A frequently used measure is cosine similarity:
\begin{equation}
sim_{cos}(a_1, a_2) = cos(\theta) = \frac{\bm{z}_{[a_1]} \cdot \bm{z}_{[a_2]}}{|| \bm{z}_{[a_1]} || \; || \bm{z}_{[a_2]} ||}
\end{equation}
whereby $\bm{z}_{[a_1]}$ and $\bm{z}_{[a_2]}$ are the vector representations of terms $a_1$ and $a_2$ respectively, $|| \bm{z}_{[a_1]} ||$ and $|| \bm{z}_{[a_2]} ||$ is the length of these vectors as computed by the Euclidean norm, and $\theta$ is the angle between the vectors. Cosine similarity gives the cosine of the angle between the term representation vectors $\bm{z}_{[a_1]}$ and $\bm{z}_{[a_2]}$ \citep[p.~122]{Manning:2008vf}. If the angle between the vectors equals 0$^{\circ}$, meaning that the vectors have the exact same orientation, the cosine is 1 \citep[p.~281]{Moore2013}. If the angle is 90$^{\circ}$, meaning that the vectors are orthogonal to each other, then $cos(\theta) = 0$ \citep[p.~281]{Moore2013}.\footnote{If the elements of the term vectors are non-negative, e.g.~because they indicate the (weighted) frequency with which a term occurs across the documents in the corpus, then the angle between the vectors will be in the range $[0^{\circ}, 90^{\circ}]$ and cosine similarity will be in the range $[0, 1]$. If, on the other hand, elements of term representation vectors can become negative, then the vectors can also point into opposing directions. In the extreme: If the vectors point into diametrically opposing directions, then $cos(\theta) = -1$.}

A frequently used term representation are word embeddings \citep[see e.g.][]{Diaz2016, Kuzi2016, Silva2020}. A word embedding is a real-valued vector representation of a term. Important model architectures to learn word embeddings are the continuous bag-of-words (CBOW) and the skip-gram models \citep{Mikolov2013b} as well as Global Vectors (GloVe) \citep{Pennington2014} and fastText \citep{Bojanowski2017}. In learning the word embedding for a target term $a_t$, these architectures make use of the words that occur in a context window around $a_t$ (\citealp[p.~4]{Mikolov2013b}; \citealp[p.~1533-1535]{Pennington2014}). In doing so, these procedures for learning word embeddings implicitly draw on the distributional hypothesis \citep{Firth1957} stating that the meaning of a word can be deduced from the words it typically co-occurs with \citep[p.~102]{Rodriguez2022}. This in turn implies that semantically or syntactically similar terms are likely to have similar word embedding vectors that point into a similar direction (\citealp[p.~1139-1140]{Bengio2003}; \citealp{Mikolov2013a}). 

In similarity-based query expansion techniques, terms that are closest to the query terms are used as query expansion terms. The number of terms added varies from approach to approach between five to a few hundred \citep[p.~1714]{Azad2019}. 
In \citet{Silva2020}, for example, the original query is represented by a single vector that is computed by taking the weighted average of the word embeddings of all terms in the original query. Then the five terms whose embeddings have the highest cosine similarity with the embedding of the query are selected for expansion.

To sum up, researchers that implement query expansion methods require a data source for expansion, a way to compute a measure that captures the relatedness between terms, and a procedure that determines which and how many terms are added via which process. If they plan to represent terms as word embeddings, then either pretrained word embeddings are required or the embeddings have to be learned. Consequently, considerable resources and expertise is needed. Yet, whereas individuals due to inhibitory processes may fail to create a comprehensive list of search terms, query expansion methods can uncover terms that denote the entity of interest and are used in the corpus at hand. As query expansion techniques have the potential to expand the initial query with synonymous and related terms, recall is likely to increase \citep[p.~193]{Manning:2008vf}. Precision, however, may decrease---especially if the added terms are homonyms or polysemes (i.e.~terms that have different meanings; whereby the meanings can be conceptually distinct (homonyms) or related (polysemes)) (\citealp[p.~110]{Manning1999}; \citealp[p.~193]{Manning:2008vf}). It thus may be advantageous to use as a data source for query expansion a corpus or thesaurus that is specific to the domain of the retrieval task rather than a global corpus or general thesaurus \citep[p.~193]{Manning:2008vf}. Moreover, query expansion techniques require researchers to a priori come up with an initial set of query terms (which will encode the researchers' assumptions) and there is no guarantee the expansion starting from the initial set will capture all different denominations of the entity. For example, there is no guarantee that query expansion will succeed in moving from \emph{`Biden'} to \emph{`Sleepy Joe'}. Finally, if the entity of interest is also referred to with multi-term expressions (e.g.~\emph{`United States'}), then these only can be extracted if the term representations used by the expansion procedure also cover multi-term expressions. Word embeddings would have to be learned or be available also for bigrams and trigrams. This increases the methods' complexity, the computational resources required and limits the availability of external globally pretrained word embeddings.\footnote{Note that beside the similarity-based automatic query expansion approaches discussed so far, there are further expansion methods. Most prominently there are query language modeling and operations based on relevance feedback from the user or pseudo-relevant feedback (\citealp{Lavrenko2001}; \citeauthor{Manning:2008vf}, \citeyear{Manning:2008vf}, p.~177-188; \citeauthor{Azad2019}, \citeyear{Azad2019}, p.~1709-1713).}


\subsection{Topic Model-Based Classification Rules} \label{sec:tmbcr}

Recently \citet{Baden2020} have proposed a procedure in which documents are categorized based on classification rules that are built by researchers on the basis of topics estimated by a topic model. \citet{Baden2020} call their procedure Hybrid Content Analysis. The idea is to assign those documents to a pre-defined category that are estimated to be comprised to a considerable share of topics that the researchers deem to be related to the category \citep{Baden2020}. 
Whilst \citet{Baden2020} formulate their method for multi-class or multi-label classification tasks in a descriptive manner, here the procedure is presented with precise mathematical expressions and the focus is exclusively on the binary classification task of retrieving relevant documents.

\begin{figure}[p]
\begin{center}
\footnotesize
\includegraphics[width=0.8\textwidth]{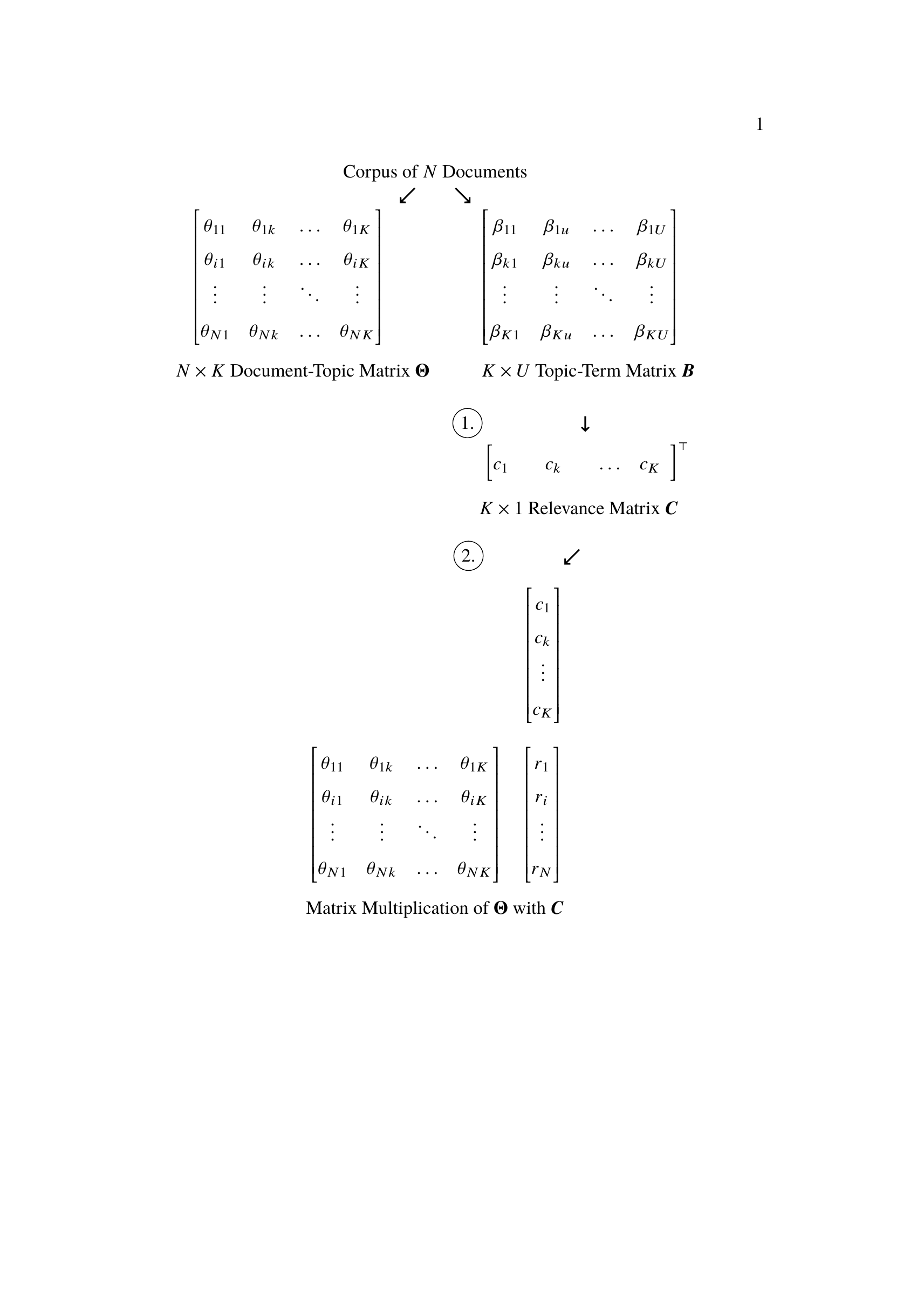}
\caption[Building Topic Model-Based Classification Rules]{\textbf{Building Topic Model-Based Classification Rules.} \footnotesize{Classification rules can be built from any topic model that on the basis of a corpus comprising $N$ documents estimates a latent topic structure characterized by two matrices: $N \times K$ document-topic matrix $\bm{\Theta}$ and $K \times U$ topic-term matrix $\bm{B}$. $\beta_{ku}$ is the estimated probability for the $u$th term to occur given topic $k$. $\theta_{ik}$ is the estimated share assigned to topic $k$ in the $i$th document. The topic model-based classification rule procedure proceeds as follows: Step 1: Researchers inspect matrix $\bm{B}$, determine which topics are relevant and create $K \times 1$ relevance matrix $\bm{C}$. Step 2: Matrix multiplication of $\bm{\Theta}$ with $\bm{C}$ yields the resulting vector $\bm{r}$. Step 3~(not shown): Documents with $r_i >=$ threshold $\xi \in [0,1]$ are retrieved.}}
\label{fig:tm}
\end{center}
\end{figure}

The family of topic models most widely applied in social science are Bayesian hierarchical mixed membership models that estimate a latent topic structure based on observed word frequencies in text documents (\citealp[p.~993, 995-997]{Blei:2003hl}; \citealp[p.~18]{Blei2007a}; \citealp[p.~988]{Roberts2016}; \citealp[p.~4713-4714]{Zhao2021a}). These topic models (which are here simply referred to as \emph{topic models}) assume that each topic is a distribution over the terms in the corpus and each document is characterized by a distribution over topics (\citealp[p.~995-997]{Blei:2003hl}; \citealp[p.~18]{Blei2007a}). Given a corpus of $N$ documents, topic models estimate a latent topic structure defined by $N \times K$ document-topic matrix $\bm{\Theta}$ and $K \times U$ topic-term matrix $\bm{B}$ (see Figure \ref{fig:tm}). 
Topic-term matrix $\bm{B} = [ \bm{\beta}_{1}, \dots, \bm{\beta}_{k}, \dots, \bm{\beta}_{K}]^\top$ gives for each topic, $k \in \{1, \dots, K\}$, the estimated probability mass function across the $U$ unique terms in the vocabulary: $\bm{\beta}_k = [\beta_{k1}, \dots, \beta_{ku}, \dots, \beta_{kU}]$; whereby $\beta_{ku}$ is the probability for the $u$th term to occur given topic $k$. Document-topic matrix $\bm{\Theta} = [\bm{\theta}_{1}, \dots, \bm{\theta}_{i}, \dots, \bm{\theta}_{N}]^\top$ contains for each document $d_i$ the estimated proportion assigned to each of $K$ latent topics: $\bm{\theta}_i = [\theta_{i1}, \dots, \theta_{ik}, \dots, \theta_{iK}]$, with $\theta_{ik}$ being the estimated share of document $d_i$ assigned to topic $k$.

Given the estimated latent topic structure characterized by $K \times U$ topic-term matrix $\bm{B}$ and $N \times K$ document-topic matrix $\bm{\Theta}$, the topic model-based classification rule building procedure proceeds as follows (see Figure \ref{fig:tm}) \citep[p.~171-174]{Baden2020}:

\begin{enumerate}
\item Based on $K \times U$ topic-term matrix $\bm{B}$ the researcher inspects for each topic the most characteristic terms, e.g.~the terms that are most likely to occur in a topic and the terms that are the most exclusive for a topic.\footnote{The \emph{most likely} terms are the terms with the highest occurrence probabilities, $\beta_{ku}$, for a given topic $k$. The \emph{most exclusive} terms refer to highly discriminating terms whose probability to occur is high for topic $k$ but low for all or most other topics. Exclusivity can be measured as: $exclusivity_{ku} = \beta_{ku}/\sum_{j=1}^K \beta_{ju}$ \citep[see for example][p.~12]{stm2019}.} Given these terms that inform about the content of each topic, the researcher determines which topics refer to the entity of interest. The researcher then creates relevance matrix $\bm{C}$ of size $K \times 1$ whose elements are 1 if the topic is considered relevant and are 0 otherwise.
\item Then $N \times K$ document-topic matrix $\bm{\Theta}$ is multiplied with $\bm{C}$. The resulting vector $\bm{r} = [r_1, \dots, r_i, \dots, r_N]^\top$ gives for each document the sum over those topic shares that refer to relevant topics. $r_i$ can be interpreted as the share of words in document $d_i$ that come from relevant topics.
\item A threshold value $\xi \in [0,1]$ is set. All documents for which $r_i >= \xi$ are considered to be relevant.
\end{enumerate}

The procedure utilizes a topic model as an unsupervised tool to uncover information about the latent topic structure of a corpus. Leveraging this information for the retrieval of relevant documents allows researchers to operate without a set of explicit keywords. Rather than having to come up with information about to be retrieved documents a priori, researchers merely have to recognize topics that refer to relevant entities.  As topic models are well known and frequently developed and applied in social science \citep[e.g.][]{Quinn:2010jl, Grimmer:2013df, Roberts:2014es, Bauer:2017kr, Maier2018, Baerg2020, Eshima2021, Schulze2021} and furthermore are implemented in corresponding software packages \citep[e.g.][]{topicmodels:2011gh, stm2019}, the procedure of building classification rules based on topic models seems easily accessible to the social science community. 

Yet, estimating a topic model in the first place induces costs. Especially the number of topics $K$ has to be set a priori. To set a useful value for $K$ typically several topic models with varying $K$ are estimated and after a manual inspection of the most likely and most exclusive terms for a topic as well as the computation of performance metrics (e.g.~held-out likelihood), researchers decide on a topic number \citep[p.~60-62]{Roberts2016wg}. Moreover, as topic models are unsupervised there is no way for researchers---beyond setting parameters as $K$---to guide the estimation process such that the results are related to the concepts of interest. Ideally one would like to have a topic model that produces one or several topics that refer to the entity of interest and are characterized by high semantic coherence as well as exclusivity. A coherent topic referring to the entity of interest would have high occurrence probabilities for frequently co-occurring terms that refer to the entity (\citeauthor{Roberts:2014es}, \citeyear{Roberts:2014es}, p.~1069; \citeauthor{stm2019}, \citeyear{stm2019}, p.~10). It would be clearly about the entity of interest rather than being a fuzzy topic without a nameable content. An exclusive topic would solely refer to the entity of interest and would not refer to any other entities.

It is not guaranteed, however, that there is a topic that distinctly covers the relevant entity. 
Additionally, topic models can generate topics that relate to several entities rather than a single entity. By selecting each topic that refers to the relevant entity but also relates to several non-relevant entities, a researcher will construct a topic model-based classification rule that will be characterized by high recall but low precision. For this reason \citet[p.~173]{Baden2020} suggest to set $K$ to a rather high value such that topics are fine-grained. Yet, whether this will work out in a given application is unclear as the latent topic structure uncovered by the topic model cannot be forced to neatly separate topics referring to relevant entities from topics referring to non-relevant entities. And topic models also cannot be forced to produce coherent topics referring to the entity of interest at all. 

\subsection{Passive and Active Supervised Learning}

Supervised learning algorithms are trained on the basis of a training data set. The training data set contains a set of documents and corresponding classes or values. In the context of retrieving relevant documents, a training set document is assigned to the relevant class if it refers to the entity of interest and is assigned to the irrelevant class otherwise. Central to the supervised learning process is the loss function. The loss function returns a cost-signifying value which is a function of the predicted and the true values of the training set documents. In an optimization process, the parameters of the supervised learning algorithm are moved toward values for which the loss function reaches a (local) minimum.

Supervised learning methods have the advantage that they come with supervision: the separation between relevant and irrelevant documents is encoded in the training data set and then learned by the model. This is a considerable advantage over automatic query expansion methods and topic model-based approaches. In the former, researchers cannot be entirely sure that the expansion really will include terms related to the initial query terms and in the latter it is unclear whether there will be coherent and exclusive topics referring to the entity of interest.

Moreover, as the true class assignments for the training set documents are known, supervised learning approaches allow researchers to use resampling techniques (e.g.~cross-validation) in order to assess how well the retrieval of relevant documents works. The values for precision and recall not only provide information about the performance of the retrieval method but also indicate the nature of the (mis)classifications. (Is the model lenient in assigning documents to the positive relevant class and therefore most of the relevant documents are retrieved (high recall) but there are many false positives among the retrieved documents (low precision) or is it rather the other way round?) 

Furthermore, just as the topic model-based approach, supervised learning techniques depend on recognizing rather than recalling: When creating the training data set, coders read the training documents and assign them to the relevant vs.~irrelevant class as specified in coding instructions. Hence, supervised learning techniques require the coders to merely recognize relevant documents rather than creating information on relevant documents from scratch.

Supervised learning methods, however, also come with disadvantages. First, the labeling of training documents by human coders is extremely costly. Precise coding instructions have to be formulated, the coders have to be trained and paid, and the intercoder reliability (e.g.~measured by Krippendorff's $\alpha$ \citep[p.~277-294]{Krippendorff2013}) has to be assessed. Reading an adequately large sample of documents and labeling each as relevant vs.~irrelevant (or having this being done by trained coders) takes time.

Second, in the context of retrieving relevant documents, it is likely that the share of relevant documents is small and thus further problems arise: If the training set documents are randomly sampled from the entire corpus from which relevant documents are to be retrieved and only a small share of documents refer to the entity of interest, then a large number of training documents have to be sampled, read, and coded such that the training data set contains a sufficiently large number of documents falling into the positive relevant class for the supervised method to effectively learn the distinctions between the relevant and the irrelevant class. If, for example, $3 \%$ of documents are relevant, then after coding 1,000 randomly sampled training documents only about 30 documents will be assigned to the relevant category.\footnote{Note that research suggests that it is rather the number of training examples in the positive relevant class than the number of all documents in the training set that affects the amount of information provided to the learning method \citep{Wang2020a}.}

What is more: If no adjustments are made, then each training set document has the same weight in the calculation of the value of the loss function. This is, the optimization algorithm attaches the same importance to the correct classification of each training set document.
Yet, in a retrieval situation characterized by imbalance, researchers typically care more about the correct classification of relevant training documents than irrelevant documents (see also argumentation in Section \ref{sec:imblearn} above) \citep[p.~2-4]{Branco2016}. Or put differently, missing a truly relevant document (false negative) is considered more problematic than falsely predicting an irrelevant document to be relevant (false positive) \citep{Brownlee2020b}. 
So, there is the question of what to do to make the supervised learning algorithm focus on correctly detecting relevant documents. 

The statistical learning community has devised a large spectrum of approaches to deal with imbalanced classification problems \citep[for an overview see][]{Branco2016}. 
Among the most common and most easily applicable procedures that are employed to make the optimization algorithm put more weight on the correct classification of instances that are part of the relevant minority class are techniques that adjust the distribution of training set instances \citep[p.~7-15, 21-27]{Branco2016}. This set of techniques comprises procedures such as random oversampling, random undersampling and the synthetic minority oversampling technique (SMOTE) \citep{Chawla2002} \citep[p.~22]{Branco2016}.\footnote{SMOTE \citep{Chawla2002} is a well-known technique in which the minority class is enlarged by adding synthetically generated minority class training examples. A synthetic training instance is created by the following process: For each feature, a feature value is randomly drawn from the line joining the feature value of a randomly sampled minority class instance and the feature value of one of its $Q$ nearest neighbors \citep[p.~328-329]{Chawla2002}. This implies that SMOTE is ``operating in `feature space' rather than `data space'''\citep[p.~328]{Chawla2002} of data that are represented in tabular form \citep{Brownlee2021a}. (Thus, SMOTE can be applied on a bag-of-words-based document-feature matrix but not original sequential text data.) In contrast to a simple random oversampling procedure, SMOTE adds new instances rather than exact copies to the training data and thereby reduces the risk of overfitting \citep[p.~860]{Chawla2005}
. (Note that SMOTE typically is combined with random undersampling \citep[p.~330]{Chawla2002}. There are various modifications of SMOTE or combinations of SMOTE with other techniques and models \citep[p.~25-26]{Branco2016}.)} In random oversampling instances of the minority class are randomly resampled with replacement and appended as duplicates to the training data set \citep[p.~9833]{Wang2020a}. In random undersampling, randomly selected instances of the majority class are removed from the training set \citep[p.~9831]{Wang2020a}. 
Both resampling techniques typically are applied until a user-specified distribution of class labels is reached (e.g.~until the minority class contains as many instances as the majority class) \citep{Brownlee2021}. Thereby, both resampling strategies make the training set more balanced and thus put more weight on the minority class than in the original training set distribution. As random oversampling implies that resampled minority instances are added as exact duplicates, random oversampling can lead to overfitting on the training data and reduced generalization performance on the test data \citep[p.~22]{Branco2016}. Moreover, oversampling implies higher computational costs \citep[p.~22]{Branco2016}. In random undersampling, on the other hand, information from removed majority class instances is lost \citep{Brownlee2021}.\footnote{Beside these random resampling techniques mentioned here, there are also methods that perform oversampling or undersampling in an informed way; e.g.~based on distance criteria \citep[see][p.~23-24]{Branco2016}.} 

Beside these techniques that adjust the training set distribution, a second set of methods to address imbalanced classification problems is the usage of cost-sensitive algorithms \citep[p.~27 ff.]{Branco2016}. There are specifically developed modifications of algorithms that allow for incorporating higher costs for misclassifying instances of the minority class \citep[for an overview over these special-purpose methods see][p.~27-29]{Branco2016}. A more general method, however, is to set up a cost matrix that specifies which cell in the confusion matrix (see Table \ref{tab:precrecall} in Section \ref{sec:imblearn}) is associated with which cost \citep{Elkan2001, Brownlee2020b}. During training, the loss of each training instance takes into account the respective cost depending on which cell the instance is in \citep[p.~973]{Elkan2001}. In this way, higher costs can be specified for false negatives than for false positives and be directly incorporated into the training process.

The idea of the cost matrix also underlies the techniques that modify the distribution of training instances \citep[p.~975]{Elkan2001}. The undersampling rates for the majority class or the oversampling rates for the minority class ideally should reflect the cost induced by misclassifying an instance from the respective class \citep{Brownlee2020b}. For example, if falsely predicting an instance from the positive minority class to be negative is considered 10 times more costly than falsely predicting an instance from the negative majority class to be positive, then the cost of a false negative is 10 and the cost for a false positive 1 (and true positives and true negatives induce no costs) \citep[p.~36]{Branco2016}. Positive minority class instances then 
could be randomly oversampled such that their number increases by a factor of 10, or the majority class instances could be undersampled such that their number decreases by a factor of 1/10 \citep[p.~36]{Branco2016}.\footnote{Note that the outlined relationship between cost ratios and over- or undersampling rates only holds if the threshold at which the classifier considers an instance to fall into the positive rather than the negative class is at $p = 0.5$ \citep[p.~975]{Elkan2001}. Note furthermore that although it would be good practice for resampling rates to reflect an underlying distribution of misclassification costs as specified in the cost matrix, resampling with rates reflecting misclassification costs will not yield the same results as incorporating misclassification costs into the learning process \citep[p.~36]{Branco2016}. One reason, for example, is that in random undersampling instances are removed entirely \citep[p.~36]{Branco2016}. For information on the relationship between oversampling/undersampling, cost-sensitive learning, and domain adaptation see \citet[p.~4-5, 7]{Kouw2019}.}

In practice, however, all discussed techniques suffer from the problem that researchers often cannot specify precise values for misclassification costs \citep{Brownlee2020b}. In the context of the task of retrieving relevant documents, researchers may be able to say that false negatives are more costly than false positives but how much so is likely to be highly difficult to specify (\citeauthor{Branco2016}, \citeyear{Branco2016}, p.~3; \citeauthor{Brownlee2020b}, \citeyear{Brownlee2020b}).

The focus of the so far mentioned methods for imbalanced classification problems has been on the difference in the misclassification costs associated with instances from the positive minority vs.~negative majority class. Yet, there are other types of cost that also should be considered: As elaborated above, the annotation of training documents is costly due to the resources required. And in the context of imbalanced classification problems annotating a random sample of documents is inefficient as a disproportionately large number of documents has to be annotated until an acceptably number of instances form the minority class is labeled. These training set annotation costs are the focus of active learning strategies.

Active learning refers to learning techniques in which the learning algorithm itself indicates which training instances should be labeled next \citep[p.~4]{Settles2010}. The idea is to let the learning algorithm select instances for labeling that are likely to be informative for the learning process \citep[p.~5]{Settles2010}. Such instances could be, for example, those instances whose prediction the learner is most uncertain about \citep[p.~5]{Settles2010}. The underlying hypothesis is that by letting the learner actively select the instances from which it seeks to learn, a high as possible prediction accuracy can be achieved with a small as possible number of annotated training instances \citep[p.~4, 5]{Settles2010}. Active learning stands in contrast to the usual supervised learning procedure in which the training set instances are being randomly sampled, annotated and then handed over to the learning algorithm. When juxtaposing active learning to this usual supervised learning procedure, the latter sometimes is called passive learning \citep[p.~534]{Miller2020}.

Active learning is useful in situations in which unlabeled training instances are abundant but the labeling process is costly \citep[p.~4]{Settles2010}. There are several different scenarios in which active learning can be applied \citep[see][p.~8-12]{Settles2010}. In this study, the focus is on pool-based sampling. In pool-based sampling a large collection of instances has been collected from some data distribution in one step \citep[p.~11]{Settles2010}. At the start of the learning algorithm, labels are obtained only for a very small set of instances, denoted $\mathcal{I}$, whilst the other instances are part of the large pool of unlabeled instances $\mathcal{U}$ \citep[p.~11]{Settles2010}. 
In each iteration of the active learning algorithm, the algorithm is trained on instances in the labeled set $\mathcal{I}$ and makes predictions for all instances in pool $\mathcal{U}$ (\citeauthor{Lewis1994}, \citeyear{Lewis1994}, p.~4; \citeauthor{Settles2010}, \citeyear{Settles2010}, p.~6, 11). The instances in pool $\mathcal{U}$ then are ranked according to how much information the learner would gather from an instance if it were labeled \citep[p.~11-12]{Settles2010}. 
Then the most informative instances in $\mathcal{U}$ are selected and labeled (e.g.~by human coders) \citep[p.~6]{Settles2010}. The newly labeled instances are added to set $\mathcal{I}$ and a new iteration starts \citep[p.~6]{Settles2010}.\footnote{Ideally, a single instance is selected and labeled in each iteration \citep[p.~4]{Lewis1994}. Yet, re-training a model often is costly and time consuming. An economic alternative is batch-mode active learning \citep[p.~35]{Settles2010}. Here a batch of instances is selected and labeled in each iteration \citep[p.~35]{Settles2010}. When selecting a batch of instances, there is the question of which instances to select. Selecting the $K$ most informative instances is one strategy that, however, ignores the homogeneity of the selected instances \citep[p.~35]{Settles2010}. Alternative approaches that seek to increase the heterogeneity among the selected instances have been developed \citep[see][p.~35]{Settles2010}.}

In the active learning community several different strategies of how the informativeness of an instance is defined and how the most informative instances are selected have been developed \citep[for an overview see][p.~12 ff.]{Settles2010}. These strategies are termed query strategies \citep[p.~12]{Settles2010}. Here, the ``[p]erhaps [...] simplest and most commonly used query framework'' \citep[p.~12]{Settles2010} will be presented: uncertainty sampling \citep{Lewis1994}. In uncertainty sampling those instances are considered to be the most informative about which the learning algorithm expresses the highest uncertainty \citep[p.~4]{Lewis1994}. 
In the context of the binary document retrieval classification task, the uncertainty could be said to be highest for instances for which the predicted probability to belong to the relevant class is closest to 0.5 \citep[p.~4]{Lewis1994}.\footnote{Note that in multi-class classification tasks it is less straight forward to operationalize uncertainty. Here one can distinguish between least confident sampling, margin sampling, and entropy-based sampling \citep[for precise definitions see][p.~12-13]{Settles2010}.} The usage of such a definition of uncertainty and informativeness only is possible for learning methods that return predicted probabilities \citep[p.~12]{Settles2010}. For methods that do not, other uncertainty-based sampling strategies have been developed \citep[see][p.~14-15]{Settles2010}. With regard to SVMs, \citet{Tong2001} have introduced three theoretically motivated query strategies. In their Simple Margin strategy the data point that is closest to the hyperplane is selected to be labeled next \citep[p.~53-54]{Tong2001}.

One important aspect to be kept in mind when applying active learning techniques is that because the training instances are not sampled randomly from the underlying corpus but are purposefully selected, the distribution of the class labels in training data set $\mathcal{I}$ and in unlabeled pool $\mathcal{U}$ is different from the distribution of labels in the entire corpus \citep[p.~539]{Miller2020}. 
If the expected generalization error is to be estimated, then one option is to randomly sample a set of instances from the corpus at the very start of the analysis (\citeauthor{Tong2001}, \citeyear{Tong2001}, p.~57; \citeauthor{Miller2020}, \citeyear{Miller2020}, p.~539, 541). This set then is annotated and set aside such that it neither can become part of set $\mathcal{I}$ nor set $\mathcal{U}$ (\citeauthor{Tong2001}, \citeyear{Tong2001}, p.~57; \citeauthor{Miller2020}, \citeyear{Miller2020}, p.~539, 541). After each learning iteration or a fixed number of iterations, the performance of the active learning algorithm then can be evaluated on this independent test set (\citeauthor{Tong2001}, \citeyear{Tong2001}, p.~57; \citeauthor{Miller2020}, \citeyear{Miller2020}, p.~539, 541).

Empirically one can say that in a majority of published works active learning reaches the same level of prediction accuracy with fewer training instances than supervised learning with random sampling of training instances (passive learning) \citep{Lewis1994, Tong2001, Ertekin2007, Settles2010, Miller2020}. 
Especially if data sets are imbalanced, active learning tends to reach the same level of classification performance with a substantively smaller number of labeled training instances than passive learning  (\citeauthor{Ertekin2007}, \citeyear{Ertekin2007}, p.~131; \citeauthor{EinDor2020}, \citeyear{EinDor2020}, p.~7954; \citeauthor{Miller2020}, \citeyear{Miller2020}, p.~543-544). Closer inspections show that during the learning process, the training set $\mathcal{I}$, which is selected by the active learning algorithm, is more balanced than the original data distribution  (\citeauthor{Ertekin2007}, \citeyear{Ertekin2007}, p.~133-134; \citeauthor{Miller2020}, \citeyear{Miller2020}, p.~545). 
One likely reason for this observation is that as active learning algorithms tend to pick instances for labeling from the uncertain region between the classes and in this region of the feature space the class distribution tends to be more balanced, the class distribution among instances an active learning algorithm tends to select is likely to be more balanced \citep[p.~129, 133-134]{Ertekin2007}. A more balanced distribution implies that more weight is given to the minority class instances. Another likely reason is that because active learning algorithms tend to pick instances close to the boundary between the classes, they are able to learn the class boundary with a smaller number of training instances \citep[p.~28]{Settles2010}.

\section{Comparison} \label{sec:comparison} 

In the following section, retrieving documents via keyword lists is compared to a query expansion technique, topic model-based classification rules and active as well as passive supervised learning on the basis of three retrieval tasks. The code to replicate the analysis can be accessed via figshare at https://doi.org/10.6084/m9.figshare.19699840.v1. The analysis is conducted in R \citep{RCoreTeam2020} and Python \citep{vanRossum2009}. For the analyses pertaining to active and passive supervised learning with the pretrained language representation model BERT (standing for Bidirectional Encoder Representations from Transformers), the Python code is run in Google Colab \citep{GoogleColab2020} in order to have access to a GPU.\footnote{The employed R packages are data.table \citep{datatable2020}, dplyr \citep{dplyr2021}, facetscales \citep{facetscales2021}, ggplot2 \citep{ggplottwo:2016hw}, ggridges \citep{ggridges2021}, lsa \citep{lsa2020}, plot3D \citep{plot3D2019}, quanteda \citep{Benoit2018}, RcppParallel \citep{RcppParallel2020}, rstudioapi \citep{rstudioapi2020}, stm \citep{stm2019}, stringr \citep{stringr2019}, text2vec \citep{text2vec2020}, and xtable \citep{xtable2019}. The used Python packages and libraries are Beautiful Soup \citep{bs42020}, gdown \citep{Kentaro2020}, imbalanced-learn \citep{Lemaitre2017}, matplotlib \citep{Hunter2007}, NumPy \citep{Oliphant2006}, pandas \citep{McKinney2010}, seaborn \citep{Waskom2020}, scikit-learn \citep{sklearn2011}, PyTorch \citep{Paszke2019}, watermark \citep{Raschka2020}, and HuggingFace's Transformers \citep{Wolf2019}. If a GPU was used, an NVIDIA Tesla P100-PCIE-16GB was employed.}

\subsection{Data} \label{sec:data} 

{\bfseries\sffamily{Twitter}}: The first inspected retrieval task operates on a corpus comprising 24,420 German tweets. These tweets are a random sample of all tweets in German language in a larger collection of tweets that has been collected by \citet{Barbera2016}. \citet{Linder2017} sampled 24,420 German tweets and used CrowdFlower workers to label the sampled tweets. For each tweet, the label indicates whether the tweet refers to refugees, refugee policies, and the refugee crisis and thus is considered relevant or not \citep[p.~23-24]{Linder2017}. The task of retrieving the relevant tweets from this corpus indeed is an imbalanced classification problem as only 727 out of the 24,420 tweets (2.98\%) are labeled to be about the refugee topic. 

{\bfseries\sffamily{SBIC}}: The aim of the second retrieval task 
is to extract all posts from the Social Bias Inference Corpus (SBIC) \citep{Sap2020} that have been labeled to be offensive toward mentally or physically disabled people. The SBIC includes 44,671 potentially toxic and offensive posts from Reddit, Twitter and three websites of online hate communities \citep[p.~5480]{Sap2020}.\footnote{For a detailed elaboration about the exact composition of the SBIC see \citet[p.~5480]{Sap2020}} The SBIC was collected with the aim of studying implied---rather than explicitly stated---social biases \citep[p.~5477]{Sap2020}. The subreddits and websites selected to be included in the SBIC constitute intentionally offensive online communities \citep[p.~5480]{Sap2020}. The additionally included reddit comments and tweet data sets were collected such that there is an increased likelihood that the content of the collected posts is offensive (e.g.~by selecting tweets that include hashtags known to be racist or sexist) \citep[p.~5480]{Sap2020}. \citet{Sap2020} used Amazon Mechanical Turk for the annotation of the posts. For each post the coder indicated, amongst others, whether the post is offensive and if so, whether the target is an individual (meaning that the post is a personal insult) or a group (implying that the post offends a social group, e.g.~women, people of color) \citep[p.~5479-5480]{Sap2020}. If one or several groups were targeted, the coders were asked to name the targeted group or groups \citep[p.~5479-5480]{Sap2020}. The authors merge the 1,414 targeted groups into seven larger group categories \citep[p.~5481]{Sap2020}. One of these group categories are mentally or physically disabled people. $2.15\%$ of the 44,671 posts are annotated as being offensive toward the disabled.\footnote{Note that each post was annotated by three independent coders and that the data shared by \citet{Sap2020} lists each annotation separately. Here the SBIC is preprocessed such that the post is considered to be offensive toward a group category if at least on annotator indicated that a group falling into this category was targeted.} 
The category of disabled people is selected as the focus of this study because this group category is the most coherent capturing a well-defined group of people.

{\bfseries\sffamily{Reuters}}: The third retrieval task is to identify all newspaper articles in the Reuters-21578 corpus \citep{Reuters1997} that refer to the topic surrounding crude oil. Reuters-21578 \citep{Reuters1997} is a widely used corpus for evaluating retrieval approaches \citep{Tong2001, Ertekin2007, Huggingface2021}. The corpus contains 21,578 newspaper articles that were published on the Reuters financial newswire service in 1987 \citep{Reuters1997, Huggingface2021}. 10,377 articles are assigned to one or several out of 135 economic subject categories called topics \citep{Reuters1997}. These categories are e.g. \emph{`gold'}, \emph{`grain'}, \emph{`cotton'}. Here, the 10,377 topic-annotated articles are used for the analysis. The aim is to identify the $566$ ($5.45\%$) newspaper articles that are labeled to be about the crude oil topic. The topic is the fourth largest. It is large enough to possibly contain enough documents for the algorithms to learn from and at the same time is small enough such that the identification of crude oil articles can be considered an imbalanced classification problem. 

The three data sets employed here are selected with the aim to achieve and represent various types of retrieval tasks common in social science. Tweets, posts from online platforms, and newspaper articles are types of documents that are often analyzed in social science and whose analysis typically involves some preliminary retrieval step \citep[see e.g.][]{King2013, Beauchamp:2017cp, Baum2018, Stier2018, FogelDror2019, Zhang2019, Watanabe2020, Muchlinski2021}. The entities of interest in social science studies vary widely with regard to their nature and their level of abstraction. \citet{Zhang2019} study collective action events, \citet{Baum2018} focus on rape incidents, \citet{Puglisi2011} retrieve information on persons involved in political scandals, \citet{Uyheng2020} extracts tweets referring to the COVID-19 pandemic, \citet{Jungherr2016} examine parties, candidates, and campaign events during an election campaign, and the entities of interest for \citet{FogelDror2019} are Israel and the Palestinian Authority. In this study, the entities of interest range from a multi-dimensional topic that includes abstract policies, occurrences as well as a social group (refugee policies, refugee crisis, refugees), to a one-dimensional topic about a single economic product (crude oil), to a specific social group (disabled people) that is referred to in a specific (namely: offending) way. Moreover, the corpora from which documents are retrieved in social science can be thematically highly heterogeneous (as is the case with the corpus of German tweets here and with the Weibo posts studied by \citet{Zhang2019}) or---due to the nature of the source---be more homogeneous with regard to topics, linguistic style, or attitudes (see e.g.~the corpus of speeches from leaders of EU institutions and member states employed by \citet{Rauh2019} and the SBIC corpus here). Note also that the task of retrieving posts that offend disabled people involves retrieving posts that are of a specific kind (namely: offending) and refer to a specific entity (disabled people). Such a retrieval task is common in sentiment analysis in which the aim is to extract documents that express an attitude toward a specific entity. The documents to be identified in such cases are required not only to refer to the entity of interest but also to be of a specific kind (namely: attitude expressing in contrast to being objective or fact-based).



\subsection{Approaches} \label{sec:approaches2} 

\subsubsection{Keyword Lists}
\label{seq:kl}
In order to compare the retrieval performance of keyword lists with the other discussed methods, keyword lists have to be generated for each of the three retrieval tasks. Due to what is known from research on the human construction of keyword lists, however, the keyword lists created by humans are likely to overlap very little and thus are likely to be unreliable \citep[p.~973-975]{King:2017io}. This poses a problem for the planned comparison because it would be best to have a challenging and reliable basis against which the other approaches can be compared to. To address this problem, the keyword lists are not constructed by humans but rather from the set of the most predictive keywords for the positive relevant class.

To identify predictive keywords, for each of the three studied corpora, the documents are preprocessed into a document-feature matrix.\footnote{The documents are preprocessed by tokenization into unigrams, lowercasing, removing terms that occur in less than 5 documents or less than 5 times throughout the corpus, and applying a boolean weighting on the entries of the document-feature matrix such that a 1 signals the occurrence of a term in a document and a 0 indicates the absence of the term in a document.} Then, logistic regression with regularization is applied. The regularization is introduced via the least absolute shrinkage and selection operator (LASSO; $L^1$ penalty) or ridge regression ($L^2$ penalty) depending on the outcome of hyperparameter tuning. The model is trained on the entire corpus and then the 50 most predictive terms (i.e.~the terms with the highest coefficients) are extracted. 
The extracted terms are listed in Tables \ref{tab:featimptwitter} to \ref{tab:featimpreuter} in Appendix \ref{app:mpt}. From each set of 50 most predictive terms 10 keywords are randomly sampled whereby the probability of drawing a term is proportional to the relative size of the term's coefficient. The 10 sampled keywords constitute one keyword list. The sampling of keywords from the set of predictive terms is repeated 100 times such that for each evaluated corpus there are 100 keyword lists of length 10 that serve as a basis for evaluation and comparison.\footnote{Note that the keyword lists comprising empirically highly predictive terms are not only applied on the corpora to evaluate the retrieval performance of keyword lists, but also form the basis for query expansion (see Section \ref{seq:queryexp}). The query expansion technique makes use of GloVe word embeddings \citep{Pennington2014} trained on the local corpora at hand and also makes use of externally obtained GloVe word embeddings trained on large global corpora. In the case of the locally trained word embeddings there is a learned word embedding for each predictive term. Thus, the set of extracted highly predictive terms can be directly used as starting terms for query expansion. In the case of the globally pretrained word embeddings, however, not all of the highly predictive terms have a corresponding global word embedding. Hence, for the globally pretrained embeddings the 50 most predictive terms \emph{for which a globally pretrained word embedding is available} are extracted. If a predictive term has no corresponding global embedding, the set of extracted predictive terms is enlarged with the next most predictive term until there are 50 extracted terms. Consequently, in Tables \ref{tab:featimptwitter} to \ref{tab:featimpreuter} in Appendix \ref{app:mpt} for each corpus two lists of the most predictive features are shown. Moreover, for the evaluation of the initial keyword lists of 10 predictive keywords, the local keyword lists have to be used because the global keyword lists have been adapted for the purposes of query expansion on the global word embedding space. \label{fnlabel}}

In contrast to human-constructed keyword lists for which it would be difficult to judge whether the lists perform on the higher or lower end of all lists humans would possibly generate for the posed retrieval tasks, the here constructed keyword lists mark the situation of a good start in which the selected keywords are highly indicative for the relevant class.

\subsubsection{Query Expansion}
\label{seq:queryexp}

The keyword lists serve as the starting point for query expansion. Each keyword list is expanded via the following procedure:
\begin{enumerate}
\item Take a set of trained word embeddings, here denoted by $\{\bm{z}_{1}, \dots, \bm{z}_{u}, \dots, \bm{z}_{U}\}$.\footnote{If necessary, the set of word embeddings is reduced to those embeddings whose terms occur in the corpus of interest and in the keyword list.}
\item For each keyword $s_v$ in the keyword list $\{s_1, \dots, s_V\}$:
\begin{enumerate}
\item Get the word embedding of the keyword: $\bm{z}_{[s_v]}$
\item Compute the cosine similarity between $\bm{z}_{[s_v]}$ and each word embedding $\bm{z}_{u}$ in the set $\{\bm{z}_{1}, \dots, \bm{z}_{u}, \dots, \bm{z}_{U}\}$:
\begin{equation}
sim_{cos}(s_v, z_u) = \frac{\bm{z}_{[s_v]} \cdot \bm{z}_{u}}{|| \bm{z}_{[s_v]} || \; || \bm{z}_{u} ||} 
\end{equation}
\item Take the $M$ terms that are not keyword $s_v$ itself and have the highest cosine similarity with keyword $s_v$. Add these $M$ terms to the keyword list.
\end{enumerate}
\end{enumerate}

This query expansion strategy makes use of word embedding representations and the cosine similarity as has been done in previous studies \citep[e.g.][]{Kuzi2016, Silva2020}. By not merging the keyword list into a single word vector representation but rather expanding the keyword list for each keyword separately, this expansion method allows to move into a different direction for each keyword. This might help in extracting a more varied range of linguistic denominations for the entity of interest and might be especially useful if the entity is abstract or combines several dimensions (as e.g.~is the case with the refugee topic that combines policies, occurrences, and a group of people). A similar procedure for query expansion has been studied by \citet{Kuzi2016}. 

For each evaluated retrieval task two different sets of word embeddings are used: embeddings that have been externally pretrained on large global corpora and embeddings trained locally on the corpus from which documents are to be retrieved. With regard to the globally pretrained embeddings for the English SBIC and the Reuters corpus, GloVe embeddings with 300 dimensions that have been trained on CommonCrawl data are made use of \citep{Pennington2014}.\footnote{The embeddings can be downloaded from https://nlp.stanford.edu/projects/glove/. GloVe embeddings here are used because they tend to be frequently employed in social science \citep[p.~104]{Rodriguez2022}.}
For the German Twitter data set 300-dimensional GloVe embeddings trained on the German Wikipedia are employed.\footnote{The embeddings can be downloaded from https://deepset.ai/german-word-embeddings.}

To get locally trained embeddings, on each corpus examined here, a GloVe embedding model is trained. GloVe embeddings with 300 dimensions are obtained for all unigram features that occur at least 5 times in the corpus. In training, a symmetric context window size of six tokens on either side of the target feature as well as a decreasing weighting function is used; such that a token that is $q$ tokens away from the target feature counts $1/q$ to the co-occurrence count \citep{Pennington2014}. After training, following the approach in \citet{Pennington2014}, the word embedding matrix and the context word embedding matrix are summed to yield the finally applied embedding matrix. Note that in their analysis of a large spectrum of settings for training word embeddings, \citet{Rodriguez2022} found that the here used popular setting of using 300-dimensional embeddings with a symmetric window size of six tokens tends to be a setting that yields good performances whilst at the same time being cost-effective regarding the embedding dimensions and the context window size.

The number of expansion terms $M$ is varied from 1 to 9 such that after the expansion the lists of originally 10 keywords then comprise between 20 and 100 keywords. The original as well as the expanded keyword lists are applied on the lowercased documents. Following the logic of a boolean query with the OR operator, a document is predicted to belong to the positive relevant class if it contains at least one of the keywords in the keyword list.

\subsubsection{Topic Model-Based Classification Rules}
When constructing topic model-based classification rules, there are three steps at which researchers have to make decisions that are likely to substantively affect the results. First, after having selected a specific type of topic model that is to be used, the number of to be estimated topics $K$ has to be set. Second, for the construction of a topic model-based classification rule, a researcher has to determine how many and which of the estimated topics are considered to be about the entity of interest (see Step 1 of the procedure described in Section \ref{sec:tmbcr}).
Finally, threshold value $\xi \in [0, 1]$ has to be set. If the sum of topic shares relating to relevant topics of a document is $\geq \xi$, the document is predicted to be relevant (see Step 2 of the procedure described in Section \ref{sec:tmbcr}). In each of these decision steps a researcher may be guided by expertise and/or an exploration of the results following from deciding for one or another option.

Whilst in practice a researcher has to finally settle for one of the options in each step such that a single classification rule is produced, here the aim rather is to comprehensively evaluate topic model-based classification rules and also to inspect how well topic model-based classification rules can perform if optimal decisions (w.r.t.~retrieval performance) are made. Consequently, specific values for the number of topics, the number of relevant topics and threshold values are set within reasonable ranges a priori. Then, the retrieval performance for all combinations of these values is evaluated. More precisely: On each corpus seven topic models---each with a different number of topics $K \in \{5, 15, 30, 50, 70, 90, 110\}$---are estimated. Then, for each estimated topic model with a specific topic number, initially only one topic is considered relevant, then two topics, and then three. For each number of topics considered to be relevant, all possible combinations regarding the question \emph{which} topics are considered relevant are evaluated. This implies that all ways of choosing one, two and three relevant topics (irrespective of the order in which they are selected) from the overall sets of 5, 15, 30, 50, 70, 90, and 110 topics have to be determined and evaluated. This amounts to 426,725 combinations---all of which are evaluated here.\footnote{For example, in a topic model with $K=15$ topics, there are 15 ways to select one relevant topic from 15 topics (namely: the first, the second, ..., and the 15th); and there are $\binom{15}{2} = 105$ ways of choosing two relevant topics from the set of 15 topics, and there are $\binom{15}{3} = 455$ ways to pick three topics from 15 topics.}

Finally, for each of the 426,725 combinations, four different threshold values $\xi$ are inspected: $0.1, 0.3, 0.5$, and $0.7$. Whereas $\xi = 0.7$ only considers those documents to be relevant that have 70$\%$ of the words they contain estimated to be generated by relevant topics, $\xi = 0.1$ is the most lenient solution in which all documents are classified to be relevant that have 10$\%$ of their words assigned to relevant topics. As $\xi$ increases, recall is likely to decrease and precision is likely to increase.

The type of topic model estimated here is a Correlated Topic Model (CTM) \citep{Blei2007a}. CTM extends the basic Latent Dirichlet Allocation (LDA) \citep{Blei:2003hl} by allowing topic proportions to be correlated. For more details on the CTM see \citet{Blei2007a}.\footnote{The CTM is estimated via the stm R-package \citep{stm2019} that originally is designed to estimate the Structural Topic Model (STM) \citep{Roberts2016}. The STM extends the LDA by allowing document-level variables to affect the topic proportions within a document (topical prevalence) or to affect the term probabilities of a topic (topical content) \citep[p.~989]{Roberts2016}. If no document-level variables are specified (as is done here), the STM reduces to the CTM \citep[p.~991]{Roberts2016}. In estimation, the approximate variational expectation-maximization (EM) algorithm as described in \citet[p.~992-993]{Roberts2016} is employed. This estimation procedure tends to be faster and tends to produce higher held-out log-likelihood values than the original variational approximation algorithm for the CTM presented in \citet{Blei2007a} \citep[p.~29-30]{stm2019}. The model is initialized via spectral initialization (\citeauthor{Arora2013}, \citeyear{Arora2013}; \citeauthor{Roberts2016wg}, \citeyear{Roberts2016wg}, p.~82-85; \citeauthor{stm2019}, \citeyear{stm2019}, p.~11). The model is considered to have converged if the relative change in the approximate lower bound on the marginal likelihood from one step to the next is smaller than 1e-04 (\citeauthor{Roberts2016}, \citeyear{Roberts2016}, p.~992; \citeauthor{stm2019}, \citeyear{stm2019}, p.~10, 28).}

\subsubsection{Active and Passive Supervised Learning}
\label{sec:activepassive}

Two types of supervised learning methods are employed. First, Support Vector Machines (SVMs) \citep{Boser1992, Cortes1995}, and second, BERT (standing for Bidirectional Encoder Representations from Transformers) \citep{Devlin2019}.

SVMs have been applied frequently and relatively successfully to text classification tasks in social science \citep{Diermeier2011, DOrazio2014, Baum2018, Pilny2019, Seboek2020, Erlich2021}; and also in active learning settings \citep{Miller2020}. An SVM operates on a document-feature matrix, $\bm{X}$. 
In a document-feature matrix, each document is represented as a feature vector of length $U$: $\bm x_i = (x_{i1}, \dots, x_{iu}, \dots, x_{iU})$. The information contained in the vector's entries, $x_{iu}$, typically is based on the frequency with which each of the $U$ textual features occurs in the $i$th document \citep[p.~147]{Turney2010}. Given the document feature vectors, $\{\bm{x}_i\}_{i=1}^N$, and corresponding binary class labels, $\{y_i\}_{i=1}^N$, whereby $y_i \in \{-1, +1\}$, an SVM tries to find a hyperplane, that separates the training documents as well as possible into the two classes \citep{Cortes1995}.\footnote{To create the required vector representation for each document, here the following text preprocessing steps are applied: The documents are tokenized into unigram tokens. Punctuation, symbols, numbers, and URLs are removed. The tokens are lowercased and stemmed. Subsequently terms whose mean tf-idf value across all documents in which they occur belongs to the lowest 0.1$\%$ (Twitter, SBIC) or 0.2$\%$ (Reuters) of mean tf-idf values of all terms in the corpus are discarded. Also terms that occur in only one (Twitter) or two (SBIC, Reuters) documents are removed. Finally, a boolean weighting scheme, in which only the absence (0) vs.~presence (1) of a term in a corpus is recorded, is applied on the document-feature matrix. To determine the hyperparameter values for the SVMs, hyperparameter tuning via a grid search across sets of hyperparameter values is conducted in a stratified 5-fold cross-validation setting on one fold of the training data. A linear kernel and a Radial Basis Function (RBF) kernel are tried. Moreover, for the inverse regularization parameter $\mathcal{C}$, that governs the trade-off between the slack variables and the training error, the values $\{0.1, 1.0, 10.0, 100.0\}$ (linear) and $\{0.1, 1.0, 10.0\}$ (RBF) are inspected. Additionally, for the RBF's parameter $\gamma$, that governs the training example's radius of influence, the values $\{0.001, 0.01, 0.1\}$ are evaluated.\footnote{On the precise definition of $\mathcal{C}$ and $\gamma$ see \citet{SklearnUserGuide2020a} and \citet{SklearnUserGuide2020b}.} The folds are stratified such that the share of instances falling into the relevant minority class is the same across all folds. In each cross-validation iteration, in the folds used for training, random oversampling of the minority class is conducted such that the number of relevant minority class examples increases by a factor of 5. Among the inspected hyperparameter settings, the setting that achieves the highest $F_1$-Score regarding the prediction of the relevant minority class and does not exhibit excessive overfitting is selected.} 

The document-feature matrix regards each document as a \emph{bag of words} \citep[p.~147]{Turney2010}. A bag-of-words representation only encodes information on the weighted frequency with which the terms occur in documents, but disregards word order, contextual information, and dependencies between the tokens in a document \citep[p.~147]{Turney2010}. Yet, a document is a sequence---not a bag---of tokens among which dependencies exist. Moreover, the meaning of a word often depends on the context of other words in which it is embedded in. (Take, for instance, the homonyms \emph{`bank‘} or \emph{`party'}.) In order to also use a supervised learning method that processes a document as a sequence of tokens and captures dependencies between tokens as well as context-dependent meanings of tokens, the Transformer-based language representation model BERT is additionally employed here.

BERT is a deep neural network based on the Transformer architecture \citep{Vaswani2017}. The central element of the Transformer architecture is the (self-)attention mechanism \citep{Bahdanau2015, Vaswani2017}. This mechanism allows the representation of each token to include information from the representations of other tokens (in the same sequence) \citep[p.~6001-6002]{Vaswani2017}; thereby enabling the model to produce token representations that encode contextual information and token dependencies.

BERT typically is applied in a sequential transfer learning setting \citep[p.~4175, 4179]{Devlin2019}. In sequential transfer learning, a model first is pretrained on a source task \citep[p.~64]{Ruder2019}. In pretraining, the aim is to learn model parameters such that the model can function as a well-generalizing input to a large range of different target tasks \citep[p.~64]{Ruder2019}. Then, in the following adaptation phase, the pretrained model (with its pretrained parameters) serves as the input for the training process on the target task \citep[p.~64]{Ruder2019}. The transferral of information (in the form of pretrained model parameters) to the learning process of a target task tends to reduce the number of training instances required to reach the same level of prediction performance than when not applying transfer learning and training the model from scratch \citep[p.~334]{Howard2018}. 

This potential of pretrained deep language representation models to reduce the number of required training instances is highly important for the application of deep neural networks in practice: In text classification tasks, deep neural networks tend to outperform conventional machine learning methods (such as SVMs) that often are applied on bag-of-words representations \citep{Socher2013, Ruder2020c}. But deep neural networks have a much higher number of parameters to learn than conventional models and thus require much more training instances. In situations in which the annotation of training instances is expensive or inefficient---such as in the context of retrieval with a strong imbalance between the relevant vs.~irrelevant class---applying a deep neural network from scratch may become prohibitively expensive. In a transfer learning setting, however, an already pretrained deep language representation model merely has to be fine-tuned to the target task at hand. If the pretrained model generalizes well, the number of training instances required to reach the same level of performance as a deep neural network that is not used in a transfer learning setting is reduced by several times \citep[p.~334]{Howard2018}. This allows deep neural networks to be applied to natural language processing tasks for which only relatively few training instances are available. Moreover, \citet{EinDor2020} show that especially in imbalanced classification settings active learning strategies can further improve the prediction performance of BERT such that even fewer training instances are needed for the same performance levels.\footnote{For an introduction to transfer learning with Transformer-based language representation models as BERT see \citet{Wankmueller2021}.}

There are two limiting factors when applying BERT: First, due to memory limitations, BERT cannot process text sequences that are longer than 512 tokens \citep[p.~4183]{Devlin2019}. This poses no problem for the Twitter corpus that has a maximum sequence length of 73 tokens. In the Reuters news corpus, however, whilst the largest share of articles is shorter than 512 tokens, there is a long tail of longer articles comprising up to around 1,500 tokens.\footnote{There is a single outlier article that is as long as 3,797 tokens.} Following the procedure by \citet{Sun2019}, Reuters news stories that exceed 512 tokens are reduced to the maximum accepted token length by keeping the first 128 and keeping the last 382 tokens whilst discarding the remaining tokens in the middle.\footnote{Note that to meet the input format required by BERT in single sequence text classification tasks, two additional special tokens, \emph{`[CLS]‘} and \emph{`[SEP]‘}, have to be added \citep[p.~4174]{Devlin2019}.} The maximum sequence length recorded for the SBIC is 354 tokens. In order to reduce the required memory capacities, the few posts that are longer than 250 tokens are shortened to 250 tokens by keeping the first 100 and the last 150 tokens.

The second limiting factor is that the prediction performance achieved by BERT after fine-tuning on the target task can vary considerably---even if the same training data set is used for fine-tuning and only the random seeds, that initialize the optimization process and set the order of the training data, differ (\citealp[p.~4176]{Devlin2019}; \citealp[p.~5-7]{Phang2019}; \citealp{Dodge2020}). Especially when the training data set is small (e.g.~smaller than 10,000 or 5,000 documents), fine-tuning with BERT has been observed to yield unstable prediction performances (\citealp[p.~4176]{Devlin2019}; \citealp[p.~5-7]{Phang2019}). Recently, \citet{Mosbach2021} established that the variance in the prediction performance of BERT models, that have been fine-tuned on the same training data set with different seeds, to a large extent are likely due to vanishing gradients in the fine-tuning optimization process. \citet[p.~5]{Mosbach2021} also note that it is not that small training data sets per se yield unstable performances but rather that if small data sets are fine-tuned for the same number of epochs than larger data sets (typically for 3 epochs), then this implies that smaller data sets are fine-tuned for a substantively smaller number of training iterations---which in turn negatively affects the learning rate schedule and the generalization ability \citep[p.~4-5]{Mosbach2021}. 
Finally, \citet[p.~2, 8-9]{Mosbach2021} show that fine-tuning with a small learning rate (in the paper: 2e-05), with warmup, bias correction, and a large number of epochs (in the paper: 20) not only tends to increase prediction performances but also significantly decreases the performance instability in fine-tuning. Here, the advice of \citet{Mosbach2021} is followed. For BERT, the AdamW algorithm \citep{Loshchilov2019} with bias correction, a warmup period lasting 10$\%$ of the training steps, and a global learning rate of 2e-05 is used. Training is conducted for 20 epochs. Dropout is set to 0.1. The batch size is set to 16.

\defcitealias{dbmdz2021}{dbmdz 2021}
For all applications, the pretrained BERT models are taken from HuggingFace's Transformers open source library \citep{Wolf2019}. The BERT model used as a pretrained input for the English applications based on SBIC and the Reuters corpus, has been pretrained on the English Wikipedia and the BooksCorpus \citep{Zhu2015} as in the original BERT paper \citep{Devlin2019}. For the data set of German tweets, a German BERT model pretrained on, amongst others, Wikipedia and CommonCrawl data by the digital library team at the Bavarian State Library is used \citep{dbmdz2021}. All BERT models are employed in the base (rather than the large) model version and operate on lowercased (rather than cased) tokens.

For both models, SVM and BERT, an active and a passive supervised learning procedure is implemented. The procedures consist of the following steps. (If the procedures differ between the active and the passive learning setting, it will be explicitly pointed out.):
\begin{itemize}
\item The data are randomly separated into 10 (SBIC, Twitter) or 5 (Reuters) equally sized folds.
\item Then, for each fold $g$ of the 10 (SBIC, Twitter) or 5 (Reuters) folds the data have been separated into the following steps are conducted:
\begin{enumerate}
\item Fold $g$ is set aside as a test set.
\item From the remaining folds, 250 instances are randomly sampled to form the initial set of labeled instances $\mathcal{I}$. The other instances constitute the pool of unlabeled instances $\mathcal{U}$.
\item The model is trained on the instances in set $\mathcal{I}$ and afterward makes predictions for all instances in pool $\mathcal{U}$ and the set aside test fold $g$. Recall, precision and the $F_1$-Score for the predictions made for pool $\mathcal{U}$ and test fold $g$ are separately recorded. During training in the passive learning setting, random oversampling of the instances falling into the positive relevant class is conducted such that the number of positive relevant instances increases by a factor of 5---thereby reflecting a cost matrix in which the cost of a false negative prediction is set to 5 and the cost of a false positive prediction is set to 1. In the active learning setting, no random oversampling is conducted.
\item A batch of 50 instances from pool $\mathcal{U}$ is added to the set of labeled instances in set $\mathcal{I}$. In passive learning, these 50 instances are randomly sampled from pool $\mathcal{U}$. In active learning, the following query strategies are applied: In the active learning setting with BERT, the 50 instances whose predicted probability to fall into the positive relevant class is closest to 0.5 are selected. When applying an SVM for active learning, the 50 instances with the smallest perpendicular distance to the hyperplane are retrieved and added to $\mathcal{I}$.
\item Steps 3 and 4 are repeated for 15 iterations, i.e.~until set $\mathcal{I}$ comprises 1,000 labeled instances.
\end{enumerate}
\end{itemize}
Hence, passive supervised learning with random oversampling and pool-based active learning with uncertainty sampling are applied. As the described learning procedures are repeated for 10 (SBIC, Twitter) or 5 (Reuters) times and are evaluated on each of the 10 (SBIC, Twitter) or 5 (Reuters) folds the data have been separated into, this allows taking the mean of the $F_1$-Scores across the 10 (SBIC, Twitter) or 5 (Reuters) test folds as an estimate of the expected generalization error of the applied models. 

\subsection{Results} \label{sec:results} 

The results are presented in Figures \ref{fig:queryres} to \ref{fig:allresreuters} and Tables \ref{tab:expansion} to \ref{tab:restopic3reuters}.

\subsubsection{Keyword Lists and Query Expansion}

\begin{sidewaysfigure}
   \centering
\begin{tabular}{ccc}
Twitter-local &
SBIC-local&
Reuters-local\\
\includegraphics[height=0.24\textwidth]{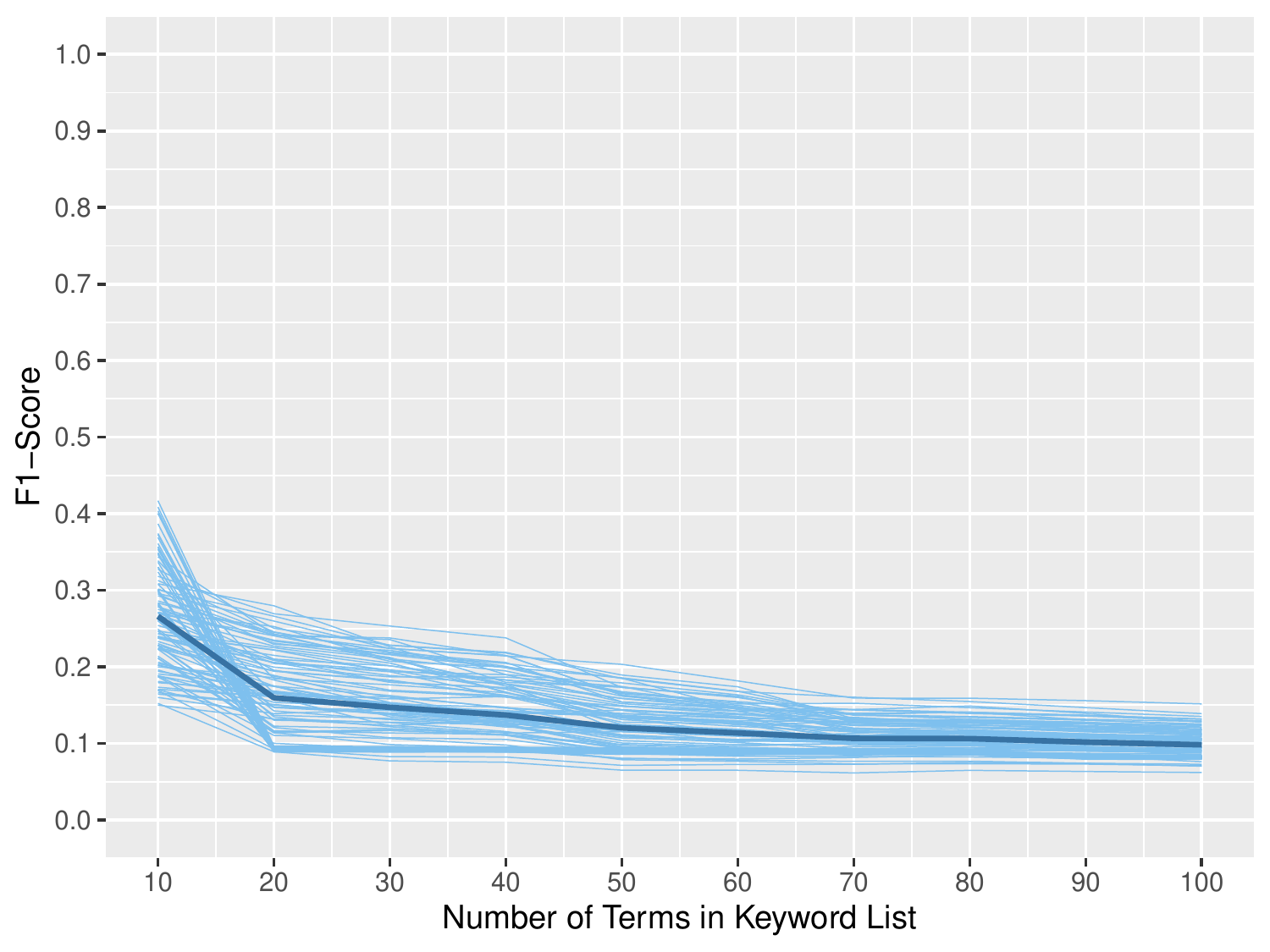}&
\includegraphics[height=0.24\textwidth]{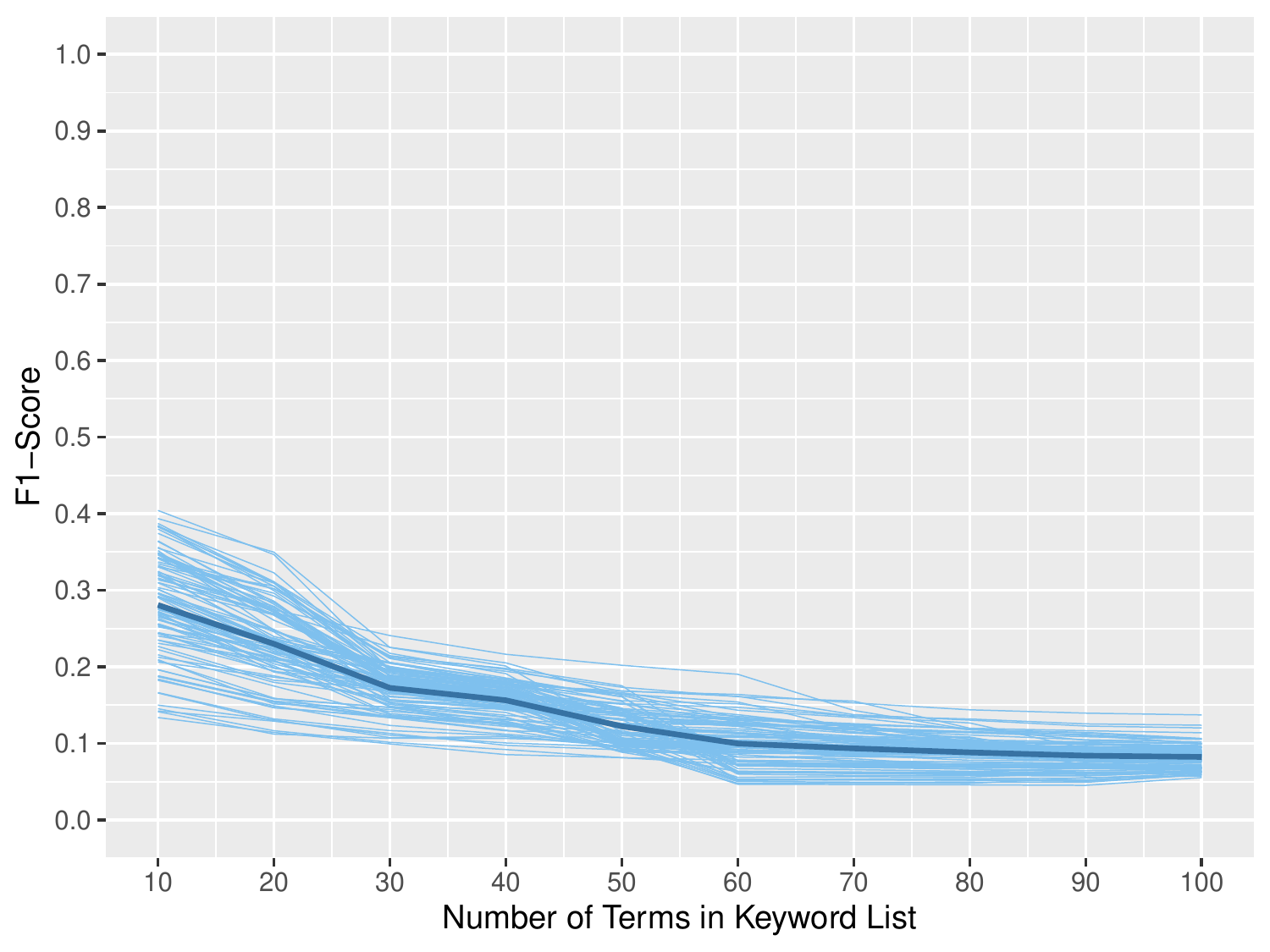}&
\includegraphics[height=0.24\textwidth]{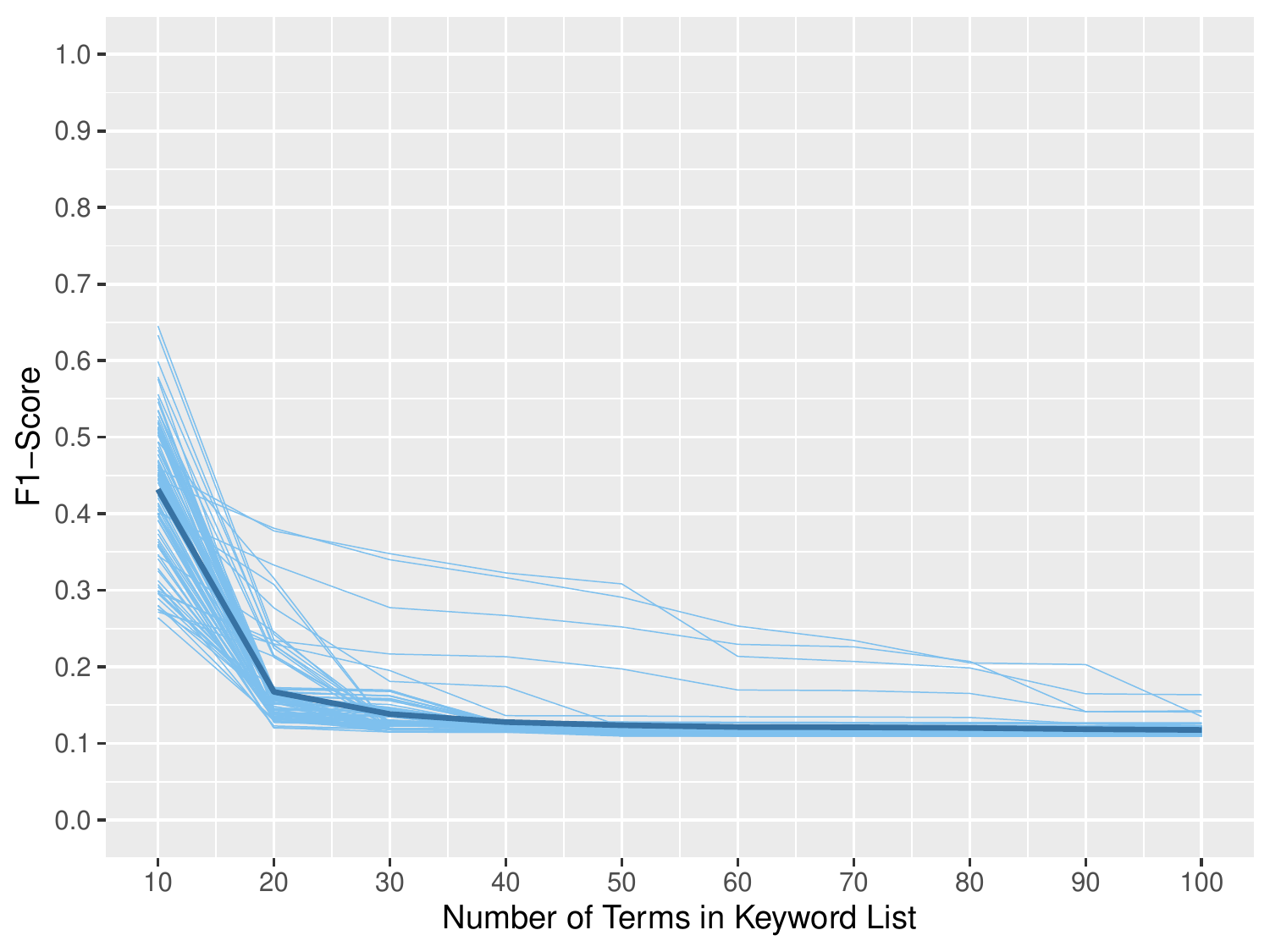}\\
& & \\
Twitter-global&
SBIC-global&
Reuters-global\\
\includegraphics[height=0.24\textwidth]{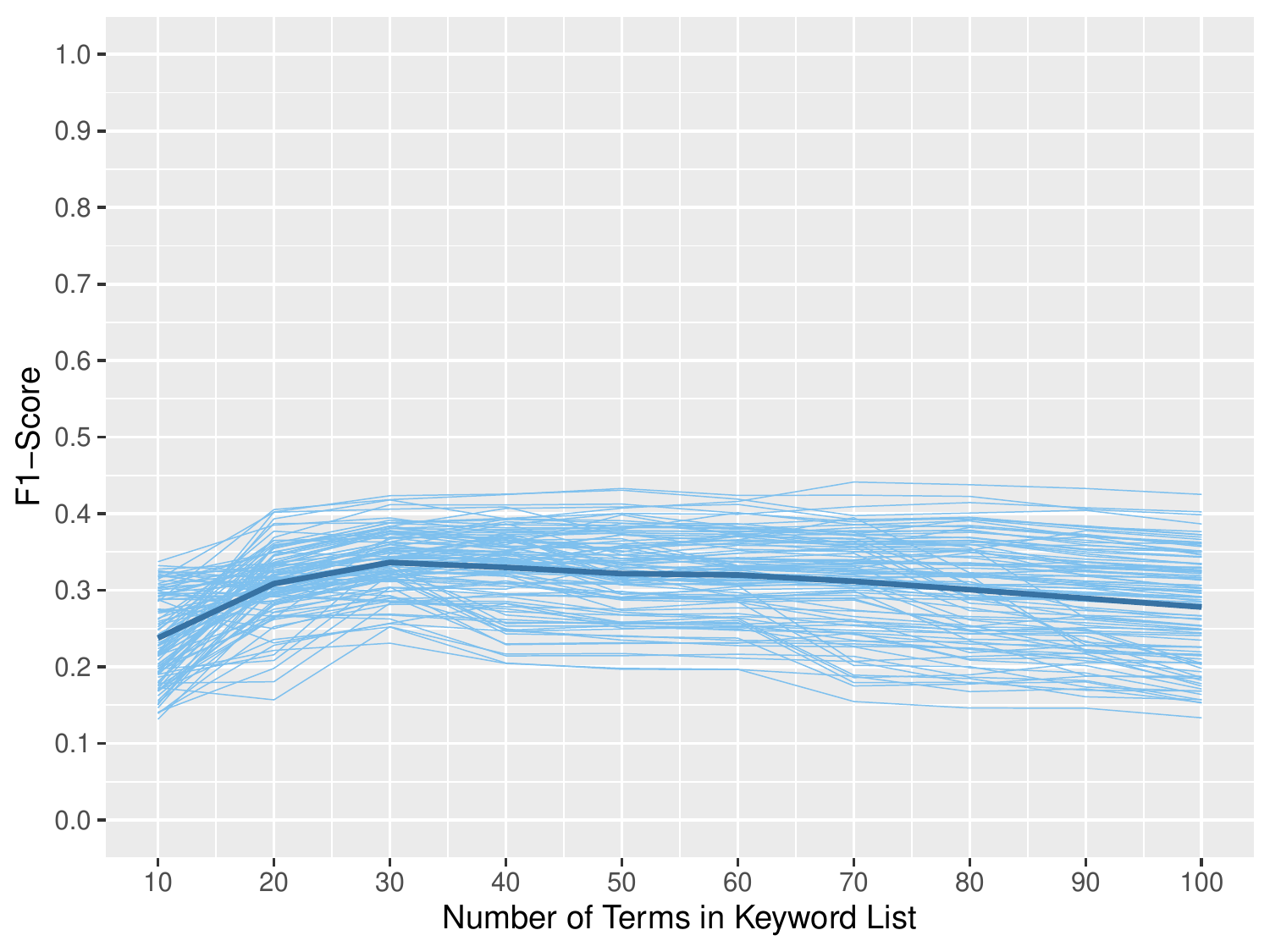}&
\includegraphics[height=0.24\textwidth]{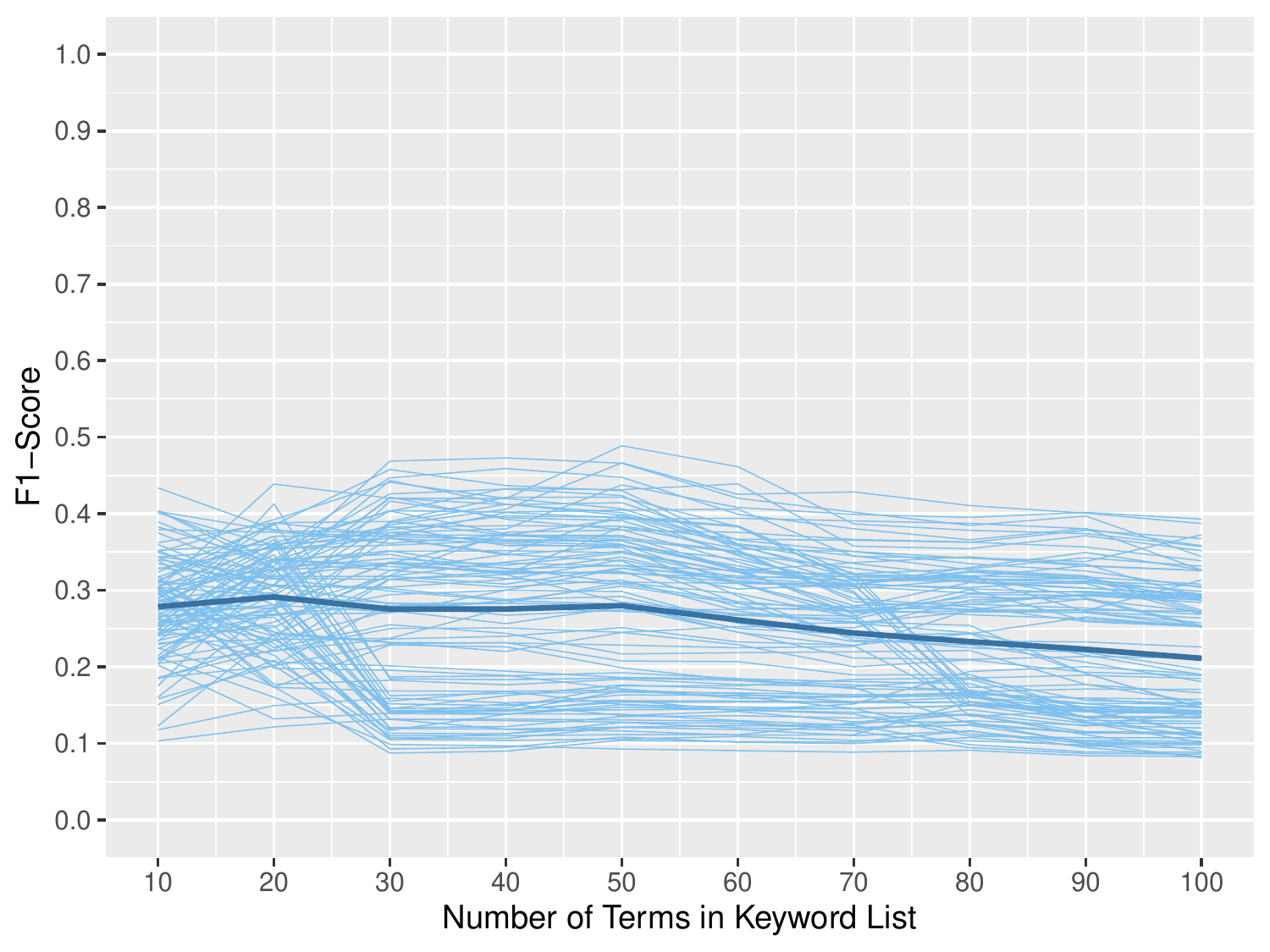}&
\includegraphics[height=0.24\textwidth]{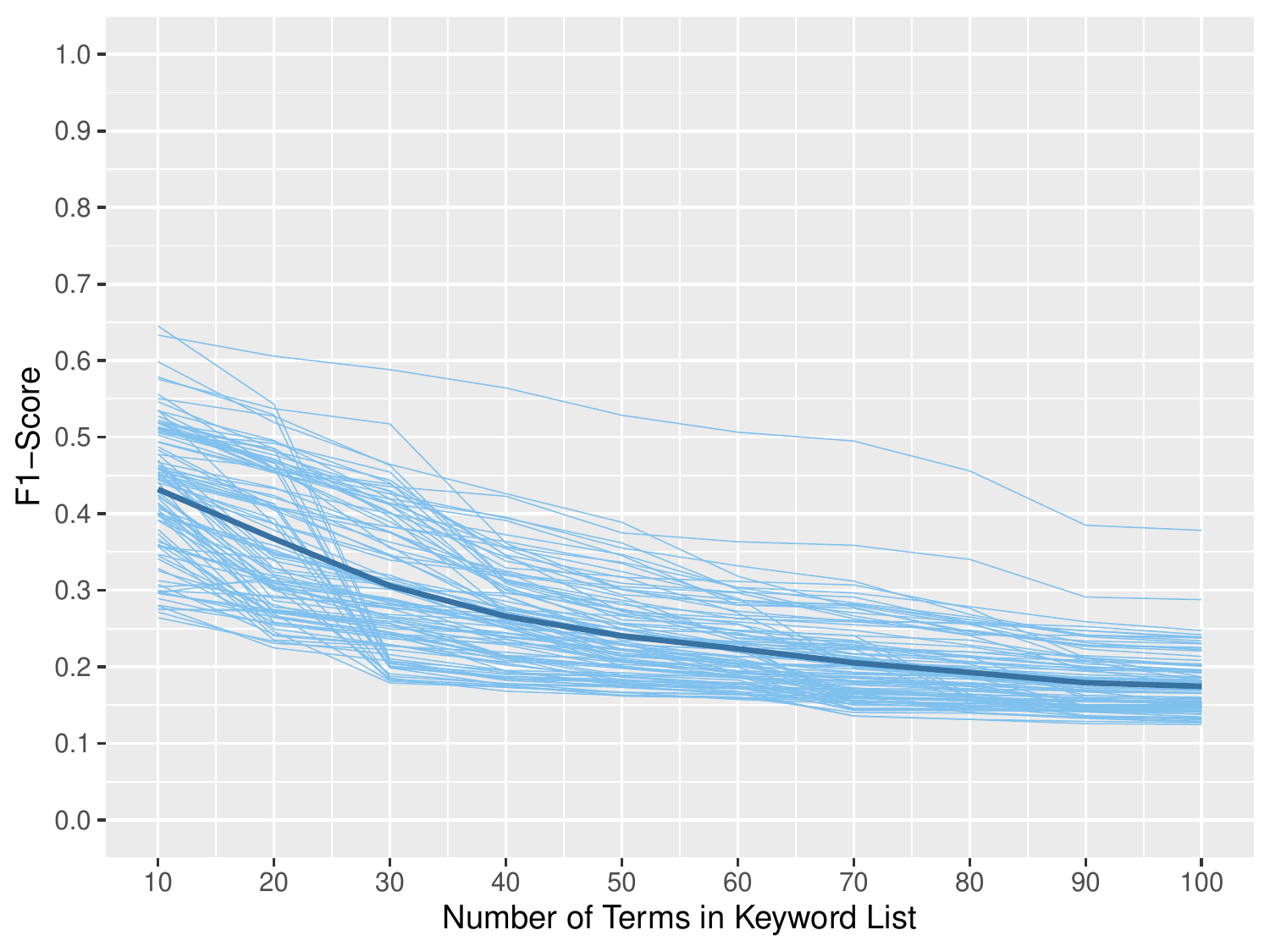}\\
\end{tabular}
\caption[$F_1$-Scores for Retrieving Relevant Documents with Keyword Lists and Query Expansion]{\textbf{$F_1$-Scores for Retrieving Relevant Documents with Keyword Lists and Query Expansion.} \footnotesize{This plot shows the $F_1$-Scores resulting from the application of the keyword lists of 10 highly predictive terms as well as the evolution of the $F_1$-Scores across the query expansion procedure based on locally trained GloVe embeddings (top row) and globally trained GloVe embeddings (bottom row). For each of the sampled 100 keyword lists that then are expanded, one light blue line is plotted. The thick dark blue line gives the mean over the 100 lists.}}
\label{fig:queryres} 
\end{sidewaysfigure}

Figure \ref{fig:queryres} visualizes for each of the three studied retrieval tasks (Twitter, SBIC, Reuters) the $F_1$-Scores resulting from the application of the 100 keyword lists of 10 highly predictive terms as well as the evolution of the $F_1$-Scores across the query expansion procedure based on locally trained GloVe embeddings (top row) and globally trained GloVe embeddings (bottom row). 

In general, the retrieval performances of the initial keyword lists of 10 predictive keywords are mediocre. Only the initial keyword lists for the Reuters corpus achieve what could be called acceptable performance levels. The maximum $F_1$-Scores reached by the initial lists of 10 predictive keywords are 0.417 (Twitter), 0.404 (SBIC), and 0.645 (Reuters).\footnote{Note that the local lists of 10 predictive keywords have to be used for evaluating the retrieval performance of keyword lists as the global keyword lists have been adapted for the purposes of query expansion on the global word embedding space (see Footnote \ref{fnlabel}).}
Moreover, also with the here used empirically driven procedure for the construction of keyword lists, the variation in the initial keyword lists' retrieval performances is considerable. The difference between the maximum and the minimum $F_1$-Scores is 0.267 (Twitter), 0.270 (SBIC), and 0.381 (Reuters).

Interestingly, the applied query expansion technique tends to decrease rather than increase the $F_1$-Score and only shows some improvement of the $F_1$-Score for the Twitter and SBIC data sets---and only if operating on the basis of word embeddings that are trained on large global external corpora rather than the local corpus at hand.

There are several factors that are likely to play a role here. First, when retrieving those terms that have the highest cosine similarity with an initial starting term, the terms retrieved from the global embedding space seem semantically or syntactically related to the initial term, whereas this is not the case for the local word embeddings (as an example see Table \ref{tab:expansion}). One reason why the local embedding space does not yield word embeddings that position related terms closely together could be that the three corpora used here are relatively small. 
The information provided by the context window-based co-occurrence counts of terms thus could be too little for the embeddings to be effectively trained. 

Second, in the global embedding space, terms with high cosine similarities seem to be closely related to the initial query term (see again Table \ref{tab:expansion}). Adding these related terms, nevertheless decreases the retrieval performance (as measured by the $F_1$-Score) for the Reuters corpus. In the case of the Twitter and SBIC corpora, adding these terms with the highest cosine similarities for some iterations at some points (especially at the beginning) slightly increase the $F_1$-Score, whereas at other points there are decreasing or no visible effects. Here, a second factor comes into play: As is to be expected, query expansion increases recall and decreases precision (see Figures \ref{fig:queryres2local} and \ref{fig:queryres2global} in Appendix \ref{app:resquery2}). Hence, in general query expansion is only worthwhile if---and as long as---the increase in recall outweighs the decrease in precision. 
Applying the initial set of 10 highly predictive keywords on the Twitter and SBIC data sets, yields a retrieval result that is characterized by low recall and high precision, whereas applying the initial set of 10 highly predictive keywords on the Reuters corpus, leads to very high (sometimes even perfect) recall and low precision (see Figures \ref{fig:queryres2local} and \ref{fig:queryres2global} in Appendix \ref{app:resquery2}). Whereas in the second situation, there is no room for query expansion to further improve the retrieval performance via increasing recall (and thus the $F_1$-Score for the Reuters corpus is moving downward), in the low-recall-high-precision situation of the Twitter and SBIC data sets there is at least the potential for query expansion to increase recall without causing a too strong decrease in precision. This potential is realized in some iterations at some expansion sets, but the decrease in precision more often than not tends to outweigh the increase in recall.

A further reason why query expansion does not perform very well also for global embeddings is the meaning conflation deficiency \citep[p.~60]{Pilehvar2021}: Because word embedding models such as GloVe represent one term by a single embedding vector, a polyseme or homonym is likely to have the various meanings that it refers to encoded within its single representation vector \citep[p.~1059]{Neelakantan2014}. The meanings get conflated into one representation \citep[p.~102]{Schuetze1998}. This is what also happens here. Moreover, it also seems that the conflation of meanings for the GloVe embeddings that have been pretrained on large, global corpora proceeds unequally: the global embedding space tends to position polysemous or homonymous terms close to terms that are semantically or syntactically related to the most common and general meaning of the polysemous or homonymous term (see for example the term \emph{`vegetables'} in Table \ref{tab:expansion}). Query expansion in the global embedding space thus fails if an initial query term is a polyseme or homonym and its intended meaning is highly context-specific.

 \begin{table}[h!]
 \begin{center}
\scriptsize
 \begin{tabular}{p{0.9cm}p{1.1cm}p{11.4cm}}
 \hline
 \textbf{Embed- dings} & \textbf{Initial term} & \textbf{Terms with the highest cosine similarity to the initial term} \\
   \hline
   local & retard  &  anymore, blend, float, sex, arguments, college, 93, meanjokes, fever   \\
   local &   vegetables &  100,000,     name,        U+1f407,  knew,        combination, traveled,    pulled,      strip,       developed   \\
   local &  epileptic  & oj,          blond,       include,     tactics,     crown,       tampons,     demands,     prostitutes, newspapers     \\
   global & retard  & retards,  retarded, dumbass, moron, idiot,  faggot,  fuckin,   stfu,  stupid   \\      
   global &  vegetables & veggies,    fruits,     vegetable,  potatoes,   carrots,    tomatoes,   meats,      onions,     cooked  \\
   global &  epileptic  &   seizures,      seizure,       psychotic,     schizophrenic, epileptics,    fainting,      migraine,      spasms,        disorder   \\
    \hline
 \end{tabular}
  \end{center}
 \caption[Example SBIC Expansion Terms]{\textbf{Example SBIC Expansion Terms.} \footnotesize{This table gives for each of the highly predictive terms \emph{`retard'}, \emph{`vegetables'}, and \emph{`epileptic'} the nine terms with the highest cosine similarity in the local and global embedding spaces.}}
\label{tab:expansion}
 \end{table}

\subsubsection{Topic Model-Based Classification Rules}

\begin{sidewaysfigure}
   \centering
\begin{tabular}{ccc}
Twitter &
\hspace{-1.8cm} SBIC&
\hspace{-1.8cm} Reuters\\
\includegraphics[height=0.38\textwidth]{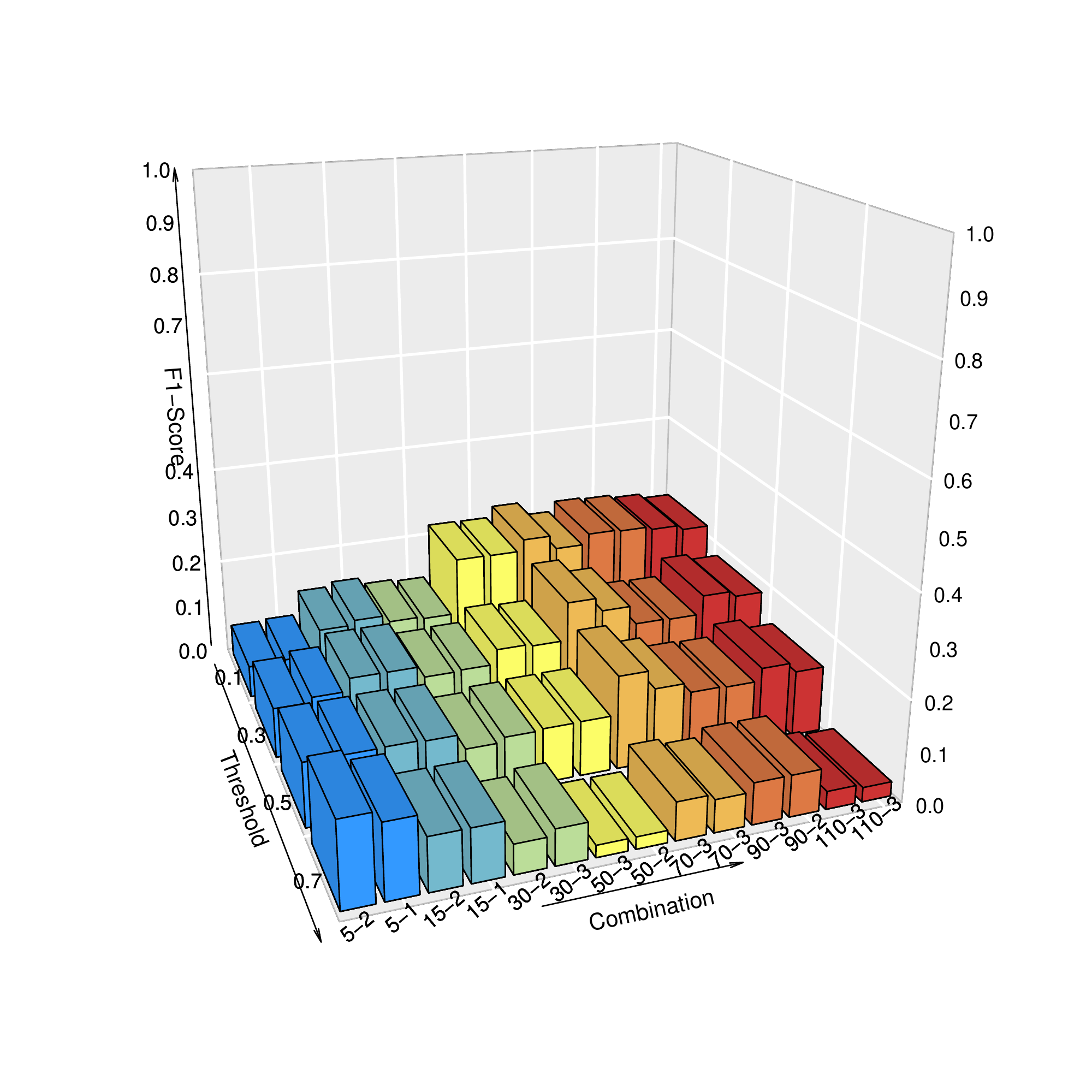} &
\hspace{-1.8cm} \includegraphics[height=0.38\textwidth]{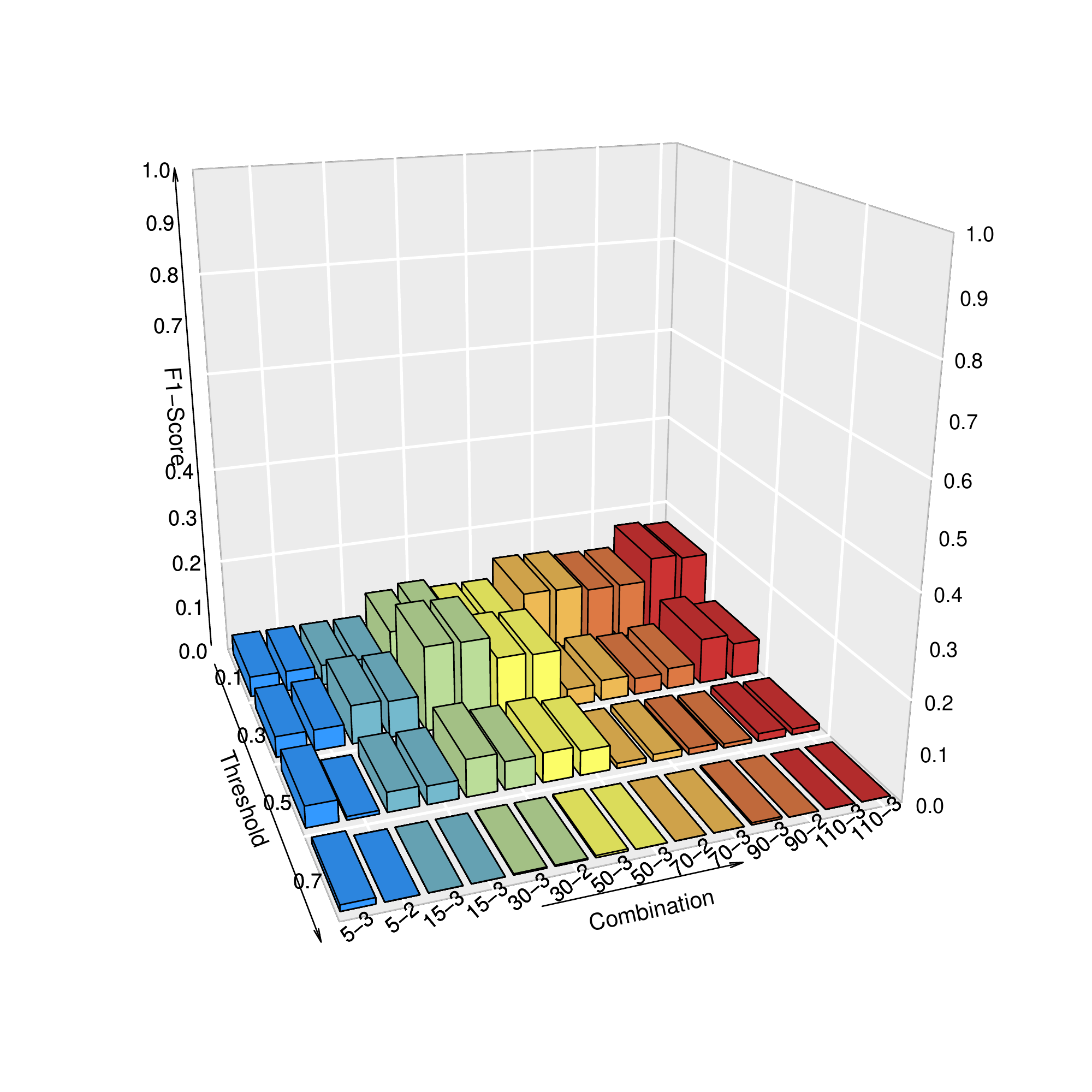}&
\hspace{-1.8cm} \includegraphics[height=0.38\textwidth]{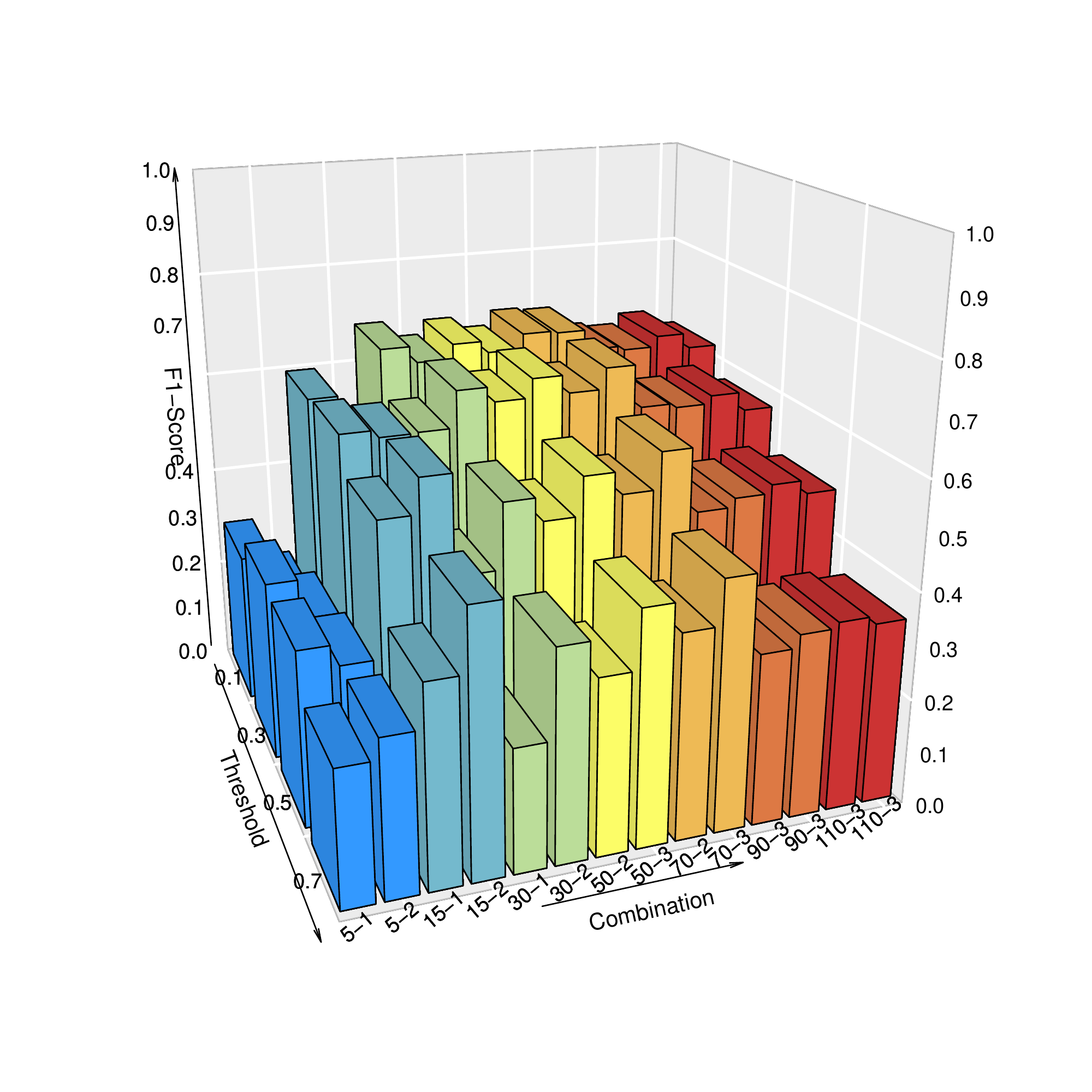}\\
\end{tabular}
\caption[$F_1$-Scores for Retrieving Relevant Documents with Topic Model-Based Classification Rules]{\textbf{$F_1$-Scores for Retrieving Relevant Documents with Topic Model-Based Classification Rules.} \footnotesize{The height of a bar indicates the $F_1$-Score resulting from the application of a topic model-based classification rule. For each number of topics $K \in \{5, 15, 30, 50, 70, 90, 110\}$, those two combinations out of all explored combinations regarding the question how many and which topics are considered relevant are shown that reach the highest $F_1$-Score for the given topic number. The labels of the combinations are such that the first number indicates the number of topics $K$ and the second number denotes the number of topics considered to be relevant by the combination. For example: In the Twitter data set the two best performing combinations for a topic model with $K = 70$ topics both regard three topics as being related to the Twitter topic. So the label for both is 70-3. (The two combinations, of course, differ regarding \emph{which} three topics they assume to be relevant.) For each combination, the $F_1$-Score for each evaluated threshold value $\xi \in \{0.1, 0.3, 0.5, 0.7 \}$ is given. Classification rules that assign none of the documents to the positive relevant class have a recall value of 0 and an undefined value for precision and the $F_1$-Score. Undefined values here are visualized by the value 0.}}
\label{fig:topicres} 
\end{sidewaysfigure}

Figure \ref{fig:topicres} presents the $F_1$-Scores reached by topic model-based classification rules. 
The most notable aspect is that the retrieval performance of topic model-based classification rules is low for the Twitter and SBIC corpora and relatively high for the Reuters corpus. The highest $F_1$-Score reached in the Twitter retrieval task is 0.253 
and regarding the SBIC is 0.175, 
whereas on the Reuters corpus a score of 0.685 is achieved. 

To better understand this result, the terms with the highest occurrence probabilities and the terms with the highest FREX-Score are inspected. The FREX metric is the weighted harmonic mean of a term's occurrence probability $ \beta_{ku}$ and a term's exclusivity (which is given by $\beta_{ku}/\sum_{j=1}^K \beta_{ju}$) \citep[p.~993]{Roberts2016}:
\begin{equation}
FREX_{ku} = \Bigg ( \frac{\omega}{ECDF(\beta_{ku}/\sum_{j=1}^K \beta_{ju})} + \frac{1-\omega}{ECDF(\beta_{ku})} ) \Bigg )^{-1}
\end{equation}
whereby $ECDF$ stands for empirical cumulative distribution function and $\omega$ is the weight balancing the two measures. Here $\omega$ is set to $0.5$.

This inspection (see Tables \ref{tab:restopic3twitter}, \ref{tab:restopic3sbic}, and \ref{tab:restopic3reuters}) reveals that whether and in how far there are exclusive and coherent topics that relate to the entity of interest likely determines whether a topic model-based classification rule can effectively retrieve relevant documents or not.

\newpage

\begin{table}[H]
\begin{center}
\scriptsize
{\renewcommand{\arraystretch}{0.99}
\begin{tabular}{p{0.8cm}p{2.3cm}p{2.2cm}p{2.3cm}p{2.2cm}p{2.4cm}}
  \hline
 & Topic 1 & Topic 2 & Topic 3 & Topic 4 & Topic 5 \\ 
  \hline
 {\textbf{Prob.}} & @therealliont & alt & zeit & klein & ne \\ 
   & facebook & stund & seid & zwei & woch \\ 
   & lass & los & scheiss & \#pegida & nach \\ 
   {\textbf{FREX}}  & gepostet & fertig & zeit & \#pegida & wert \\ 
   & facebook & cool & \#emabiggestfans1d & \#nopegida & passt \\ 
   & monday-giveaway & wahrschein & gonn & setz & woch \\ 
\end{tabular}

\begin{tabular}{p{0.8cm}p{2.3cm}p{2.2cm}p{2.3cm}p{2.2cm}p{2.4cm}}
 & \\
  \hline
 & Topic 6 & Topic 7 & Topic 8 & Topic 9 & Topic 10 \\ 
  \hline
 {\textbf{Prob.}} & toll & fall & best & the & eigent \\ 
   & richtig & nochmal & nacht & of & sonntag \\ 
   & dabei & h & onlin & hatt & eig \\ 
     {\textbf{FREX}}  & toll & nochmal & schulmobel & team & eigent \\ 
   & wenigst & fall & videos & artikel & lag \\ 
   & manchmal & \#breslau & onlin & \#techjob & sonntag \\ 
\end{tabular}

\begin{tabular}{p{0.8cm}p{2.3cm}p{2.2cm}p{2.3cm}p{2.2cm}p{2.4cm}}
 & \\
  \hline
 & Topic 11 & Topic 12 & Topic 13 & Topic 14 & Topic 15 \\ 
  \hline
 {\textbf{Prob.}} & wurd & xd & sich & gern & @youtube-playlist \\ 
   & viel & mensch & suss & sammelt & hinzugefugt \\ 
   & lohnt & frag & vielleicht & kind & @youtub \\ 
     {\textbf{FREX}}  & bordelldatenbank.eu & wunderschon & reinhard\_4711 & sammelt & \#younow \\ 
   & erforscht & frag & mallorcamagazin & zeig & nightcor \\ 
   & publiziert & geil & sich & glucklich & vs \\ 
\end{tabular}
%
%

\begin{tabular}{p{0.8cm}p{2.3cm}p{2.2cm}p{2.3cm}p{2.2cm}p{2.4cm}}
 & \\
  \hline
 & Topic 16 & Topic 17 & Topic 18 & Topic 19 & Topic 20 \\ 
  \hline
 {\textbf{Prob.}}  & \#iphonegam & \#iphon & weiss & lang & oh \\ 
   & fahrt & steh & welt & veranstalt & schnell \\ 
   & hotel & gruss & end & event & nix \\ 
      {\textbf{FREX}}  & antalya & \#iphon & kompakt & lkr & mag \\ 
   & erendiz & \#blondin & deintraum & veranstalt & aufgeregt \\ 
   & sightseeing & \#blondinenwitz & haus & event & sing \\ 
\end{tabular}

\begin{tabular}{p{0.8cm}p{2.3cm}p{2.2cm}p{2.3cm}p{2.2cm}p{2.4cm}}
 & \\
  \hline
 & Topic 21 & Topic 22 & Topic 23 & Topic 24 & Topic 25 \\ 
  \hline
 {\textbf{Prob.}}  & endlich & bitt & folg & beim & war \\ 
   & gross & abgeschloss & warum & spass & grad \\ 
   & brauch & frau & \#votesami & seit & sowas \\ 
      {\textbf{FREX}} & \#immortalis & mission & erfullt & abonni & geschlecht \\ 
   & immortalis & bitt & belohn & total & mutt \\ 
   & pvp-gefecht & aufgab & international & herzlich & star \\ 
\end{tabular}

\begin{tabular}{p{0.8cm}p{2.3cm}p{2.2cm}p{2.3cm}p{2.2cm}p{2.4cm}}
 & \\
  \hline
 & Topic 26 & Topic 27 & Topic 28 & Topic 29 & Topic 30 \\ 
  \hline
 {\textbf{Prob.}}  & \#kca & komm & find & retweet & halt \\ 
   & twitt & steht & voll & leut & zuruck \\ 
   & \#votedagi & klar & mach & bett & uhr \\ 
      {\textbf{FREX}}  & gezuchtet & \#hamburg & wahr & @ischtaris & anschau \\ 
   & ratselhaft & geplant & find & bett & zuruck \\ 
   & \#votedagi & hamburg & zeigt & \#kca & uberhaupt \\ 
   \hline
\end{tabular}}
\end{center}
\caption[Twitter: Terms with the Highest Probability and the Highest FREX-Score]{\textbf{Twitter: Terms with the Highest Probability and the Highest FREX-Score.} \footnotesize{For each topic in the CTM with 30 topics estimated on the Twitter corpus, this Table presents the 3 terms with the highest probability (Prob.) and the 3 terms with the highest FREX-Score (FREX). See Topic 4 for the here only moderately coherent Pegida topic. Note that German umlauts here are removed as the preprocessing procedure for the CTM involved stemming---which here also implied removing umlauts.}}
\label{tab:restopic3twitter} 
\end{table}

The entity of interest in the Twitter data set is multi-dimensional. It includes refugees as a social group, refugee policies as well as actions and occurrences revolving around the refugee crisis. When examining the most likely and exclusive terms for topic models estimated on the Twitter corpus, it becomes clear that not each aspect of this multi-dimensional refugee topic is captured in a coherent and exclusive topic (see for example Table \ref{tab:restopic3twitter}). In each model for $K \geq 30$ there is one relatively coherent topic on Pegida, an anti-Islam and anti-immigration movement that held many demonstrations in the context of the refugee crisis. Beside that there are further more or less integrated topics that touch refugees and refugee policies without, however, being exclusively about these entities. Thus, the topic models do not offer a set of topics that, taken together, cover all dimensions of the refugee topic in an exclusive manner. 

Regarding the SBIC, the situation is even more disadvantageous for the application of topic model-based classification rules. Across all CTMs estimated on the SBIC there is no topic that identifiably relates to disabled people in a disrespectful way (as an example see Table \ref{tab:restopic3sbic}). The CTMs with higher topic numbers include some topics that very slightly touch disabilities, but these topics are not coherent. Applying topic model-based classification rules in this situation is futile. Among all evaluated 426,725 $\times$ 4 = 1,706,900 settings, an $F_1$-Score of 0.175 is as good as it maximally gets.

\begin{table}[H]
\scriptsize
\begin{center}
{\renewcommand{\arraystretch}{0.99}
 \begin{tabular}{p{0.8cm}p{1.2cm}p{1.2cm}p{2cm}p{1.2cm}p{1.2cm}p{1.2cm}p{1.2cm}p{1.2cm}}
  \hline
 & Topic 1 & Topic 2 & Topic 3 & Topic 4 & Topic 5 & Topic 6 & Topic 7 & Topic 8 \\ 
  \hline
 {\textbf{Prob.}}  & U+1F602 & go & now & more & sinc & as & time & bitch \\ 
   & rt & we & right & take & move & year & back & got \\ 
   & bad & out & left & doe & fire & had & actual & hoe \\ 
       {\textbf{FREX}}  & U+1F602 & go & \#releasethememo & mani & season & ago & comput & U+1F612 \\ 
   &  U+1F62D & let & now & take & U+1F643 & best & hitler & these \\ 
   & bad & tonight & hillari & wors & move & almost & finish & retard \\ 
\end{tabular}

 \begin{tabular}{p{0.8cm}p{1.2cm}p{1.2cm}p{2cm}p{1.2cm}p{1.2cm}p{1.2cm}p{1.2cm}p{1.2cm}}
    & \\
  \hline
 & Topic 9 & Topic 10 & Topic 11 & Topic 12 & Topic 13 & Topic 14 & Topic 15 & Topic 16 \\ 
  \hline
 {\textbf{Prob.}}  & her & day & by & there & don't & dark & man & guy \\ 
   & she & today & children & eat & know & movi & into & gay \\ 
   & make & play & hit & lot & women & an & ask & posit \\ 
       {\textbf{FREX}}  & girlfriend & april & = & hide & sexist & chees & bar & gay \\ 
   & her & fool & bottl & eat & women & prostitut & walk & fag \\ 
   & cri & humid & mosquito & space & don't & humor & into & straight \\ 
\end{tabular}

 \begin{tabular}{p{0.8cm}p{1.2cm}p{1.2cm}p{2cm}p{1.2cm}p{1.2cm}p{1.2cm}p{1.2cm}p{1.2cm}}
    & \\
  \hline
 & Topic 17 & Topic 18 & Topic 19 & Topic 20 & Topic 21 & Topic 22 & Topic 23 & Topic 24 \\ 
  \hline
 {\textbf{Prob.}}  & well & love & hate & your & our & black & muslim & he \\ 
   & made & shit & who & will & white & call & ` & was \\ 
   & friend & i'm & r & their & us & between & red & his \\ 
       {\textbf{FREX}}  & oh & dirti & reason & your & immigr & pizza & ` & kid \\ 
   & god & love & crime & peac & russia & black & rose & dad \\ 
   & \verb|^| & yo & asshol & educ & nation & common & ice & father \\ 
\end{tabular}

 \begin{tabular}{p{0.8cm}p{1.2cm}p{1.2cm}p{2cm}p{1.2cm}p{1.2cm}p{1.2cm}p{1.2cm}p{1.2cm}}
    & \\
  \hline
 & Topic 25 & Topic 26 & Topic 27 & Topic 28 & Topic 29 & Topic 30 & & \\ 
  \hline
 {\textbf{Prob.}}  & look & tri & off & see & $>$ & would  & &\\ 
   & good & start & im & here & $<$ & one \\ 
   & incel & stori & done & post & s & can \\ 
       {\textbf{FREX}}  & hair & case & piss & post & $>$ & would \\ 
   & look & ethiopia & youtub & see & $<$ & never \\ 
   & normal & touch & im & pictur & number & one  & &\\ 
   \hline
\end{tabular}}
\end{center}
\caption[SBIC: Terms with the Highest Probability and the Highest FREX-Score]{\textbf{SBIC: Terms with the Highest Probability and the Highest FREX-Score.} \footnotesize{For each topic in the CTM with 30 topics estimated on the SBIC, this Table presents the 3 terms with the highest probability (Prob.) and the 3 terms with the highest FREX-Score (FREX). See Topic 8 for a non-coherent and non-exclusive topic that slightly touches disrespectful posts about disabled people.}}
\label{tab:restopic3sbic} 
\end{table}

\begin{table}[H]
 \scriptsize
 \begin{center}
 {\renewcommand{\arraystretch}{0.99}
 \begin{tabular}{p{0.8cm}p{1.3cm}p{1.4cm}p{1.3cm}p{1.4cm}p{1.3cm}p{1.3cm}p{1.3cm}p{1.3cm}}
  \hline
 & Topic 1 & Topic 2 & Topic 3 & Topic 4 & Topic 5 & Topic 6 & Topic 7 & Topic 8 \\ 
  \hline
 {\textbf{Prob.}} & cts & profit & pct & share & export & bank & ec & unit \\ 
   & avg & oper & growth & pct & nil & rate & european & subsidiari \\ 
   & shrs & gain & rise & stock & coffe & pct & communiti & agreement \\ 
 {\textbf{FREX}}  & shrs & profit & gnp & smc & seamen & 9-13 & ec & mhi \\ 
   & avg & pretax & economi & ucpb & prev & bank & communiti & cetus \\ 
   & cts & extraordinari & growth & calmat & ibc & 9-7 & ecus & squibb \\ 
\end{tabular}

 \begin{tabular}{p{0.8cm}p{1.3cm}p{1.4cm}p{1.3cm}p{1.4cm}p{1.3cm}p{1.3cm}p{1.3cm}p{1.3cm}}
   & \\
  \hline
 & Topic 9 & Topic 10 & Topic 11 & Topic 12 & Topic 13 & Topic 14 & Topic 15 & Topic 16 \\ 
  \hline
 {\textbf{Prob.}}  & pct & vs & trade & franc & gulf & oil & offer & dollar \\ 
   & billion & loss & u. & french & ship & price & share & rate \\ 
   & januari & rev & japan & group & u. & gas & sharehold & currenc \\ 
   {\textbf{FREX}}  & unadjust & rev & lyng & ferruzzi & missil & opec & caesar & louvr \\ 
   & januari & vs & chip & cgct & warship & herrington & sosnoff & miyazawa \\ 
   & fell & mths & miti & cpc & tehran & bpd & cyacq & poehl \\ 
\end{tabular}

 \begin{tabular}{p{0.8cm}p{1.3cm}p{1.4cm}p{1.3cm}p{1.4cm}p{1.3cm}p{1.3cm}p{1.3cm}p{1.3cm}}
  & \\
  \hline
 & Topic 17 & Topic 18 & Topic 19 & Topic 20 & Topic 21 & Topic 22 & Topic 23 & Topic 24 \\ 
  \hline
 {\textbf{Prob.}}  & share & billion & tonn & ltd & price & food & analyst & canadian \\ 
   & stock & trade & wheat & plc & contract & beef & earn & canada \\ 
   & dividend & reserv & export & pct & cent & philippin & market & credit \\ 
   {\textbf{FREX}}  & dividend & fed & beet & csr & octan & satur & analyst & card \\ 
   & payabl & surplus & cane & transcanada & kwacha & diseas & ibm & canadian \\ 
   & payout & taiwan & rapese & monier & sulphur & nppc & rumor & nova \\ 
\end{tabular}

 \begin{tabular}{p{0.8cm}p{1.3cm}p{1.4cm}p{1.3cm}p{1.4cm}p{1.3cm}p{1.3cm}p{1.3cm}p{1.3cm}}
  & \\
  \hline
 & Topic 25 & Topic 26 & Topic 27 & Topic 28 & Topic 29 & Topic 30 & & \\ 
  \hline
 {\textbf{Prob.}} & quarter & price & chemic & court & gold & debt  & & \\ 
   & first & produc & u. & file & mine & payment \\ 
   & earn & stock & busi & general & ton & loan \\ 
   {\textbf{FREX}}  & fourth & cocoa & gaf & gencorp & assay & debt \\ 
   & quarter & buffer & hanson & afg & uranium & payment \\ 
   & earn & icco & borg-warn & court & gold & repay  & &\\ 
   \hline
\end{tabular}}
\end{center}
\caption[Reuters: Terms with the Highest Probability and the Highest FREX-Score]{\textbf{Reuters: Terms with the Highest Probability and the Highest FREX-Score.} \footnotesize{For each topic in the CTM with 30 topics estimated on the Reuters corpus, this Table presents the 3 terms with the highest probability (Prob.) and the 3 terms with the highest FREX-Score (FREX). See Topic 14 for the crude oil topic and see Topic 13 for the military topic that at times touches crude oil.}}
\label{tab:restopic3reuters} 
\end{table}

The situation is entirely different for the crude oil topic. For $K \geq 30$ each estimated topic model contains at least one coherent topic that clearly refers to aspects of crude oil (e.g.~\emph{`opec'}, \emph{`bpd'}, \emph{`oil'}; see Table \ref{tab:restopic3reuters}). These coherent crude oil topics are not completely but relatively exclusive. Some of the crude oil topics also cover another energy source (namely: \emph{`gas'}) and there is one reappearing conflict topic that refers to military aspects but also touches crude oil (\emph{`gulf'}, \emph{`missil'}, \emph{`warship'}, \emph{`oil'}). Other than that, no other entities are substantially covered by crude oil topics. Building topic model-based classification rules on the basis of these crude oil topics yields relatively high recall and precision values.

Hence, topic model-based classification rules can be a useful tool---but only if the estimated topics coherently and exclusively cover the entity of interest in all its aspects.

In all three applications, and as is to be expected, high recall and low precision values tend to be achieved for topic models with smaller number of topics and lower values for threshold $\xi$, whereas low recall and high precision values tend to result from topic models with a higher number of topics and higher values for $\xi$ (see Figure \ref{fig:restopic2} in Appendix \ref{app:restopic2}).\footnote{A too high $\xi$, however, at times may lead to a classification rule in which none of the documents is assigned to the positive relevant class---thereby producing an undefined value for precision and the $F_1$-Score (here visualized by 0).} Classification rules that use topic models with a higher topic number and lower threshold $\xi$ neither tend to exhibit the highest recall nor the highest precision values but they tend to strike the best balance between recall and precision and achieve the highest $F_1$-Scores (see Figure \ref{fig:topicres} here and Figure \ref{fig:restopic2} in Appendix \ref{app:restopic2}).

\subsubsection{Active and Passive Supervised Learning}

\begin{figure}[p]
\centering
\small
\hspace{-1.5cm} Twitter \\
\includegraphics[height=0.39\textwidth]{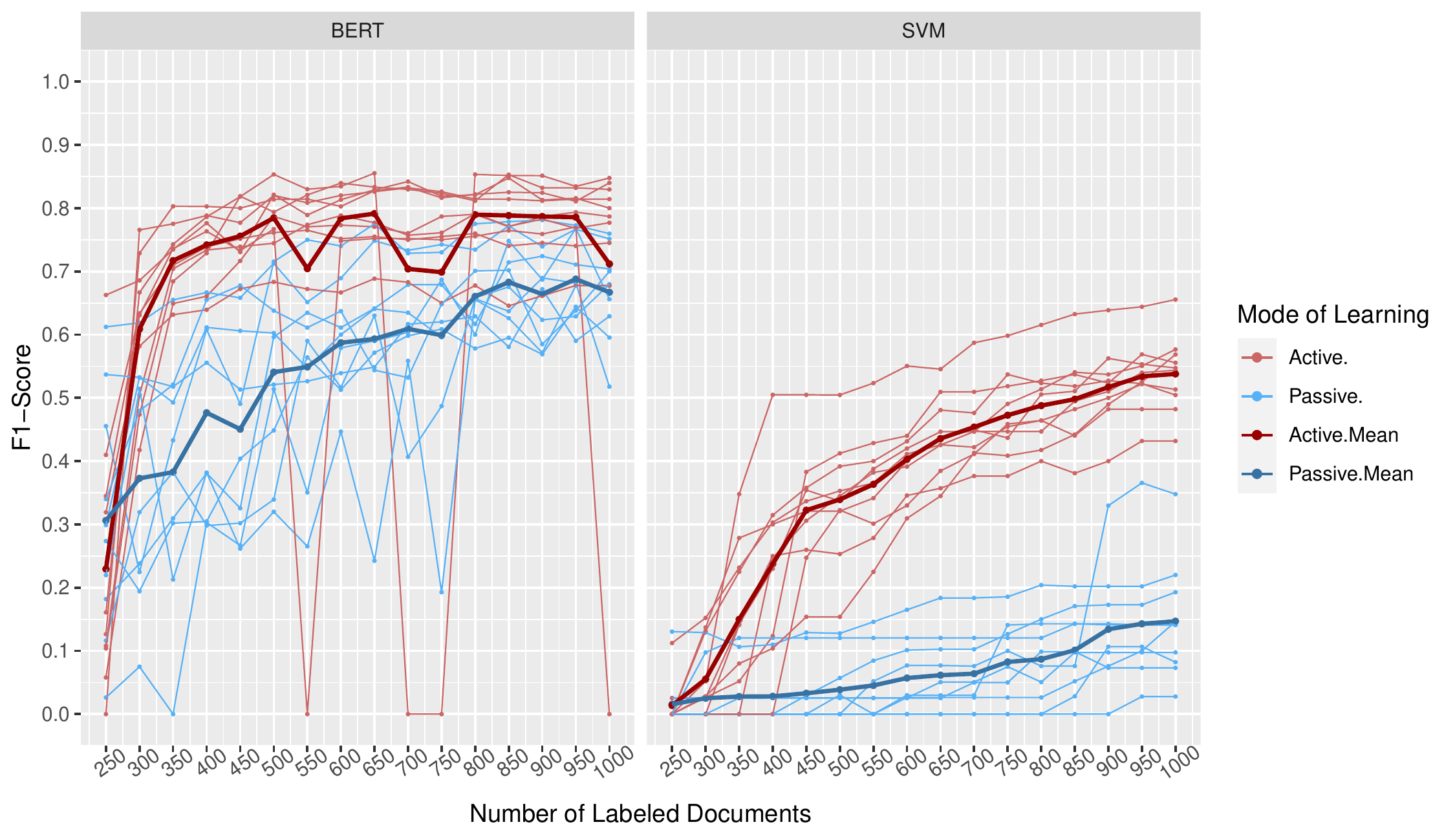}  \\
\hspace{-1.5cm} SBIC \\
\includegraphics[height=0.39\textwidth]{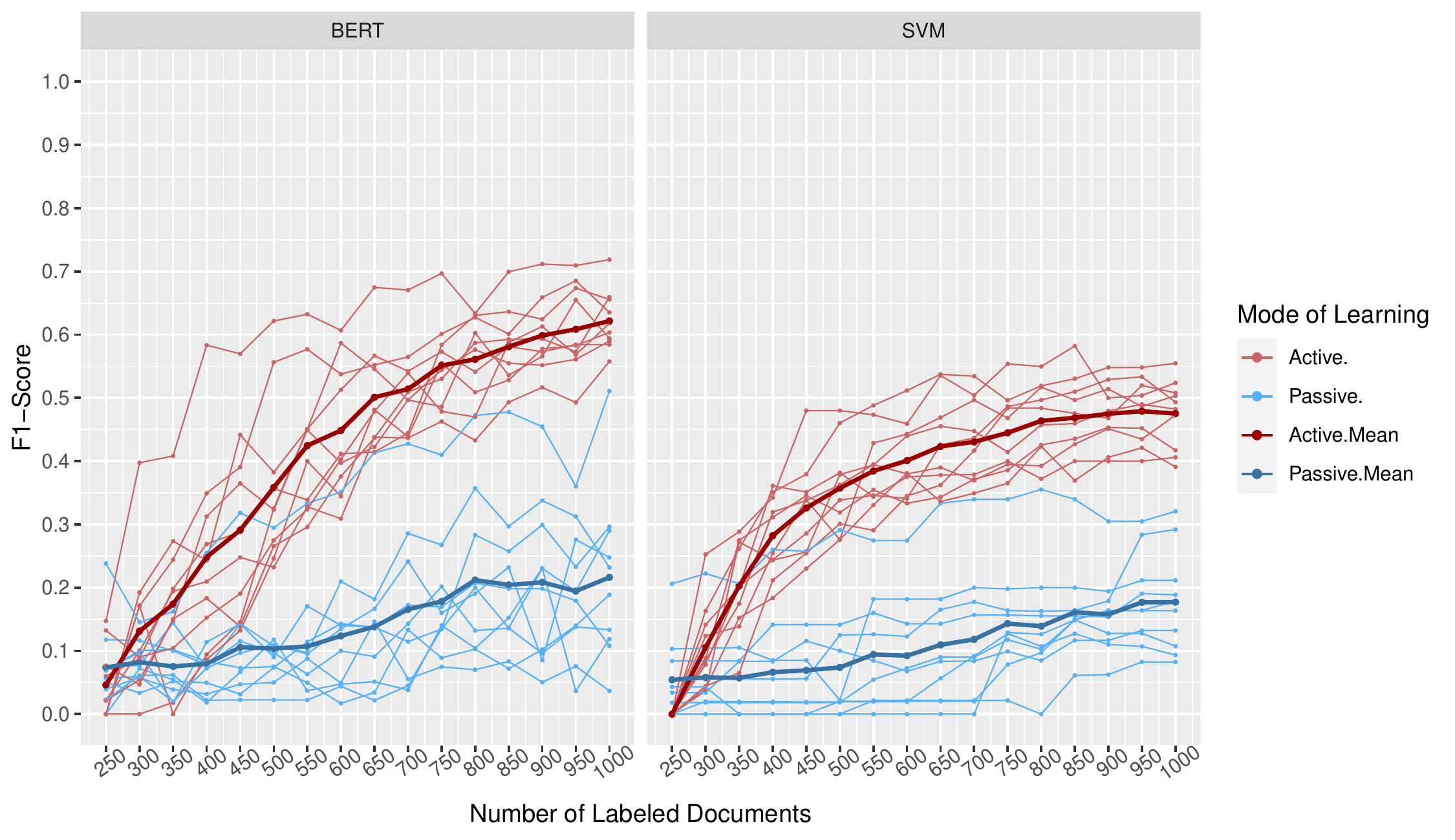}  \\
\hspace{-1.5cm} Reuters  \\
\includegraphics[height=0.39\textwidth]{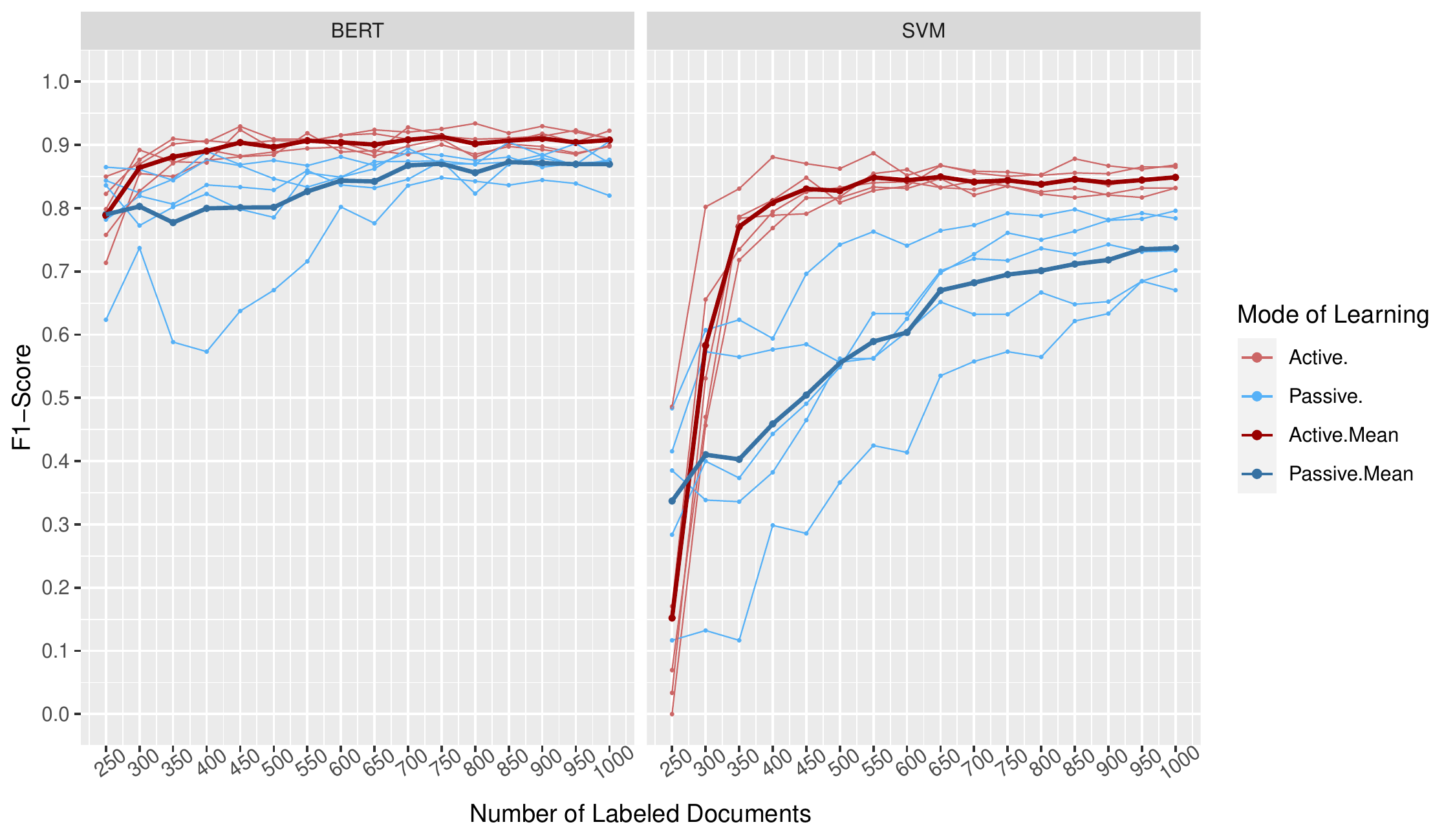}  \\
\caption[Retrieving Relevant Documents with Active and Passive Supervised Learning]{\textbf{Retrieving Relevant Documents with Active and Passive Supervised Learning.} \footnotesize{$F_1$-Scores achieved on the set aside test set as the number of unique labeled documents in set $\mathcal{I}$ increases from 250 to 1,000. Passive supervised learning results are visualized by blue lines, active learning results are given in red. For each of the 10 (Twitter, SBIC) or 5 (Reuters) conducted iterations, one light colored line is plotted. The thick and dark blue and red lines give the means across the iterations. If a trained model assigns none of the documents to the positive relevant class, then it has a recall value of 0 and an undefined value for precision and the $F_1$-Score. Undefined values here are visualized by the value 0.}}
\label{fig:superres} 
\end{figure}

\begin{figure}[p]
\centering
\small
\hspace{-1.5cm} Twitter \\
\includegraphics[height=0.39\textwidth]{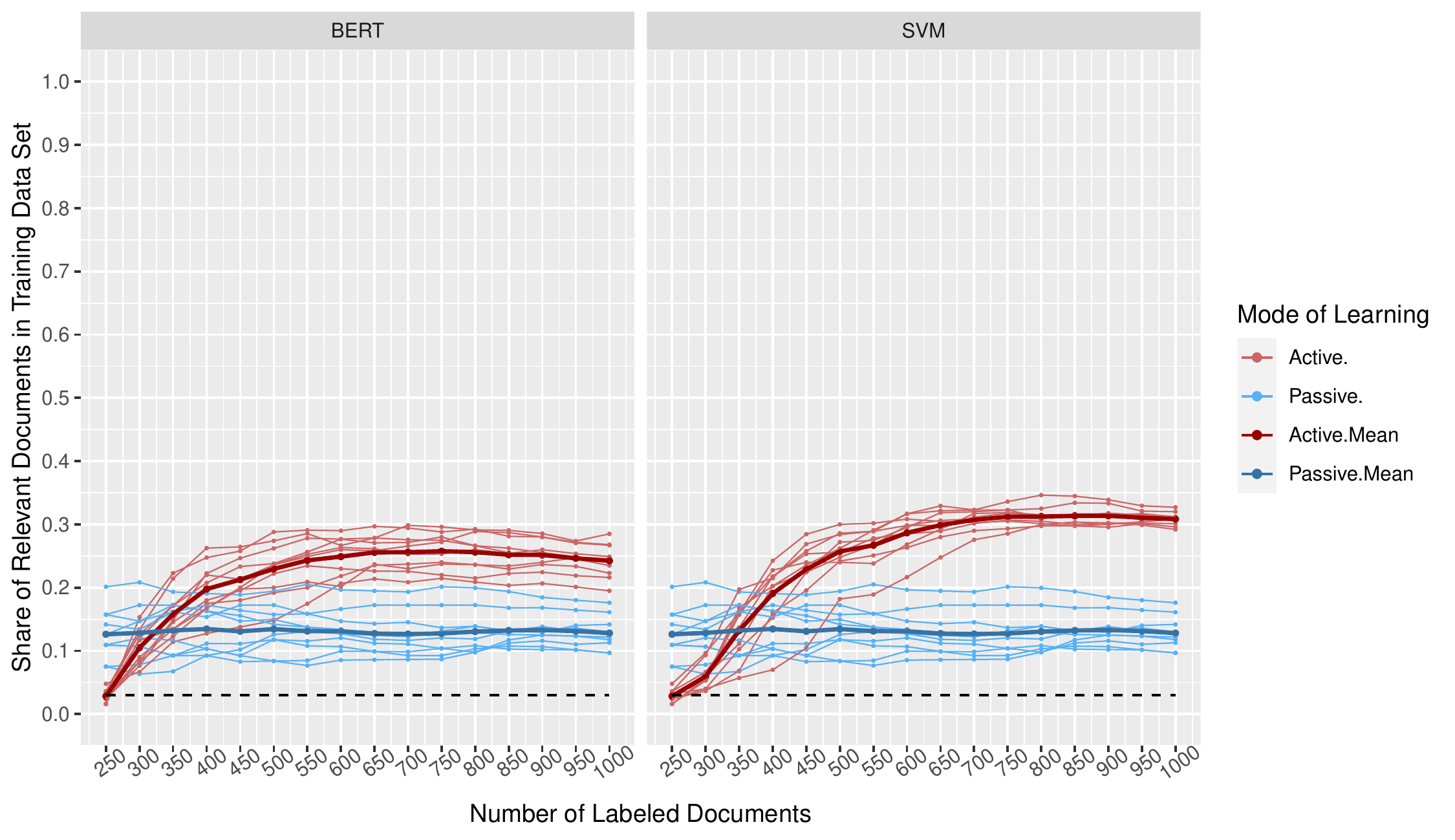}  \\
\hspace{-1.5cm} SBIC \\
\includegraphics[height=0.39\textwidth]{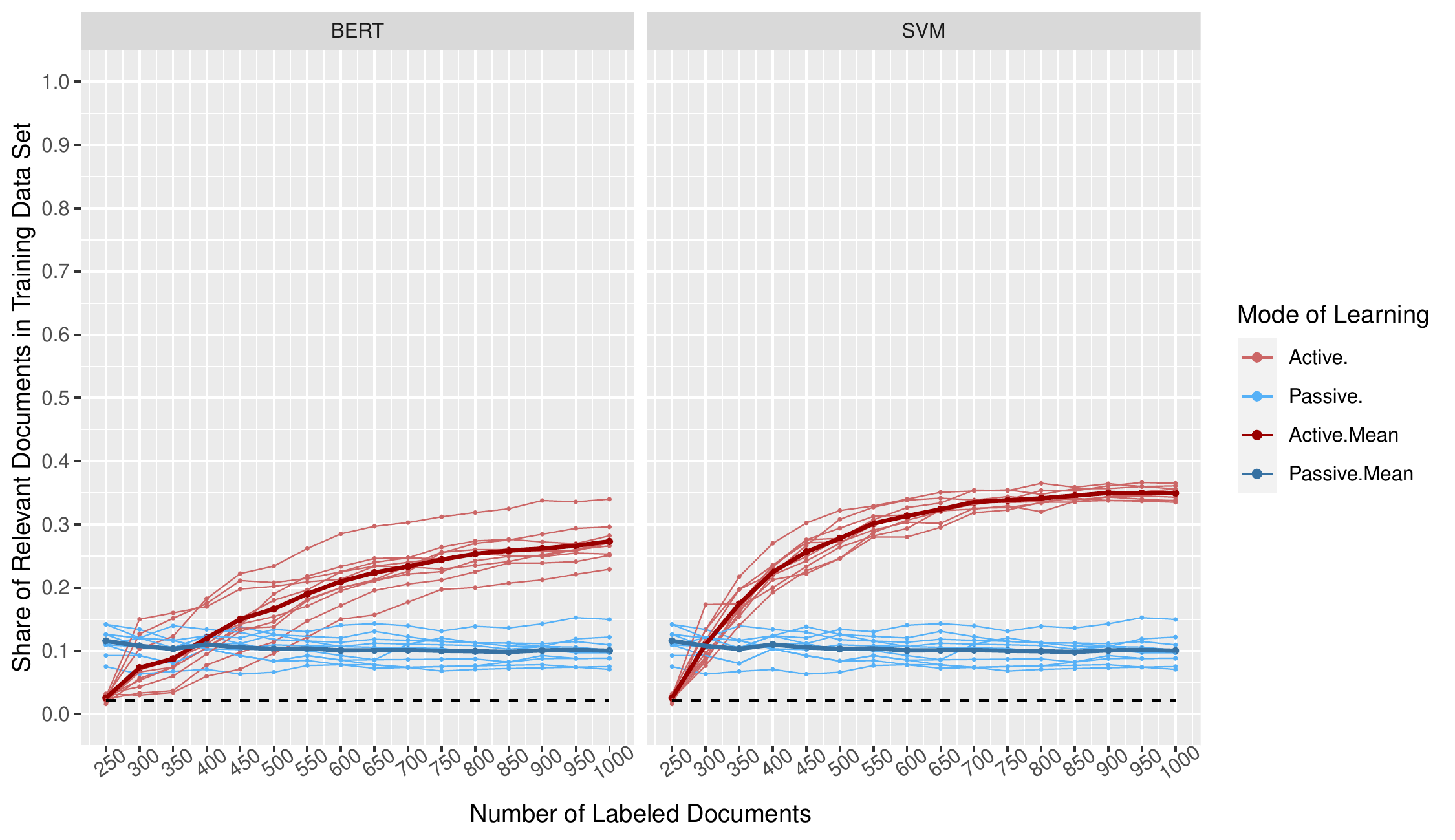}  \\
\hspace{-1.5cm} Reuters  \\
\includegraphics[height=0.39\textwidth]{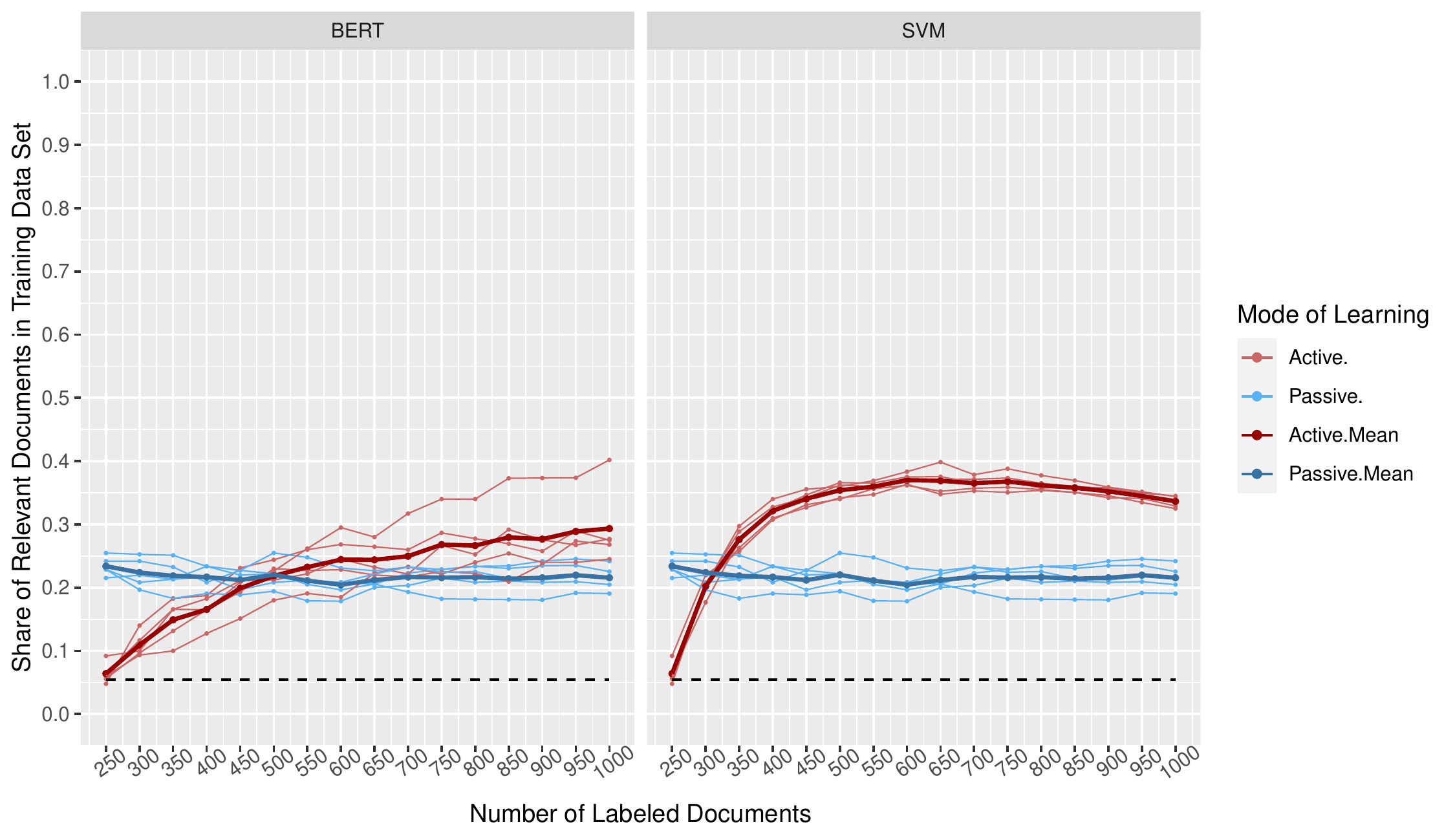}  \\
\caption[Share of Relevant Documents in the Training Set]{\textbf{Share of Relevant Documents in the Training Set.} \footnotesize{Share of documents in the training set that fall into the positive relevant class as the number of unique labeled documents in set $\mathcal{I}$ increases from 250 to 1,000. Passive supervised learning shares are visualized by blue lines, active learning shares are given in red. For each of the 10 (Twitter, SBIC) or 5 (Reuters) conducted iterations, one light colored line is plotted. The thick and dark blue and red lines give the means across the iterations. The black dashed line visualizes the share of relevant documents in the entire corpus.}}
\label{fig:superresshare} 
\end{figure}

Figure \ref{fig:superres} visualizes for each of the three studied retrieval tasks (Twitter, SBIC, Reuters), for each employed supervised learning model (BERT and SVM), for each applied learning setting (passive learning with random oversampling as well as pool-based active learning with uncertainty sampling), and for each of the 10 (Twitter, SBIC) or 5 (Reuters) conducted iterations, the $F_1$-Score achieved on the set aside test set (fold $g$) as the number of labeled training documents in set $\mathcal{I}$ increases from 250 to 1,000. Passive supervised learning results are visualized by blue lines, active learning results are given in red. The thick and dark blue and red lines give the mean $F_1$-Scores across the iterations. They visualize the estimate of the expected generalization error.\footnote{Note that the x-axis denotes the number of unique labeled instances in set $\mathcal{I}$. As in passive learning with random oversampling the documents from the relevant minority class in set $\mathcal{I}$ are randomly resampled with replacement to then form the training set on which the model is trained, in passive learning the size of the training set is larger than the size of set $\mathcal{I}$. Yet, the size of $\mathcal{I}$ indicates the number of unique documents on which training is performed and---as only unique documents have to be annotated---it indicates the annotation costs.}

Across all three applications and for BERT as well as SVM, active learning with uncertainty sampling tends to dominate passive learning with random oversampling. Passive learning with random oversampling on average only shows a similar or higher $F_1$-Score for the first learning iteration (i.e.~at the start when training is conducted on the randomly sampled training set of 250 labeled instances). Then, however, the active learning retrieval performance strongly increases such that for the same number of labeled training instances active learning, on average, produces a higher $F_1$-Score than passive learning.

One likely reason for this difference between passive and active learning is revealed in Figure \ref{fig:superresshare} that contains the same information as Figure \ref{fig:superres}; except that on the y-axis not the $F_1$-Score but the share of documents from the positive relevant class in the training set is shown. The black dashed line visualizes the share of relevant documents in the entire corpus and thus would be the expected share of relevant documents in a randomly sampled training set if neither random oversampling nor active learning were conducted. In passive learning with random oversampling (shown in blue) the 50 training instances, that are added in each step to the set of labeled training instances $\mathcal{I}$, are randomly sampled from pool $\mathcal{U}$. Then, the relevant instances in set $\mathcal{I}$ are randomly oversampled such that their number increases by a factor of 5. For this reason, the share of positive training instances in the passive learning setting is higher than in the corpus (black dashed line) but remains relatively constant across the training steps. In active learning (shown in red), no random oversampling is conducted---which is why at the beginning the share of relevant documents in at about equals the share of relevant documents in the corpus. Then, however, active learning at each step selects the 50 instances the algorithm is most uncertain about. As has been observed in other studies before (\citealp[p.~133-134]{Ertekin2007}; \citealp[p.~545]{Miller2020}), this implies that disproportionately many instances from the relevant minority class are selected into set $\mathcal{I}$. The share of positive training instances increases substantively---which in turn tends to increase generalization performance on the test set as shown in Figure \ref{fig:superres}. 

When decomposing the $F_1$-Score into recall and precision (see Figure \ref{fig:superres2} in Appendix \ref{app:superres2}), it is revealed that the supervised models' recall values gradually improve as the number of training instances increases. The precision values early reach higher levels and exhibit a more volatile path. The observed retrieval performance enhancements hence particularly are caused by the models from step to step becoming better at identifying a larger share of the truly relevant documents from the corpora. Models trained in active rather than passive learning mode tend to yield higher recall values. 

\begin{figure}[p]
   \centering
   \small
\begin{tabular}{cc}
\hspace{0.8cm} Twitter & \hspace{-0.9cm} Twitter  \\
\includegraphics[height=0.35\textwidth]{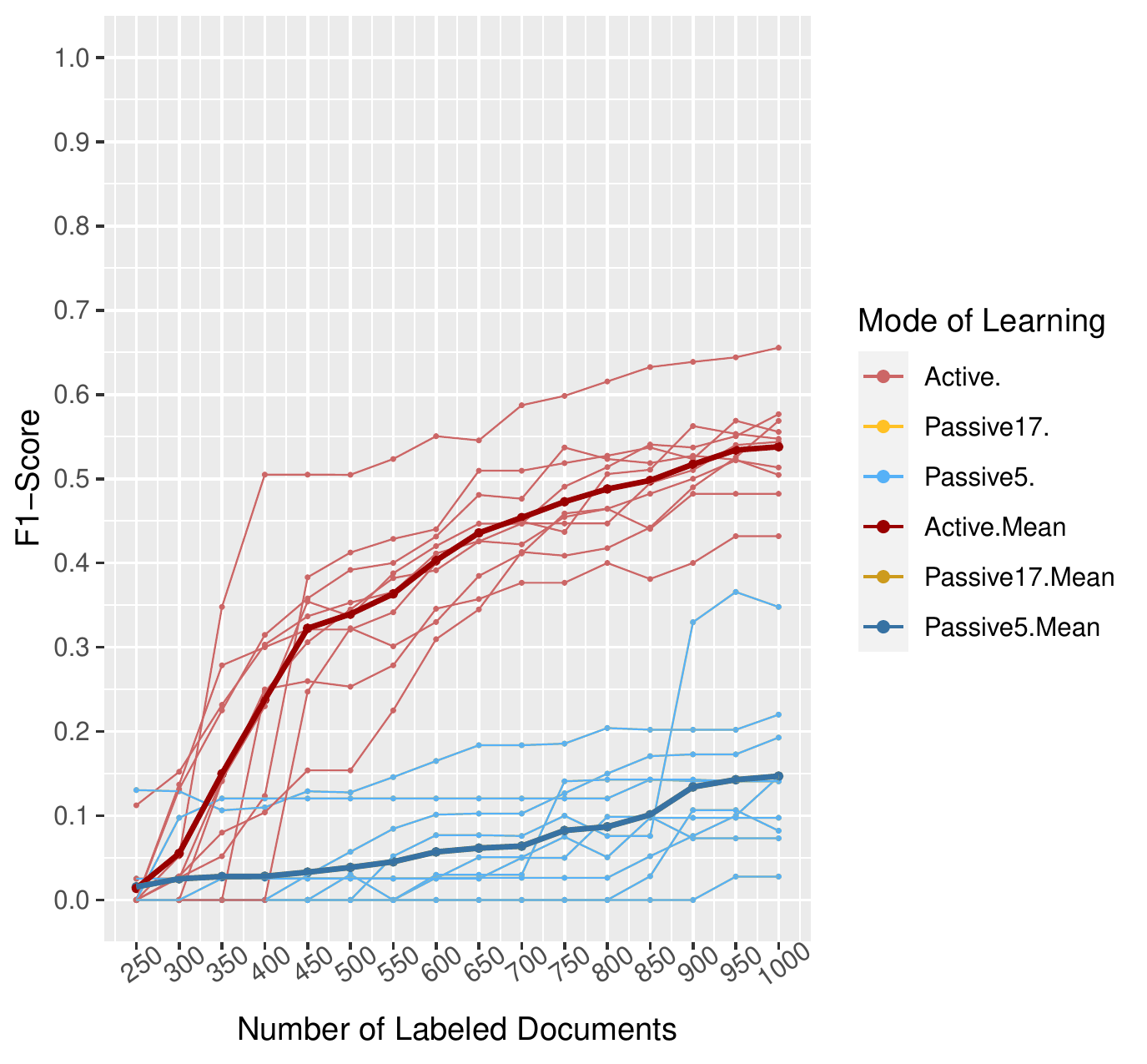}  \hspace{-2cm} &
\includegraphics[height=0.35\textwidth]{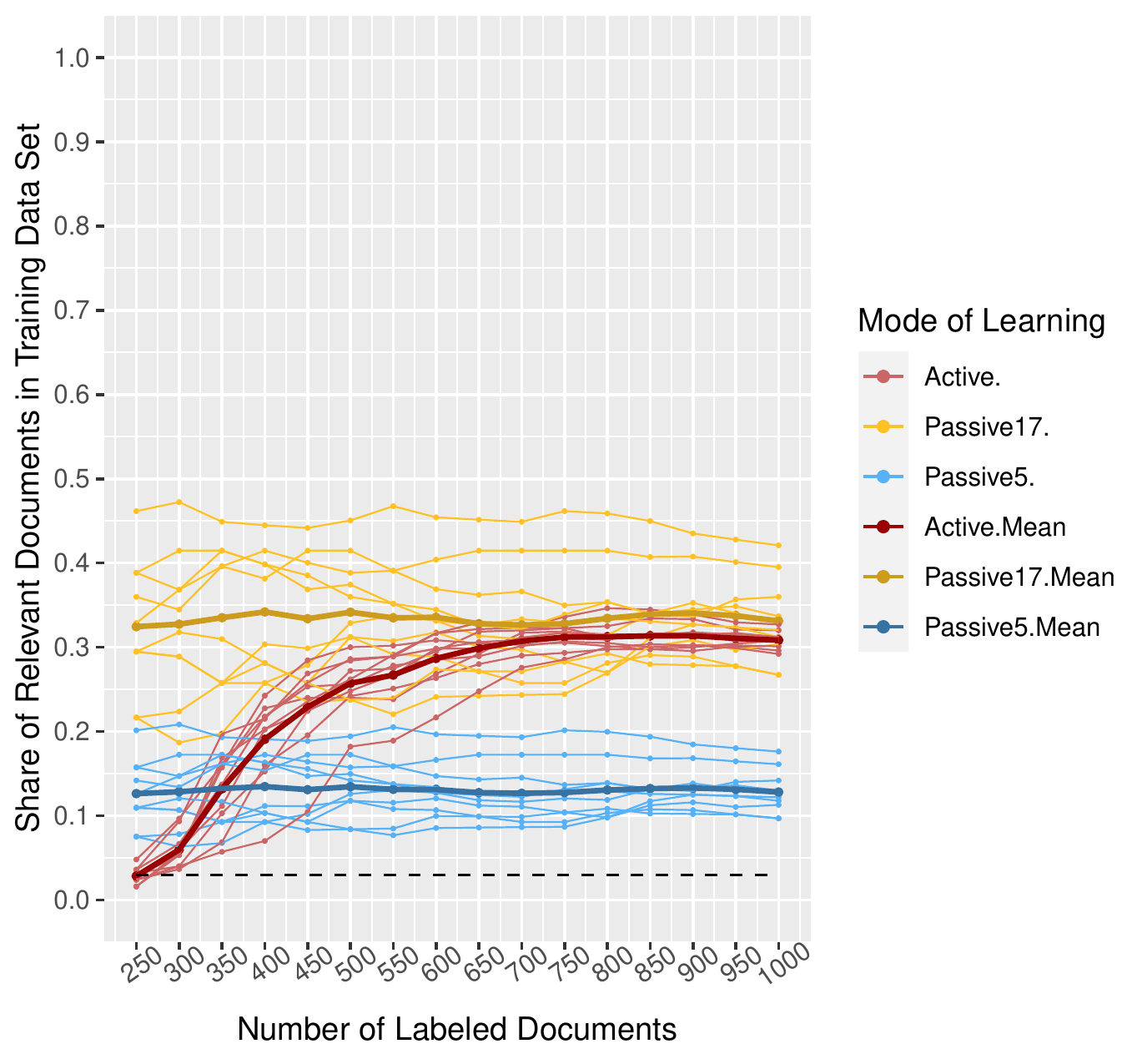} \\
\hspace{0.8cm} SBIC & \hspace{-0.9cm} SBIC \\
\includegraphics[height=0.35\textwidth]{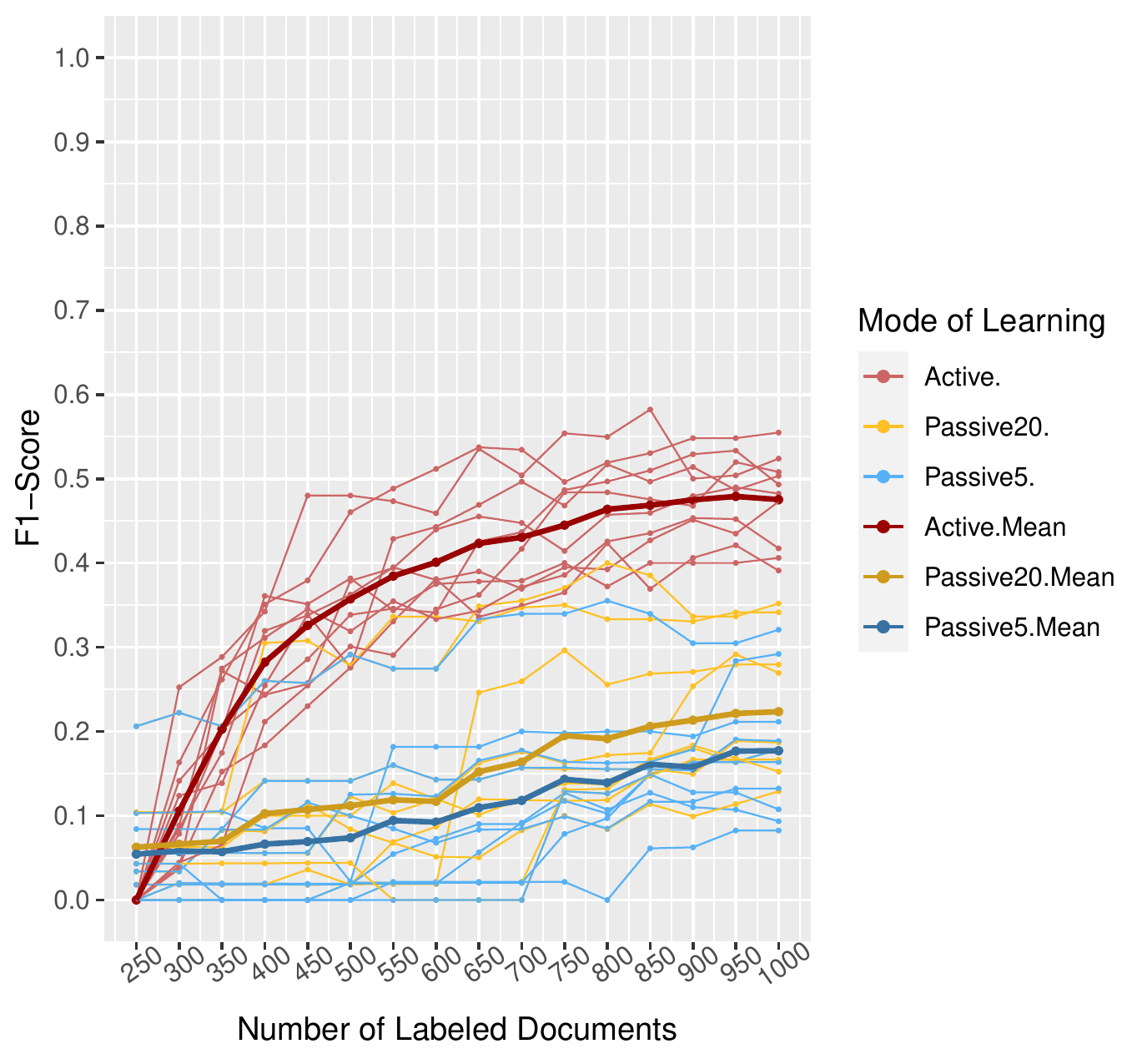} \hspace{-2cm} &
\includegraphics[height=0.35\textwidth]{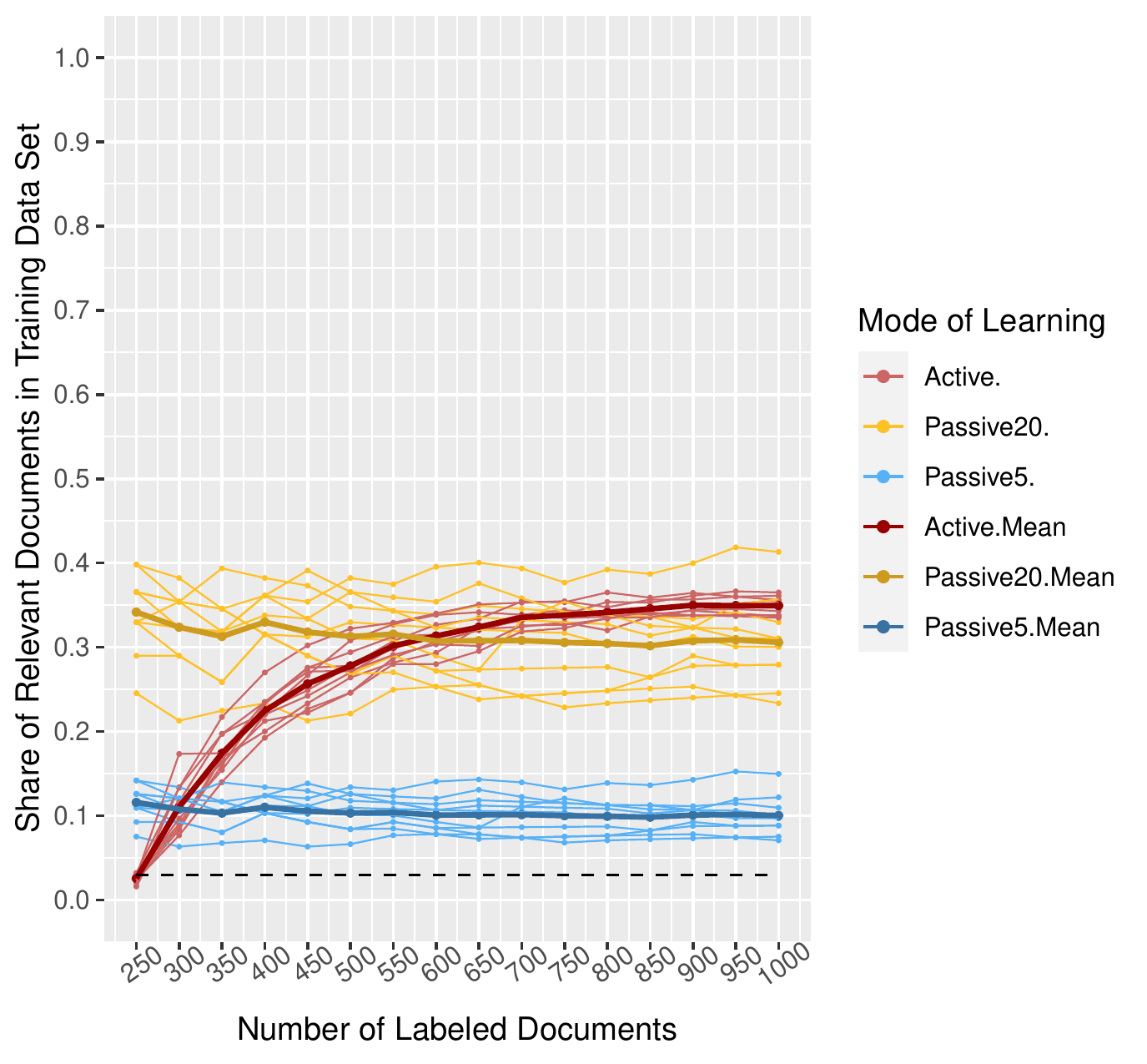} \\
\hspace{0.8cm} Reuters & \hspace{-0.9cm} Reuters  \\
\includegraphics[height=0.35\textwidth]{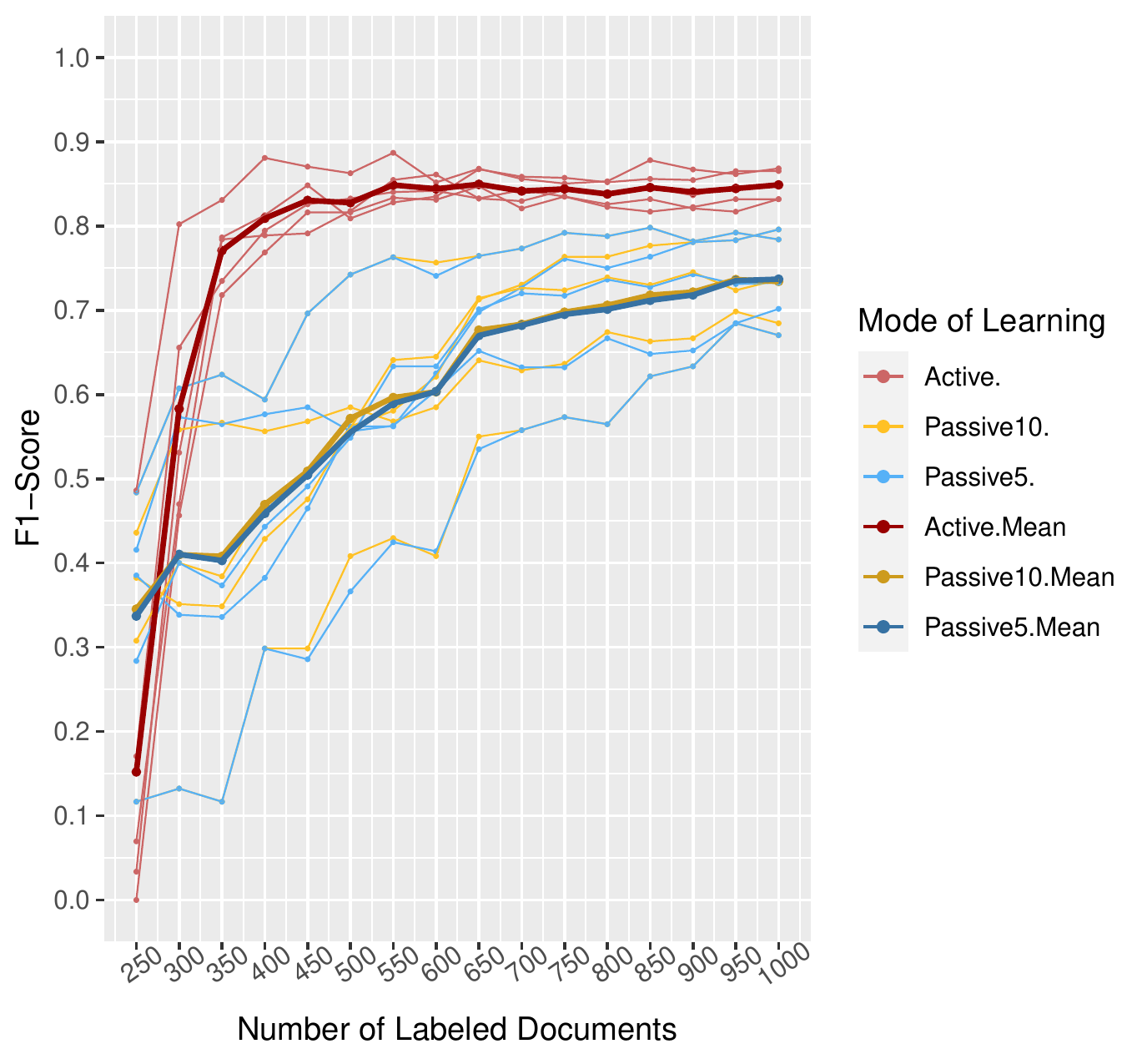}  \hspace{-2cm} &
\includegraphics[height=0.35\textwidth]{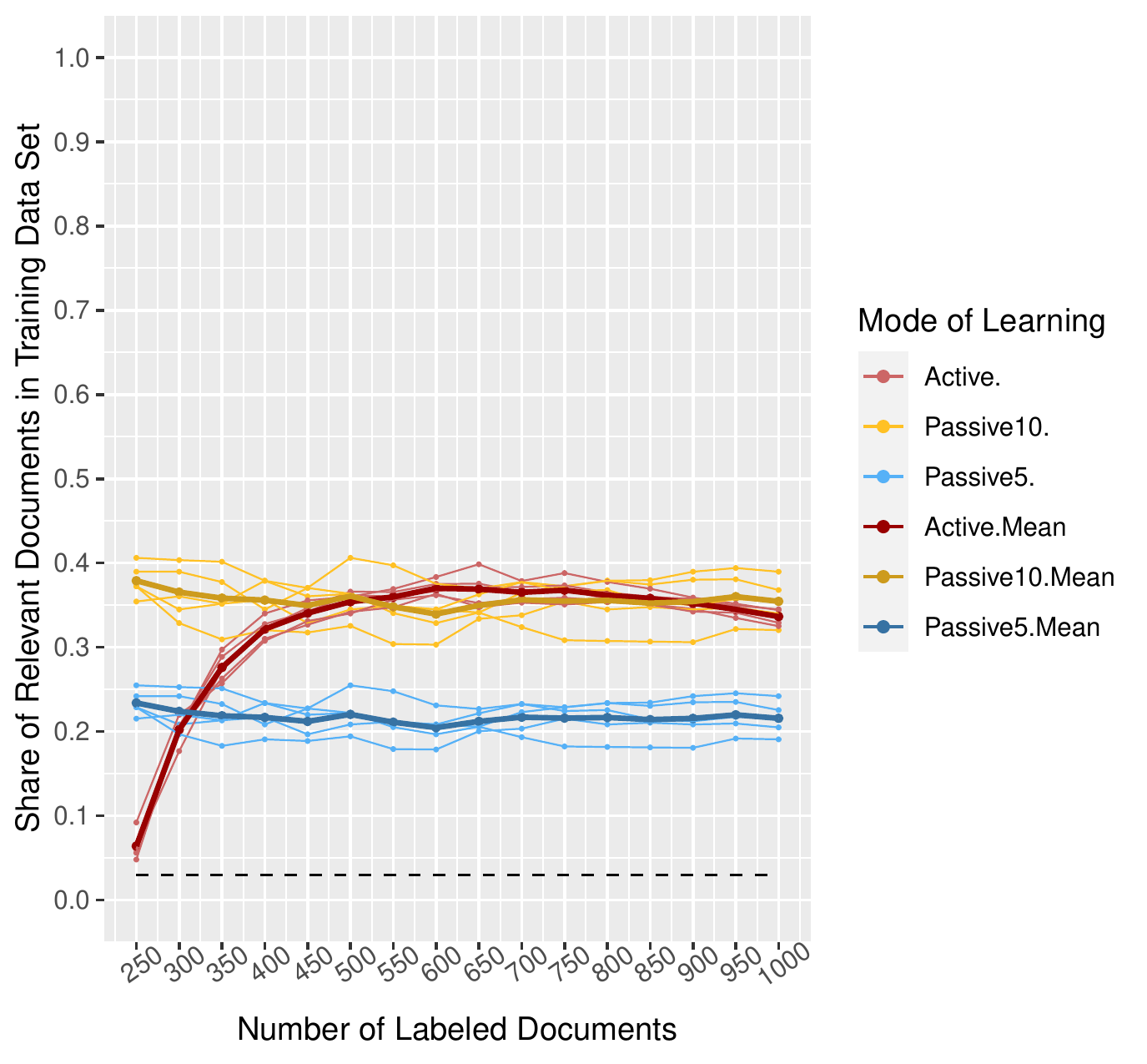}\\
\end{tabular}
\caption[SVM: Comparing Passive Learning with Two Oversampling Factors to Active Learning]{\textbf{SVM: Comparing Passive Learning with Two Oversampling Factors to Active Learning.} \footnotesize{\textbf{Left column:} $F_1$-Scores achieved by the SVMs on the set aside test set as the number of unique labeled documents in set $\mathcal{I}$ increases from 250 to 1,000. Undefined values here are visualized by the value 0. \textbf{Right column:} Share of documents in the training set that falls into the positive relevant class as the number of unique labeled documents in set $\mathcal{I}$ increases from 250 to 1,000. The black dashed line visualizes the share of relevant documents in the entire corpus. \textbf{Both columns:} The results are presented for passive learning with a random oversampling factor of 5 (blue lines), passive learning with random oversampling factors of 17 (Twitter), 20 (SBIC), and 10 (Reuters) (golden lines), as well as pool-based active learning with uncertainty sampling (red lines). The thick and dark blue, golden, and red lines give the means across the iterations.}}
\label{fig:superresdouble} 
\end{figure}

Yet, there is the question of whether active learning exhibits a superior performance to passive learning with random oversampling simply because after a certain number of training steps the share of training instances is higher for active than for passive learning or whether active learning dominates passive learning (also) because active learning, due to focusing on the uncertain region between the classes and due to operating on unique---rather than duplicated---positive training instances, learns a better generalizing class boundary with fewer training instances \citep[p.~28]{Settles2010}. To inspect this question, for the SVMs passive learning with random oversampling is repeated whereby positive relevant documents are randomly oversampled such that their number increases by a factor of 10 (Reuters), 17 (Twitter), or 20 (SBIC) (instead of a factor of 5 as before). This results in higher shares of relevant documents in the training set for passive learning (see right column in Figure \ref{fig:superresdouble}). Yet, the prediction performance on the test set as measured by the $F_1$-Score either does not or does only minimally increase compared to the situation of random oversampling by a factor of 5 (see left column in Figure \ref{fig:superresdouble}). Moreover, although the share of relevant documents in the stronger oversampled passive learning training data sets is similar to those of active learning, active learning still yields considerably higher $F_1$-Scores. This indicates that from a certain point, merely duplicating positive instances by random oversampling has no or only a small effect on the class boundary learned by the SVM. The finding also indicates that active learning improves upon passive learning because it is effectively able to select a large share of truly positive documents for training that are not duplicated but unique and because its selection of uncertain documents provides crucial information on the class boundary. 

To be more precise: When applying SVMs to imbalanced data sets, the problem in general is that the learned hyperplane tends to be positioned too close to the positive minority class instances \citep[p.~40-44]{Akbani2004}. The reason is that whereas the many negative training instances occur across the entire area \emph{belonging} to the negative class, the few positive instances only occur at few points within the area \emph{belonging} to the positive class \citep[p.~40-44]{Akbani2004}. Hence, the boundary of the negative area is well represented in the training data whereas the boundary of the positive area is not \citep[for an illustration see][p.~43-44]{Akbani2004}. The hyperplane can be moved toward the negative side by giving more weight to positive instances, e.g. by duplicating them via random oversampling or by introducing different costs for misclassifying positive vs.~negative instances \citep{Veropoulos1999}. A problem that often arises when doing so, however, is that the hyperplane tends to overfit on the positive instances. Its orientation and shape too strongly tends to reflect the positions of positive instances \citep[p.~46]{Akbani2004}. 
In active learning, in contrast, the training instances the learning algorithm requests to be labeled and added next are unique---which has a positive effect on generalization performance.\footnote{Consider the following example: Assume that the share of positive instances in a corpus is 2.5$\%$. Assume further that there are two training data sets each comprising 1,000 training instances and the share of positive instances in each is 30$\%$. This high share of positive instances has been generated by random oversampling in one training data set and by selection via active learning in the other. Then, in the training set generated with random oversampling, the 300 positive training instances are mere duplicates of around 25 unique instances. In the training data set selected by active learning, in contrast, each of the 300 positive training instances is a unique instance. Therefore, active learning provides much more information on the distribution of positive training instances than passive learning with random oversampling. There are more unique positive instances observed across the area \emph{belonging} to the positive class, thereby providing more information on the possible boundary of the positive area. In active as compared to passive learning with random oversampling, the fact that there are more unique positive training instances is likely to make the hyperplane more smooth and thus is likely to produce a better generalizing hyperplane.}

Another important observation concerns the performances' variability (see again Figure \ref{fig:superres}): For all models and learning modes, given a fixed number of labeled training instances, the $F_1$-Scores on the set aside test sets can vary considerably between iterations. Which set of documents is randomly sampled to form the (initial) training set and which set aside test fold is used for evaluation thus can have a profound effect on the measured retrieval performance.

A further observation is that BERT on average tends to outperform SVM (see also Figure \ref{fig:superrescompare} in Appendix \ref{app:superrescompare}). Hence, applying the Transformer-based pretrained language representation model BERT with its ability to learn context-dependent meanings of tokens and with the information acquired in the pretraining phase, here tends to be better able to identify the few relevant documents, 
than an SVM operating on bag-of-words representations. That BERT only can process sequences of at maximum 512 tokens is not a particularly limiting or performance reducing factor here. The performance difference between the two learning methods is distinct and relatively consistent with regard to the Twitter and Reuters retrieval tasks. It is less clear cut for the SBIC. This is rather surprising as in the SBIC the disrespectful remarks toward disabled people are implied rather than stated explicitly. 

Note also that with regard to the Twitter and SBIC retrieval tasks, BERT exhibits a higher instability from one learning step to the next as the number of labeled training instances $\mathcal{I}$ increases by a batch of 50 documents (see Figure \ref{fig:superrescomparediffs} in Appendix \ref{app:superrescompare}). On the Reuters corpus, where the retrieval task seems much more easy and all training and testing iterations with BERT early settle for similar, high performing solutions, instabilities are minimal. With regard to the more complex Twitter and SBIC retrieval tasks, however, after adding a new batch of 50 labeled training instances and fine-tuning BERT on this new, slightly expanded training data set, the $F_1$-Score achieved on the test set may not only increase but also decrease considerably. The strongest de- and increases can be observed for active learning on the Twitter data set where drops and rises of the $F_1$-Score by a value of about 0.85 occur. 

As noted in Section \ref{sec:activepassive} above, 
BERT's prediction performance can exhibit considerable variance across random initializations---even if trained on the exactly same training data set. Here, to make predictions and performances more stable, precautions have been taken by choosing a small learning rate of 2e-05 and setting the number of epochs to 20 such that many training iterations are conducted. To evaluate how the level and the stability of BERT's prediction performance changes if the hyperparameters are set to more conventional values (e.g.~combining a learning rate of 2e-05 or 3e-05 with 3 or 4 epochs), BERT is also trained using hyperparameter values in these common ranges. Figure \ref{fig:superreshyper} in Appendix \ref{app:superreshyper} visualizes the $F_1$-Scores of these models. The instability of the learning paths is only slightly to moderately higher (see Figure \ref{fig:superreshyper2} in Appendix \ref{app:superreshyper}). Yet, for the Twitter and SBIC tasks and especially as the training data sets are very small, the $F_1$-Scores of BERT models with common hyperparameter values are substantively lower than the $F_1$-Scores of BERT models that have been trained for 20 epochs. In the Twitter application, for example, the mean $F_1$-Score of active learning with a BERT model that is trained for 20 epochs on a set of 500 labeled training instances is 0.568 higher than active learning with a BERT model trained for 3 epochs on 500 training instances. Hence, although training over 20 epochs takes proportionally more computing time than training over 3 or 4 epochs,\footnote{For example, training BERT for 20 epochs on 1,000 tweets takes on average 172 seconds, whereas training BERT on 1,000 tweets for 3 epochs takes on average 26 seconds.} when applying BERT to small data sets---a scenario which is likely in the retrieval settings focused on here---training for many epochs (and thus presenting each document in the small training data set many times to the model) seems important to enhance performance.

Nevertheless, because BERT tends to exhibit a relatively high degree of variability in its performance even if training is conducted for a larger number of epochs, monitoring retrieval performance with a set aside test set seems important for researchers to detect situations in which (likely due to vanishing gradients) retrieval performance drops to low values. Such situations can be easily fixed by, for example, by choosing another random seed for initialization. 

To conclude, if a team of researchers has the resources to retrieve documents referring to their entity of interest via supervised learning and they have a fixed number of training instances they can maximally label, then active learning is likely to yield better results than passive learning. Moreover, if none or only a small share of documents exceeds the maximum number of tokens that Transformer-based language representation models as BERT can process, then applying a BERT-like model that is trained with a small learning rate for a large number of epochs is likely to achieve better results than applying conventional supervised machine learning methods (auch as SVM) on bag-of-words-based representations. Yet, the predictions made by BERT (and hence also BERT's retrieval performance) is prone to considerable variation depending on the initializing random seed, the initial training data set, and the changes to the training data set from one learning step to the next. This problem, however, is mitigated by the fact that mediocre performances can be easily detected if performance is monitored with a set aside test set.


\subsubsection{Comparison Across Approaches}
\label{sec:conclusio}

\begin{sidewaysfigure}
  \centering
\includegraphics[height=0.4\textwidth]{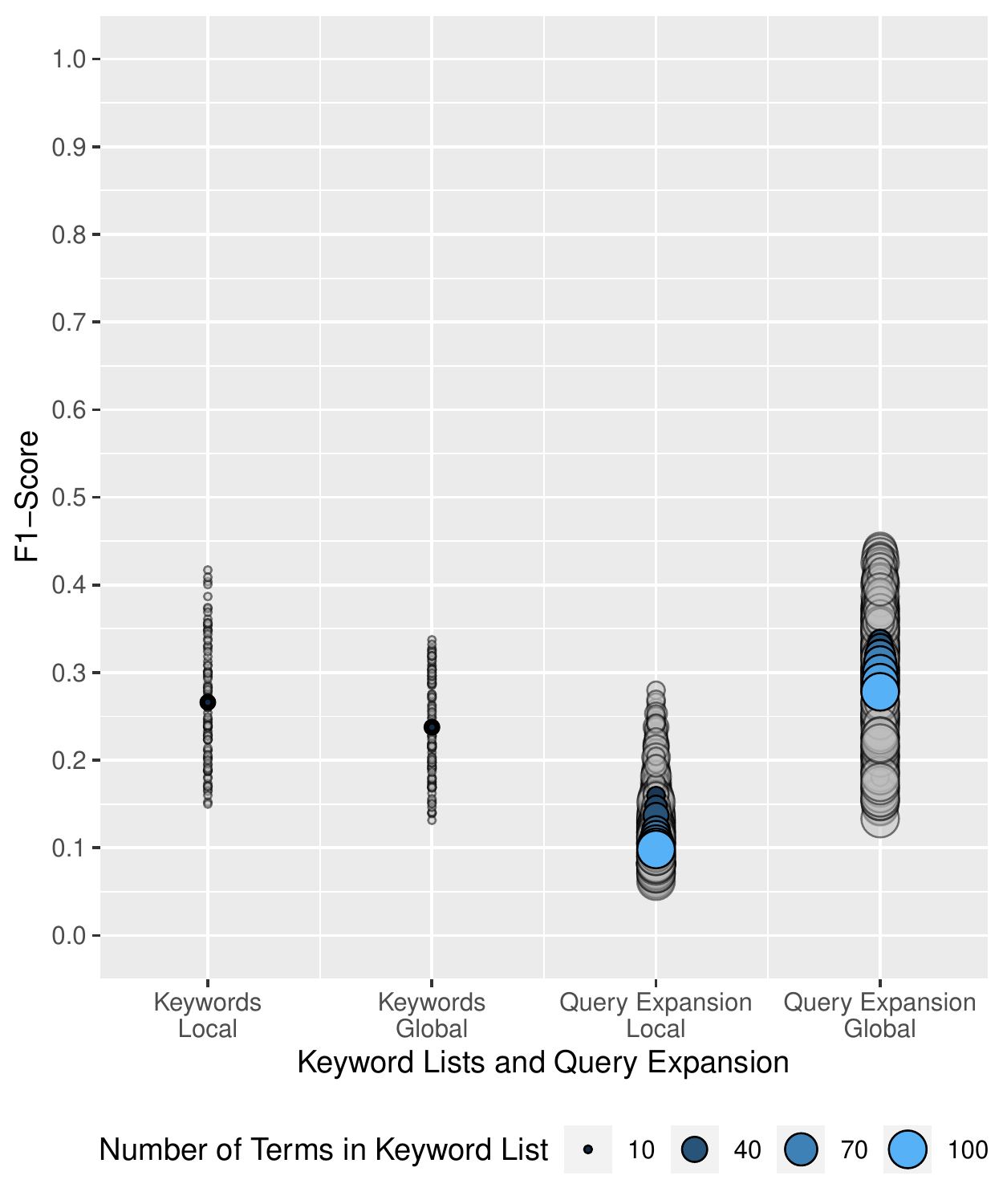} \hspace{-0.3cm} \includegraphics[height=0.4\textwidth]{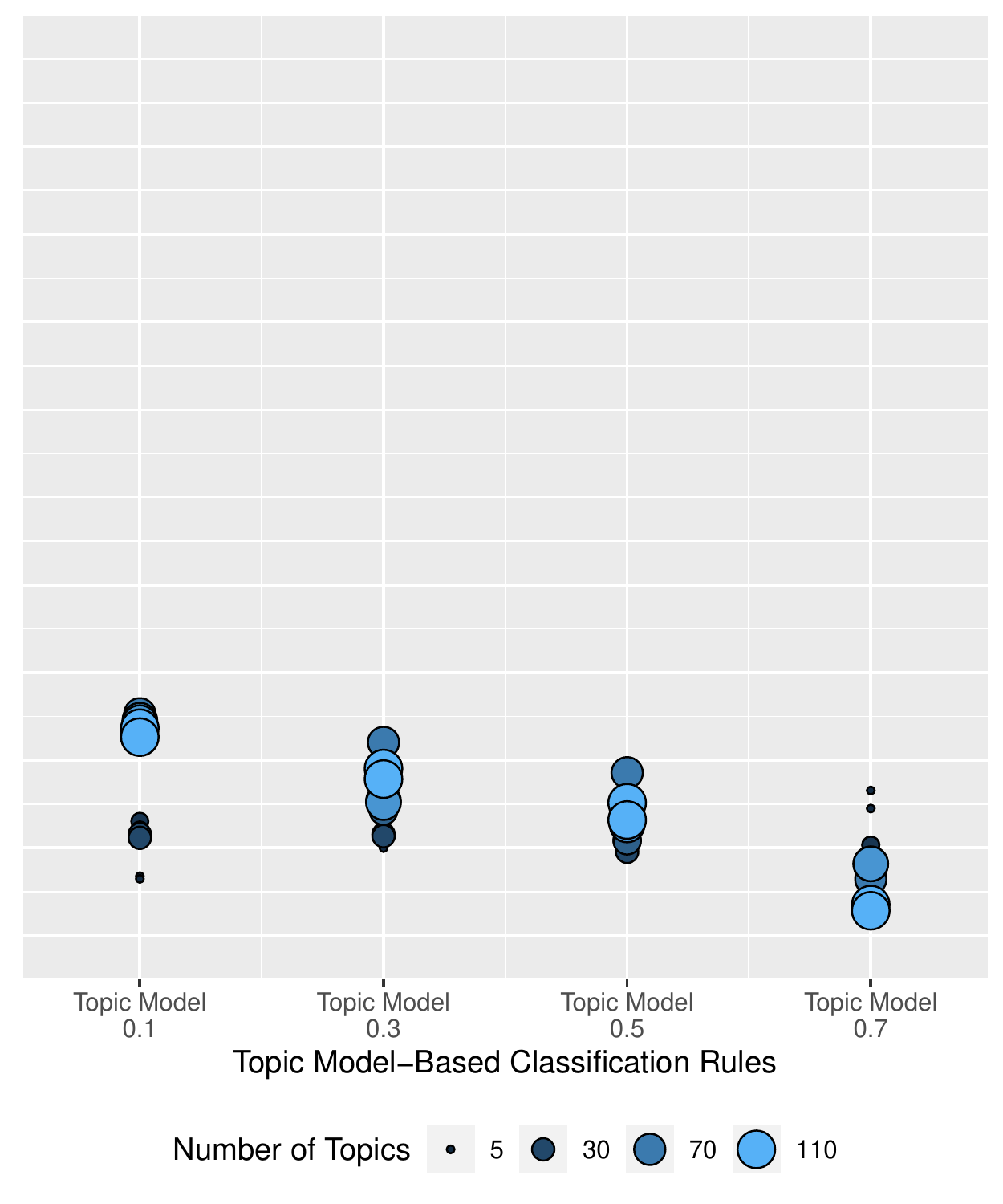}  \hspace{-0.3cm} \includegraphics[height=0.4\textwidth]{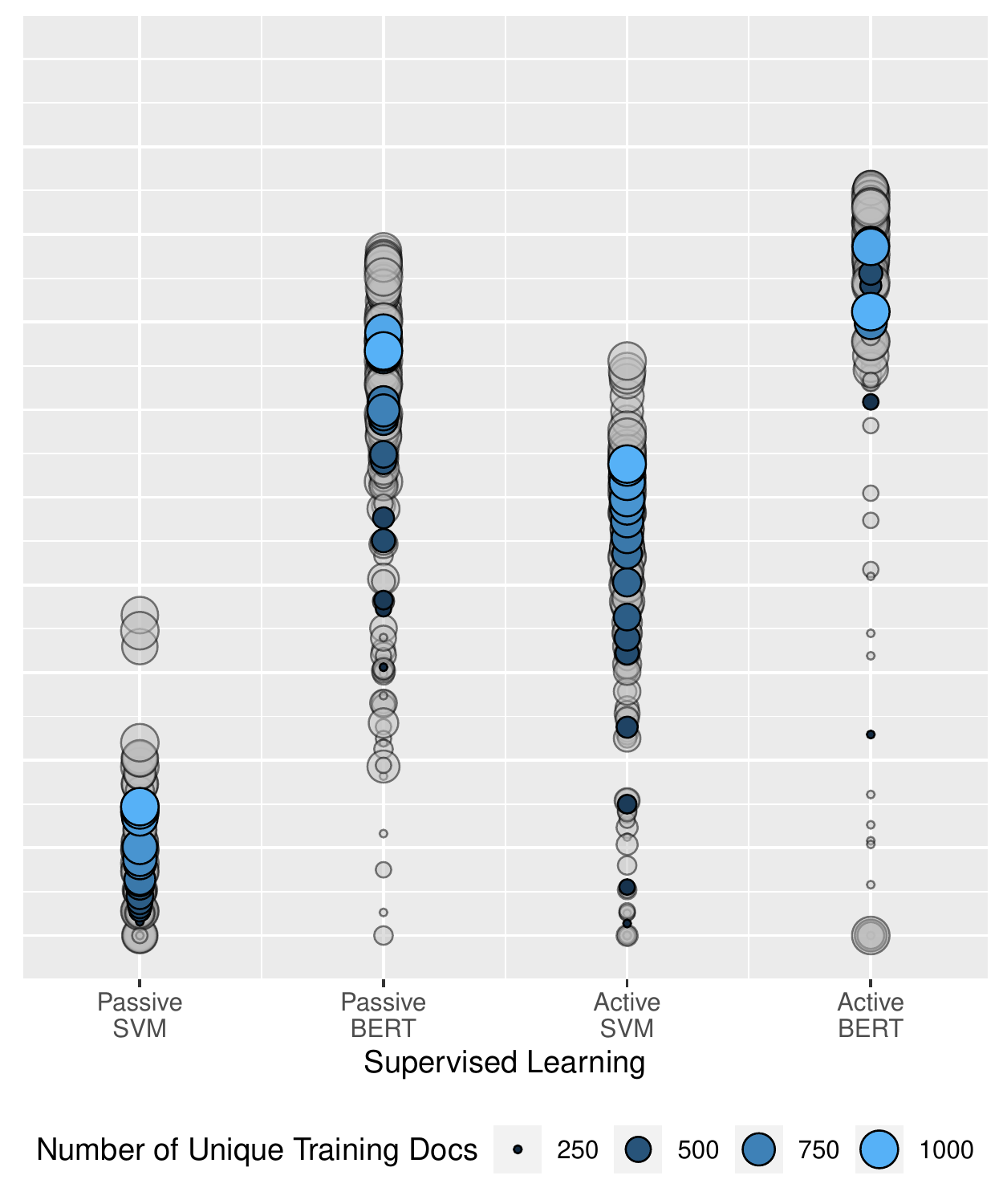}\\
\caption[Twitter: Comparison of Retrieval Methods]{\textbf{Twitter: Comparison of Retrieval Methods.} \footnotesize{This plot summarizes the retrieval performances of the here evaluated approaches on the retrieval task associated with the Twitter corpus. \textbf{Left panel:} $F_1$-Scores for the 100 lists of 10 predictive keywords that then are expanded in the local and global embedding spaces. A gray transparent dot marks the $F_1$-Score reached by a single (expanded) keyword list. A blue opaque dot marks the mean of the $F_1$-Scores across 100 (expanded) keyword lists that contain the same number of search terms. The larger the number of terms in an (expanded) keyword list, the larger the size and the lighter the color of the printed dots. \textbf{Panel at the center:} $F_1$-Scores of the topic model-based classification rules with different values for threshold $\xi$. The larger the number of topics in an estimated CTM, the larger the size and the lighter the color of the printed dots. (From all CTMs with a given topic number that have been estimated here, the best two performing combinations regarding the question how many and which topics are considered relevant are presented.) \textbf{Right panel:} $F_1$-Scores for active as well as passive supervised learning with SVM and BERT. A gray transparent dot marks the $F_1$-Score reached by a model that has been trained by a given number of unique training documents on one set aside test set $g$. A blue opaque dot marks the mean of the $F_1$-Scores of 10 such models that have been trained on the same number of unique training documents. Each of the 10 models is evaluated on another test set and together the 10 test sets constitute the entire Twitter corpus. The larger the number of unique labeled training instances, the larger the size and the lighter the color of the printed dots.}}
\label{fig:allrestwitter} 
\end{sidewaysfigure}

\begin{sidewaysfigure}
  \centering
\includegraphics[height=0.4\textwidth]{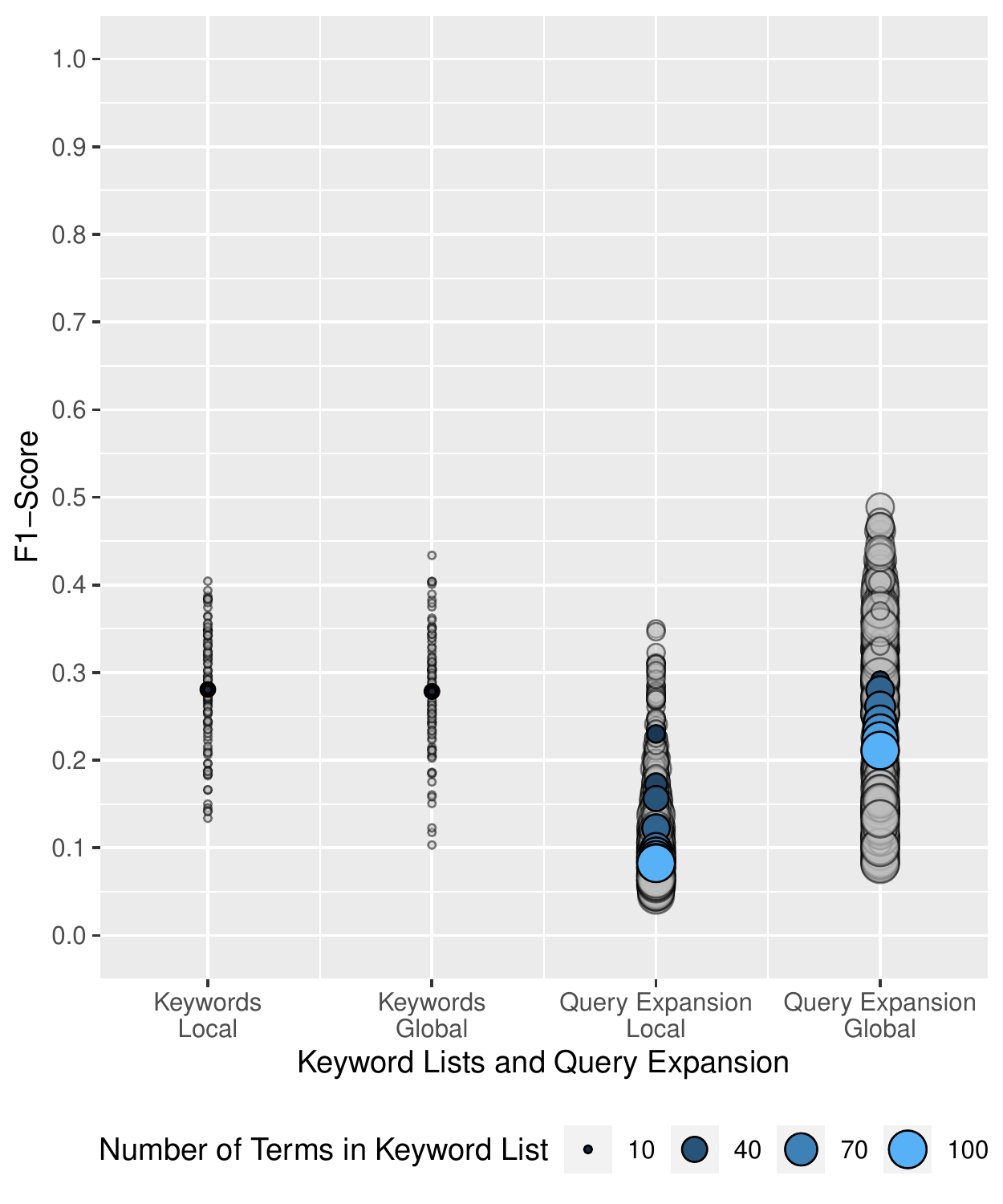} \hspace{-0.3cm} \includegraphics[height=0.4\textwidth]{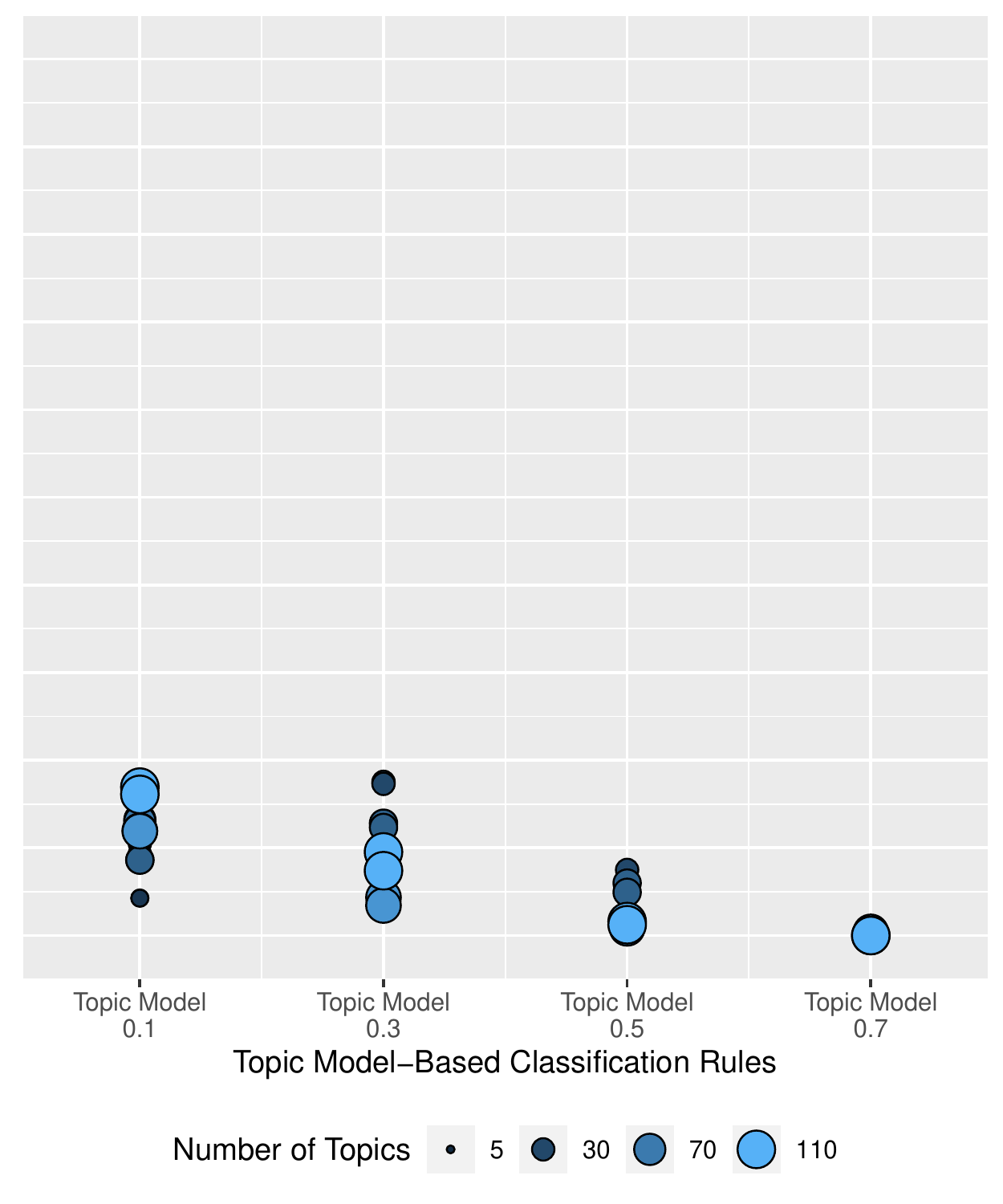}  \hspace{-0.3cm} \includegraphics[height=0.4\textwidth]{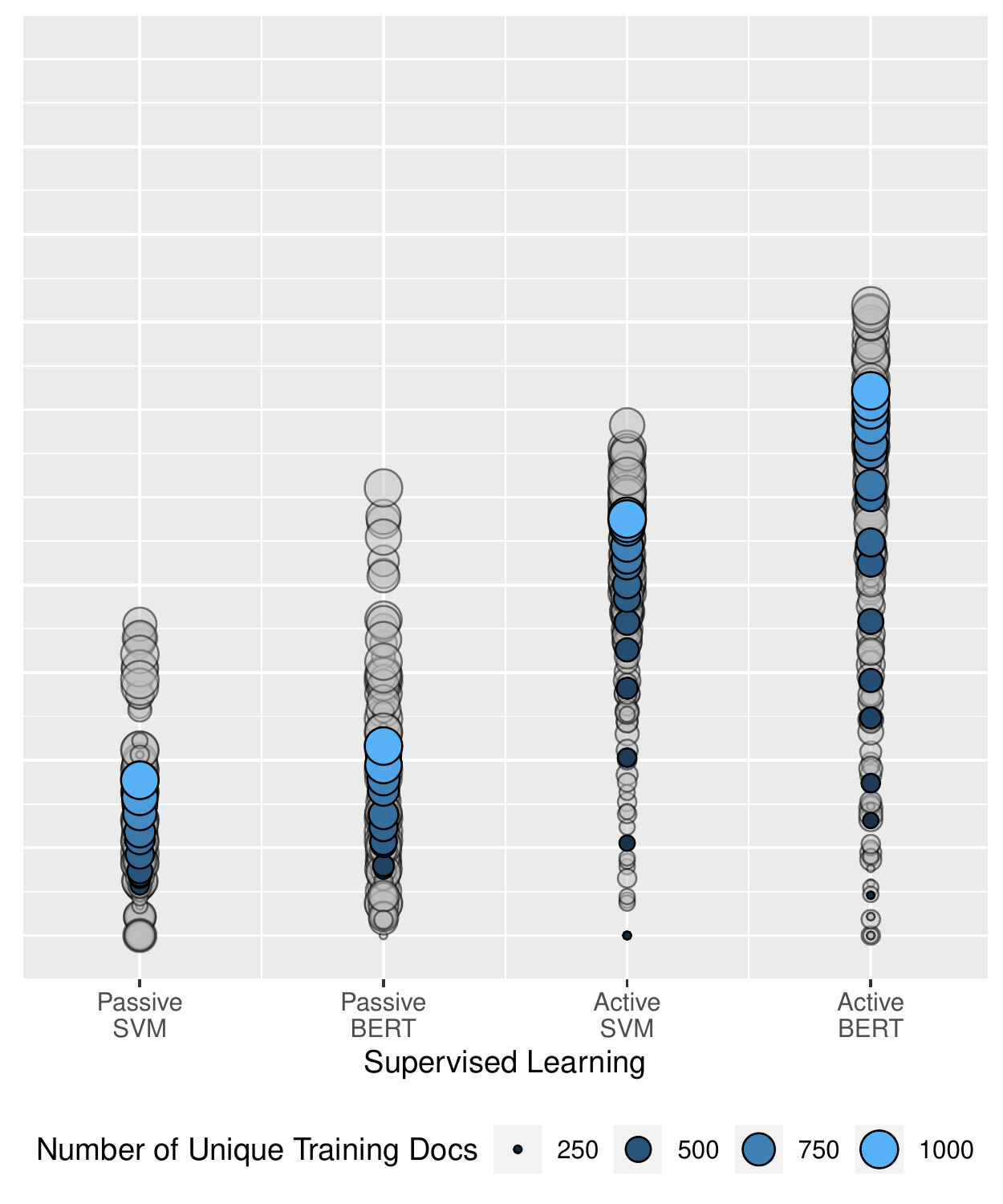}\\
\caption[SBIC: Comparison of Retrieval Methods]{\textbf{SBIC: Comparison of Retrieval Methods.} \footnotesize{This plot summarizes the retrieval performances of the here evaluated approaches on the retrieval task associated with the SBIC. \textbf{Left panel:} $F_1$-Scores for the 100 lists of 10 predictive keywords that then are expanded in the local and global embedding spaces. A gray transparent dot marks the $F_1$-Score reached by a single (expanded) keyword list. A blue opaque dot marks the mean of the $F_1$-Scores across 100 (expanded) keyword lists that contain the same number of search terms. The larger the number of terms in an (expanded) keyword list, the larger the size and the lighter the color of the printed dots. \textbf{Panel at the center:} $F_1$-Scores of the topic model-based classification rules with different values for threshold $\xi$. The larger the number of topics in an estimated CTM, the larger the size and the lighter the color of the printed dots. (From all CTMs with a given topic number that have been estimated here, the best two performing combinations regarding the question how many and which topics are considered relevant are presented.) \textbf{Right panel:} $F_1$-Scores for active as well as passive supervised learning with SVM and BERT. A gray transparent dot marks the $F_1$-Score reached by a model that has been trained by a given number of unique training documents on one set aside test set $g$. A blue opaque dot marks the mean of the $F_1$-Scores of 10 such models that have been trained on the same number of unique training documents. Each of the 10 models is evaluated on another test set and together the 10 test sets constitute the entire SBIC. The larger the number of unique labeled training instances, the larger the size and the lighter the color of the printed dots.}}
\label{fig:allressbic} 
\end{sidewaysfigure}

\begin{sidewaysfigure}
  \centering
\includegraphics[height=0.4\textwidth]{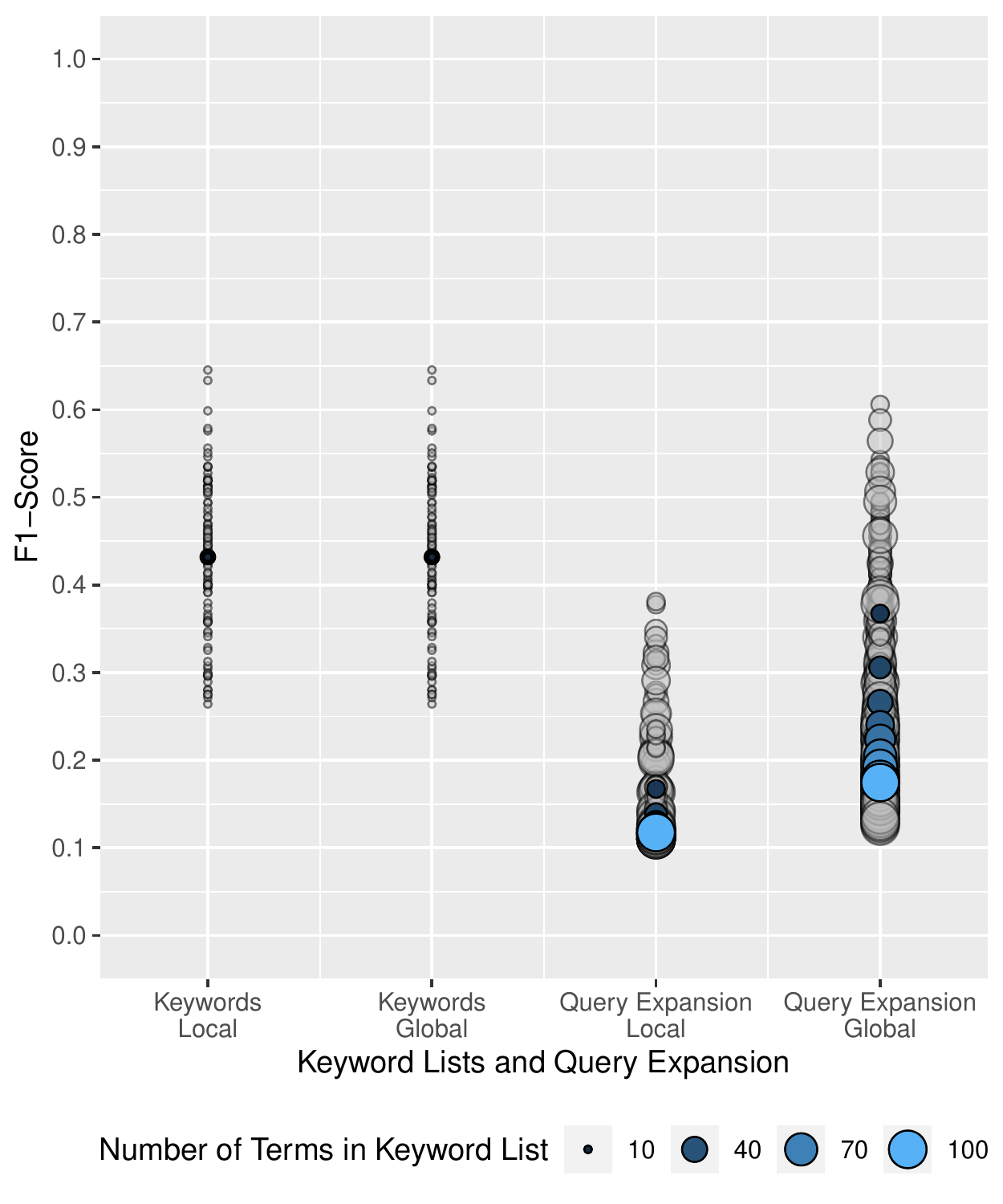} \hspace{-0.3cm} \includegraphics[height=0.4\textwidth]{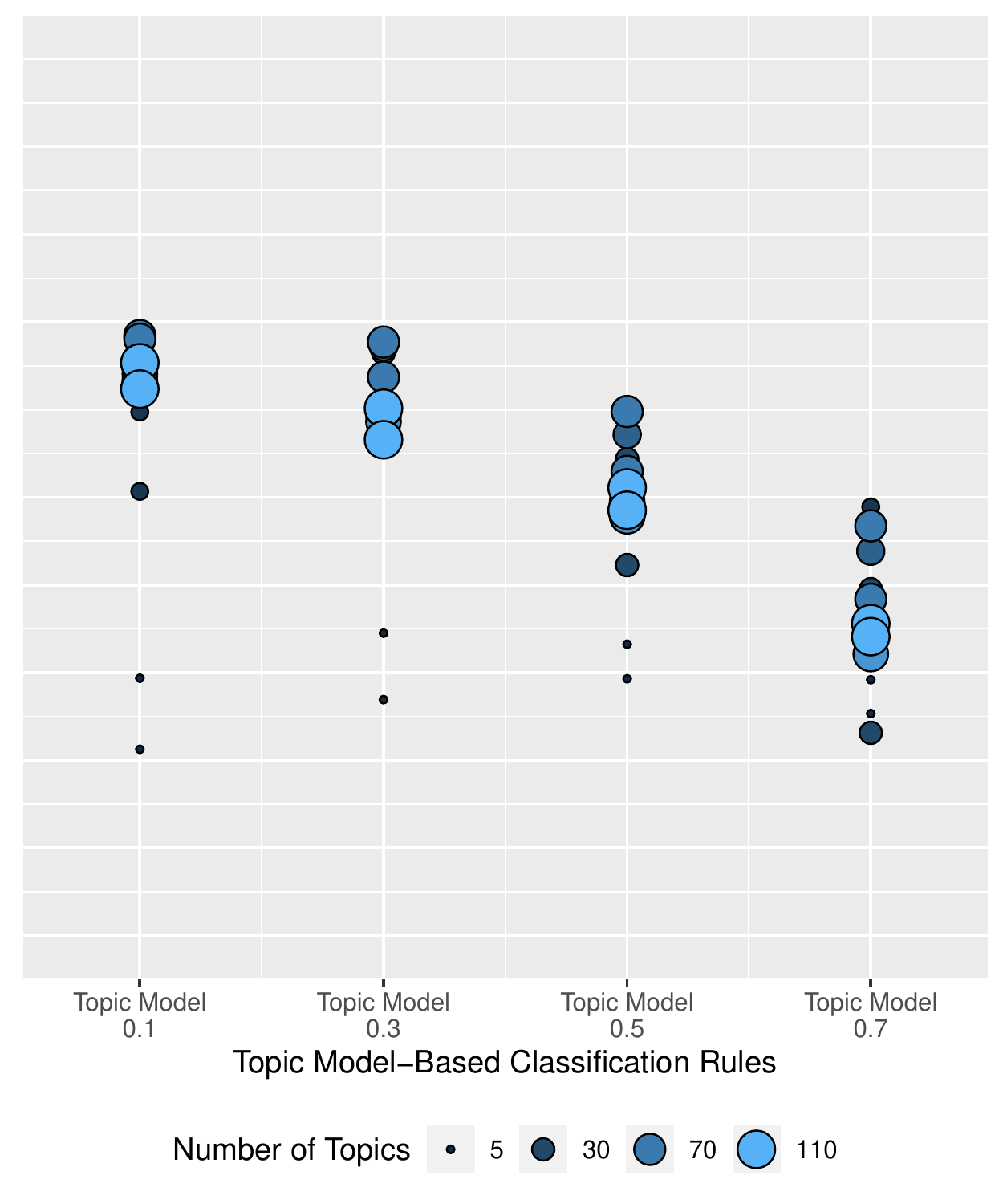}  \hspace{-0.3cm} \includegraphics[height=0.4\textwidth]{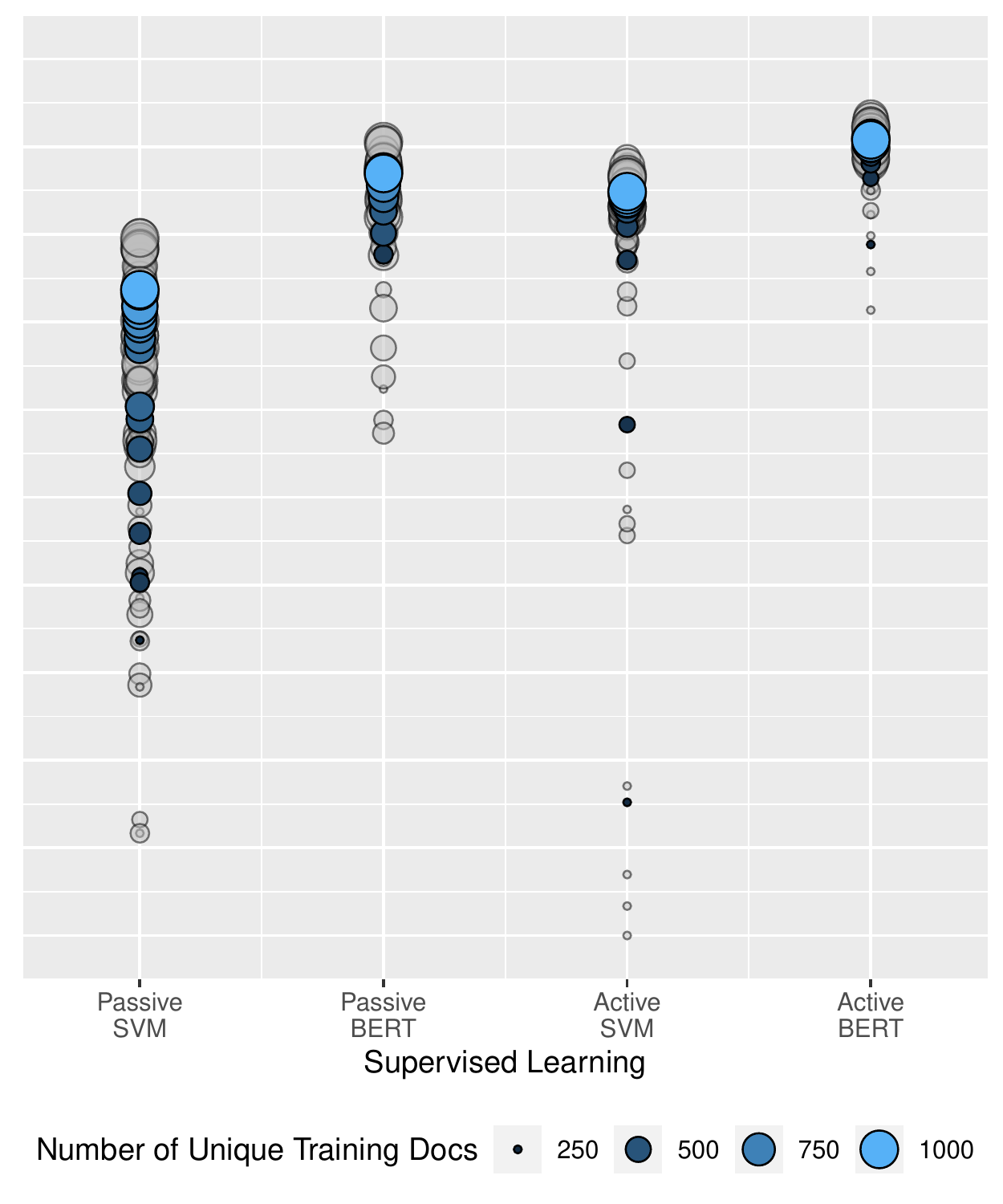}\\
\caption[Reuters: Comparison of Retrieval Methods]{\textbf{Reuters: Comparison of Retrieval Methods.} \footnotesize{This plot summarizes the retrieval performances of the here evaluated approaches on the retrieval task associated with the Reuters-21578 corpus. \textbf{Left panel:} $F_1$-Scores for the 100 lists of 10 predictive keywords that then are expanded in the local and global embedding spaces. A gray transparent dot marks the $F_1$-Score reached by a single (expanded) keyword list. A blue opaque dot marks the mean of the $F_1$-Scores across 100 (expanded) keyword lists that contain the same number of search terms. The larger the number of terms in an (expanded) keyword list, the larger the size and the lighter the color of the printed dots. \textbf{Panel at the center:} $F_1$-Scores of the topic model-based classification rules with different values for threshold $\xi$. The larger the number of topics in an estimated CTM, the larger the size and the lighter the color of the printed dots. (From all CTMs with a given topic number that have been estimated here, the best two performing combinations regarding the question how many and which topics are considered relevant are presented.) \textbf{Right panel:} $F_1$-Scores for active as well as passive supervised learning with SVM and BERT. A gray transparent dot marks the $F_1$-Score reached by a model that has been trained by a given number of unique training documents on one set aside test set $g$. A blue opaque dot marks the mean of the $F_1$-Scores of 5 such models that have been trained on the same number of unique training documents. Each of the 5 models is evaluated on another test set and together the 5 test sets constitute the entire Reuters-21578 corpus. The larger the number of unique labeled training instances, the larger the size and the lighter the color of the printed dots.}}
\label{fig:allresreuters} 
\end{sidewaysfigure}

The central question this study seeks to answer is: What, if anything, can be gained by applying more costly retrieval approaches as query expansion, topic model-based classification rules, or supervised learning instead of the relatively simple and inexpensive usage of a boolean query with a keyword list? In order to finally answer this question and compare the approaches against each other, Figures \ref{fig:allrestwitter}, \ref{fig:allressbic}, and \ref{fig:allresreuters} summarize the retrieval performance---as measured by the $F_1$-Score---of the evaluated approaches on the retrieval tasks associated with the Twitter corpus (Figure \ref{fig:allrestwitter}), the SBIC (Figure \ref{fig:allressbic}), and the Reuters-21578 corpus (Figure \ref{fig:allresreuters}). In each Figure, the left panel gives the $F_1$-Scores for the lists of 10 predictive keywords that then are expanded in the local and global embedding spaces. The middle panel shows the $F_1$-Scores of topic model-based classification rules with different values for threshold $\xi$. The right panel visualizes the $F_1$-Scores for active as well as passive supervised learning with SVM and BERT. 

In general, the direct comparison shows that, when taking keyword lists comprising 10 empirically predictive terms as the baseline, then the application of more complex and more expensive retrieval techniques does not guarantee better retrieval results. 

Query expansion techniques here rather decrease than increase the $F_1$-Score. Minimal improvements only occur sporadically in the embedding space trained on external global corpora if the increase in recall outweighs the decrease in precision. The farther the expansion, the worse the results tend to become. 

In general, it seems that the identification of newspaper articles referring to crude oil from the Reuters corpus is a more simple task than the retrieval of tweets relating to the multi-dimensional refugee topic or the extraction of posts that refer to an entity (disabled people) and are of a particular kind (here: disrespectful). All approaches show higher $F_1$-Scores on the Reuters corpus, and lower scores for the other two evaluated retrieval tasks. 

Topic model-based classification rules work relatively well for the Reuters corpus but not the other corpora. Hence, if there are no coherent and exclusive topics that cover the entity of interest in all its aspects, topic model-based classification rules exhibit rather poor retrieval performances. In the Twitter and SBIC data sets, the $F_1$-Score reached by the topic model-based classification rules are in the lower range of the values achieved by the lists of predictive keywords. If, on the other hand, coherent and exclusive topics relating to the entity of interest exist (as is the case for the Reuters corpus), acceptable retrieval results are possible. Here, gains over the best performing keyword lists are achieved for combinations with larger topic numbers and smaller values for threshold $\xi$.
The best performing topic model-based classification rule on the Reuters corpus is based on a CTM with 70 topics that considers 2 topics to be relevant and predicts documents to be relevant that have 10$\%$ of their words assigned to these two relevant topics. This topic model-based classification rule reaches an $F_1$-Score of 0.685, which is 0.04 higher than the $F_1$-Score of the best performing keyword list.

Whereas query expansion techniques and topic model-based classification rules show no or small improvements, supervised learning---if conducted in an active learning mode---has the potential to yield a substantively higher retrieval performance than a list of 10 predictive keywords. The prerequisite for this, however, is that not too few training instances are used. 
The larger the number of training instances, the higher the $F_1$-Score tends to be. Yet, as has been established above, especially for BERT this relationship is not monotonic and can exhibit considerable variability. What number of training documents is required to produce acceptable retrieval results that are better than what could be achieved with a keyword list, depends on the specifics of the retrieval task at hand and the employed learning mode and model. In the Twitter application, for example, it is likely to achieve an acceptable to good retrieval performance that improves upon keyword lists when applying active learning with BERT using $\geq 350$ training documents or applying passive learning with BERT on $\geq 800$ unique training documents (see again Figure \ref{fig:superres}).

In general, active learning is to be preferred over passive learning, as across applications and learning models, active learning tends to reach a higher retrieval performance than passive learning with the same number of training documents. Moreover, the pretrained deep neural network BERT tends to yield a higher $F_1$-Score compared to an SVM operating on bag-of-words-based document representations. BERT produces more unstable results but this behavior can be monitored with a set aside test set.

Across applications, applying active learning with BERT until 1,000 training instances have been labeled here produces a good separation of relevant and irrelevant documents that considerably improves upon the separation achieved by applying a keyword list. The mean $F_1$-Scores of BERT applied in an active learning mode with a training budget of 1,000 labeled instances are 0.712 (Twitter), 0.622 (SBIC), and 0.908 (Reuters), whereas the \emph{maximum} $F_1$-Scores reached by the empirically constructed initial keyword lists are 0.417 (Twitter), 0.404 (SBIC), and 0.645 (Reuters). Hence, the improvements in the $F_1$-Scores that are achieved by applying active learning with BERT rather than the best performing keyword list are 0.295 (Twitter), 0.218 (SBIC), and 0.263 (Reuters). If applying BERT-like models exceeds available capacities, active learning with a conventional machine learning model as an SVM still is a good and viable alternative.\footnote{If a large share of documents in the corpus at hand are longer than the 512 tokens that can be processed by BERT, Transformer-based models that can process longer sequences of tokens, e.g.~the Longformer \citep{Beltagy2020}, can be applied.} The mean $F_1$-Scores of active learning with SVM trained with 1,000 labeled documents are 0.538 (Twitter), 0.475 (SBIC), and 0.849 (Reuters). This still yields enhancements of the $F_1$-Score by 0.121 (Twitter), 0.071 (SBIC), and 0.204 (Reuters).


Note that the performance enhancements of active learning here are observed across applications. Irrespective of document length, textual style, the type of the entity of interest, and the homogeneity or heterogeneity of the corpus from which the documents are retrieved, active learning with 1,000 training documents shows superior performance to keyword lists and the other approaches.

\section{Conclusion}  \label{sec:conclusion} 

In text-based analyses researchers typically are interested to study documents referring to a particular entity. Yet, textual references to specific entities are often contained within multi-thematic corpora. In consequence, documents that contain references toward the entities of interest have to be separated from those that do not. 

A very common approach in social science to retrieve relevant documents is to apply a list of keywords. Keyword lists are inexpensive and easy to apply, but they may result in biased inferences if they systematically miss out relevant documents. 
Query expansion techniques, topic model-based classification rules and active as well as passive supervised learning constitute alternative, more expensive, more complex, and in social science rarely applied procedures for the retrieval of relevant documents. These more complex procedures theoretically can have the potential to reach a higher retrieval performance than keyword lists and thus to reduce the potential size of selection biases. So far, a systematic comparison of these approaches was lacking and therefore it was unclear, whether the employment of any of these more expensive methods would yield any improvements of retrieval performance, and if so how large and consistent across contexts the improvement would be. 

This study closed this gap. The comparison of the approaches on the basis of retrieval tasks associated with a data set of German tweets \citep{Linder2017}, the Social Bias Inference Corpus (SBIC) \citep{Sap2020}, and the Reuters-21578 corpus \citep{Reuters1997} shows that neither of the applied more complex approaches necessarily enhances the retrieval performance, as measured by the $F_1$-Score, over the application of a keyword list containing 10 empirically predictive terms. Yet, whereas across all settings and combinations evaluated for query expansion techniques and topic model-based classification rules at the very best small increases in the $F_1$-Score can be observed, active supervised learning---with the Transformer-based language representation model BERT---increases the $F_1$-Scores across application contexts substantively if the number of labeled training documents used in the active learning process is not too small. 

Thus, in terms of retrieval performance, supervised learning in an active learning mode---preferably with a pretrained deep neural network---is the procedure to be preferred to all other approaches. However, this procedure is also the most expensive of the evaluated methods. Supervised learning implies human, financial, and time resources for annotating the training documents. A training data set comprising 1,000 instances is very small for usual supervised learning settings, but the coding process also has to be monitored and coordinated. Moreover, active learning involves a dynamic labeling process in which after each iteration those documents are annotated for which a label is requested by the model. While active learning reduces the overall number of training instances for which a label is required, the dynamic labeling process may increase coordination costs or the time coders spend on coding as they wait for the model to request the next labels. 

The precise separation of documents that refer to the entity of interest and thus are relevant for the planned study at hand from documents that are irrelevant is an essential analytic step. This step defines the set of documents on which all the following core analyses are conducted. Selection biases induced by the applied retrieval method ultimately bias the study's results. Therefore, attention and care should be taken when it comes to extracting relevant documents. Compared to the creation of a set of keywords, active learning requires substantive amounts of additional resources. But given the observed considerably higher retrieval performances achieved by active learning compared to keyword lists, spending these resources is likely to be worthwhile for the quality of the study.

The aim of this study was to compare different learning approaches: keyword lists, query expansion techniques, topic model-based classification rules, and active as well as passive supervised learning. One problem that naturally arises if one seeks to compare learning approaches is that different approaches can only be compared on the basis of specific models that follow specific procedures (e.g. for query expansion), that have specific hyperparameter settings, that are trained on a specific finite set of training documents and are evaluated on a specific finite set of test set documents, and that are initialized by specific random seed values \citep{Reimers2018}. Here, care was taken to have a broad range of several specific models with different settings for each approach. With regard to keyword lists and query expansion, 100 different keyword lists were expanded in local as well as global embedding spaces (whereby the number of expansion terms was varied from 1 to 9). For topic-model-based classification rules 4 different values for threshold $\xi$ for each of 426,725 combinations were evaluated. With regard to active and passive supervised learning two different types of models (SVM and BERT) were applied 10 or 5 times with different random initializations. In each of the 10 or 5 runs, a different initial training set was used that then was enlarged by passive random sampling or active selection in 15 iterations. 
This broad evaluation setting makes it more likely that the conclusions drawn here on the set of models evaluated for each approach hold and that active learning indeed is superior to keyword lists for the studied tasks. 

Nevertheless, future studies might inspect the effect of the chosen model settings of the evaluated approaches on the obtained results: Here, for each of the evaluated approaches the most simple setting or version was used. For query expansion, GloVe embeddings that represent each term by a single vector were employed and a simple boolean query using the OR operator was conducted. More complex procedures from the field of information retrieval that make use of contextualized embeddings and pseudo-relevant feedback \citep[e.g.][]{Zheng2020} were not applied. For the estimation of the topics, the CTM rather than, for example, a neural topic model was used  \citep[for an overview over neural topic models see][]{Zhao2021a}. For passive supervised learning, simple random oversampling was employed and for active learning uncertainty sampling was applied as a query strategy rather than more complex procedures as query-by-committee or expected model change \citep[see e.g.][]{Settles2010}. These simplest versions applied here present the core idea of each approach often most clearly and also are most easy to implement for scientists that seek to use one of the approaches as a first step in their analysis. Nevertheless, future studies might explore whether applying more complex procedures changes the substantive conclusions reached here. 

Finally, there is the question of in how far the conclusions drawn here on the basis of three applications travel to further contexts. The selected data sets and tasks differ with regard to textual length and style, the heterogeneity of the corpora, the characteristics of the entities of interest, and the share of relevant documents in the corpora. The finding that active supervised learning---if applied with a not too small amount of training instances---considerably increases the $F_1$-Score compared to keyword lists holds across these applications' differences. But further studies could look more closely at which contextual factors lead to which effects on retrieval performance for which procedures.

\newpage
\begin{appendix}

\section{Most Predictive Terms}
\label{app:mpt}

Note on Tables \ref{tab:featimptwitter} to \ref{tab:featimpreuter}: The keyword lists comprising empirically highly predictive terms are not only applied on the corpora to evaluate the retrieval performance of keyword lists, but also form the basis for query expansion (see Section \ref{seq:queryexp}). The query expansion technique makes use of GloVe word embeddings \citep{Pennington2014} trained on the local corpora at hand and also makes use of externally obtained GloVe word embeddings trained on large global corpora. In the case of the locally trained word embeddings there is a learned word embedding for each predictive term. Thus, the set of extracted highly predictive terms can be directly used as starting terms for query expansion. In the case of the globally pretrained word embeddings, however, not all of the highly predictive terms have a corresponding global word embedding. Hence, for the globally pretrained embeddings the 50 most predictive terms \emph{for which a globally pretrained word embedding is available} are extracted. If a predictive term has no corresponding global embedding, the set of extracted predictive terms is enlarged with the next most predictive term until there are 50 extracted terms. In consequence, below for each corpus two lists of the most predictive features are shown.
 \begin{table}[H]
 \centering
 \begingroup\scriptsize
  {\renewcommand{\arraystretch}{0.99}
 \begin{tabular}{l|l}
 \hline
 \textbf{A} & \textbf{B} \\
   \hline
\#fl\"uchtlinge      &     migranten   \\    
migranten        &     asylanten    \\    
\#refugeeswelcome     &       fl\"uchtlingen  \\    
 \#refugee  &        asylrecht    \\        
asylanten             &asylunterkunft  \\   
fl\"uchtlingen      &      asyl    \\    
asylrecht           &     fl\"uchtlinge     \\    
\#migration        & asylbewerber        \\       
 \#fluechtlinge        &       ausl\"ander    \\    
 asylunterkunft        &       fl\"uchtlingsheim  \\    
 asyl            &         refugees         \\    
 fl\"uchtlinge         &           fl\"uchtling  \\      
migrationshintergrund    &        asylpolitik   \\    
asylbewerber        &          ungarn      \\    
ausl\"ander        &        refugee     \\    
fl\"uchtlingsheim      &  mittelmeer \\          
 fl\"uchtlingsheime     &        kritisiert      \\    
 fl\"uchtlingskrise      &     syrer   \\    
 \#schauhin         &       syrischen       \\    
 refugees              &  bamberg  \\      
 fl\"uchtling            &          brandanschlag     \\    
 asylpolitik      &         behandelt        \\    
 ungarn        &           merkels       \\      
 \#refugeecamp        &        ermittelt        \\      
 \#bloggerfuerfluechtlinge   &   zusammenhang  \\    
 \#refugees            &         innenminister   \\    
 \#fl\"uchtlingen   &           kundgebung       \\    
 fl\"uchtlingsheimen      &  pegida     \\      
 \#asyl         &        welcome        \\    
 refugee        &           unterbringung     \\    
 mittelmeer         &             migration       \\    
 kritisiert             &   ben\"otigt  \\      
 syrer            &       erfahrungen     \\    
 \@proasyl         &                   sollen  \\     
 syrischen          &         heimat        \\       
 bamberg            &    tja      \\         
 brandanschlag         &   balkanroute    \\      
 behandelt         &         rechte     \\    
 merkels            &       dort       \\    
 ermittelt       &   b\"urger    \\           
 zusammenhang      &          merkel    \\         
 innenminister      &             demo        \\    
 kundgebung           &    mehrheit    \\    
 pegida    &     letztes  \\                   
 welcome              &         geplante   \\    
 \#deutschland       &        recht     \\     
 \@refugeeswlcm\_le     &    hilfe  \\      
 unterbringung     &   europa    \\            
\@ndaktuell           &    afghanistan  \\        
migration       &     islam        \\          
    \hline
 \end{tabular}}
 \endgroup
  \caption[Most Predictive Features in the Twitter Data Set]{\textbf{Most Predictive Features in the Twitter Data Set.} \footnotesize{\textbf{A}: This is a list of the 50 terms that are extracted as the most predictive features for the relevant class of documents that refer to the refugee topic (most predictive term at the top). From this list of terms, 100 samples of 10 keywords are sampled to construct initial keyword lists that then are extended via query expansion using the locally trained GloVe word embeddings. \textbf{B}: This is a list of the 50 terms \emph{for which a globally pretrained GloVe word embedding is available} that are extracted as the most predictive features for the relevant class of documents that refer to the refugee topic (most predictive term at the top). From this list of terms, 100 samples of 10 keywords are sampled to construct initial keyword lists that then are extended via query expansion using the globally trained GloVe word embeddings.}}
\label{tab:featimptwitter}
 \end{table}
\newpage

 \begin{table}[H]
 \centering
 \begingroup\scriptsize
   {\renewcommand{\arraystretch}{0.99}
 \begin{tabular}{l|l}
 \hline
 \textbf{A} & \textbf{B} \\
   \hline
   retard     &    retard  \\    
    retards   &     retards     \\    
     retarded  &    retarded    \\    
  quadriplegic &        quadriplegic  \\        
     autistic        & autistic   \\   
   paralyzed  &     paralyzed    \\    
        schizophrenic  &    schizophrenic    \\    
   vegetables  &     vegetables        \\       
   wheelchair      &    wheelchair     \\    
     epileptic  &       epileptic    \\    
        parkinson's    &              disabled    \\    
        disabled  &    anorexic             \\      
  anorexic      &     stevie     \\    
       stevie   &         cripples       \\    
        cripples       &        paralympics    \\    
 paralympics      &   adhd   \\          
    adhd    &   paraplegic    \\    
   paraplegic    & syndrome  \\    
      syndrome   &          paralysed     \\    
         paralysed       &        midget   \\      
   midget        &      leper         \\    
     leper    &           amputee    \\    
     amputee       &        cripple      \\      
      cripple    &          tons        \\      
     tons     & handicapped   \\    
   handicapped        &   bipolar   \\    
bipolar &           wheelchairs       \\    
  wheelchairs      &    dyslexic     \\      
       dyslexic      &          chromosome      \\    
     chromosome          &        blind      \\    
       blind  &       suicidal             \\    
   suicidal            &    crippled     \\      
       crippled        &        chromosomes  \\    
     chromosomes     &       vegetable               \\     
        vegetable         &        challenged          \\       
   alzheimer's         &       special   \\         
            challenged    &       veggie      \\      
           special   &             cancer         \\    
          veggie       &     spade          \\    
        cancer      &             helen    \\           
   spade      &   jenga          \\         
    helen      &      autism      \\    
    jenga           & medication       \\    
    autism   &    deaf      \\                   
     medication           &  logan     \\    
     deaf         &      depressed        \\     
     logan      &    christopher      \\      
         depressed      &   mentally        \\            
    christopher         &  potato      \\        
 mentally        &       shouted       \\          
    \hline
 \end{tabular}}
 \endgroup
  \caption[Most Predictive Features in the SBIC]{\textbf{Most Predictive Features in the SBIC.} \footnotesize{\textbf{A}: This is a list of the 50 terms that are extracted as the most predictive features for the relevant class of documents that are offensive toward disabled people (most predictive term at the top). From this list of terms, 100 samples of 10 keywords are sampled to construct initial keyword lists that then are extended via query expansion using the locally trained GloVe word embeddings. \textbf{B}: This is a list of the 50 terms \emph{for which a globally pretrained GloVe word embedding is available} that are extracted as the most predictive features for the relevant class of documents that are offensive toward disabled people (most predictive term at the top). From this list of terms, 100 samples of 10 keywords are sampled to construct initial keyword lists that then are extended via query expansion using the globally trained GloVe word embeddings.}}
\label{tab:featimpsbic}
 \end{table}
 \newpage

 \begin{table}[H]
 \centering
 \begingroup\scriptsize
   {\renewcommand{\arraystretch}{0.99}
 \begin{tabular}{l|l}
 \hline
 \textbf{A} & \textbf{B} \\
   \hline
      oil  &    oil  \\    
        crude   &      crude    \\    
     barrels   &    barrels   \\    
   barrel  &    barrel    \\        
   exploration     & exploration \\   
     energy  &       energy    \\    
         petroleum     &   petroleum    \\    
     production   &     production    \\       
    drilling     &  drilling      \\    
     bpd     &       bpd    \\    
          refinery  &     refinery    \\    
         opec     &       opec         \\      
    gulf       &      gulf      \\    
       tanker  &     tanker       \\    
         texas    &       texas      \\    
offshore      &   offshore   \\          
       canada   &   canada  \\    
   resources   &   resources  \\    
      sea  &      sea      \\    
     rigs     &         rigs   \\      
     out         &        out         \\    
    petrobras    &      petrobras    \\    
       rig        &      rig          \\      
       refineries     &              refineries         \\      
    agency     &   agency    \\    
   along     &    along     \\    
 depressed   &             depressed       \\    
   conoco      &    conoco      \\      
      raise      &             raise       \\    
  shelf         &        shelf    \\    
         iranian &        iranian         \\    
     platform     &       platform   \\      
          day       &       day      \\    
         maintain    &      maintain             \\     
         drill       &        drill             \\       
     total       &         total     \\         
        well     &       well          \\      
   deal     &               deal         \\    
            16          &     16          \\    
              fiscal      &         fiscal    \\           
      upon        &    upon        \\         
   postings      &      postings      \\    
   light     &        light      \\    
 blocks &     blocks   \\                   
   tuesday         &  tuesday       \\    
     about        &      about      \\     
     meters   &    meters       \\      
     daily         &   daily        \\            
  future    &  future   \\        
    equivalent       &     equivalent       \\          
    \hline
 \end{tabular}}                
 \endgroup
  \caption[Most Predictive Features in the Reuters-21578 Corpus]{\textbf{Most Predictive Features in  Reuters-21578 Corpus.} \footnotesize{\textbf{A}: This is a list of the 50 terms that are extracted as the most predictive features for the relevant class of documents that are about the crude oil topic (most predictive term at the top). From this list of terms, 100 samples of 10 keywords are sampled to construct initial keyword lists that then are extended via query expansion using the locally trained GloVe word embeddings. \textbf{B}: This is a list of the 50 terms \emph{for which a globally pretrained GloVe word embedding is available} that are extracted as the most predictive features for the relevant class of documents that are about the crude oil topic (most predictive term at the top). From this list of terms, 100 samples of 10 keywords are sampled to construct initial keyword lists that then are extended via query expansion using the globally trained GloVe word embeddings.}}
\label{tab:featimpreuter}
 \end{table}
\newpage

\section{Recall and Precision of Keyword Lists and Query Expansion}
\label{app:resquery2}

\begin{figure}[H]
   \centering
   \small
\begin{tabular}{cc}
Twitter-global-Recall & Twitter-global-Precision \\
\includegraphics[height=0.3\textwidth]{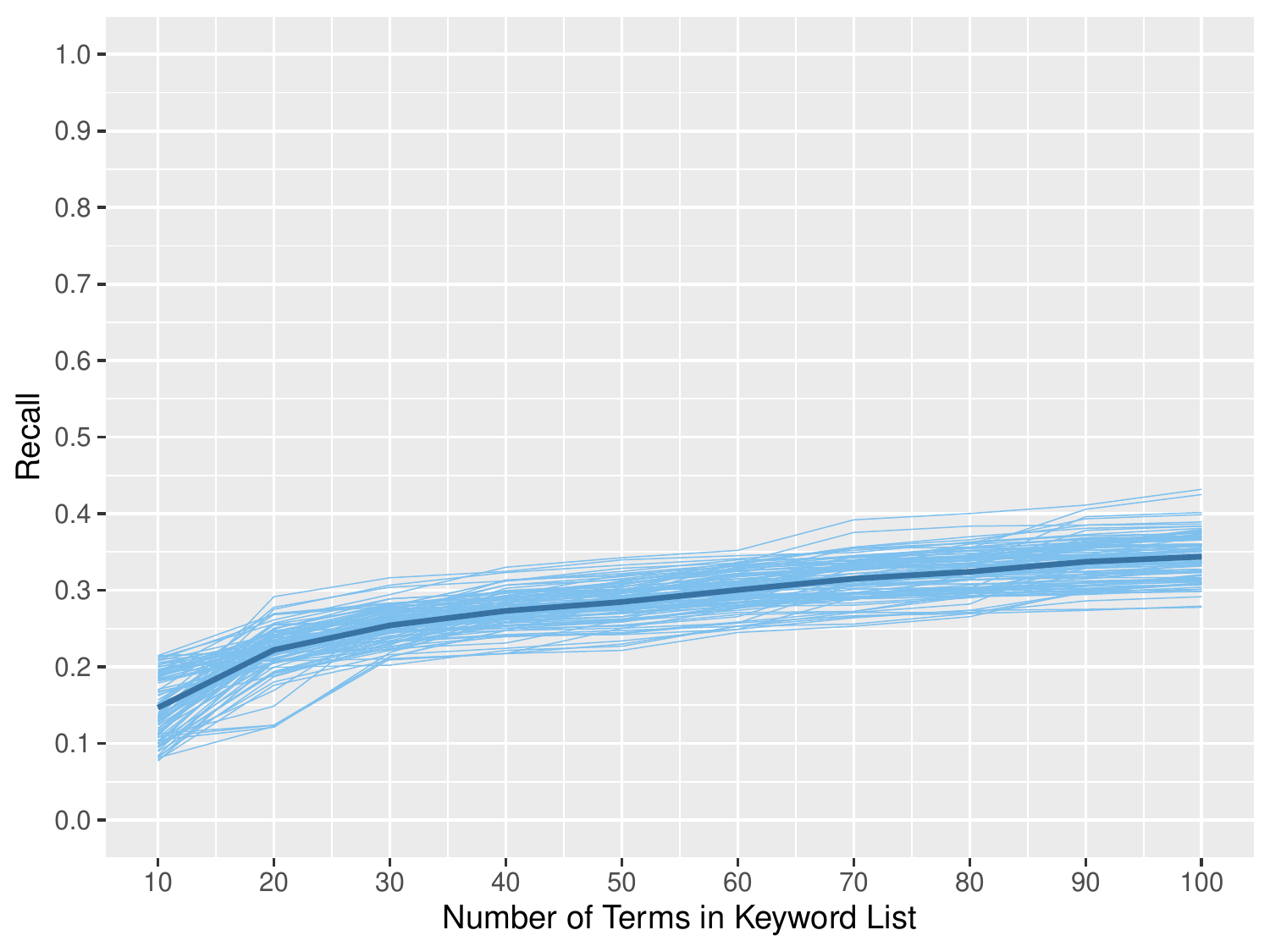}& \includegraphics[height=0.3\textwidth]{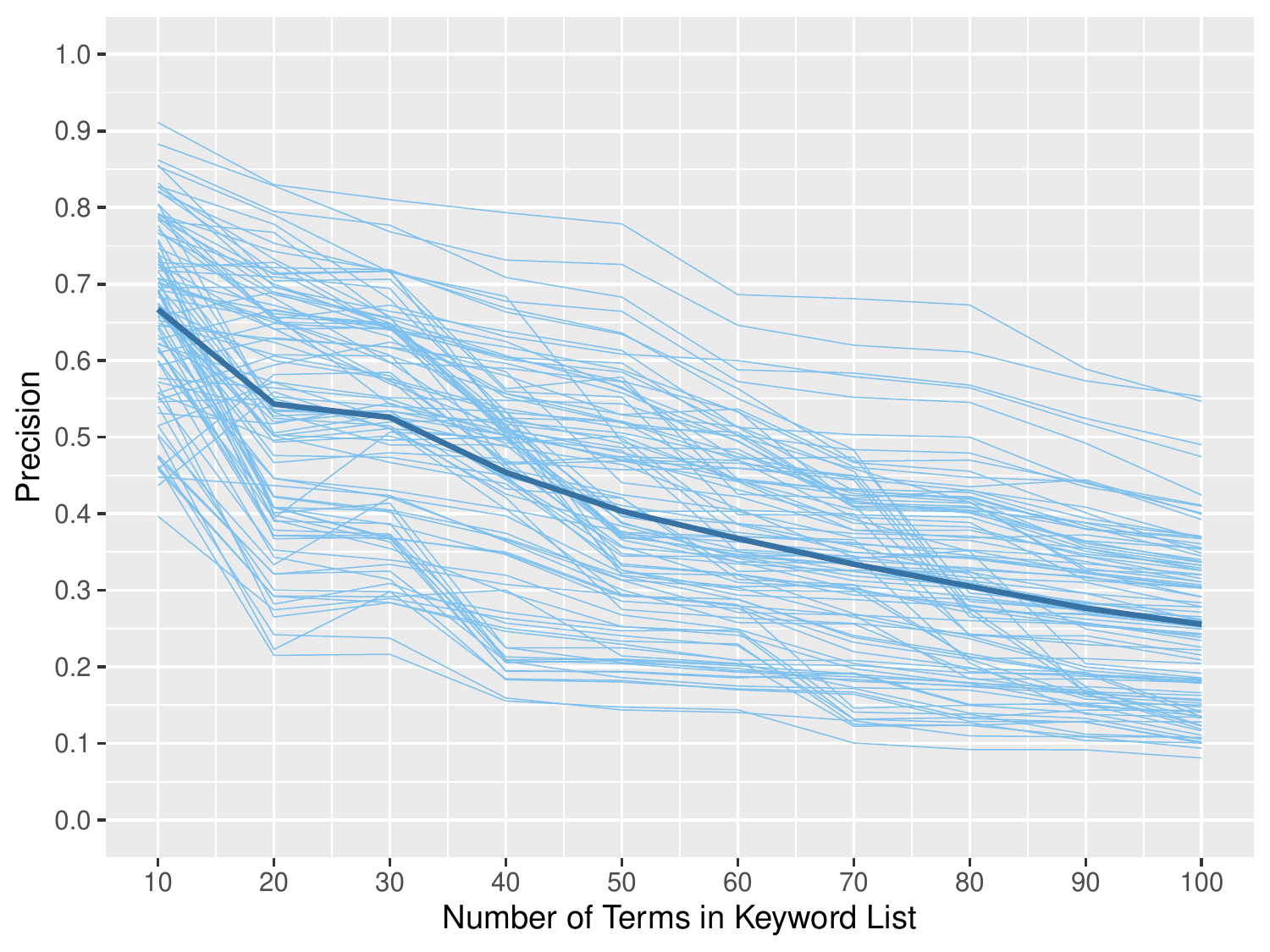} \\
SBIC-global-Recall & SBIC-global-Precision \\
\includegraphics[height=0.3\textwidth]{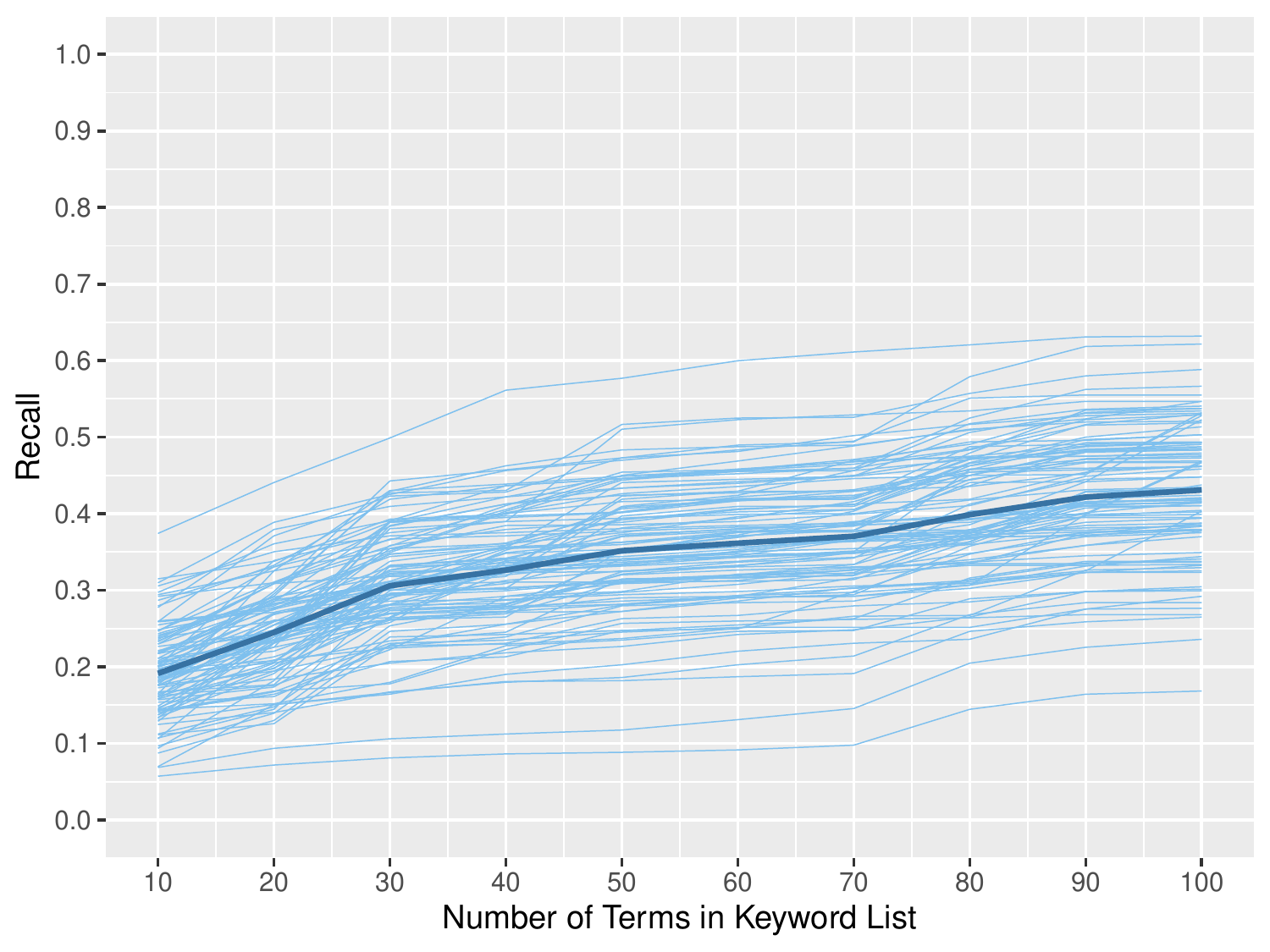}& \includegraphics[height=0.3\textwidth]{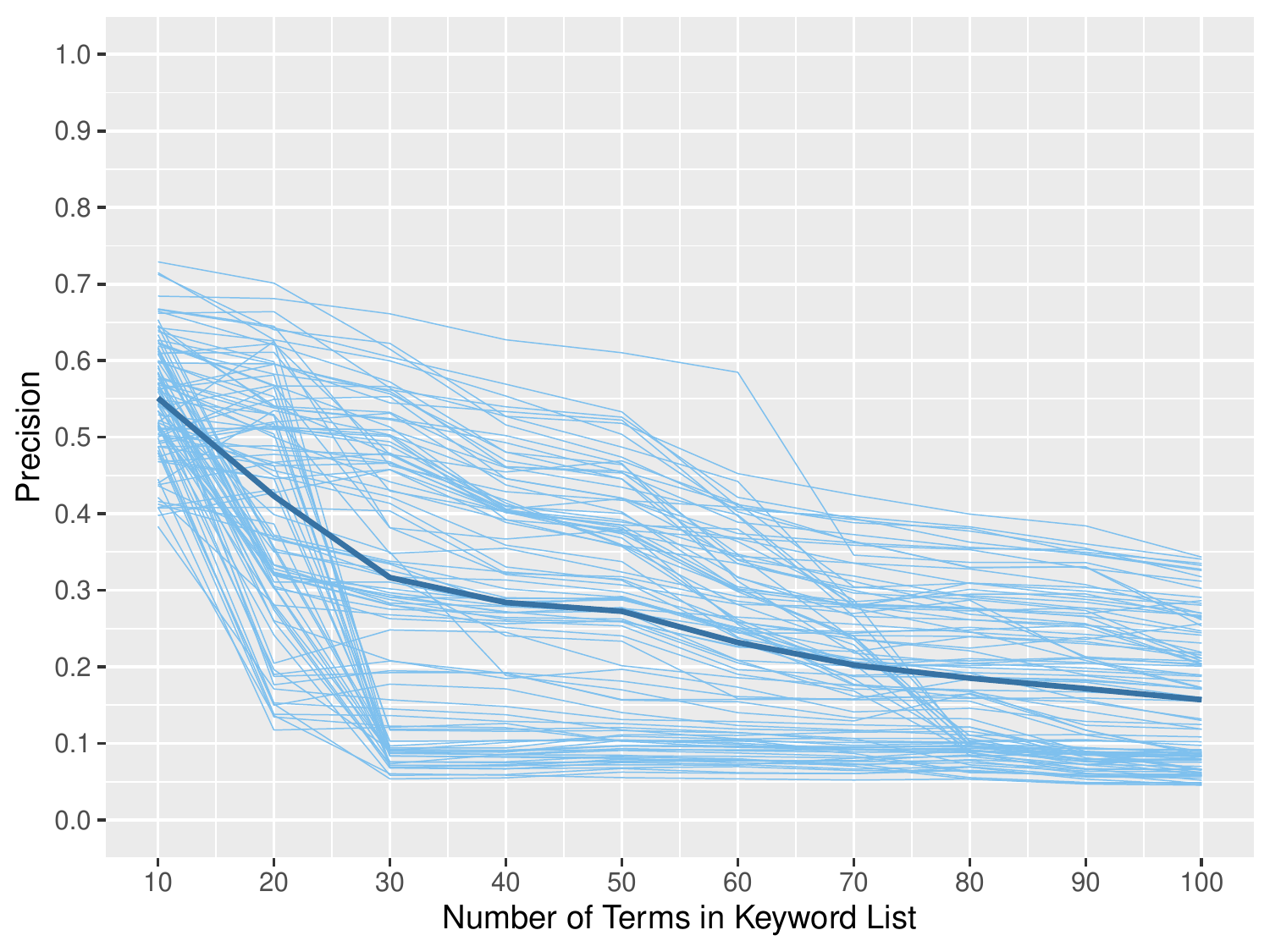} \\
Reuters-global-Recall & Reuters-global-Precision \\
\includegraphics[height=0.3\textwidth]{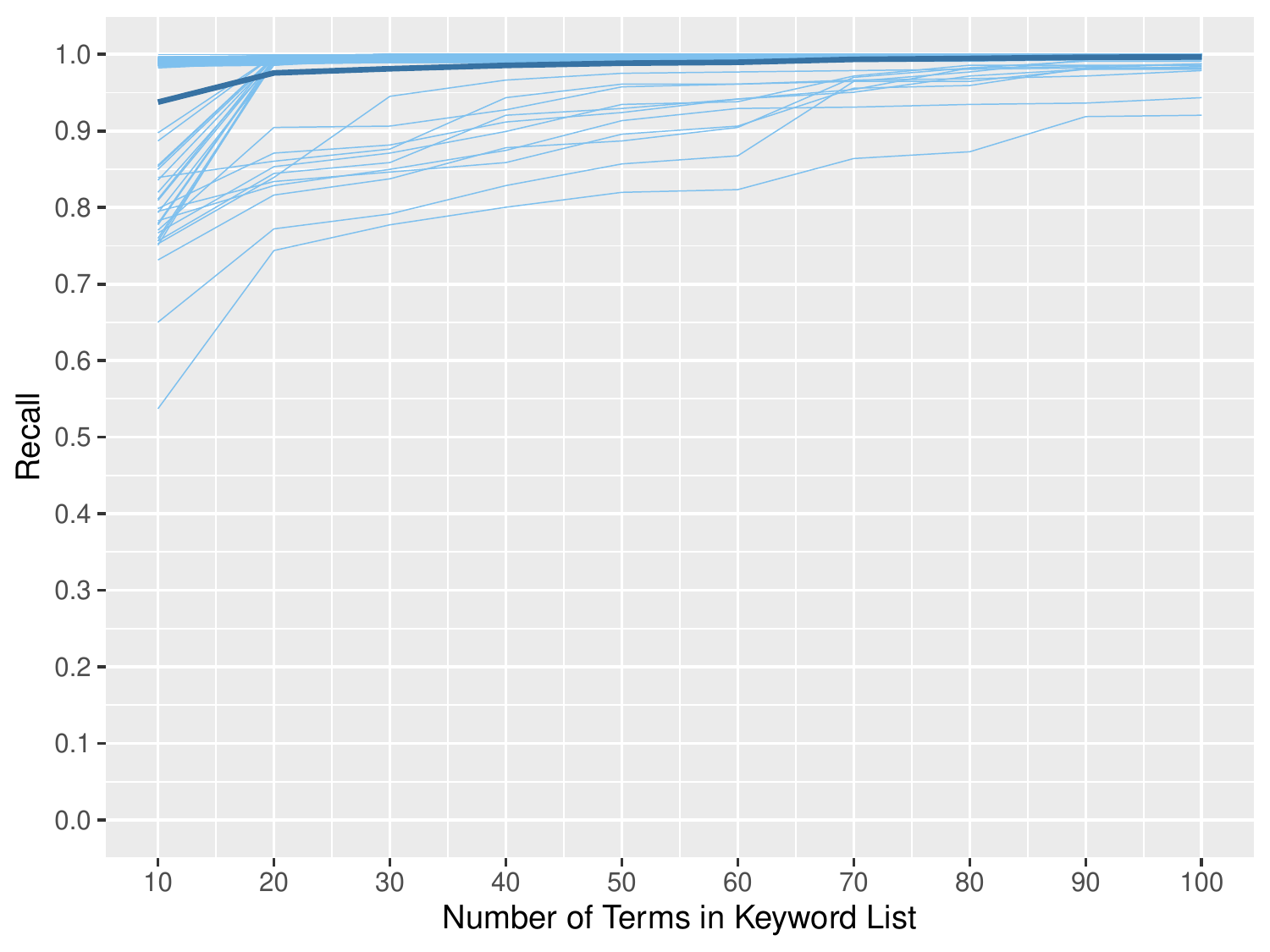} & \includegraphics[height=0.3\textwidth]{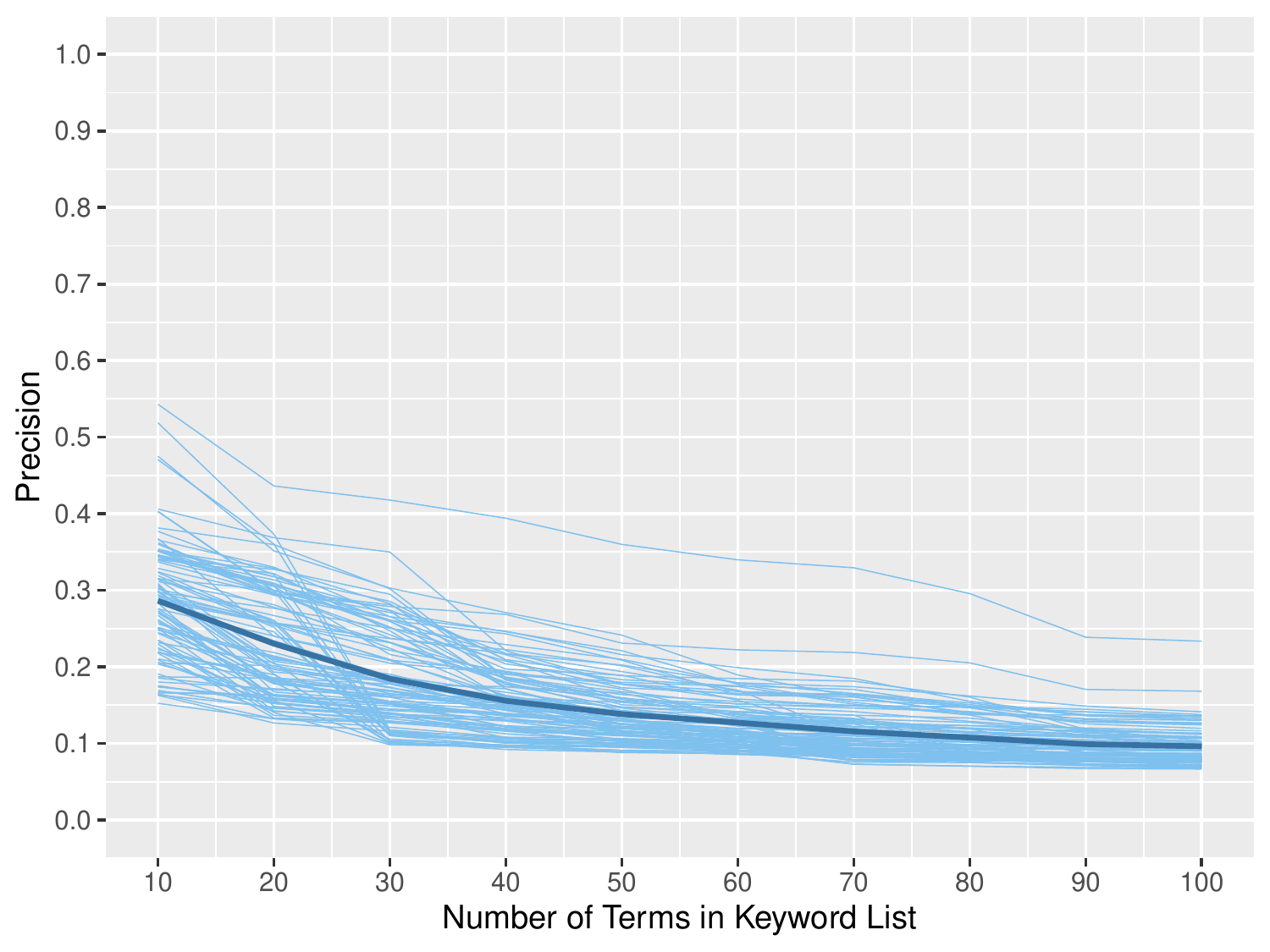}\\
\end{tabular}
\caption[Recall and Precision for Retrieving Relevant Documents with Keyword Lists and Global Query Expansion]{\textbf{Recall and Precision for Retrieving Relevant Documents with Keyword Lists and Global Query Expansion.} \scriptsize{This plot shows recall and precision scores resulting from the application of the keyword lists of 10 highly predictive terms as well as the evolution of the recall and precision scores across the query expansion procedure based on globally trained GloVe embeddings. For each of the sampled 100 keyword lists that then are expanded, one light blue line is plotted. The thick dark blue line gives the mean over the 100 lists.}}
\label{fig:queryres2global} 
\end{figure}

\begin{figure}[H]
   \centering
   \small
\begin{tabular}{cc}
Twitter-local-Recall &Twitter-local-Precision \\
\includegraphics[height=0.3\textwidth]{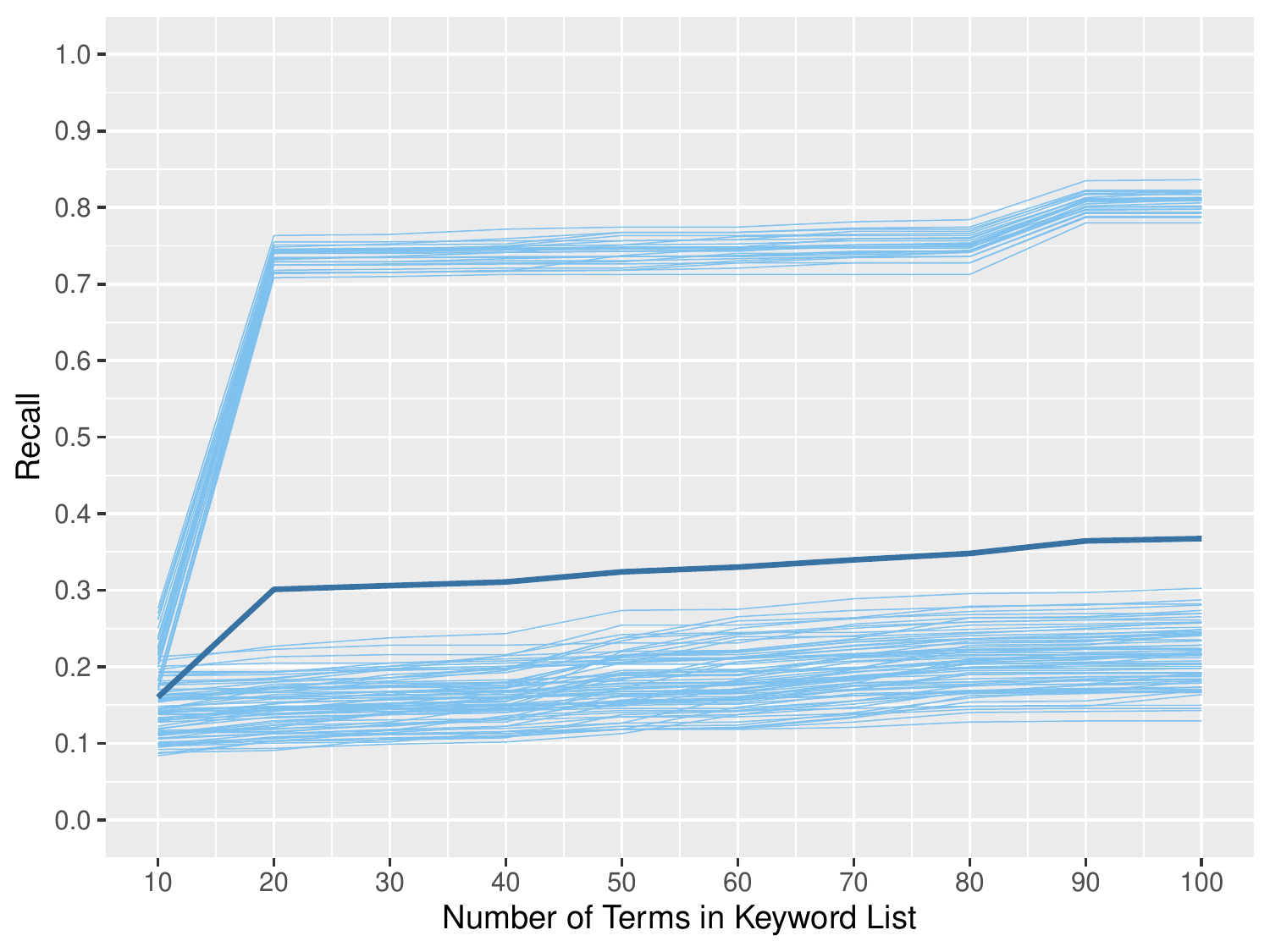}& \includegraphics[height=0.3\textwidth]{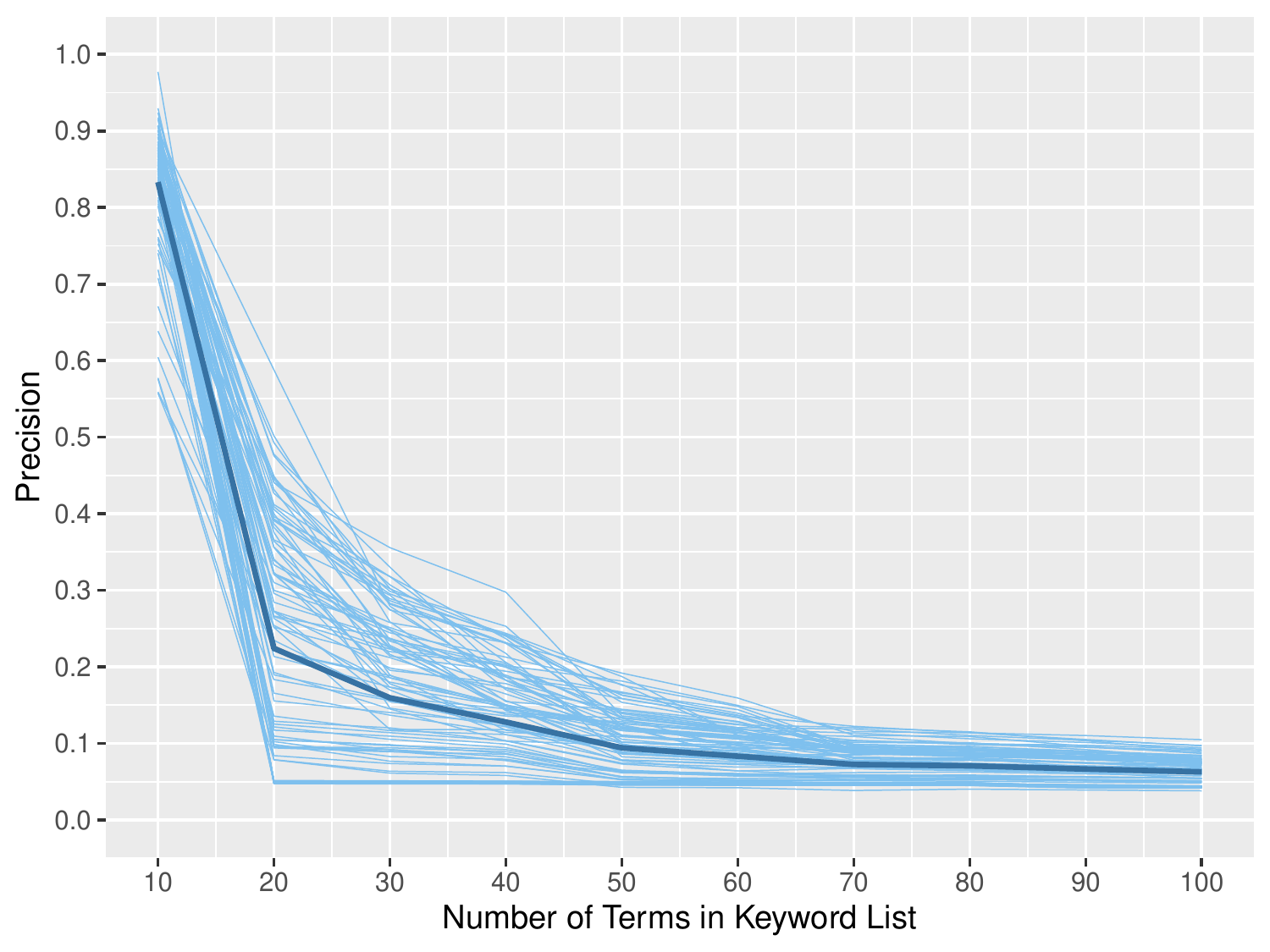} \\
SBIC-local-Recall & SBIC-local-Precision \\
\includegraphics[height=0.3\textwidth]{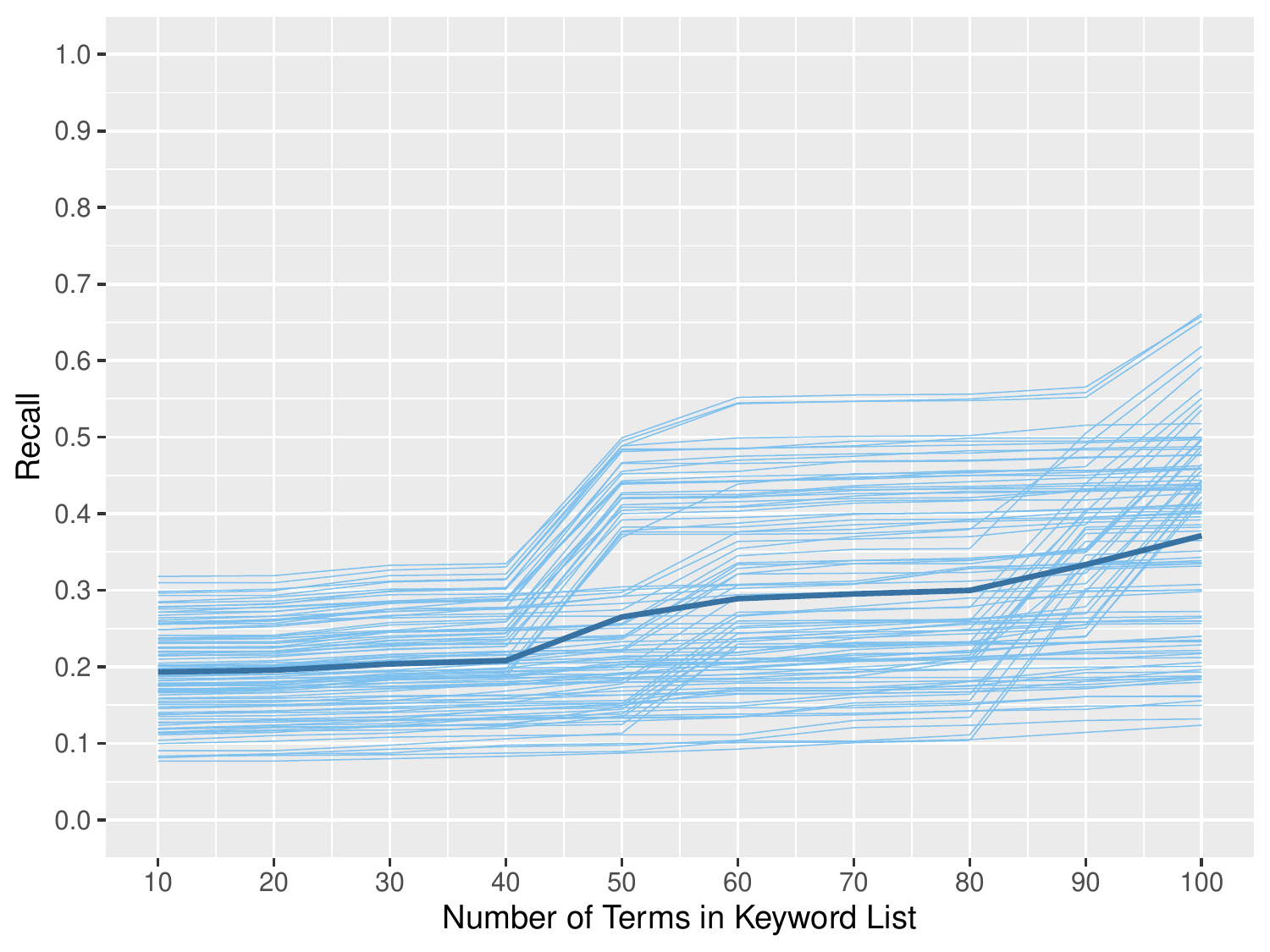}& \includegraphics[height=0.3\textwidth]{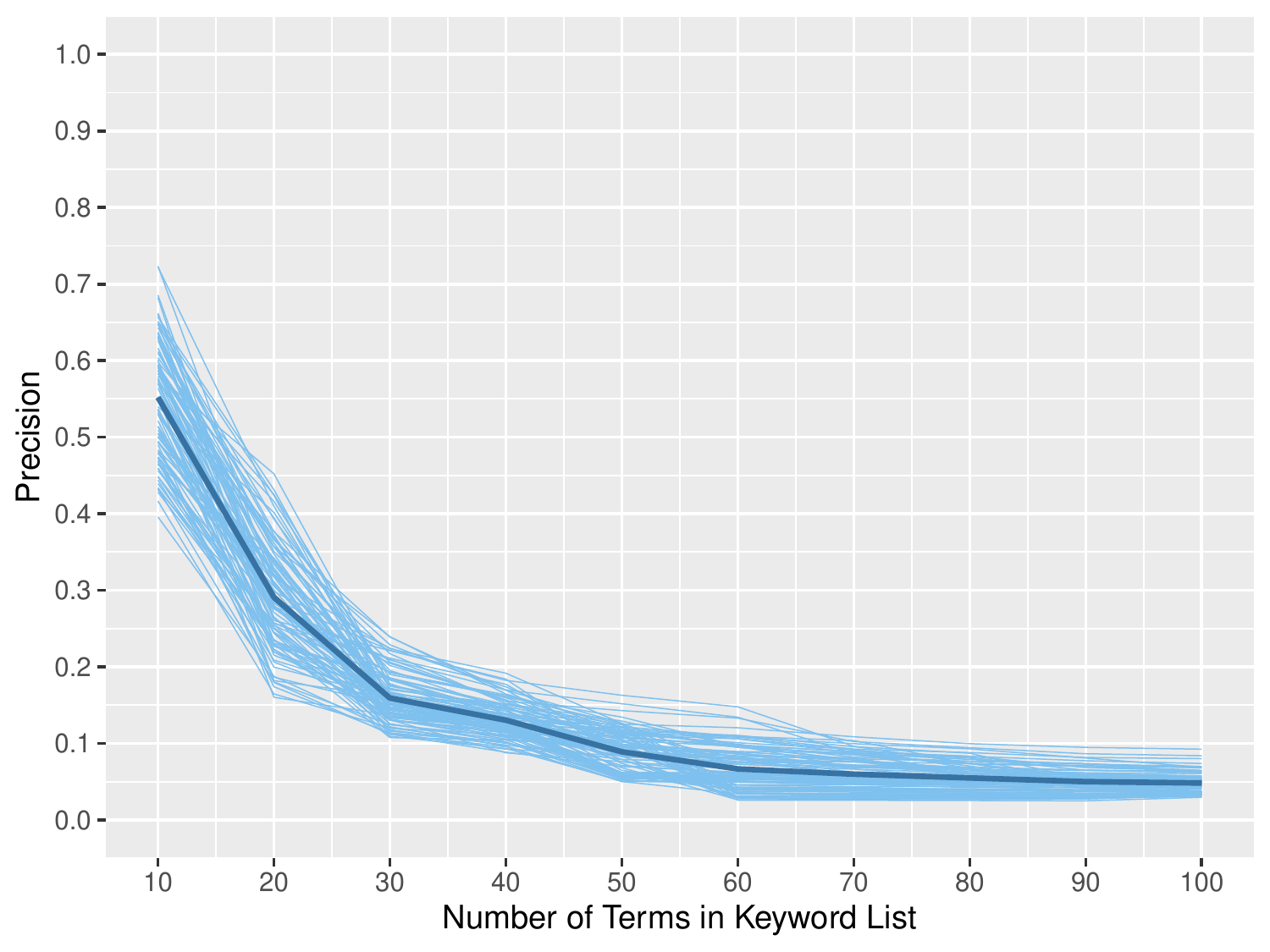} \\
Reuters-local-Recall &Reuters-local-Precision \\
\includegraphics[height=0.3\textwidth]{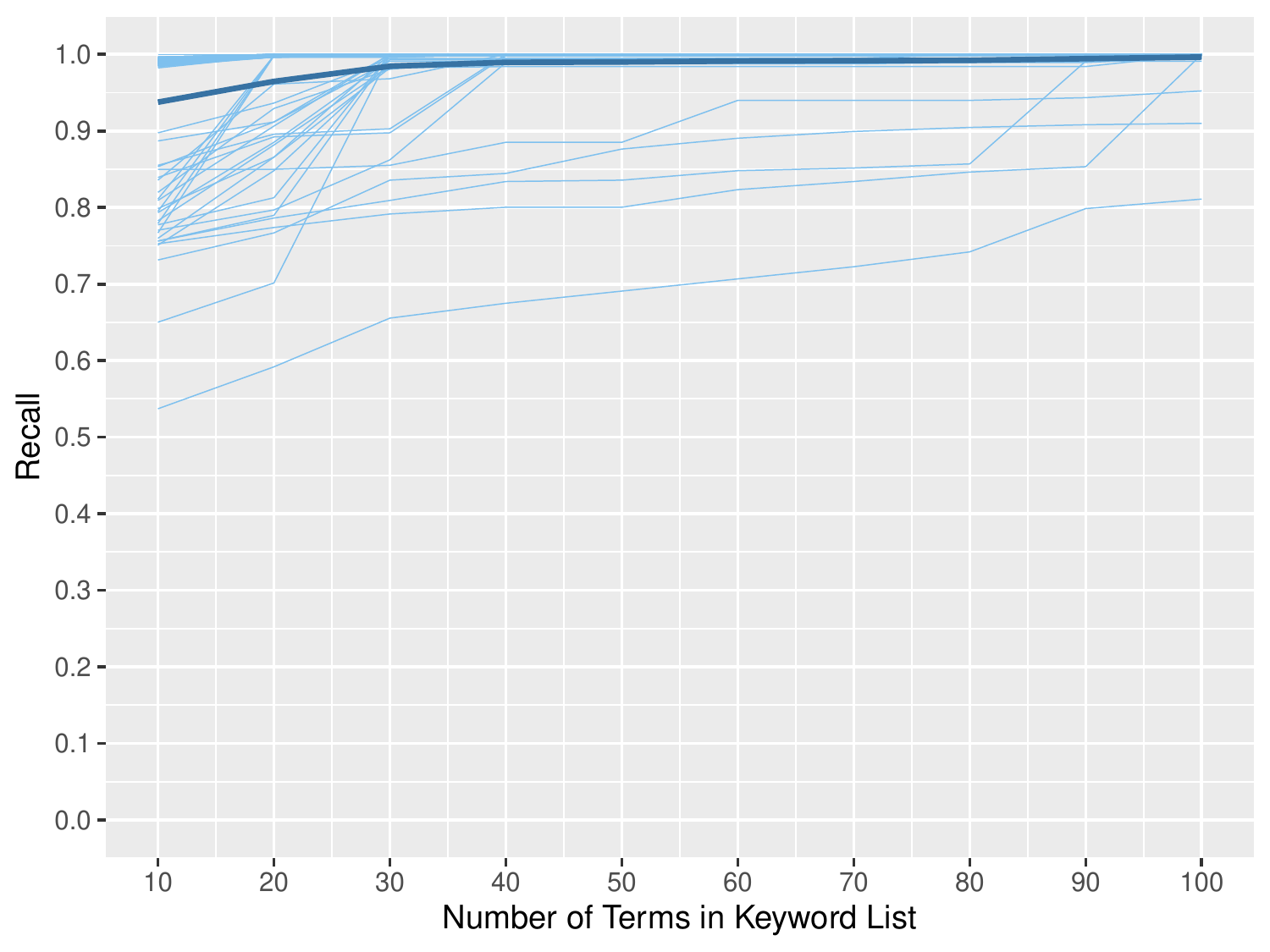} & \includegraphics[height=0.3\textwidth]{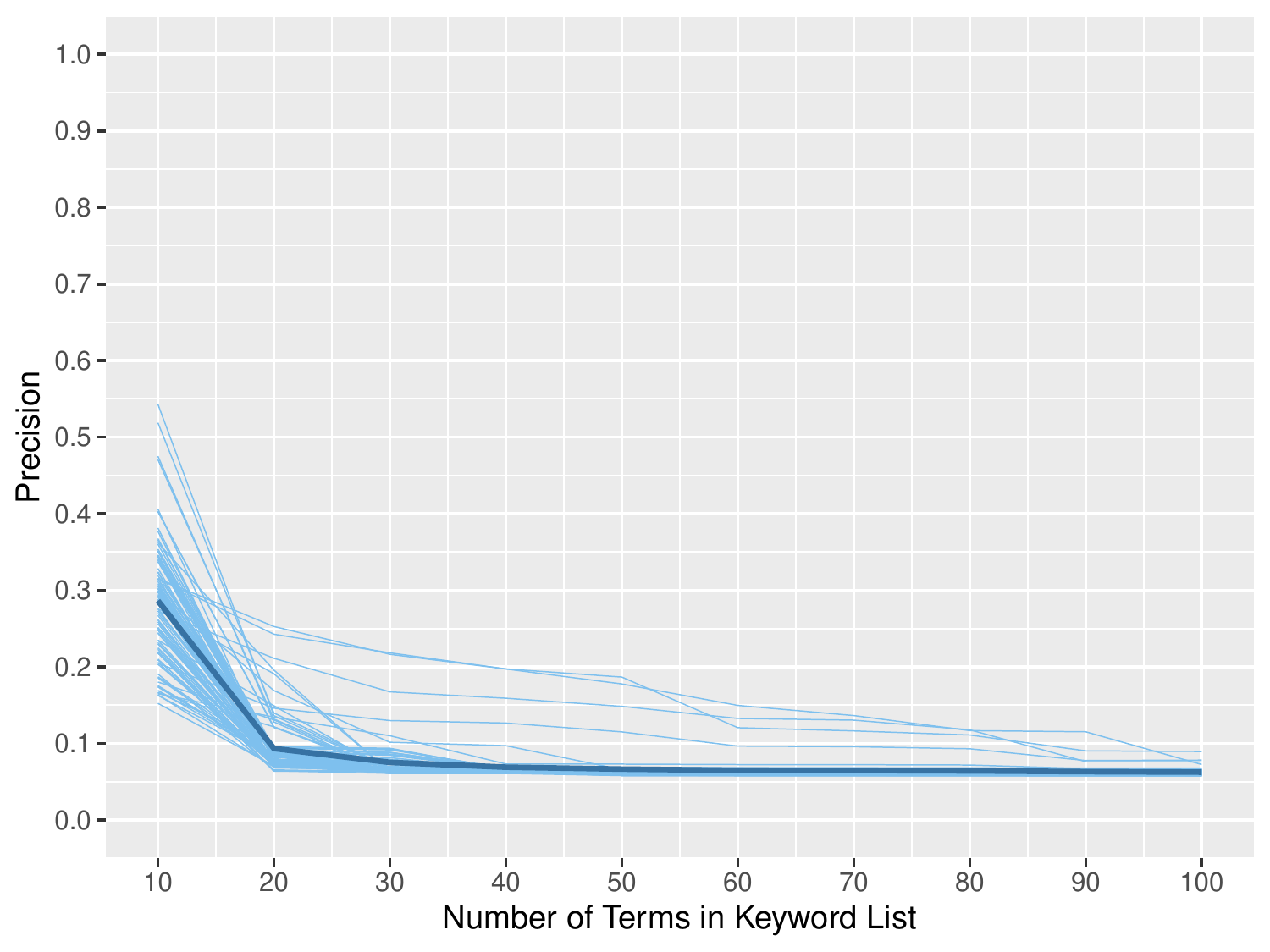}\\
\end{tabular}
\caption[Recall and Precision for Retrieving Relevant Documents with Keyword Lists and Local Query Expansion]{\textbf{Recall and Precision for Retrieving Relevant Documents with Keyword Lists and Local Query Expansion.} \scriptsize{This plot shows recall and precision scores resulting from the application of the keyword lists of 10 highly predictive terms as well as the evolution of the recall and precision scores across the query expansion procedure based on locally trained GloVe embeddings. For each of the sampled 100 keyword lists that then are expanded, one light blue line is plotted. The thick dark blue line gives the mean over the 100 lists. Note that the strong increase in recall for some keyword lists in the Twitter data set is due to the fact that the textual feature with the highest cosine similarity to the highly predictive initial term \emph{`fl\"uchtlinge'} (translation: \emph{`refugees‘}) is the colon \emph{`:'}.}}
\label{fig:queryres2local} 
\end{figure}

\section{Recall and Precision of Topic Model-Based Classification Rules}
\label{app:restopic2}

\begin{figure}[H]
\footnotesize
\centering
  {\renewcommand{\arraystretch}{0}
\begin{tabular}{cc}
Twitter-Recall &Twitter-Precision \\
\includegraphics[height=0.33\textwidth]{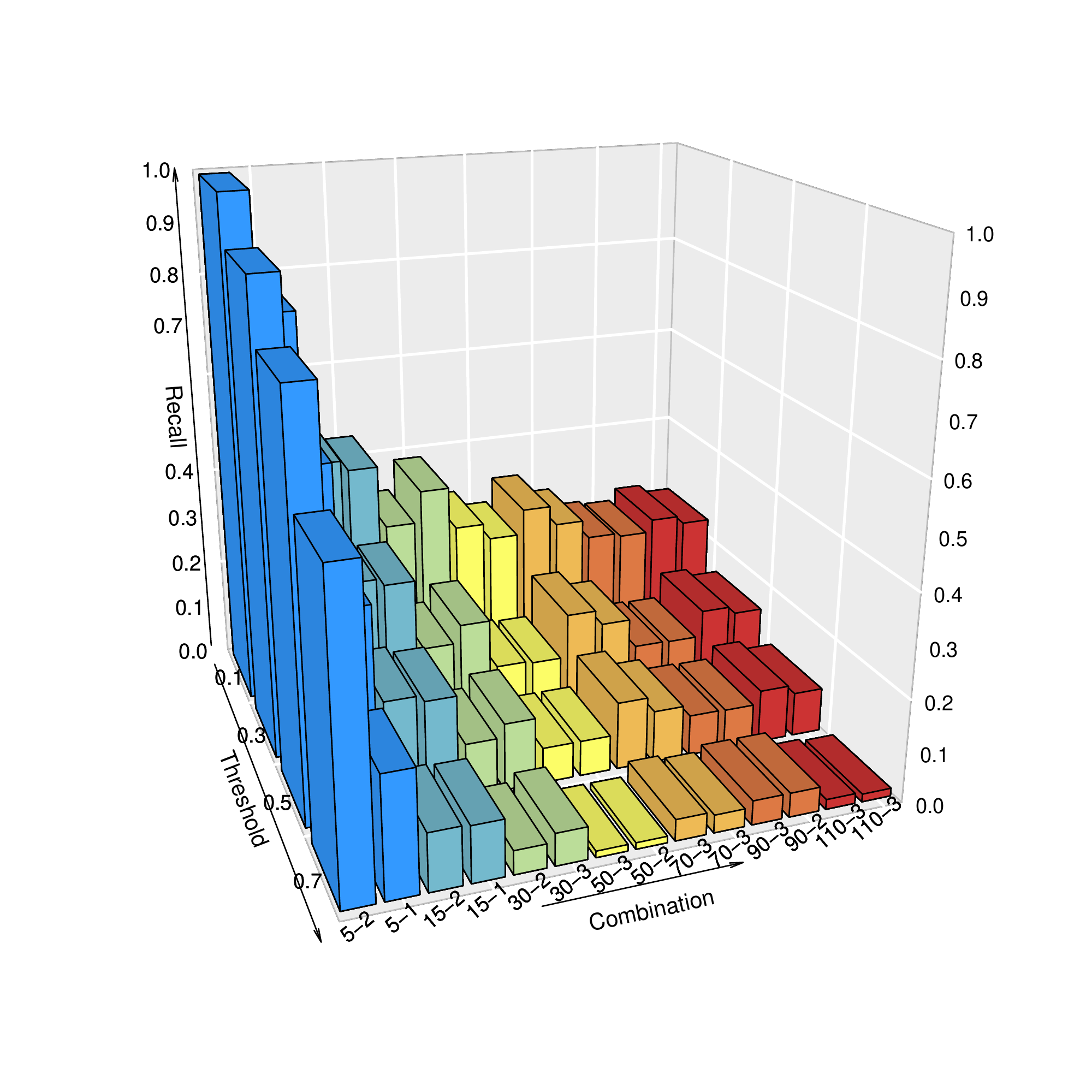}& \includegraphics[height=0.33\textwidth]{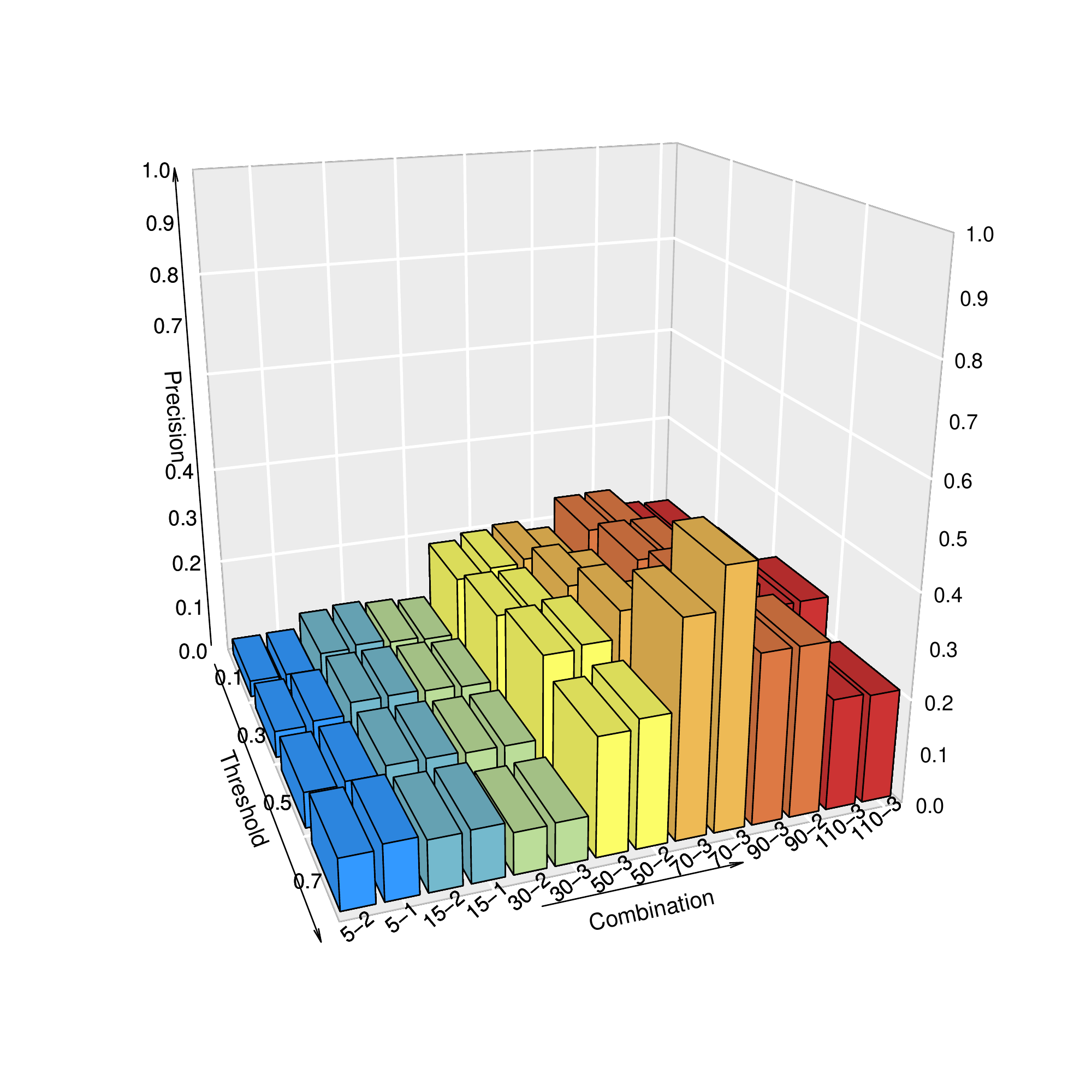} \\
SBIC-Recall & SBIC-Precision \\
\includegraphics[height=0.33\textwidth]{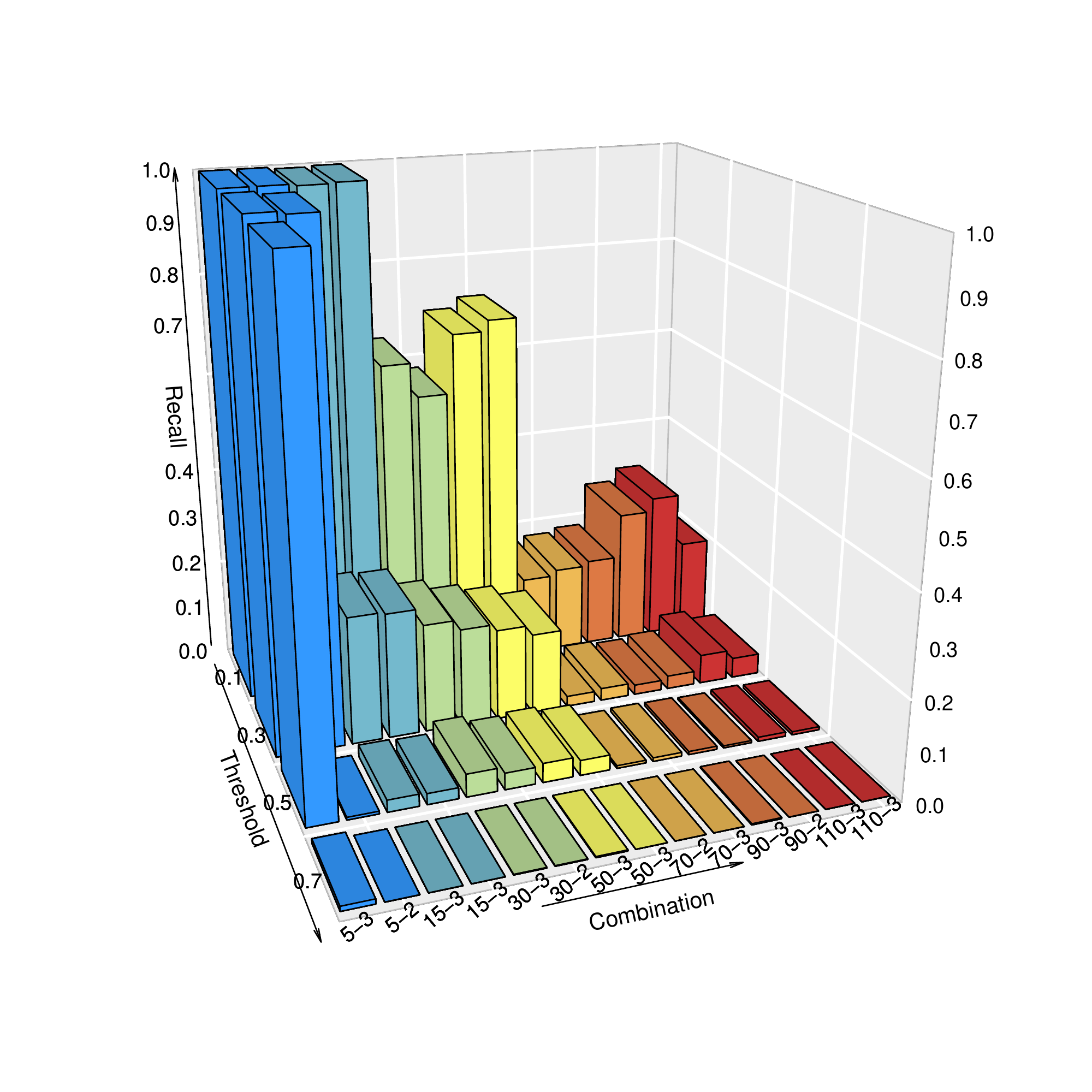}& \includegraphics[height=0.33\textwidth]{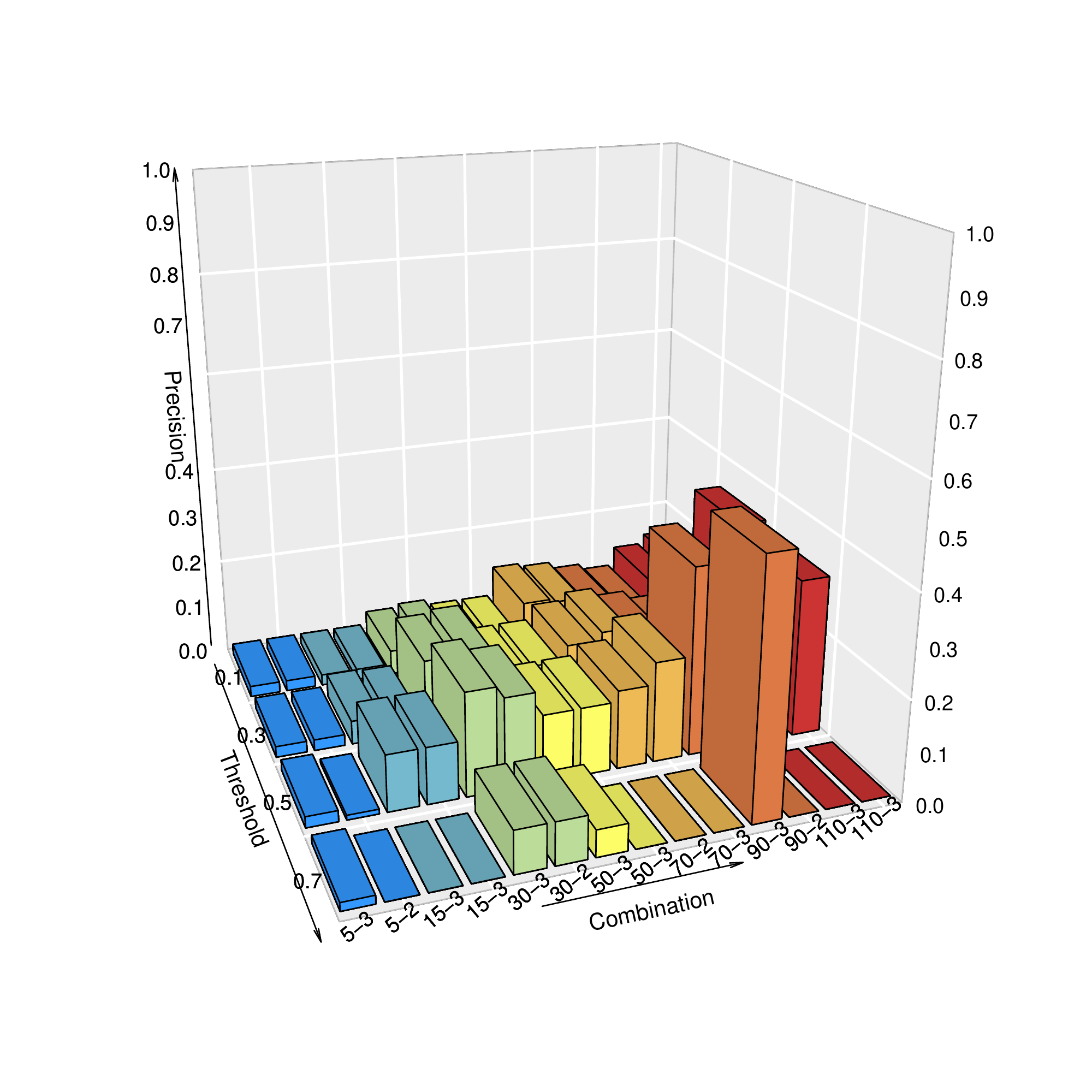} \\
Reuters-Recall & Reuters-Precision \\
\includegraphics[height=0.33\textwidth]{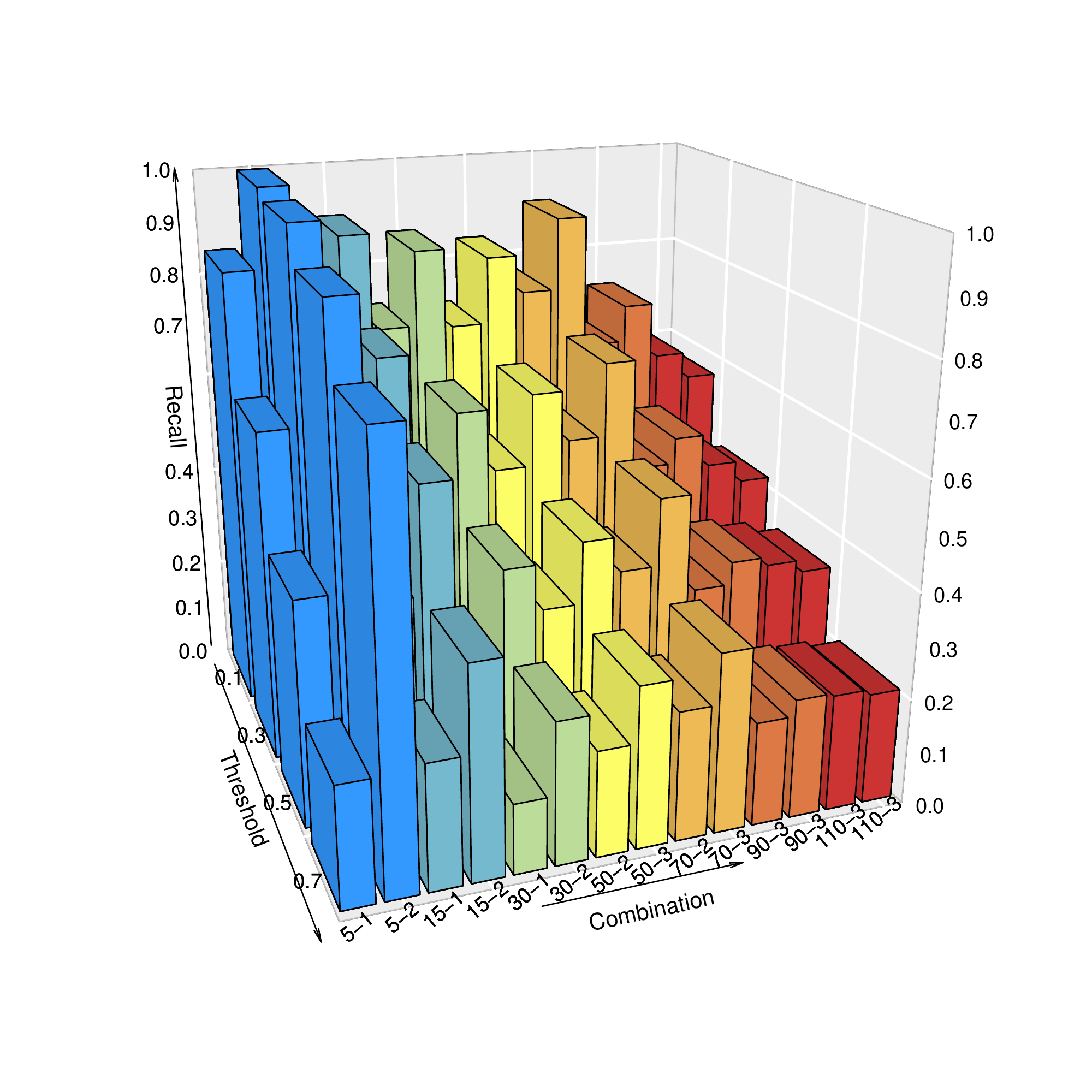} & \includegraphics[height=0.33\textwidth]{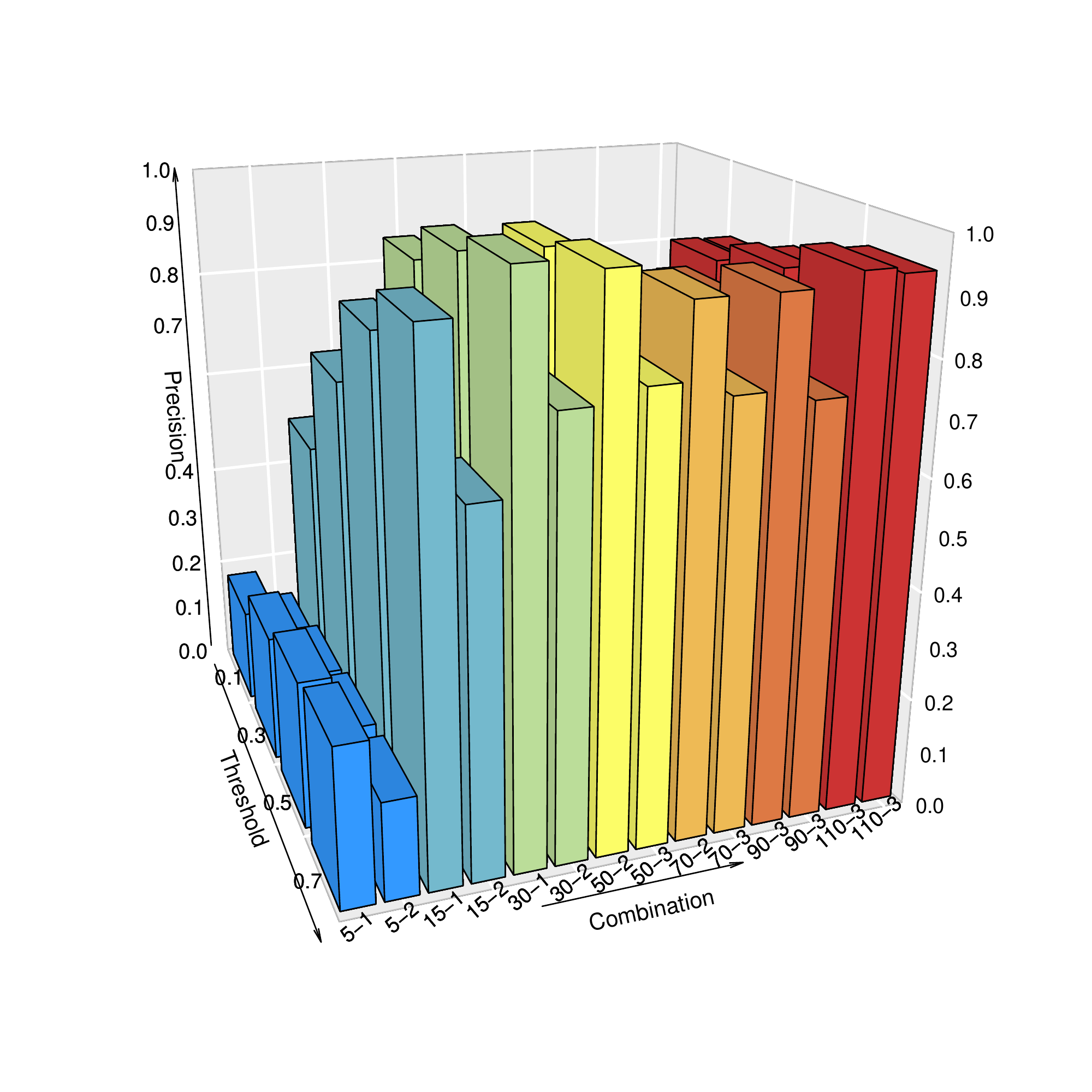} \\
\end{tabular}}
\caption[Recall and Precision of Topic Model-Based Classification Rules]{\textbf{Recall and Precision of Topic Model-Based Classification Rules.} \scriptsize{The height of a bar indicates the recall values (left column) and precision values (right column) resulting from the application of a topic model-based classification rule. For each number of topics $K \in \{5, 15, 30, 50, 70, 90, 110\}$, those two combinations out of all explored combinations regarding the question how many and which topics are considered relevant are shown that reach the highest $F_1$-Score for the given topic number. For each combination, recall and precision values for each threshold value $\xi \in \{0.1, 0.3, 0.5, 0.7 \}$ is given. Classification rules that assign none of the documents to the positive class have a recall value of 0 and an undefined value for precision and the $F_1$-Score. Undefined values here are visualized by the value 0.}}
\label{fig:restopic2} 
\end{figure}
\newpage

\section{Recall and Precision of Active and Passive Supervised Learning}
\label{app:superres2}

\begin{figure}[H]
\footnotesize
   \centering
     {\renewcommand{\arraystretch}{0.8}
\begin{tabular}{cc}
\hspace{0.8cm} Twitter-Recall & \hspace{-0.9cm} Twitter-Precision  \\
\includegraphics[height=0.32\textwidth]{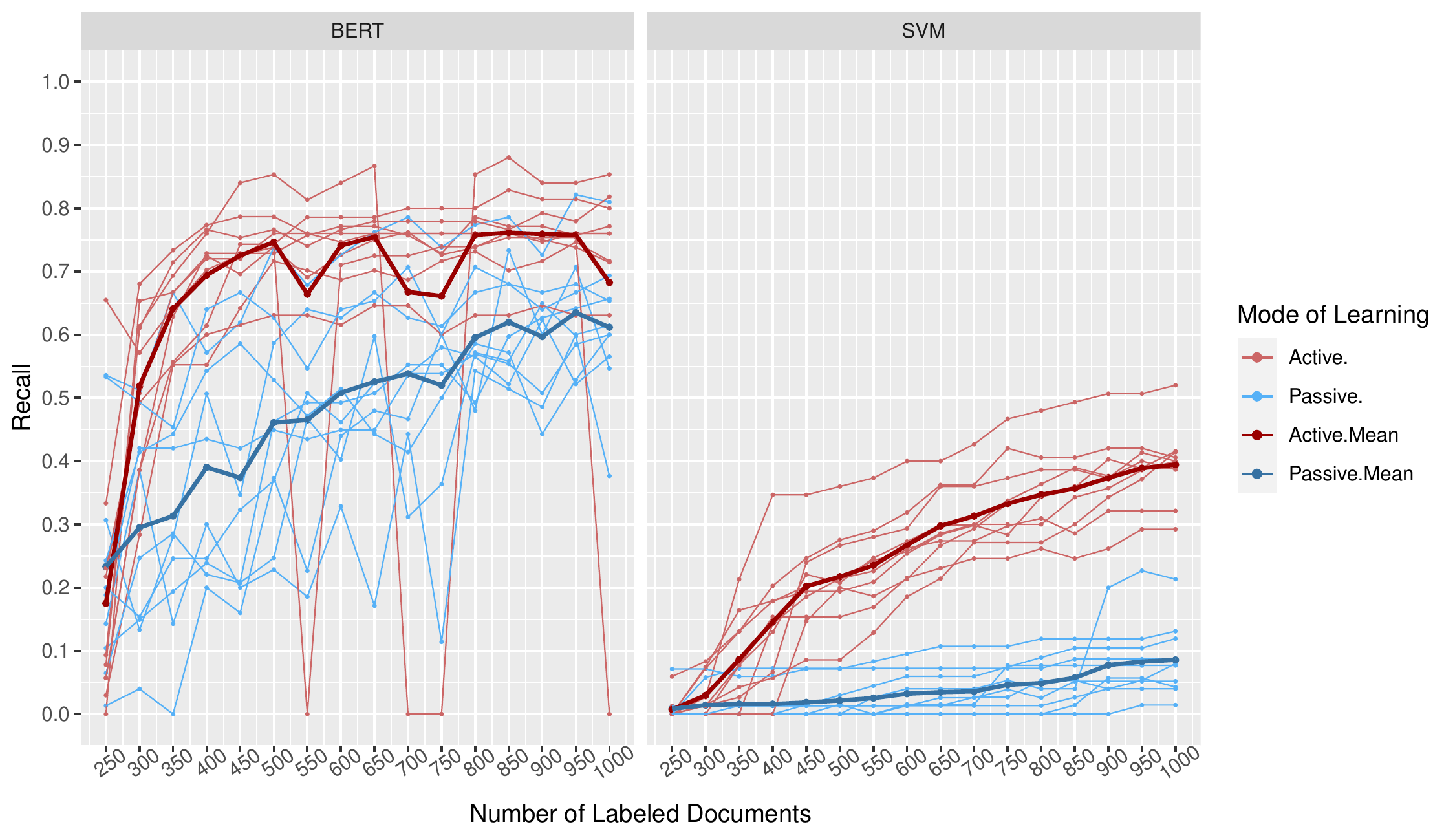}  \hspace{-1.9cm} &
\includegraphics[height=0.32\textwidth]{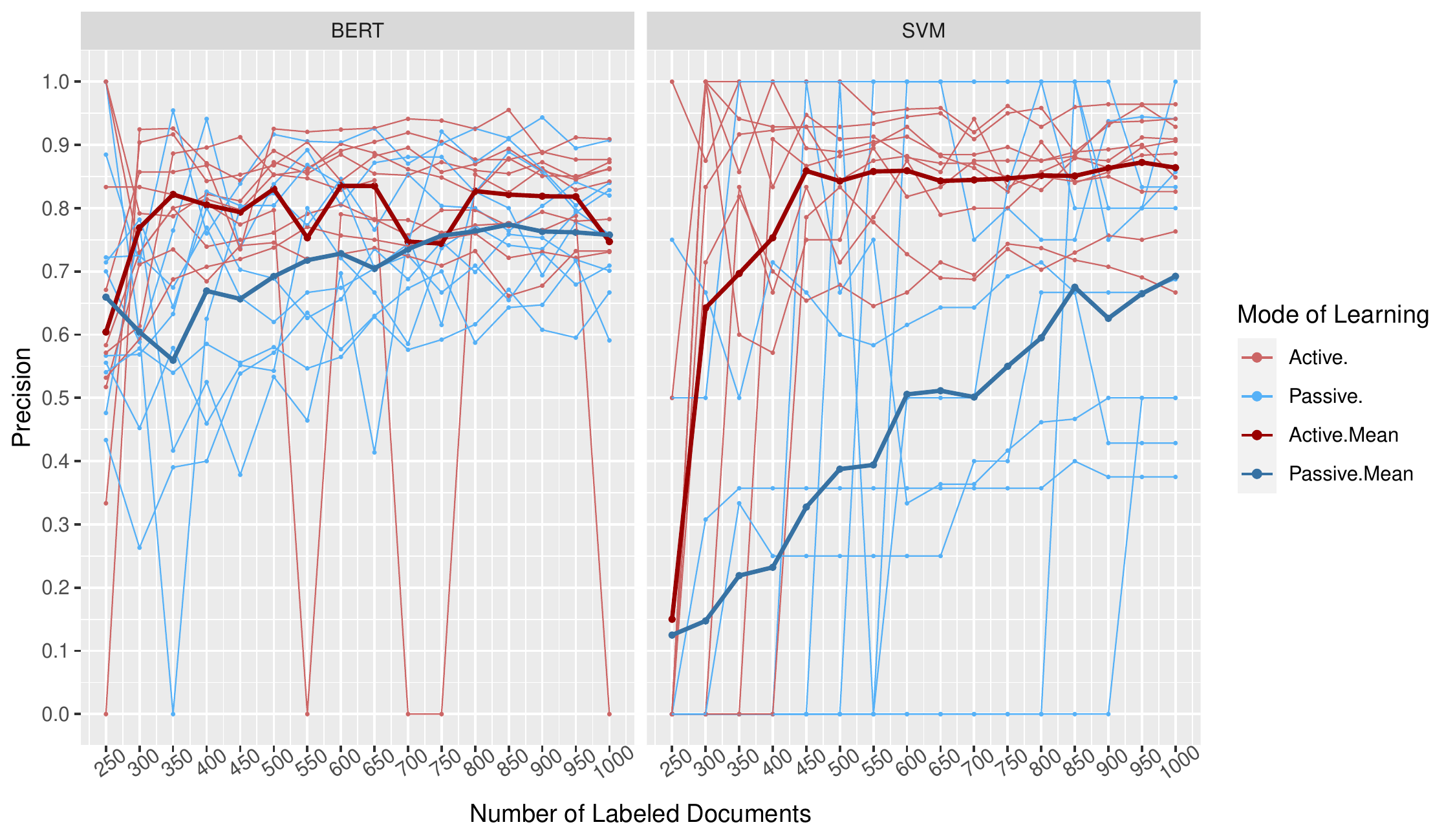} \\
\hspace{0.8cm} SBIC-Recall  & \hspace{-0.9cm} SBIC-Precision \\
\includegraphics[height=0.32\textwidth]{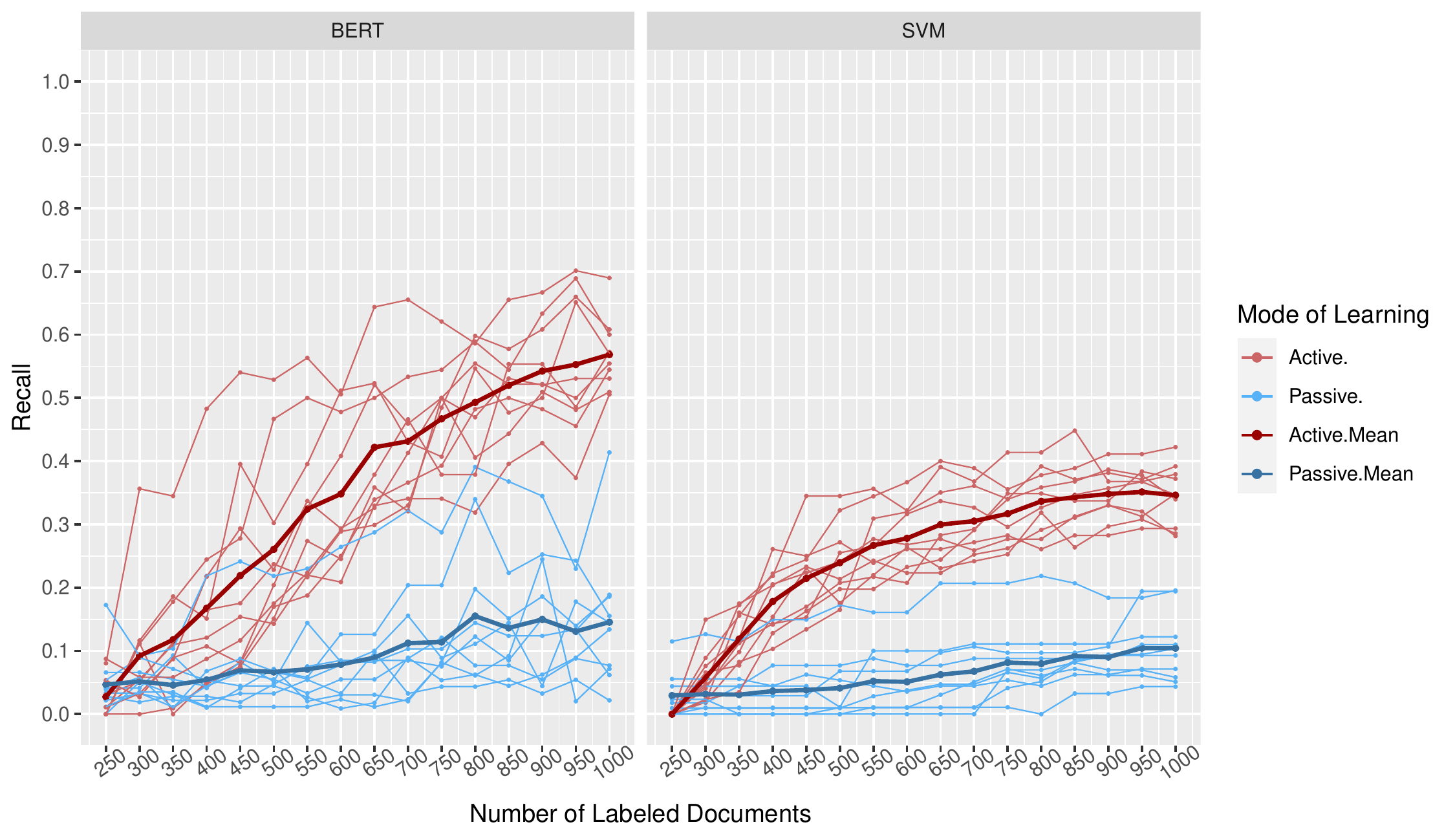} \hspace{-1.9cm} &
\includegraphics[height=0.32\textwidth]{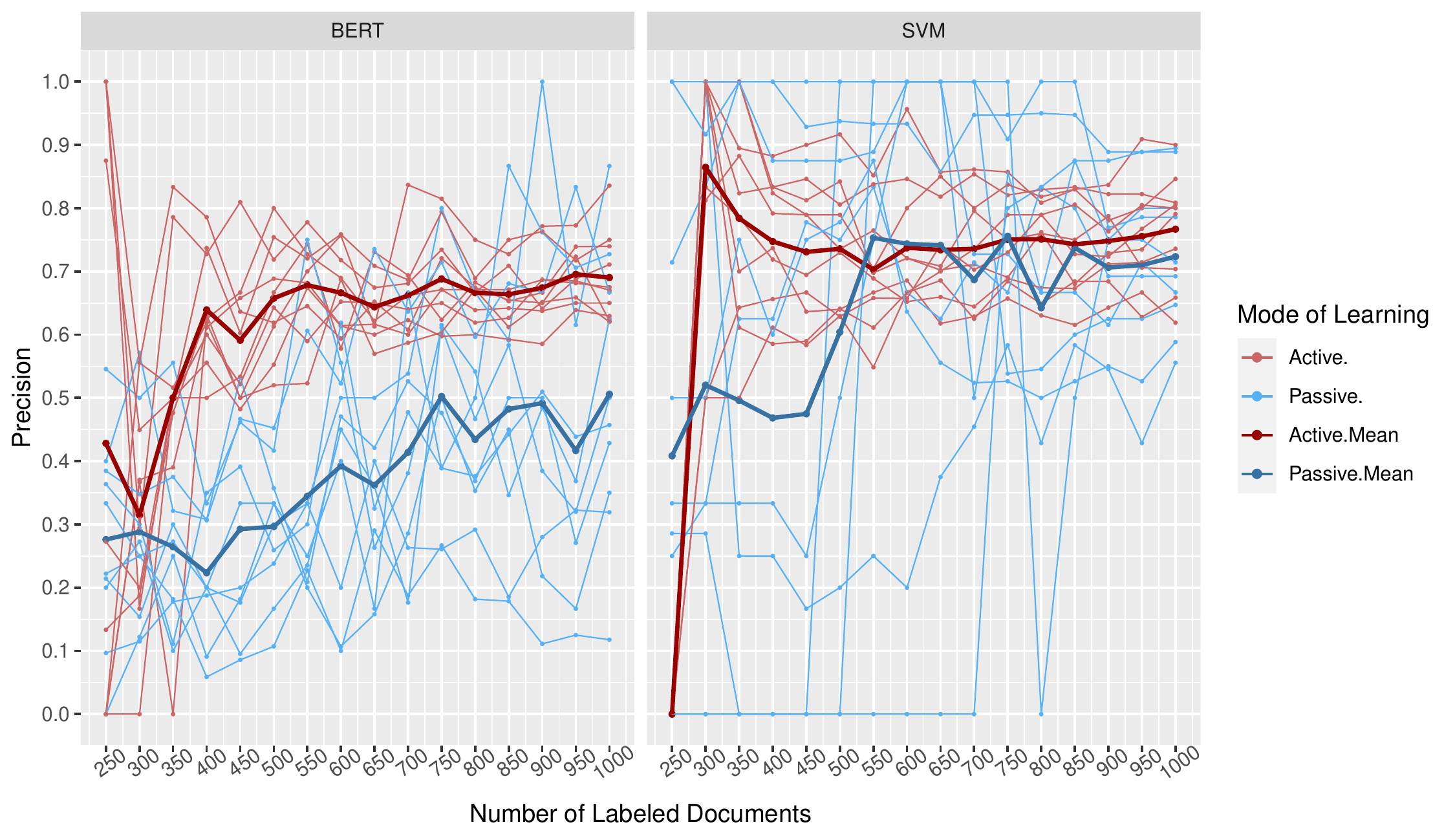} \\
\hspace{0.8cm} Reuters-Recall  & \hspace{-0.9cm} Reuters-Precision  \\
\includegraphics[height=0.32\textwidth]{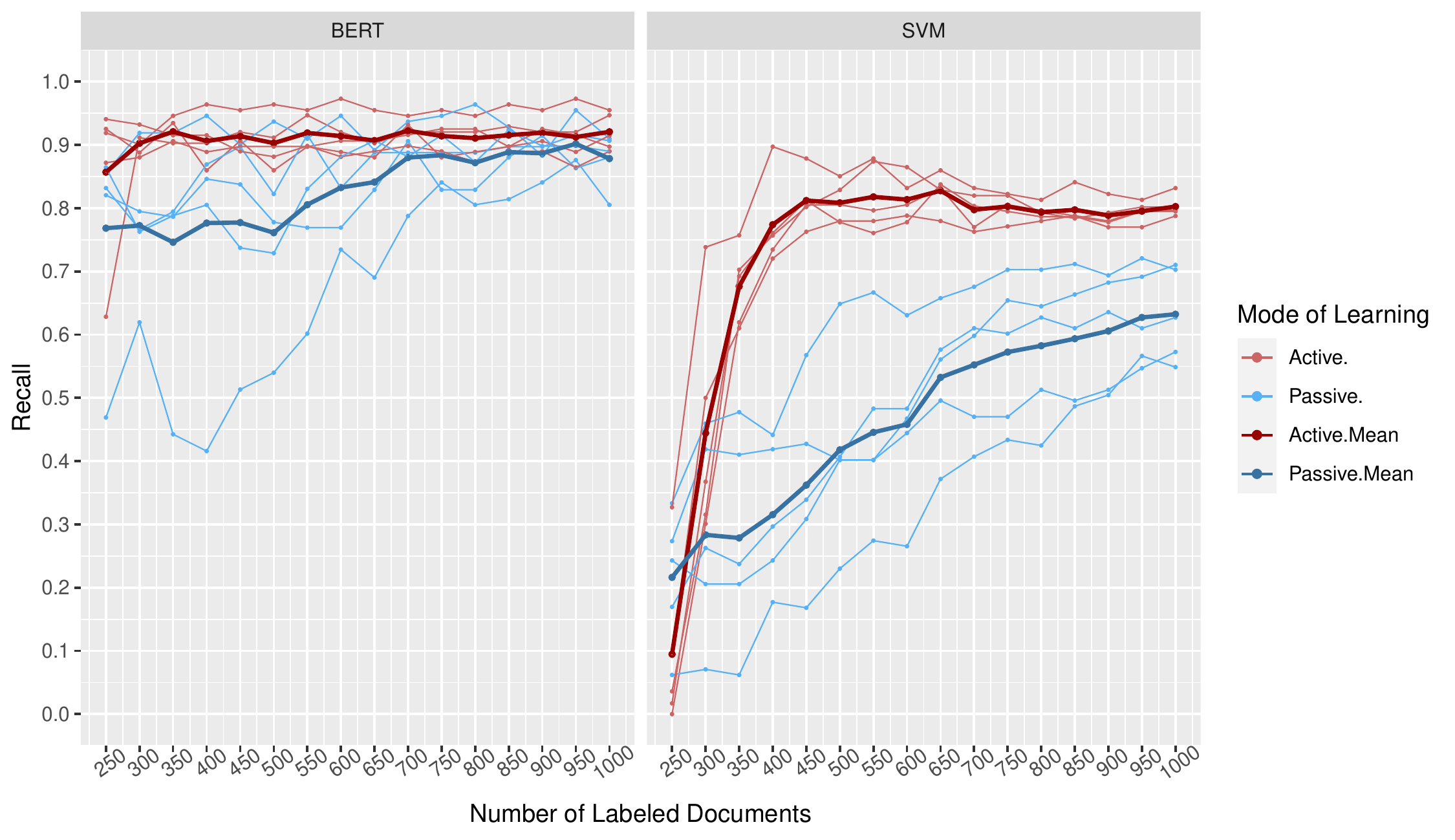}  \hspace{-1.9cm} &
\includegraphics[height=0.32\textwidth]{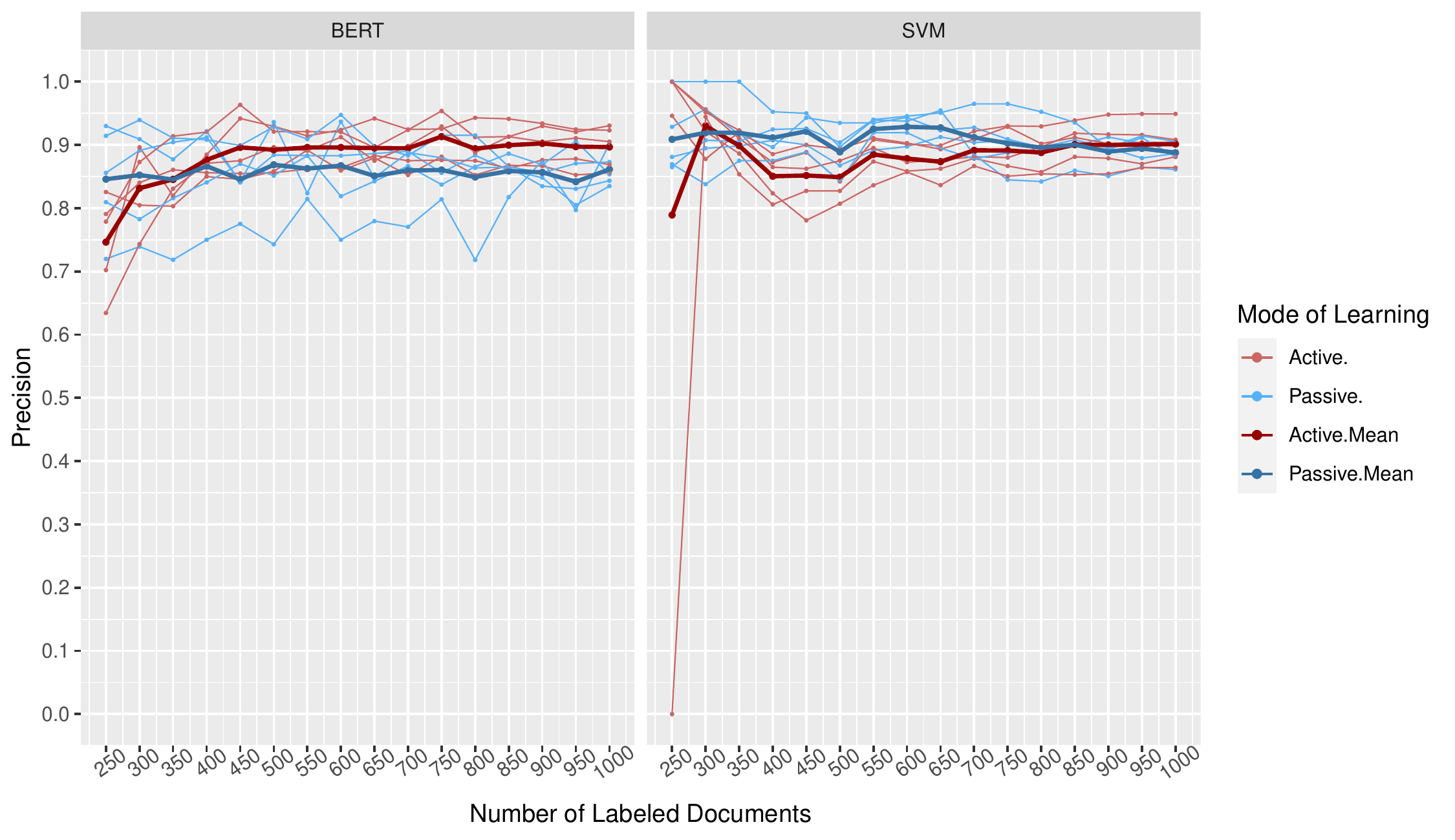}\\
\end{tabular}}
\caption[Recall and Precision of Active and Passive Supervised Learning]{\textbf{Recall and Precision of Active and Passive Supervised Learning.} \scriptsize{Recall values (left column) and precision values (right column) achieved on the set aside test set as the number of unique labeled documents in set $\mathcal{I}$ increases from 250 to 1,000. Passive supervised learning results are visualized by blue lines, active learning results are given in red. For each of the 10 (Twitter, SBIC) or 5 (Reuters) conducted iterations, one light colored line is plotted. The thick and dark blue and red lines give the means across the iterations. If a trained model assigns none of the documents to the positive relevant class, then it has a recall value of 0 and an undefined value for precision and the $F_1$-Score. Undefined values here are visualized by the value 0.}}
\label{fig:superres2} 
\end{figure}
\newpage

\section{Comparing BERT and SVM for Active and Passive Supervised Learning}
\label{app:superrescompare}

\begin{figure}[H]
   \centering
\footnotesize
     {\renewcommand{\arraystretch}{0.8}
\begin{tabular}{c}
\hspace{-1cm} Twitter   \\
\includegraphics[height=0.3\textwidth]{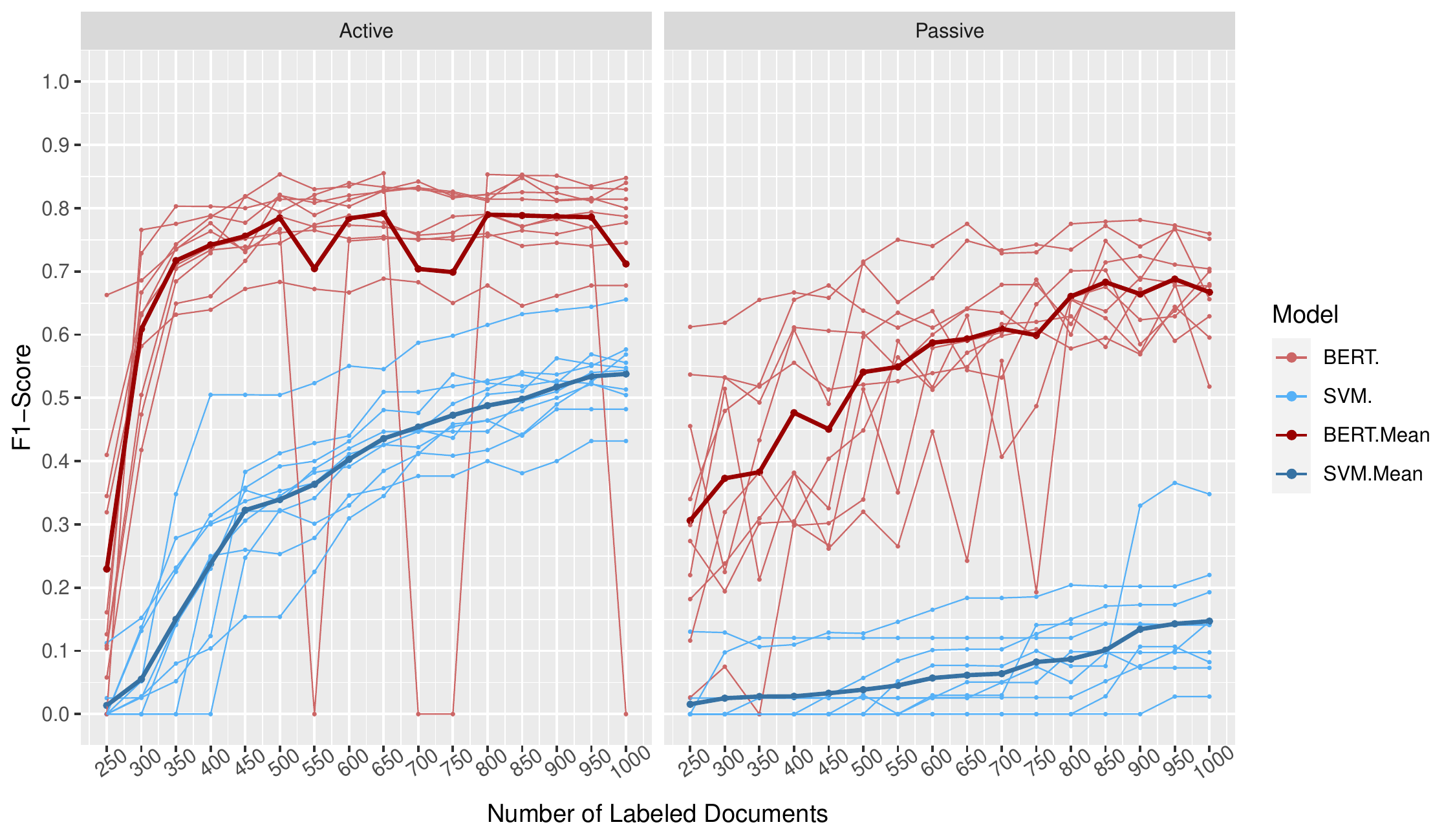} \\
\hspace{-1cm}  SBIC  \\
\includegraphics[height=0.3\textwidth]{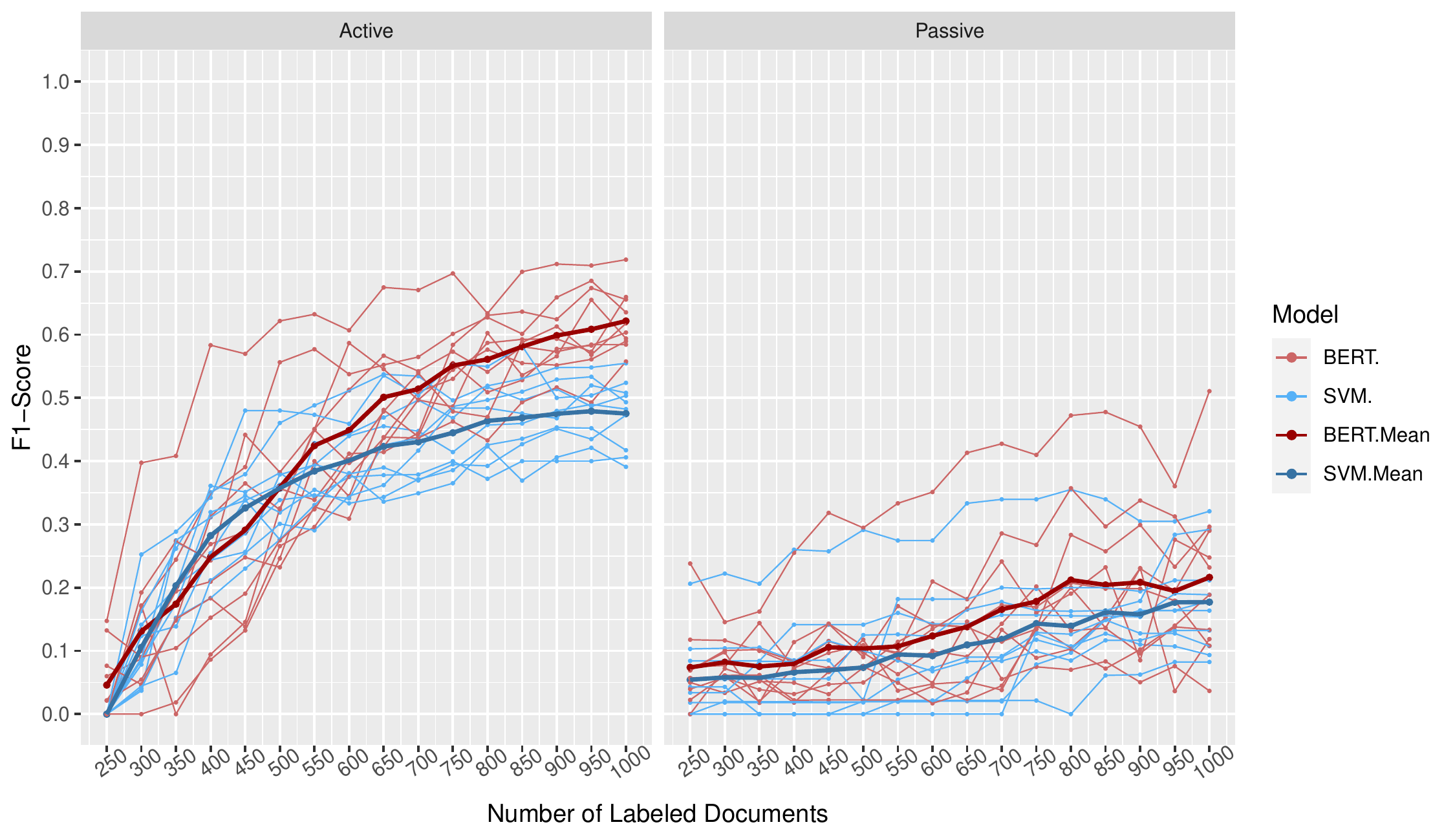}  \\
\hspace{-1cm}  Reuters   \\
\includegraphics[height=0.3\textwidth]{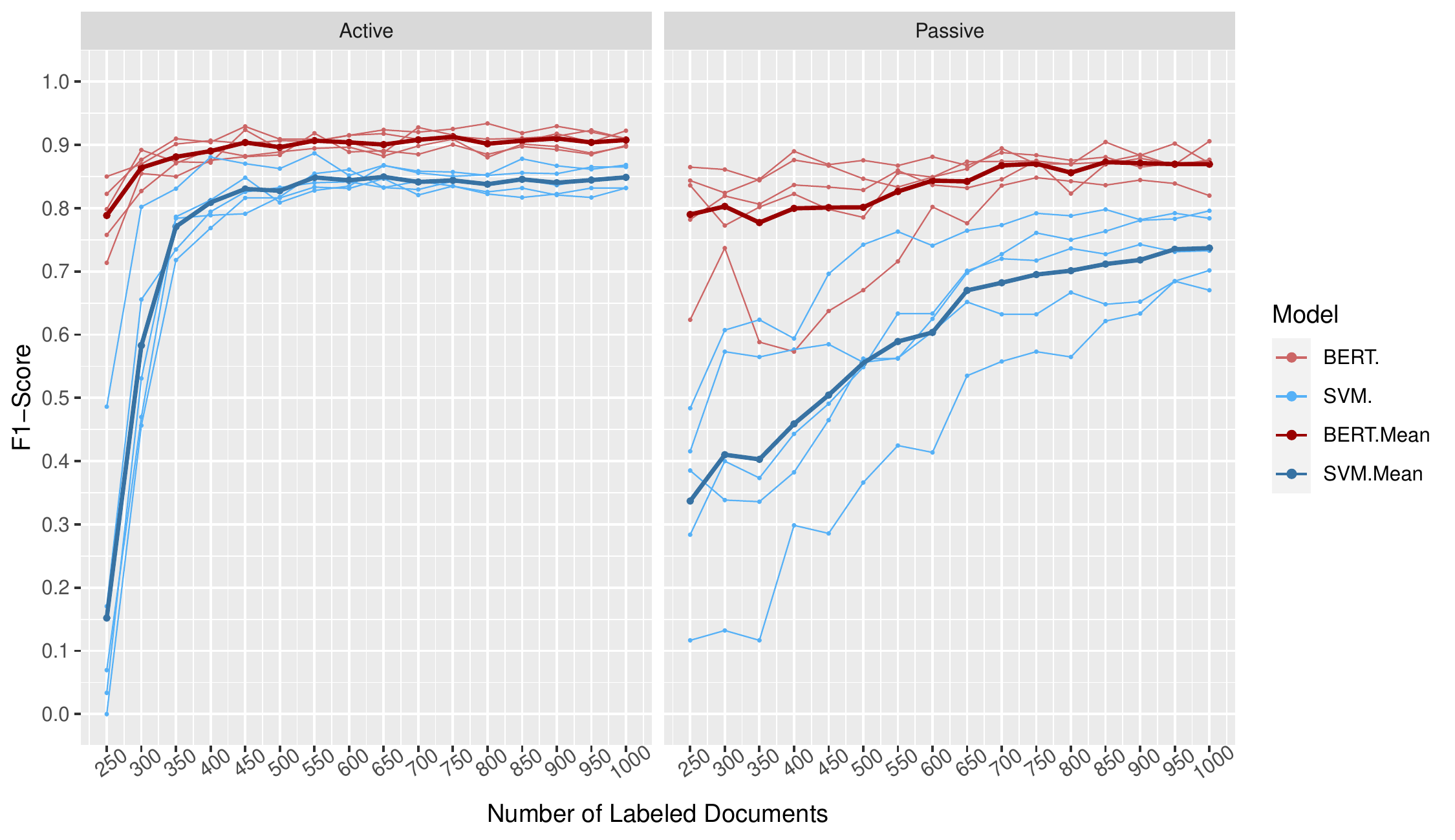} \\
\end{tabular}}
\caption[Comparing BERT and SVM for Active and Passive Supervised Learning I]{\textbf{Comparing BERT and SVM for Active and Passive Supervised Learning I.} \scriptsize{$F_1$-Scores achieved on the set aside test set as the number of unique labeled documents in set $\mathcal{I}$ increases from 250 to 1,000. Active learning results are visualized in the left panels, passive learning results are given in the right panels. $F_1$-Scores of the SVMs are visualized by blue lines, BERT performances are given in red. For each of the 10 (Twitter, SBIC) or 5 (Reuters) conducted iterations, one light colored line is plotted. The thick and dark blue and red lines give the mean across the iterations. If a trained model assigns none of the documents to the positive relevant class, then it has a recall value of 0 and an undefined value for precision and the $F_1$-Score. Undefined values here are visualized by the value 0.}}
\label{fig:superrescompare} 
\end{figure}
\newpage

 \clearpage
\begin{sidewaysfigure}
   \centering
\begin{tabular}{ccc}
Twitter & SBIC & Reuters  \\
\includegraphics[height=0.32\textwidth]{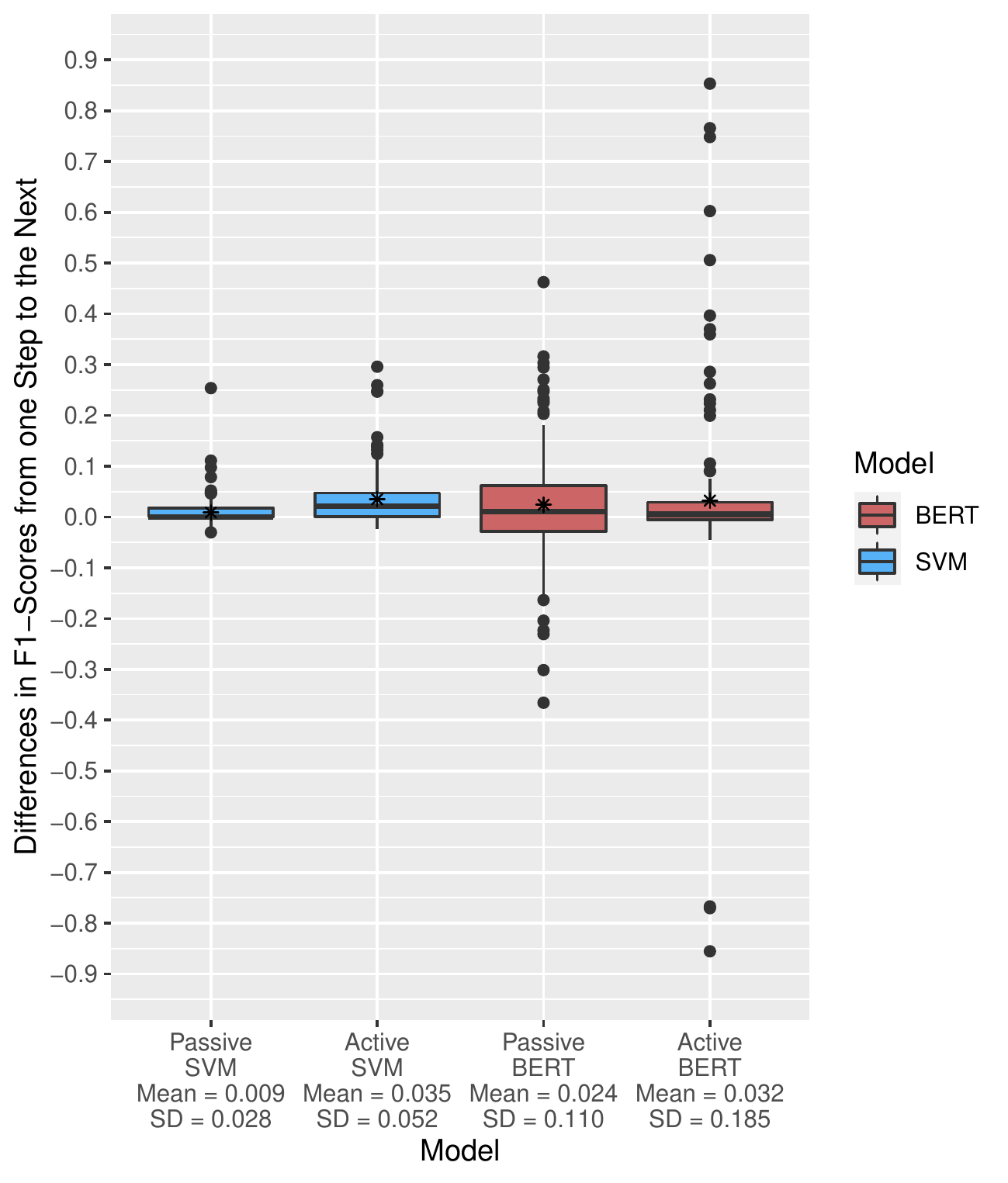} &
\includegraphics[height=0.32\textwidth]{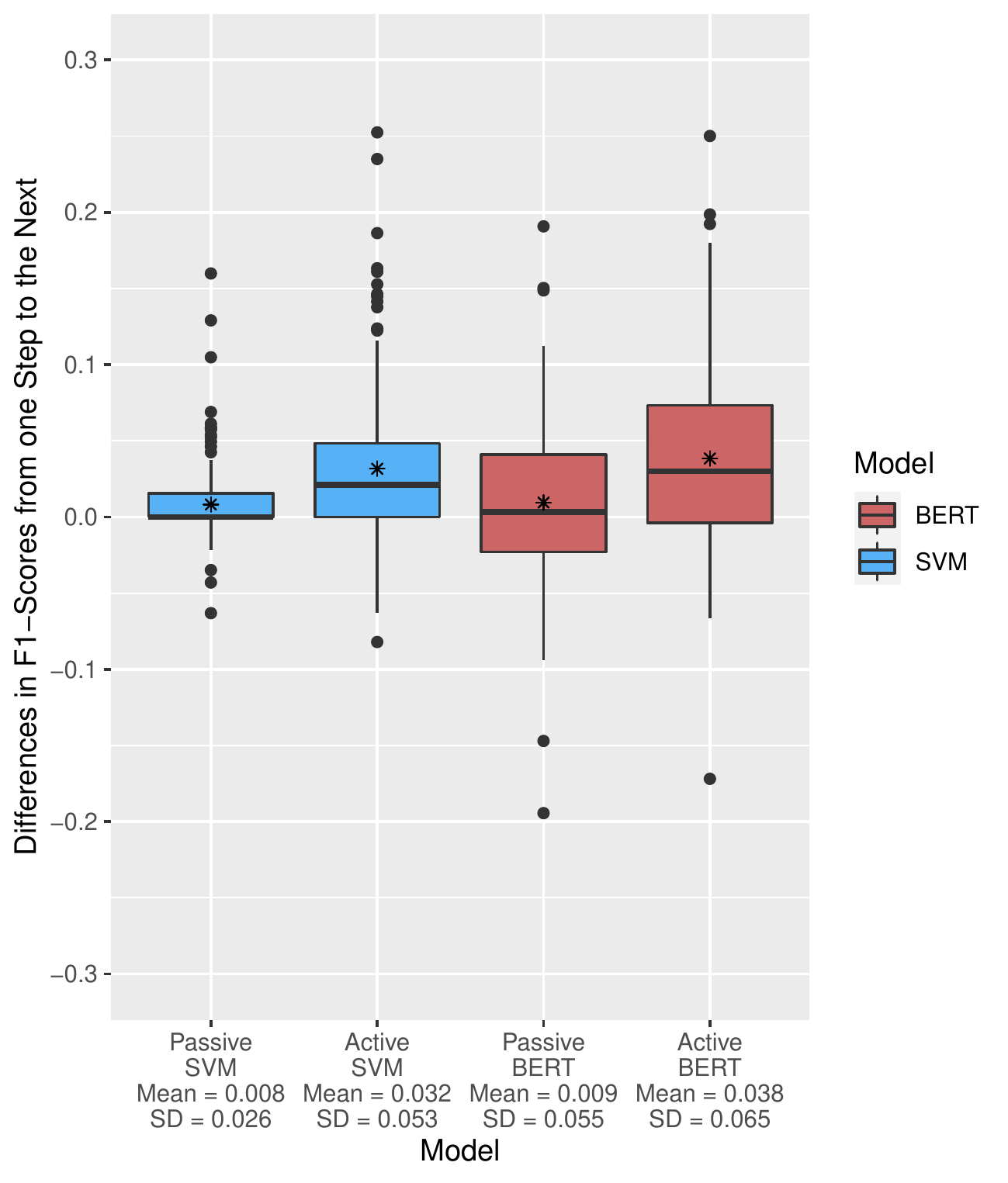}  &
\includegraphics[height=0.32\textwidth]{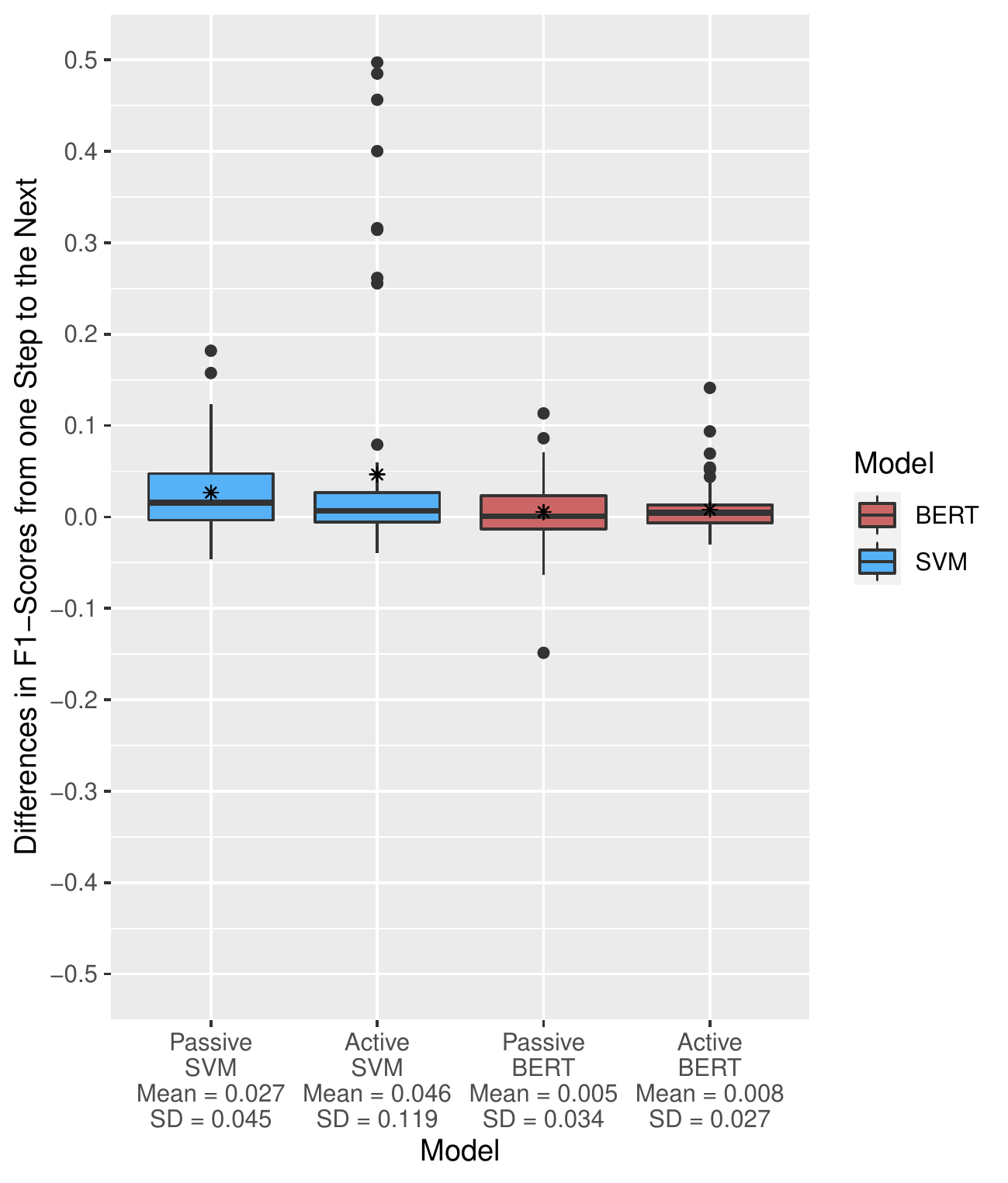} \\
\end{tabular}
\caption[Comparing BERT and SVM for Active and Passive Supervised Learning II]{\textbf{Comparing BERT and SVM for Active and Passive Supervised Learning II.} \footnotesize{Distribution of the differences in the $F_1$-Scores achieved on the set aside test set as the number of unique labeled documents in set $\mathcal{I}$ increases from one training step to the next by a batch of 50 documents. Boxplots visualizing the distribution of differences in $F_1$-Scores of the SVMs are presented in blue. $F_1$-Score differences for BERT are given in red. The mean is visualized by a star dot. The value of the mean as well as the standard deviation (SD) are given below the respective boxplots.}}
\label{fig:superrescomparediffs} 
\end{sidewaysfigure}
 \clearpage

\section{Comparing BERT with Different Hyperparameter Values}
\label{app:superreshyper}

The hyperparameter values for BERT models trained with hyperparameter values in conventional value ranges are determined via hyperparameter tuning. As for the SVMs, hyperparameter tuning via a grid search across sets of hyperparameter values is conducted in a stratified 5-fold cross-validation setting on one of the folds of the data. The AdamW algorithm \citep{Loshchilov2019} with a warmup period lasting 6$\%$ of the training steps is used. Dropout is set to 0.1. The batch size is set to 16. The inspected hyperparameter values for the global learning rate are $\{$2e-05, 3e-05$\}$, and for the number of epochs are $\{2, 3, 4, 5\}$. The folds are stratified such that the share of instances falling into the relevant minority class is the same across all folds. In each cross-validation iteration, in the folds used for training, random oversampling of the minority class is conducted such that the number of relevant minority class examples increases by a factor of 5. Among the inspected hyperparameter settings, the setting that achieves the highest $F_1$-Score regarding the prediction of the relevant minority class and does not exhibit excessive overfitting is selected.
\vfill
\newpage

\begin{figure}[H]
   \centering
\begin{tabular}{c}
\hspace{-1.5cm} Twitter   \\
\includegraphics[height=0.33\textwidth]{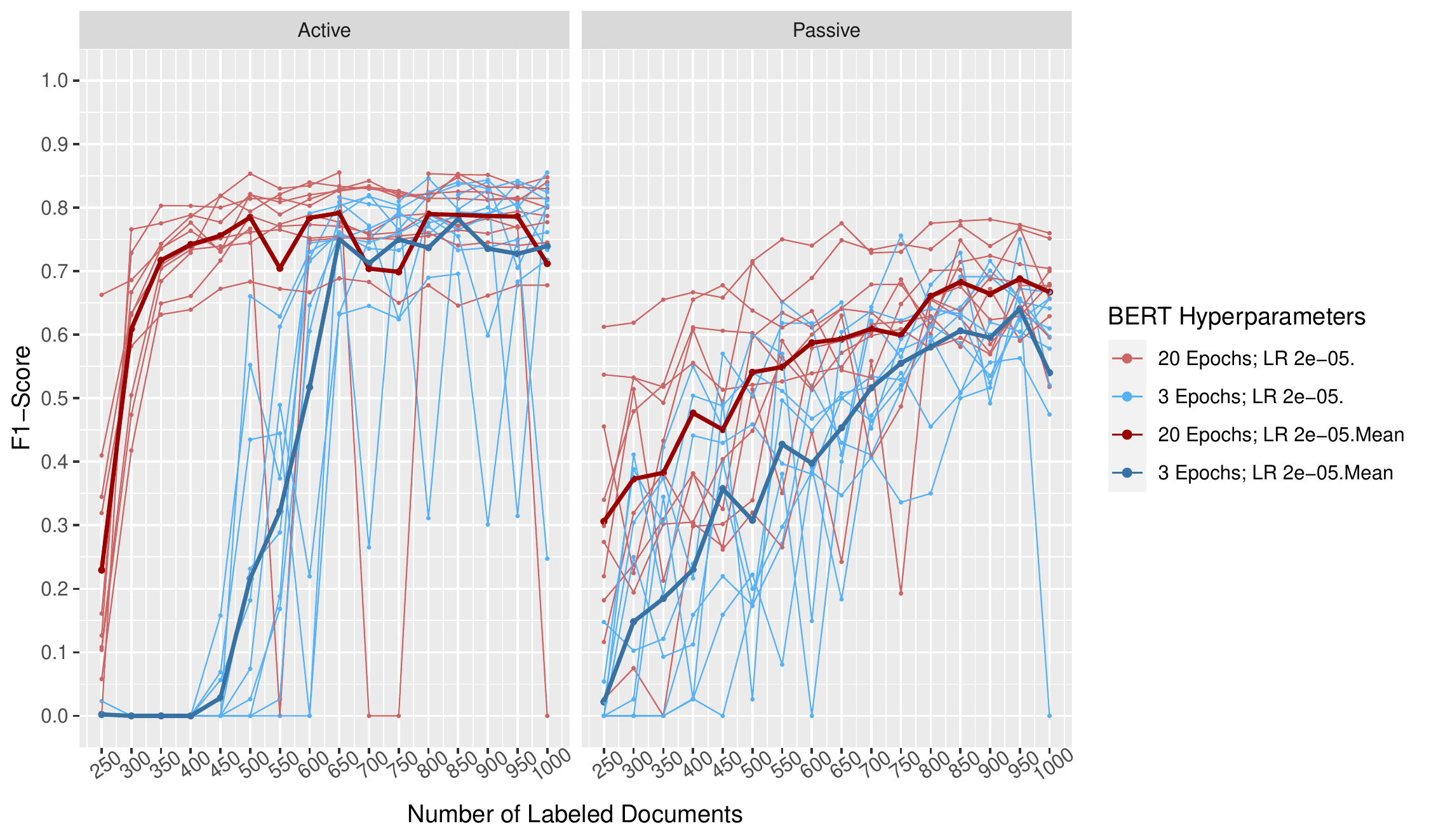} \\
 \hspace{-1.5cm} SBIC  \\
\includegraphics[height=0.33\textwidth]{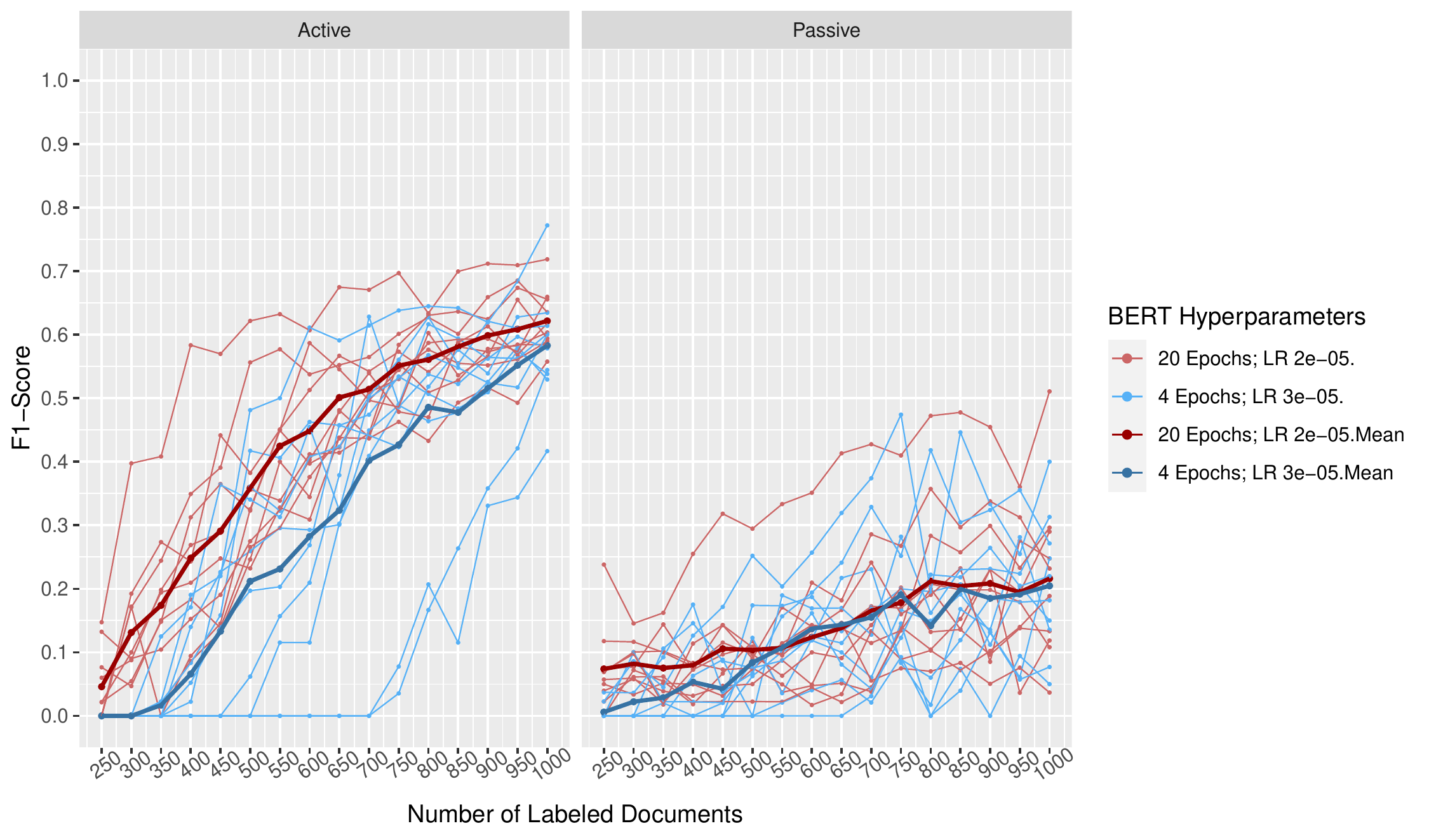}  \\
\hspace{-1.5cm}  Reuters   \\
\includegraphics[height=0.33\textwidth]{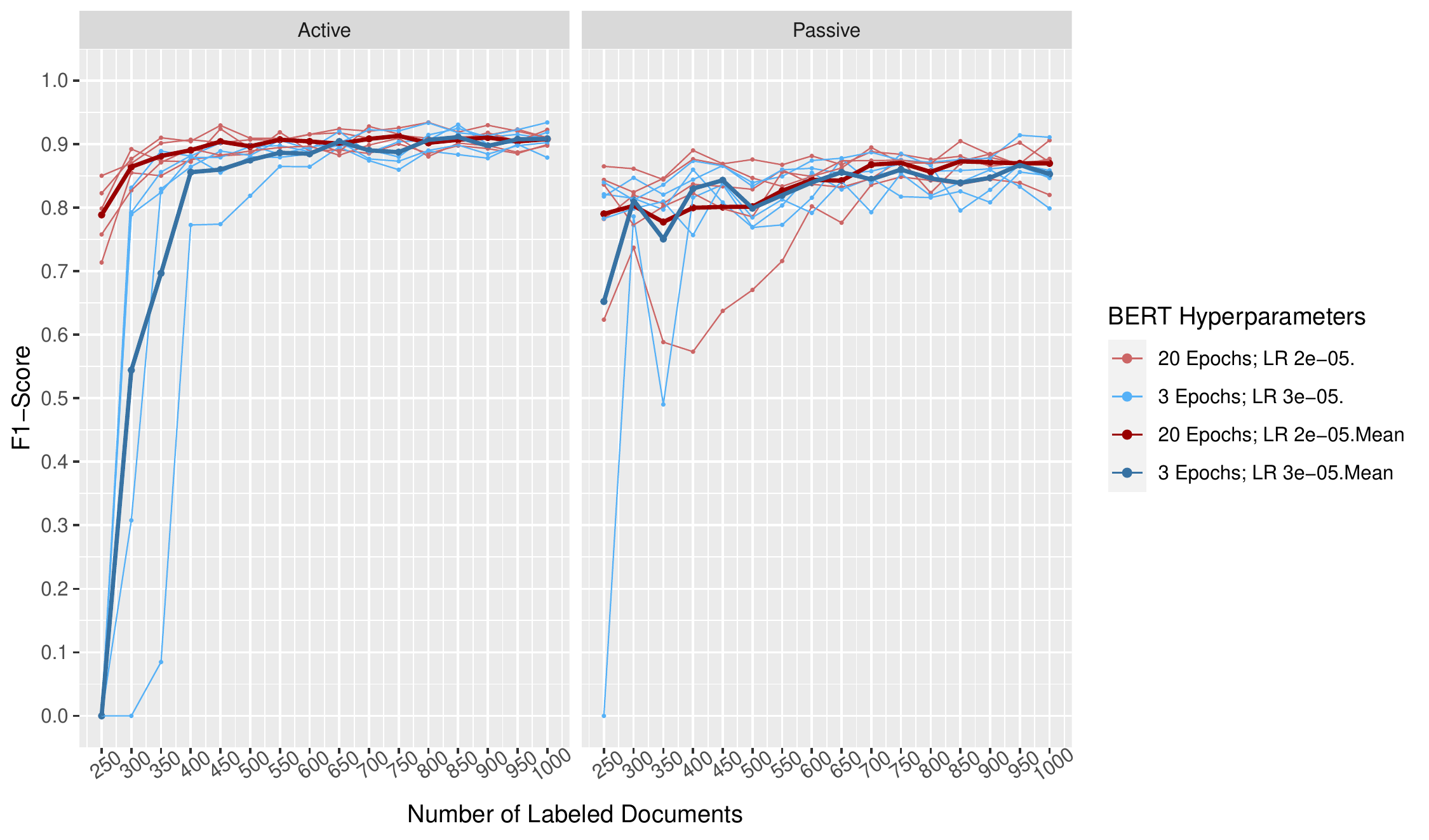} \\
\end{tabular}
\caption[Comparing BERT with Different Hyperparameter Values I]{\textbf{Comparing BERT with Different Hyperparameter Values I.} \footnotesize{$F_1$-Scores achieved by BERT models on the set aside test set as the number of unique labeled documents in set $\mathcal{I}$ increases from 250 to 1,000. Active learning results are visualized in the left panels, passive learning results are given in the right panels. $F_1$-Scores of BERT models trained with a global learning rate of 2e-05 for 20 epochs are given in red, performances of BERT models trained with hyperparameter values in conventional value ranges are visualized by blue lines. The precise values for the number of epochs and the learning rates are specified in the legends beside the plots. The thick and dark blue and red lines give the means across the iterations. If a trained model assigns none of the documents to the positive relevant class, then it has a recall value of 0 and an undefined value for precision and the $F_1$-Score. Undefined values here are visualized by the value 0.}}
\label{fig:superreshyper} 
\end{figure}

 \clearpage
\begin{sidewaysfigure}
   \centering
\begin{tabular}{ccc}
\hspace{-1.2cm} Twitter & \hspace{-1.2cm}  SBIC & \hspace{-1.2cm} Reuters  \\
\includegraphics[height=0.28\textwidth]{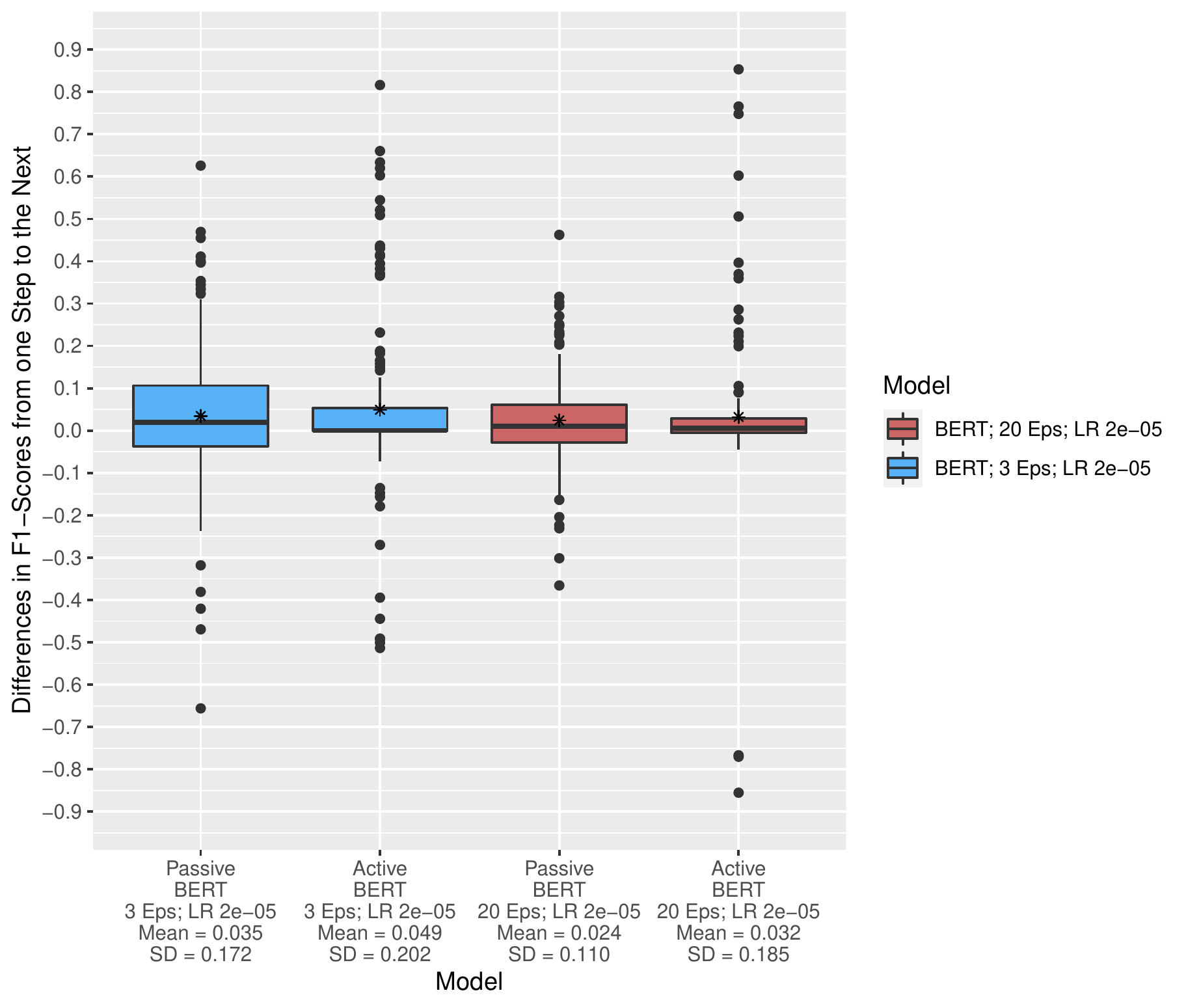} &
\hspace{-0.5cm} \includegraphics[height=0.28\textwidth]{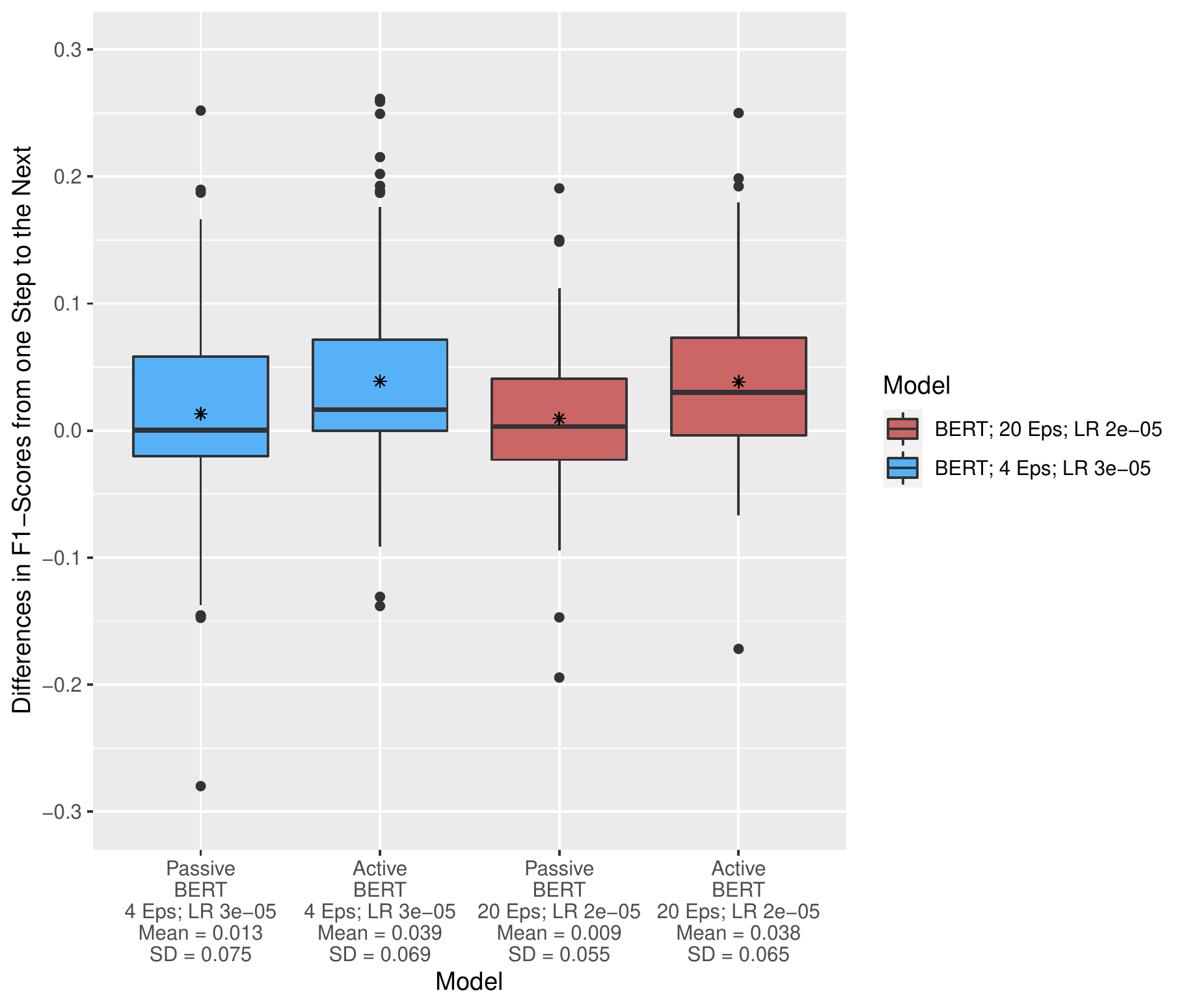}  &
\hspace{-0.5cm} \includegraphics[height=0.28\textwidth]{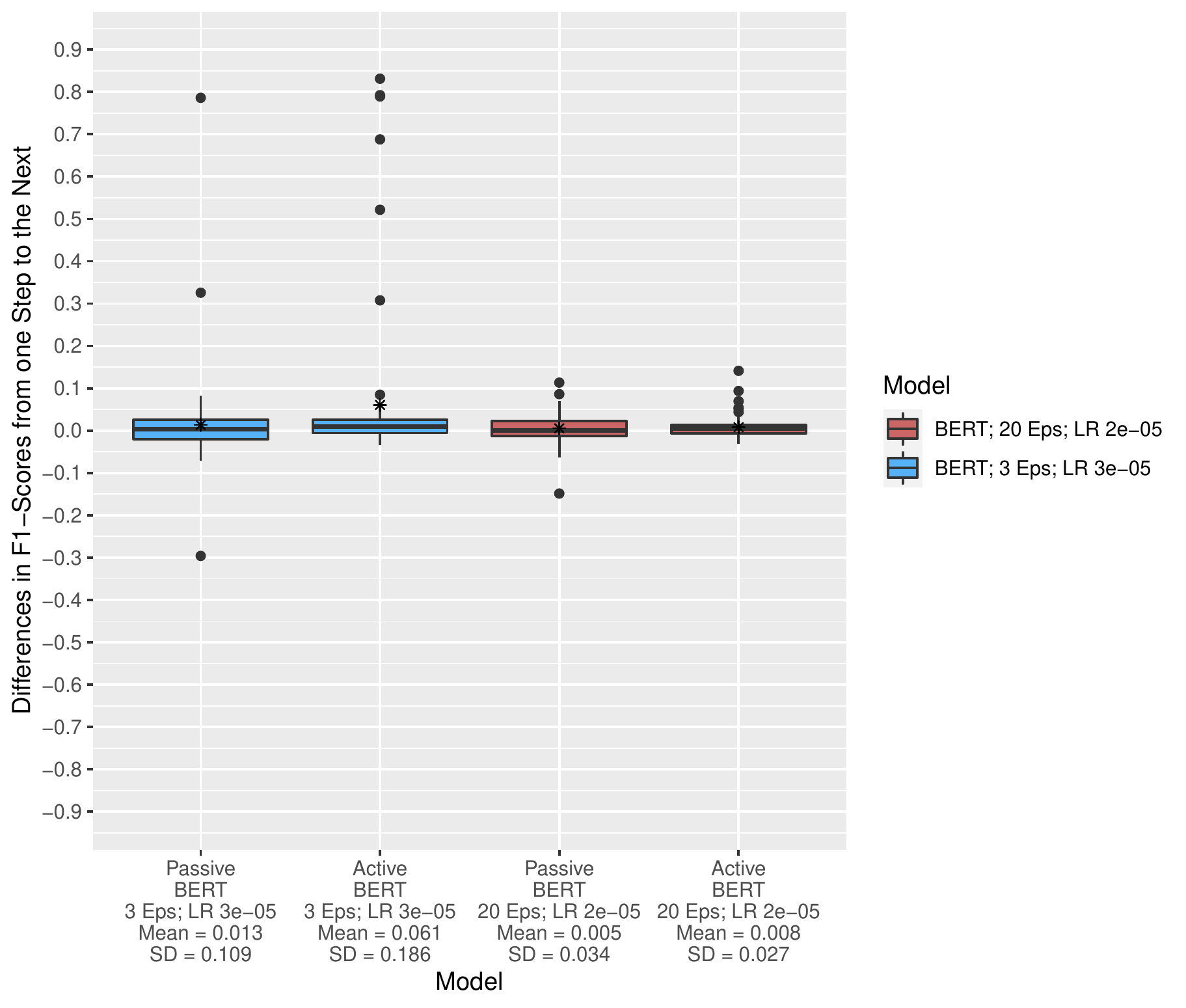} \\
\end{tabular}
\caption[Comparing BERT with Different Hyperparameter Values II]{\textbf{Comparing BERT with Different Hyperparameter Values II.} \footnotesize{Distributions of the differences in the $F_1$-Scores achieved on the set aside test set as the number of unique labeled documents in set $\mathcal{I}$ increases from one training step to the next by a batch of 50 documents. Boxplots visualizing the distribution of differences in $F_1$-Scores of BERT models trained with hyperparameter values in conventional value ranges are presented in blue. $F_1$-Score differences for BERT trained with a global learning rate (LR) of 2e-05 for 20 epochs (Eps) are given in red. The mean is visualized by a star dot. The value of the mean as well as the standard deviation (SD) are given below the respective boxplots.}}
\label{fig:superreshyper2} 
\end{sidewaysfigure}
 \clearpage


\end{appendix}

\newpage
\bibliography{libgenerale}

\begin{thebibliography}{}

\bibitem[Akbani et~al., 2004]{Akbani2004}
Akbani, R., Kwek, S., and Japkowicz, N. (2004).
\newblock Applying support vector machines to imbalanced datasets.
\newblock In Boulicaut, J.-F., Esposito, F., Giannotti, F., and Pedreschi, D.,
  editors, {\em Machine Learning: ECML 2004}, pages 39--50. Springer.

\bibitem[Allaire et~al., 2020]{RcppParallel2020}
Allaire, J.~J., Francois, R., Ushey, K., Vandenbrouck, G., Geelnard, M., and
  {Intel} (2020).
\newblock {\em {RcppParallel}: Parallel programming tools for {`Rcpp'}
  {(Version 5.0.2)}}.
\newblock [R package]. CRAN. https://CRAN.R-project.org/package=RcppParallel.

\bibitem[ALMasri et~al., 2013]{ALMasri2013}
ALMasri, M., Berrut, C., and Chevallet, J.-P. (2013).
\newblock Wikipedia-based semantic query enrichment.
\newblock In Bennett, P.~N., Gabrilovich, E., Kamps, J., and Karlgren, J.,
  editors, {\em Proceedings of the Sixth International Workshop on Exploiting
  Semantic Annotations in Information Retrieval {(ESAIR '13)}}, pages 5--8.
  Association for Computing Machinery.

\bibitem[Arora et~al., 2013]{Arora2013}
Arora, S., Ge, R., Halpern, Y., Mimno, D., Moitra, A., Sontag, D., Wu, Y., and
  Zhu, M. (2013).
\newblock A practical algorithm for topic modeling with provable guarantees.
\newblock In Dasgupta, S. and McAllester, D., editors, {\em Proceedings of the
  30th International Conference on Machine Learning}, pages 280--288.
  Proceedings of Machine Learning Research.

\bibitem[Azad and Deepak, 2019]{Azad2019}
Azad, H.~K. and Deepak, A. (2019).
\newblock Query expansion techniques for information retrieval: {A} survey.
\newblock {\em Information Processing and Management}, 56(5):1698--1735.
\newblock https://doi.org/10.1016/j.ipm.2019.05.009.

\bibitem[Azar, 2009]{Azar2009}
Azar, E.~E. (2009).
\newblock {\em {Conflict and Peace Data Bank (COPDAB), 1948-1978}}.
\newblock [Data set]. Inter-University Consortium for Political and Social
  Research. https://doi.org/10.3886/ICPSR07767.v4.

\bibitem[Baden et~al., 2020]{Baden2020}
Baden, C., Kligler-Vilenchik, N., and Yarchi, M. (2020).
\newblock {Hybrid Content Analysis}: Toward a strategy for the theory-driven,
  computer-assisted classification of large text corpora.
\newblock {\em Communication Methods and Measures}, 14(3):165--183.
\newblock https://doi.org/10.1080/19312458.2020.1803247.

\bibitem[Baerg and Lowe, 2020]{Baerg2020}
Baerg, N. and Lowe, W. (2020).
\newblock A textual {Taylor} rule: Estimating central bank preferences
  combining topic and scaling methods.
\newblock {\em Political Science Research and Methods}, 8(1):106--122.
\newblock https://doi.org/10.1017/psrm.2018.31.

\bibitem[Bahdanau et~al., 2015]{Bahdanau2015}
Bahdanau, D., Cho, K., and Bengio, Y. (2015).
\newblock Neural machine translation by jointly learning to align and
  translate.
\newblock In Bengio, Y. and LeCun, Y., editors, {\em 3rd International
  Conference on Learning Representations ({ICLR} 2015)}, pages 1--15.

\bibitem[Barber{\'a}, 2016]{Barbera2016}
Barber{\'a}, P. (2016).
\newblock Less is more? {H}ow demographic sample weights can improve public
  opinion estimates based on twitter data.
\newblock Manuscript.
\newblock Retrieved June 4, 2021, from,
  http://pablobarbera.com/static/less-is-more.pdf.

\bibitem[Bauer et~al., 2017]{Bauer:2017kr}
Bauer, P.~C., Barber{\'a}, P., Ackermann, K., and Venetz, A. (2017).
\newblock Is the left-right scale a valid measure of ideology?
\newblock {\em Political Behavior}, 39(3):553--583.
\newblock https://doi.org/10.1007/s11109-016-9368-2.

\bibitem[Baum et~al., 2018]{Baum2018}
Baum, M., Cohen, D.~K., and Zhukov, Y.~M. (2018).
\newblock Does rape culture predict rape? {E}vidence from {U.S.} newspapers,
  2000-2013.
\newblock {\em Quarterly Journal of Political Science}, 13(3):263--289.
\newblock http://dx.doi.org/10.1561/100.00016124.

\bibitem[B{\"a}uml, 2007]{Baeuml2007}
B{\"a}uml, K.-H. (2007).
\newblock Making memories unavailable: The inhibitory power of retrieval.
\newblock {\em {Journal of Psychology}}, 215(1):4--11.
\newblock https://doi.org/10.1027/0044-3409.215.1.4.

\bibitem[Beauchamp, 2017]{Beauchamp:2017cp}
Beauchamp, N. (2017).
\newblock Predicting and interpolating state-level polls using {Twitter}
  textual data.
\newblock {\em American Journal of Political Science}, 61(2):490--503.
\newblock https://doi.org/10.1111/ajps.12274.

\bibitem[Beltagy et~al., 2020]{Beltagy2020}
Beltagy, I., Peters, M.~E., and Cohan, A. (2020).
\newblock Longformer: The long-document {Transformer}.
\newblock arXiv.
\newblock https://arxiv.org/abs/2004.05150.

\bibitem[Bengio et~al., 2003]{Bengio2003}
Bengio, Y., Ducharme, R., Vincent, P., and Janvin, C. (2003).
\newblock A neural probabilistic language model.
\newblock {\em {Journal of Machine Learning Research}}, 3:1137--1155.

\bibitem[Benoit et~al., 2018]{Benoit2018}
Benoit, K., Watanabe, K., Wang, H., Nulty, P., Obeng, A., M\"uller, S., and
  Matsuo, A. (2018).
\newblock {quanteda}: An {R} package for the quantitative analysis of textual
  data.
\newblock {\em Journal of Open Source Software}, 3(30):774.
\newblock https://doi.org/10.21105/joss.00774.

\bibitem[Blei and Lafferty, 2007]{Blei2007a}
Blei, D.~M. and Lafferty, J.~D. (2007).
\newblock A correlated topic model of science.
\newblock {\em The Annals of Applied Statistics}, 1(1):17--35.
\newblock https://doi.org/10.1214/07-AOAS114.

\bibitem[Blei et~al., 2003]{Blei:2003hl}
Blei, D.~M., Ng, A.~Y., and Jordan, M.~I. (2003).
\newblock Latent {D}irichlet {A}llocation.
\newblock {\em Journal of Machine Learning Research}, 3:993--1022.

\bibitem[Bojanowski et~al., 2017]{Bojanowski2017}
Bojanowski, P., Grave, E., Joulin, A., and Mikolov, T. (2017).
\newblock Enriching word vectors with subword information.
\newblock {\em Transactions of the Association for Computational Linguistics},
  5:135--146.
\newblock https://doi.org/10.1162/tacl\_a\_00051.

\bibitem[Boser et~al., 1992]{Boser1992}
Boser, B.~E., Guyon, I.~M., and Vapnik, V.~N. (1992).
\newblock A training algorithm for optimal margin classifiers.
\newblock In Haussler, D., editor, {\em Proceedings of the Fifth Annual
  Workshop on Computational Learning Theory ({COLT '92})}, pages 144--152.
  Association for Computing Machinery.

\bibitem[Branco et~al., 2016]{Branco2016}
Branco, P., Torgo, L., and Ribeiro, R.~P. (2016).
\newblock A survey of predictive modeling on imbalanced domains.
\newblock {\em ACM Computing Surveys}, 49(2):1--50.
\newblock https://doi.org/10.1145/2907070.

\bibitem[Brownlee, 2020]{Brownlee2020b}
Brownlee, J. (2020).
\newblock Cost-sensitive learning for imbalanced classification.
\newblock {\em Machine Learning Mastery}.
\newblock Retrieved June 9, 2021, from
  https://machinelearningmastery.com/cost-sensitive-learning-for-imbalanced-classification/.

\bibitem[Brownlee, 2021a]{Brownlee2021}
Brownlee, J. (2021a).
\newblock Random oversampling and undersampling for imbalanced classification.
\newblock {\em Machine Learning Mastery}.
\newblock Retrieved June 8, 2021, from
  https://machinelearningmastery.com/random-oversampling-and-undersampling-for-imbalanced-classification/.

\bibitem[Brownlee, 2021b]{Brownlee2021a}
Brownlee, J. (2021b).
\newblock {SMOTE} for imbalanced classification with {Python}.
\newblock {\em Machine Learning Mastery}.
\newblock Retrieved June 8, 2021, from
  https://machinelearningmastery.com/smote-oversampling-for-imbalanced-classification/.

\bibitem[Burnap et~al., 2016]{Burnap:2016kv}
Burnap, P., Gibson, R., Sloan, L., Southern, R., and Williams, M. (2016).
\newblock 140 characters to victory?: {U}sing {Twitter} to predict the {UK 2015
  General Election}.
\newblock {\em Electoral Studies}, 41:230--233.
\newblock https://doi.org/10.1016/j.electstud.2015.11.017.

\bibitem[Chawla, 2005]{Chawla2005}
Chawla, N.~V. (2005).
\newblock Data mining for imbalanced datasets: An overview.
\newblock In Maimon, O. and Rokach, L., editors, {\em The Data Mining and
  Knowledge Discovery Handbook}, pages 853--867. Springer.

\bibitem[Chawla et~al., 2002]{Chawla2002}
Chawla, N.~V., Bowyer, K.~W., Hall, L.~O., and Kegelmeyer, W.~P. (2002).
\newblock {SMOTE}: Synthetic minority over-sampling technique.
\newblock {\em Journal of Artificial Intelligence Research}, 16(1):321--357.

\bibitem[Cortes and Vapnik, 1995]{Cortes1995}
Cortes, C. and Vapnik, V. (1995).
\newblock Support-vector networks.
\newblock {\em Machine Learning}, 20(3):273--297.

\bibitem[Dahl et~al., 2019]{xtable2019}
Dahl, D.~B., Scott, D., Roosen, C., Magnusson, A., and Swinton, J. (2019).
\newblock {\em {xtable}: Export Tables to {LaTeX} or {HTML} {(Version 1.8-4)}}.
\newblock [R package]. CRAN.
  https://cran.r-project.org/web/packages/xtable/index.html.

\bibitem[Devlin et~al., 2019]{Devlin2019}
Devlin, J., Chang, M.-W., Lee, K., and Toutanova, K. (2019).
\newblock {BERT}: Pre-training of deep bidirectional {T}ransformers for
  language understanding.
\newblock In Burstein, J., Doran, C., and Solorio, T., editors, {\em
  Proceedings of the 2019 Conference of the North {A}merican Chapter of the
  Association for Computational Linguistics: Human Language Technologies},
  pages 4171--4186. Association for Computational Linguistics.
\newblock https://doi.org/10.18653/v1/N19-1423.

\bibitem[Diaz et~al., 2016]{Diaz2016}
Diaz, F., Mitra, B., and Craswell, N. (2016).
\newblock Query expansion with locally-trained word embeddings.
\newblock In Erk, K. and Smith, N.~A., editors, {\em {Proceedings of the 54th
  Annual Meeting of the Association for Computational Linguistics}}, pages
  367--377. Association for Computational Linguistics.
\newblock https://doi.org/10.18653/v1/P16-1035.

\bibitem[Diermeier et~al., 2011]{Diermeier2011}
Diermeier, D., Godbout, J.-F., Yu, B., and Kaufmann, S. (2011).
\newblock Language and ideology in {Congress}.
\newblock {\em British Journal of Political Science}, 42(1):31--55.
\newblock https://doi.org/10.1017/S0007123411000160.

\bibitem[Dodge et~al., 2020]{Dodge2020}
Dodge, J., Ilharco, G., Schwartz, R., Farhadi, A., Hajishirzi, H., and Smith,
  N. (2020).
\newblock Fine-tuning pretrained language models: Weight initializations, data
  orders, and early stopping.
\newblock arXiv. arXiv:2002.06305v1 [cs.CL].
  https://arxiv.org/abs/2002.06305v1.

\bibitem[D'Orazio et~al., 2014]{DOrazio2014}
D'Orazio, V., Landis, S.~T., Palmer, G., and Schrodt, P. (2014).
\newblock Separating the wheat from the chaff: Applications of automated
  document classification using {Support Vector Machines}.
\newblock {\em Political Analysis}, 22(2):224--242.
\newblock https://doi.org/10.1093/pan/mpt030.

\bibitem[Dowle and Srinivasan, 2020]{datatable2020}
Dowle, M. and Srinivasan, A. (2020).
\newblock {\em {data.table}: Extension of {`data.frame`} {(Version 1.13.0)}}.
\newblock [R package]. CRAN. https://CRAN.R-project.org/package=data.table.

\bibitem[Durrell, 2008]{Durrell2008}
Durrell, M. (2008).
\newblock Linguistic variable - linguistic variant.
\newblock In Ammon, U., Dittmar, N., Mattheier, K.~J., and Trudgill, P.,
  editors, {\em Sociolinguistics}, pages 195--200. De Gruyter Mouton.
\newblock https://doi.org/10.1515/9783110141894.1.2.195.

\bibitem[Ein-Dor et~al., 2020]{EinDor2020}
Ein-Dor, L., Halfon, A., Gera, A., Shnarch, E., Dankin, L., Choshen, L.,
  Danilevsky, M., Aharonov, R., Katz, Y., and Slonim, N. (2020).
\newblock {A}ctive learning for {BERT}: {A}n empirical study.
\newblock In Webber, B., Cohn, T., He, Y., and Liu, Y., editors, {\em
  Proceedings of the 2020 Conference on Empirical Methods in Natural Language
  Processing (EMNLP)}, pages 7949--7962. Association for Computational
  Linguistics.
\newblock https://doi.org/10.18653/v1/2020.emnlp-main.638.

\bibitem[Elkan, 2001]{Elkan2001}
Elkan, C. (2001).
\newblock The foundations of cost-sensitive learning.
\newblock In {\em Proceedings of the 17th International Joint Conference on
  Artificial Intelligence ({IJCAI '01})}, pages 973--978. Morgan Kaufmann
  Publishers Inc.

\bibitem[Erlich et~al., 2021]{Erlich2021}
Erlich, A., Dantas, S.~G., Bagozzi, B.~E., Berliner, D., and Palmer-Rubin, B.
  (2021).
\newblock Multi-label prediction for political text-as-data.
\newblock {\em Political Analysis}, pages 1--18.
\newblock https://doi.org/10.1017/pan.2021.15.

\bibitem[Ertekin et~al., 2007]{Ertekin2007}
Ertekin, S., Huang, J., Bottou, L., and Giles, L. (2007).
\newblock Learning on the border: Active learning in imbalanced data
  classification.
\newblock In {\em Proceedings of the Sixteenth ACM Conference on Information
  and Knowledge Management ({CIKM '07})}, pages 127--136. Association for
  Computing Machinery.
\newblock https://doi.org/10.1145/1321440.1321461.

\bibitem[Eshima et~al., 2021]{Eshima2021}
Eshima, S., Imai, K., and Sasaki, T. (2021).
\newblock Keyword assisted topic models.
\newblock arXiv. arXiv:2004.05964v2 [cs.CL].
  https://arxiv.org/abs/2004.05964v2.

\bibitem[Firth, 1957]{Firth1957}
Firth, J.~R. (1957).
\newblock {\em Studies in Linguistic Analysis}.
\newblock Publications of the Philological Society. Blackwell.

\bibitem[Fogel-Dror et~al., 2019]{FogelDror2019}
Fogel-Dror, Y., Shenhav, S.~R., Sheafer, T., and Atteveldt, W.~V. (2019).
\newblock Role-based association of verbs, actions, and sentiments with
  entities in political discourse.
\newblock {\em Communication Methods and Measures}, 13(2):69--82.
\newblock https://doi.org/10.1080/19312458.2018.1536973.

\bibitem[Gessler and Hunger, 2021]{Gessler2021}
Gessler, T. and Hunger, S. (2021).
\newblock How the refugee crisis and radical right parties shape party
  competition on immigration.
\newblock {\em Political Science Research and Methods}, pages 1--21.
\newblock https://doi.org/10.1017/psrm.2021.64.

\bibitem[{Google Colaboratory}, 2020]{GoogleColab2020}
{Google Colaboratory} (2020).
\newblock {\em {Google Colaboratory Frequently Asked Questions}}.
\newblock Google Colaboratory.
\newblock Retrieved October 28, 2020, from
  https://research.google.com/colaboratory/faq.html.

\bibitem[Grimmer, 2013]{Grimmer:2013df}
Grimmer, J. (2013).
\newblock Appropriators not position takers: The distorting effects of
  electoral incentives on {C}ongressional representation.
\newblock {\em American Journal of Political Science}, 57(3):624--642.
\newblock https://doi.org/10.1111/ajps.12000.

\bibitem[Gr\"un and Hornik, 2011]{topicmodels:2011gh}
Gr\"un, B. and Hornik, K. (2011).
\newblock {topicmodels}: An {R} package for fitting topic models.
\newblock {\em Journal of Statistical Software}, 40(13):1--30.
\newblock https://doi.org/10.18637/jss.v040.i13.

\bibitem[Howard and Ruder, 2018]{Howard2018}
Howard, J. and Ruder, S. (2018).
\newblock Universal language model fine-tuning for text classification.
\newblock In Gurevych, I. and Miyao, Y., editors, {\em Proceedings of the 56th
  Annual Meeting of the Association for Computational Linguistics}, pages
  328--339. Association for Computational Linguistics.
\newblock https://doi.org/10.18653/v1/P18-1031.

\bibitem[{HuggingFace}, 2021]{Huggingface2021}
{HuggingFace} (2021).
\newblock {\em Dataset card for reuters21578}.
\newblock Retrieved May 19, 2021, from
  https://huggingface.co/datasets/reuters21578.

\bibitem[Hunter, 2007]{Hunter2007}
Hunter, J.~D. (2007).
\newblock Matplotlib: A {2D} graphics environment.
\newblock {\em {Computing in Science \& Engineering}}, 9(3):90--95.
\newblock https://doi.org/10.1109/MCSE.2007.55.

\bibitem[Jungherr et~al., 2016]{Jungherr2016}
Jungherr, A., Schoen, H., and J{\"u}rgens, P. (2016).
\newblock The mediation of politics through {T}witter: An analysis of messages
  posted during the campaign for the {German Federal Election 2013}.
\newblock {\em Journal of Computer-Mediated Communication}, 21(1):50--68.
\newblock https://doi.org/10.1111/jcc4.12143.

\bibitem[Katagiri and Min, 2019]{Katagiri2019}
Katagiri, A. and Min, E. (2019).
\newblock The credibility of public and private signals: A document-based
  approach.
\newblock {\em American Political Science Review}, 113(1):156--172.
\newblock https://doi.org/10.1017/S0003055418000643.

\bibitem[Kentaro, 2020]{Kentaro2020}
Kentaro, W. (2020).
\newblock {\em {gdown}: Download a Large File from {Google Drive}}.
\newblock [Python package]. GitHub. https://github.com/wkentaro/gdown.

\bibitem[King et~al., 2017]{King:2017io}
King, G., Lam, P., and Roberts, M.~E. (2017).
\newblock Computer-assisted keyword and document set discovery from
  unstructured text.
\newblock {\em American Journal of Political Science}, 61(4):971--988.
\newblock https://doi.org/10.1111/ajps.12291.

\bibitem[King et~al., 2013]{King2013}
King, G., Pan, J., and Roberts, M.~E. (2013).
\newblock How censorship in {C}hina allows government criticism but silences
  collective expression.
\newblock {\em American Political Science Review}, 107(2):326--343.
\newblock https://doi.org/10.1017/S0003055413000014.

\bibitem[Kouw and Loog, 2019]{Kouw2019}
Kouw, W.~M. and Loog, M. (2019).
\newblock A review of domain adaptation without target labels.
\newblock arXiv.
\newblock https://arxiv.org/abs/1901.053353.

\bibitem[Krippendorff, 2013]{Krippendorff2013}
Krippendorff, K. (2013).
\newblock {\em Content Analysis: An Introduction to Its Methodology}.
\newblock Sage Publications, 3rd edition.

\bibitem[Kuzi et~al., 2016]{Kuzi2016}
Kuzi, S., Shtok, A., and Kurland, O. (2016).
\newblock Query expansion using word embeddings.
\newblock In {\em Proceedings of the 25th ACM International on Conference on
  Information and Knowledge Management ({CIKM '16})}, pages 1929--1932.
  Association for Computing Machinery.
\newblock https://doi.org/10.1145/2983323.2983876.

\bibitem[Lavrenko and Croft, 2001]{Lavrenko2001}
Lavrenko, V. and Croft, W.~B. (2001).
\newblock Relevance based language models.
\newblock In {\em Proceedings of the 24th Annual International ACM SIGIR
  Conference on Research and Development in Information Retrieval (SIGIR '01)},
  pages 120--127. Association for Computing Machinery.
\newblock https://doi.org/10.1145/383952.383972.

\bibitem[Lema{{\^i}}tre et~al., 2017]{Lemaitre2017}
Lema{{\^i}}tre, G., Nogueira, F., and Aridas, C.~K. (2017).
\newblock Imbalanced-learn: A {Python} toolbox to tackle the curse of
  imbalanced datasets in machine learning.
\newblock {\em Journal of Machine Learning Research}, 18(17):1--5.

\bibitem[Lewis, 1997]{Reuters1997}
Lewis, D.~D. (1997).
\newblock {\em {Reuters-21578 (Distribution 1.0)}}.
\newblock [Data set].
  http://www.daviddlewis.com/resources/testcollections/reuters21578/.

\bibitem[Lewis and Gale, 1994]{Lewis1994}
Lewis, D.~D. and Gale, W.~A. (1994).
\newblock A sequential algorithm for training text classifiers.
\newblock In Croft, B.~W. and van Rijsbergen, C.~J., editors, {\em Proceedings
  of the 17th Annual International ACM SIGIR Conference on Research and
  Development in Information Retrieval (SIGIR '94)}, pages 3--12. Springer.

\bibitem[Linder, 2017]{Linder2017}
Linder, F. (2017).
\newblock Reducing bias in online text datasets: Query expansion and active
  learning for better data from keyword searches.
\newblock SSRN.
\newblock http://dx.doi.org/10.2139/ssrn.3026393.

\bibitem[Loshchilov and Hutter, 2019]{Loshchilov2019}
Loshchilov, I. and Hutter, F. (2019).
\newblock Decoupled weight decay regularization.
\newblock In {\em 7th International Conference on Learning Representations
  ({ICLR} 2019)}. OpenReview.net.

\bibitem[Maier et~al., 2018]{Maier2018}
Maier, D., Waldherr, A., Miltner, P., Wiedemann, G., Niekler, A., Keinert, A.,
  Pfetsch, B., Heyer, G., Reber, U., Häussler, T., Schmid-Petri, H., and Adam,
  S. (2018).
\newblock Applying {LDA} topic modeling in communication research: Toward a
  valid and reliable methodology.
\newblock {\em Communication Methods and Measures}, 12(2-3):93--118.
\newblock https://doi.org/10.1080/19312458.2018.1430754.

\bibitem[Manning et~al., 2008]{Manning:2008vf}
Manning, C.~D., Raghavan, P., and Sch{\"u}tze, H. (2008).
\newblock {\em {Introduction to Information Retrieval}}.
\newblock Cambridge University Press.

\bibitem[Manning and Sch{\"u}tze, 1999]{Manning1999}
Manning, C.~D. and Sch{\"u}tze, H. (1999).
\newblock {\em {Foundations of Statistical Natural Language Processing}}.
\newblock MIT Press, Cambridge.
\newblock
  https://mitpress.mit.edu/books/foundations-statistical-natural-language-processing.

\bibitem[McKinney, 2010]{McKinney2010}
McKinney, W. (2010).
\newblock Data structures for statistical computing in {Python}.
\newblock In {S}t\'efan van~der {W}alt and {J}arrod {M}illman, editors, {\em
  Proceedings of the 9th Python in Science Conference (SciPy 2010)}, pages
  56--61.
\newblock https://doi.org/10.25080/Majora-92bf1922-00a.

\bibitem[{Michael Waskom and Team}, 2020]{Waskom2020}
{Michael Waskom and Team} (2020).
\newblock {\em {Seaborn}}.
\newblock [Python package]. Zenodo. {https://zenodo.org/record/4379347}.

\bibitem[Mikolov et~al., 2013a]{Mikolov2013b}
Mikolov, T., Chen, K., Corrado, G., and Dean, J. (2013a).
\newblock Efficient estimation of word representations in vector space.
\newblock arXiv.
\newblock arXiv:1301.3781v3 [cs.CL]. https://arxiv.org/abs/1301.3781.

\bibitem[Mikolov et~al., 2013b]{Mikolov2013a}
Mikolov, T., Yih, W.-t., and Zweig, G. (2013b).
\newblock Linguistic regularities in continuous space word representations.
\newblock In Vanderwende, L., Daum{\'e}~III, H., and Kirchhoff, K., editors,
  {\em Proceedings of the 2013 Conference of the North {A}merican Chapter of
  the Association for Computational Linguistics: Human Language Technologies},
  pages 746--751. Association for Computational Linguistics.

\bibitem[Miller et~al., 2020]{Miller2020}
Miller, B., Linder, F., and Mebane, W.~R. (2020).
\newblock Active learning approaches for labeling text: Review and assessment
  of the performance of active learning approaches.
\newblock {\em Political Analysis}, 28(4):532--551.
\newblock https://doi.org/10.1017/pan.2020.4.

\bibitem[Moore and Siegel, 2013]{Moore2013}
Moore, W.~H. and Siegel, D.~A. (2013).
\newblock {\em {A Mathematics Course for Political and Social Research}}.
\newblock Princeton University Press.

\bibitem[Mosbach et~al., 2021]{Mosbach2021}
Mosbach, M., Andriushchenko, M., and Klakow, D. (2021).
\newblock On the stability of fine-tuning {BERT}: Misconceptions, explanations,
  and strong baselines.
\newblock In {\em International Conference on Learning Representations (ICLR
  2021)}. OpenReview.net.

\bibitem[Muchlinski et~al., 2021]{Muchlinski2021}
Muchlinski, D., Yang, X., Birch, S., Macdonald, C., and Ounis, I. (2021).
\newblock We need to go deeper: Measuring electoral violence using
  {Convolutional Neural Networks} and social media.
\newblock {\em Political Science Research and Methods}, 9(1):122--139.
\newblock https://doi.org/10.1017/psrm.2020.32.

\bibitem[{M\"unchener Digitalisierungszentrum der Bayerischen Staatsbibliothek
  (dbmdz)}, 2021]{dbmdz2021}
{M\"unchener Digitalisierungszentrum der Bayerischen Staatsbibliothek (dbmdz)}
  (2021).
\newblock {\em Model card for bert-base-german-uncased from dbmdz}.
\newblock Retrieved May 19, 2021, from
  https://huggingface.co/dbmdz/bert-base-german-uncased.

\bibitem[Neelakantan et~al., 2014]{Neelakantan2014}
Neelakantan, A., Shankar, J., Passos, A., and McCallum, A. (2014).
\newblock Efficient non-parametric estimation of multiple embeddings per word
  in vector space.
\newblock In Moschitti, A., Pang, B., and Daelemans, W., editors, {\em
  Proceedings of the 2014 Conference on Empirical Methods in Natural Language
  Processing ({EMNLP})}, pages 1059--1069, Stroudsburg, PA, USA. Association
  for Computational Linguistics.
\newblock https://doi.org/10.3115/v1/D14-1113.

\bibitem[Oliphant, 2006]{Oliphant2006}
Oliphant, T.~E. (2006).
\newblock {\em {A Guide to NumPy}}.
\newblock Trelgol Publishing USA.

\bibitem[{Oller Moreno}, 2021]{facetscales2021}
{Oller Moreno}, S. (2021).
\newblock {\em {facetscales}: {facet grid} with different scales per {facet}
  {(Version 0.1.0.9000)}}.
\newblock [R package]. GitHub. https://github.com/zeehio/facetscales.

\bibitem[Paszke et~al., 2019]{Paszke2019}
Paszke, A., Gross, S., Massa, F., Lerer, A., Bradbury, J., Chanan, G., Killeen,
  T., Lin, Z., Gimelshein, N., Antiga, L., Desmaison, A., Kopf, A., Yang, E.,
  DeVito, Z., Raison, M., Tejani, A., Chilamkurthy, S., Steiner, B., Fang, L.,
  Bai, J., and Chintala, S. (2019).
\newblock {PyTorch}: An imperative style, high-performance deep learning
  library.
\newblock In Wallach, H., Larochelle, H., Beygelzimer, A., d'Alch\'{e} Buc, F.,
  Fox, E., and Garnett, R., editors, {\em Advances in Neural Information
  Processing Systems 32}, pages 8024--8035. Curran Associates, Inc.

\bibitem[Pedregosa et~al., 2011]{sklearn2011}
Pedregosa, F., Varoquaux, G., Gramfort, A., Michel, V., Thirion, B., Grisel,
  O., Blondel, M., Prettenhofer, P., Weiss, R., Dubourg, V., Vanderplas, J.,
  Passos, A., Cournapeau, D., Brucher, M., Perrot, M., and Duchesnay, E.
  (2011).
\newblock Scikit-learn: Machine learning in {P}ython.
\newblock {\em Journal of Machine Learning Research}, 12:2825--2830.

\bibitem[Pennington et~al., 2014]{Pennington2014}
Pennington, J., Socher, R., and Manning, C. (2014).
\newblock {G}lo{V}e: Global vectors for word representation.
\newblock In Moschitti, A., Pang, B., and Daelemans, W., editors, {\em
  {Proceedings of the 2014 Conference on Empirical Methods in Natural Language
  Processing ({EMNLP})}}, pages 1532--1543. Association for Computational
  Linguistics.
\newblock https://doi.org/10.3115/v1/D14-1162.

\bibitem[Phang et~al., 2019]{Phang2019}
Phang, J., Févry, T., and Bowman, S.~R. (2019).
\newblock Sentence encoders on {STILT}s: Supplementary training on intermediate
  labeled-data tasks.
\newblock arXiv.
\newblock {arXiv:1811.01088v2 [cs.CL].} https://arxiv.org/abs/1811.01088v2.

\bibitem[Pilehvar and Camacho-Collados, 2020]{Pilehvar2021}
Pilehvar, M.~T. and Camacho-Collados, J. (2020).
\newblock {\em Embeddings in Natural Language Processing: {T}heory and Advances
  in Vector Representations of Meaning}.
\newblock Morgan {\&} Claypool Publishers, San Rafael, CA, USA.
\newblock https://doi.org/10.2200/S01057ED1V01Y202009HLT047.

\bibitem[Pilny et~al., 2019]{Pilny2019}
Pilny, A., McAninch, K., Slone, A., and Moore, K. (2019).
\newblock Using supervised machine learning in automated content analysis: An
  example using relational uncertainty.
\newblock {\em Communication Methods and Measures}, 13(4):287--304.
\newblock https://doi.org/10.1080/19312458.2019.1650166.

\bibitem[Puglisi and Snyder, 2011]{Puglisi2011}
Puglisi, R. and Snyder, J.~M. (2011).
\newblock Newspaper coverage of political scandals.
\newblock {\em The Journal of Politics}, 73(3):931--950.
\newblock https://doi.org/10.1017/s0022381611000569.

\bibitem[Quinn et~al., 2010]{Quinn:2010jl}
Quinn, K.~M., Monroe, B.~L., Colaresi, M., Crespin, M.~H., and Radev, D.~R.
  (2010).
\newblock How to analyze political attention with minimal assumptions and
  costs.
\newblock {\em American Journal of Political Science}, 54(1):209--228.
\newblock https://doi.org/10.1111/j.1540-5907.2009.00427.x.

\bibitem[{R Core Team}, 2020]{RCoreTeam2020}
{R Core Team} (2020).
\newblock {\em {R}: A Language and Environment for Statistical Computing}.
\newblock R Foundation for Statistical Computing, Vienna, Austria.
\newblock https://www.R-project.org/.

\bibitem[Raschka, 2020]{Raschka2020}
Raschka, S. (2020).
\newblock {\em {watermark}}.
\newblock [Python package]. GitHub. {https://github.com/rasbt/watermark}.

\bibitem[Rauh et~al., 2020]{Rauh2019}
Rauh, C., Bes, B.~J., and Schoonvelde, M. (2020).
\newblock Undermining, defusing or defending {E}uropean integration? assessing
  public communication of {E}uropean executives in times of {EU}
  politicisation.
\newblock {\em European Journal of Political Research}, 59:397--423.
\newblock https://doi.org/10.1111/1475-6765.12350.

\bibitem[Reda et~al., 2021]{AbdulReda2021}
Reda, A.~A., Sinanoglu, S., and Abdalla, M. (2021).
\newblock Mobilizing the masses: Measuring resource mobilization on twitter.
\newblock {\em Sociological Methods \& Research}, pages 1--40.

\bibitem[Reimers and Gurevych, 2018]{Reimers2018}
Reimers, N. and Gurevych, I. (2018).
\newblock Why comparing single performance scores does not allow to draw
  conclusions about machine learning approaches.
\newblock arXiv.
\newblock http://arxiv.org/abs/1803.09578.

\bibitem[Richardson, 2020]{bs42020}
Richardson, L. (2020).
\newblock {\em {Beautiful Soup 4}}.
\newblock [Python library]. Crummy.
  https://www.crummy.com/software/BeautifulSoup/.

\bibitem[Roberts et~al., 2016a]{Roberts2016}
Roberts, M.~E., Stewart, B.~M., and Airoldi, E.~M. (2016a).
\newblock A model of text for experimentation in the social sciences.
\newblock {\em Journal of the American Statistical Association},
  111(515):988--1003.
\newblock https://doi.org/10.1080/01621459.2016.1141684.

\bibitem[Roberts et~al., 2016b]{Roberts2016wg}
Roberts, M.~E., Stewart, B.~M., and Tingley, D. (2016b).
\newblock Navigating the local modes of big data: The case of topic models.
\newblock In Alvarez, R.~M., editor, {\em {Computational Social Science:
  Discovery and Prediction}}, pages 51--97. Cambridge University Press.

\bibitem[Roberts et~al., 2019]{stm2019}
Roberts, M.~E., Stewart, B.~M., and Tingley, D. (2019).
\newblock {stm}: An {R} package for {Structural Topic Models}.
\newblock {\em Journal of Statistical Software}, 91(2):1--40.
\newblock https://doi.org/10.18637/jss.v091.i02.

\bibitem[Roberts et~al., 2014]{Roberts:2014es}
Roberts, M.~E., Stewart, B.~M., Tingley, D., Lucas, C., Luis, J.~L., Gadarian,
  S.~K., Albertson, B., and Rand, D.~G. (2014).
\newblock {Structural Topic Models} for open-ended survey responses.
\newblock {\em American Journal of Political Science}, 58(4):1064--1082.
\newblock https://doi.org/10.1111/ajps.12103.

\bibitem[Rodriguez and Spirling, 2022]{Rodriguez2022}
Rodriguez, P.~L. and Spirling, A. (2022).
\newblock Word embeddings: What works, what doesn’t, and how to tell the
  difference for applied research.
\newblock {\em The Journal of Politics}, 84(1):101--115.
\newblock https://doi.org/10.1086/715162.

\bibitem[Ruder, 2019]{Ruder2019}
Ruder, S. (2019).
\newblock {\em Neural Transfer Learning for Natural Language Processing}.
\newblock PhD thesis, National University of Ireland, Galway.

\bibitem[Ruder, 2020]{Ruder2020c}
Ruder, S. (2020).
\newblock {\em NLP-Progress}.
\newblock Retrieved June 21, 2021, from
  https://nlpprogress.com/english/text\_classification.html.

\bibitem[Sap et~al., 2020]{Sap2020}
Sap, M., Gabriel, S., Qin, L., Jurafsky, D., Smith, N.~A., and Choi, Y. (2020).
\newblock Social bias frames: Reasoning about social and power implications of
  language.
\newblock In Jurafsky, D., Chai, J., Schluter, N., and Tetreault, J., editors,
  {\em Proceedings of the 58th Annual Meeting of the Association for
  Computational Linguistics}, pages 5477--5490. Association for Computational
  Linguistics.
\newblock https://doi.org/10.18653/v1/2020.acl-main.486.

\bibitem[Schulze et~al., 2021]{Schulze2021}
Schulze, P., Wiegrebe, S., Thurner, P.~W., Heumann, C., A{\ss}enmacher, M., and
  Wankm{\"u}ller, S. (2021).
\newblock Exploring topic-metadata relationships with the {STM}: A {B}ayesian
  approach.
\newblock arXiv.
\newblock {arXiv:2104.02496v1 [cs.CL]}. https://arxiv.org/abs/2104.02496.

\bibitem[Sch{\"u}tze, 1998]{Schuetze1998}
Sch{\"u}tze, H. (1998).
\newblock Automatic word sense discrimination.
\newblock {\em Computational Linguistics}, 24(1):97--123.
\newblock https://aclanthology.org/J98-1004.

\bibitem[{scikit-learn Developers}, 2020a]{SklearnUserGuide2020b}
{scikit-learn Developers} (2020a).
\newblock {\em {1.4. Support Vector Machines}}.
\newblock Retrieved November 23, 2020, from
  https://scikit-learn.org/stable/modules/svm.html.

\bibitem[{scikit-learn Developers}, 2020b]{SklearnUserGuide2020a}
{scikit-learn Developers} (2020b).
\newblock {\em {RBF SVM Parameters}}.
\newblock Retrieved November 23, 2020, from
  https://scikit-learn.org/stable/auto\_examples/svm/plot\_rbf\_parameters.html.

\bibitem[Seb{\H{o}}k and Kacsuk, 2020]{Seboek2020}
Seb{\H{o}}k, M. and Kacsuk, Z. (2020).
\newblock The multiclass classification of newspaper articles with machine
  learning: The hybrid binary snowball approach.
\newblock {\em Political Analysis}, pages 1--14.
\newblock https://doi.org/10.1017/pan.2020.27.

\bibitem[Selivanov et~al., 2020]{text2vec2020}
Selivanov, D., Bickel, M., and Wang, Q. (2020).
\newblock {\em {text2vec}: Modern Text Mining Framework for {R}}.
\newblock [R package]. CRAN. https://CRAN.R-project.org/package=text2vec.

\bibitem[Settles, 2010]{Settles2010}
Settles, B. (2010).
\newblock {\em Active Learning Literature Survey}.
\newblock Computer Sciences Technical Report 1648. University of
  Wisconsin--Madison.
\newblock http://burrsettles.com/pub/settles.activelearning.pdf.

\bibitem[Silva and Mendoza, 2020]{Silva2020}
Silva, A. and Mendoza, M. (2020).
\newblock Improving query expansion strategies with word embeddings.
\newblock In {\em Proceedings of the ACM Symposium on Document Engineering 2020
  (DocEng '20)}, pages 1--4. Association for Computing Machinery.
\newblock https://doi.org/10.1145/3395027.3419601.

\bibitem[Socher et~al., 2013]{Socher2013}
Socher, R., Perelygin, A., Wu, J., Chuang, J., Manning, C.~D., Ng, A., and
  Potts, C. (2013).
\newblock Recursive deep models for semantic compositionality over a sentiment
  treebank.
\newblock In Yarowsky, D., Baldwin, T., Korhonen, A., Livescu, K., and Bethard,
  S., editors, {\em {Proceedings of the 2013 Conference on Empirical Methods in
  Natural Language Processing}}, pages 1631--1642. Association for
  Computational Linguistics.

\bibitem[Soetaert, 2019]{plot3D2019}
Soetaert, K. (2019).
\newblock {\em {plot3D}: Plotting Multi-Dimensional Data {(Version 1.3)}}.
\newblock [R package]. CRAN. https://CRAN.R-project.org/package=plot3D.

\bibitem[Stier et~al., 2018]{Stier2018}
Stier, S., Bleier, A., Bonart, M., M\"orsheim, F., Bohlouli, M., Nizhegorodov,
  M., Posch, L., Maier, J., Rothmund, T., and Staab, S. (2018).
\newblock {\em Systematically Monitoring Social Media: the Case of the {German
  Federal Election 2017}}.
\newblock GESIS - Leibniz-Institut f\"ur Sozialwissenschaften.
\newblock https://doi.org/10.21241/ssoar.56149.

\bibitem[Sun et~al., 2019]{Sun2019}
Sun, C., Qiu, X., Xu, Y., and Huang, X. (2019).
\newblock How to fine-tune {BERT} for text classification?
\newblock arXiv.
\newblock {arXiv:1905.05583v3 [cs.CL]}. https://arxiv.org/abs/1905.05583v3.

\bibitem[Tong and Koller, 2002]{Tong2001}
Tong, S. and Koller, D. (2002).
\newblock {Support Vector Machine} active learning with applications to text
  classification.
\newblock {\em Journal of Machine Learning Research}, 2:45--66.
\newblock https://doi.org/10.1162/153244302760185243.

\bibitem[Turney and Pantel, 2010]{Turney2010}
Turney, P.~D. and Pantel, P. (2010).
\newblock From frequency to meaning: Vector space models of semantics.
\newblock {\em Journal of Artificial Intelligence Research}, 37:141--188.
\newblock https://doi.org/10.1613/jair.2934.

\bibitem[Ushey et~al., 2020]{rstudioapi2020}
Ushey, K., Allaire, J., Wickham, H., and Ritchie, G. (2020).
\newblock {\em {rstudioapi}: Safely Access the {RStudio API} {(Version 0.11)}}.
\newblock [R package]. CRAN. https://CRAN.R-project.org/package=rstudioapi.

\bibitem[Uyheng and Carley, 2020]{Uyheng2020}
Uyheng, J. and Carley, K.~M. (2020).
\newblock Bots and online hate during the {COVID-19} pandemic: Case studies in
  the {United States and the Philippines}.
\newblock {\em Journal of Computational Social Science}, 3:445--468.
\newblock https://doi.org/10.1007/s42001-020-00087-4.

\bibitem[van Atteveldt et~al., 2017]{vanAtteveldt2017}
van Atteveldt, W., Sheafer, T., Shenhav, S.~R., and Fogel-Dror, Y. (2017).
\newblock Clause analysis: Using syntactic information to automatically extract
  source, subject, and predicate from texts with an application to the
  {2008--2009 Gaza War}.
\newblock {\em Political Analysis}, 25(2):207--222.
\newblock https://doi.org/10.1017/pan.2016.12.

\bibitem[van Rijsbergen, 2000]{vanRijsbergen2000}
van Rijsbergen, C.~J. (2000).
\newblock {\em {Information Retrieval --- Session 1: Introduction to
  Information Retrieval}}.
\newblock [Lecture notes]. Universit{\"a}t Duisburg Essen.
  https://www.is.inf.uni-due.de/courses/dortmund/lectures/ir\_ws00-01/folien/keith\_intro.ps.

\bibitem[Van~Rossum and Drake, 2009]{vanRossum2009}
Van~Rossum, G. and Drake, F.~L. (2009).
\newblock {\em {Python 3} Reference Manual}.
\newblock CreateSpace.

\bibitem[Vaswani et~al., 2017]{Vaswani2017}
Vaswani, A., Shazeer, N., Parmar, N., Uszkoreit, J., Jones, L., Gomez, A.~N.,
  Kaiser, L., and Polosukhin, I. (2017).
\newblock Attention is all you need.
\newblock In Guyon, I., Luxburg, U.~V., Bengio, S., Wallach, H., Fergus, R.,
  Vishwanathan, S., and Garnett, R., editors, {\em Advances in Neural
  Information Processing Systems 30}, pages 5998--6008. Curran Associates, Inc.

\bibitem[Veropoulos et~al., 1999]{Veropoulos1999}
Veropoulos, K., Campbell, C., and Cristianini, N. (1999).
\newblock Controlling the sensitivity of {Support Vector Machines}.
\newblock In Dean, T., editor, {\em Proceedings of the International Joint
  Conference on Artificial Intelligence (IJCAI '99)}.

\bibitem[Wang, 2020]{Wang2020a}
Wang, H. (2020).
\newblock Logistic regression for massive data with rare events.
\newblock In Daum{\'e}~III, H. and Singh, A., editors, {\em Proceedings of the
  37th International Conference on Machine Learning}, pages 9829--9836.
  Proceedings of Machine Learning Research.

\bibitem[Wankm{\"u}ller, 2021]{Wankmueller2021}
Wankm{\"u}ller, S. (2021).
\newblock Neural transfer learning with {T}ransformers for social science text
  analysis.
\newblock arXiv.
\newblock {arXiv:2102.02111v1 [cs.CL]}. https://arxiv.org/abs/2102.02111.

\bibitem[Watanabe, 2021]{Watanabe2020}
Watanabe, K. (2021).
\newblock {Latent Semantic Scaling}: A semisupervised text analysis technique
  for new domains and languages.
\newblock {\em Communication Methods and Measures}, 15(2):81--102.
\newblock https://doi.org/10.1080/19312458.2020.1832976.

\bibitem[Wickham, 2016]{ggplottwo:2016hw}
Wickham, H. (2016).
\newblock {\em {ggplot2}: Elegant Graphics for Data Analysis}.
\newblock Springer.

\bibitem[Wickham, 2019]{stringr2019}
Wickham, H. (2019).
\newblock {\em {stringr}: Simple, Consistent Wrappers for Common String
  Operations {(Version 1.4.0)}}.
\newblock [R package]. CRAN. https://CRAN.R-project.org/package=stringrCRAN.

\bibitem[Wickham et~al., 2021]{dplyr2021}
Wickham, H., François, R., Henry, L., and M{\"u}ller, K. (2021).
\newblock {\em {dplyr}: A Grammar of Data Manipulation {(Version 1.0.6)}}.
\newblock [R package]. CRAN. https://CRAN.R-project.org/package=dplyr.

\bibitem[Wild, 2020]{lsa2020}
Wild, F. (2020).
\newblock {\em {lsa: Latent Semantic Analysis} {(Version 0.73.2)}}.
\newblock [R package]. CRAN. https://CRAN.R-project.org/package=lsa.

\bibitem[Wilke, 2021]{ggridges2021}
Wilke, C.~O. (2021).
\newblock {\em {ggridges}: Ridgeline Plots in {'ggplot2'} {(Version 0.5.3)}}.
\newblock [R package]. CRAN. https://CRAN.R-project.org/package=ggridgesCRAN.

\bibitem[Wolf et~al., 2020]{Wolf2019}
Wolf, T., Debut, L., Sanh, V., Chaumond, J., Delangue, C., Moi, A., Cistac, P.,
  Rault, T., Louf, R., Funtowicz, M., Davison, J., Shleifer, S., von Platen,
  P., Ma, C., Jernite, Y., Plu, J., Xu, C., Scao, T.~L., Gugger, S., Drame, M.,
  Lhoest, Q., and Rush, A.~M. (2020).
\newblock {HuggingFace's Transformers}: State-of-the-art natural language
  processing.
\newblock arXiv.
\newblock {arXiv:1910.03771v5 [cs.CL]}. https://arxiv.org/abs/1910.03771v5.

\bibitem[Zhang and Pan, 2019]{Zhang2019}
Zhang, H. and Pan, J. (2019).
\newblock {CASM}: A deep-learning approach for identifying collective action
  events with text and image data from social media.
\newblock {\em Sociological Methodology}, 49(1):1--57.
\newblock https://doi.org/10.1177/0081175019860244.

\bibitem[Zhao et~al., 2021]{Zhao2021a}
Zhao, H., Phung, D., Huynh, V., Jin, Y., Du, L., and Buntine, W. (2021).
\newblock Topic modelling meets deep neural networks: {A} survey.
\newblock In Zhou, Z.-H., editor, {\em Proceedings of the Thirtieth
  International Joint Conference on Artificial Intelligence, {IJCAI-21}}, pages
  4713--4720. International Joint Conferences on Artificial Intelligence
  Organization.
\newblock https://doi.org/10.24963/ijcai.2021/638.

\bibitem[Zheng et~al., 2020]{Zheng2020}
Zheng, Z., Hui, K., He, B., Han, X., Sun, L., and Yates, A. (2020).
\newblock {BERT-QE}: {C}ontextualized {Q}uery {E}xpansion for {D}ocument
  {R}e-ranking.
\newblock In Cohn, T., He, Y., and Liu, Y., editors, {\em Findings of the
  Association for Computational Linguistics: EMNLP 2020}, pages 4718--4728.
  Association for Computational Linguistics.
\newblock https://doi.org/10.18653/v1/2020.findings-emnlp.424.

\bibitem[Zhu et~al., 2015]{Zhu2015}
Zhu, Y., Kiros, R., Zemel, R., Salakhutdinov, R., Urtasun, R., Torralba, A.,
  and Fidler, S. (2015).
\newblock Aligning books and movies: Towards story-like visual explanations by
  watching movies and reading books.
\newblock In {\em Proceedings of the 2015 IEEE International Conference on
  Computer Vision (ICCV '15)}, pages 19--27. IEEE Computer Society.
\newblock https://doi.org/10.1109/ICCV.2015.11.

\end{thebibliography}
\bibliographystyle{apalike}

\end{document}